\documentclass[aps,prb,twocolumn,longbibliography,notitlepage,superscriptaddress,nofootinbib]{revtex4-2}

\usepackage{comment}

\usepackage{ifthen}
\usepackage{amsmath}\allowdisplaybreaks
\usepackage{amsfonts}
\usepackage{amssymb}
\usepackage{dsfont}
\usepackage{amsmath}
\usepackage{epsfig}
\usepackage{amsfonts}
\usepackage{savesym}
\usepackage{braket}
\usepackage{graphicx}
\usepackage{longtable}
\usepackage{afterpage}
\graphicspath{ {./figs/} }
\usepackage{epsfig}
\usepackage[scr=boondox]{mathalpha}
\usepackage{cancel}

\usepackage{epstopdf}
\usepackage{pst-all}
\usepackage[off]{auto-pst-pdf}

\usepackage{mathtools}
\usepackage{tikz}
\usetikzlibrary{intersections,calc,3d,arrows,decorations.markings}
\usepackage{bigdelim}
\usepackage{color}
\usepackage{footnote}
\usepackage[linktocpage]{hyperref}
\hypersetup{colorlinks=true,citecolor=blue,linkcolor=blue, urlcolor=blue, breaklinks=true}

\usepackage{slashed}
\def\H{\mathcal{H}}

\def\Z{\mathbb{Z}}
\def\R{\mathbb{R}}

\def\A{\mathcal{A}}
\def\C{\mathcal{C}}

\newcommand{\ZZ}{{\mathbb Z}}
\newcommand{\RR}{{\mathbb R}}

\newcommand{\bx}{\boldsymbol{x}}
\newcommand{\by}{\boldsymbol{y}}

\newcommand{\om}{{\omega_2}}

\newcommand{\lr}[1]{ \langle {#1} \rangle}

\newcommand{\ra}{\rightarrow}

\def\l@subsubsection#1#2{}

\usepackage{environ}
\NewEnviron{eqs}{%
\begin{equation}\begin{split}
    \BODY
\end{split}\end{equation}
}

\DeclarePairedDelimiter\floor{\lfloor}{\rfloor}
\begin{document}
\title{Classification of (2+1)D invertible fermionic topological phases with symmetry}

\author{Maissam Barkeshli}
\affiliation{Department of Physics, Condensed Matter Theory Center, and Joint Quantum Institute, University of Maryland, College Park, Maryland 20742, USA}

\author{Yu-An Chen}
\affiliation{Department of Physics, Condensed Matter Theory Center, and Joint Quantum Institute, University of Maryland, College Park, Maryland 20742, USA}

\author{Po-Shen Hsin}
\affiliation{Walter Burke Institute for Theoretical Physics, California Institute of Technology, Pasadena, CA 91125, USA}

\author{Naren Manjunath}
\affiliation{Department of Physics, Condensed Matter Theory Center, and Joint Quantum Institute, University of Maryland, College Park, Maryland 20742, USA}

\begin{abstract}
We provide a classification of invertible topological phases of interacting fermions with symmetry in two spatial dimensions for general fermionic symmetry groups $G_f$ and general values of the chiral central charge $c_-$. Here $G_f$ is a central extension of a bosonic symmetry group $G_b$ by fermion parity, $(-1)^F$, specified by a second cohomology class $[\omega_2] \in \mathcal{H}^2(G_b, \mathbb{Z}_2)$. Our approach proceeds by gauging fermion parity and classifying the resulting $G_b$ symmetry-enriched topological orders while keeping track of certain additional data and constraints. We perform this analysis through two perspectives, using $G$-crossed braided tensor categories and Spin$(2c_-)_1$ Chern-Simons theory coupled to a background $G$ gauge field. These results give a way to characterize and classify invertible fermionic topological phases in terms of a concrete set of data and consistency equations, which is more physically transparent and computationally simpler than the more abstract methods using cobordism theory and spectral sequences. Our results also generalize and provide a different approach to the recent classification of fermionic symmetry-protected topological phases by Wang and Gu, which have chiral central charge $c_- = 0$. We show how the 10-fold way classification of topological insulators and superconductors fits into our scheme, along with general non-perturbative constraints due to certain choices of $c_-$ and $G_f$. Mathematically, our results also suggest an explicit general parameterization of deformation classes of (2+1)D invertible topological quantum field theories with $G_f$ symmetry. 
\end{abstract}
\maketitle

\addtocontents{toc}{\protect\setcounter{tocdepth}{1}}
\tableofcontents

\section{Introduction}

The discovery of the integer quantum Hall (IQH) effect and, later, topological insulators and superconductors have revolutionized our understanding of phases of matter \cite{hasan2010,qi2011}. These phases are now understood to be special cases of a general class of phases of matter called invertible topological phases. An invertible topological phase of matter with symmetry group $G$ is an equivalence class of gapped systems that possess a unique $G$-symmetric ground state on any closed spatial manifold\footnote{Invertible phases are sometimes also referred to as short-range entangled phases, although not all authors use this phrase the same way.}\cite{Chen2013,Senthil2015SPT,Kapustin2014,Kapustin:2014dxa,Freed:2016rqq,yonekura2019,gaiotto2017}. Two such systems are defined to be in the same invertible phase if and only if they can be adiabatically connected without closing the bulk energy gap. 
Importantly, the concept of invertible topological phases of matter applies to systems with arbitrarily strong interactions among the constituent degrees of freedom, and thus is distinct from topological band theory \cite{hasan2010,qi2011,kitaev2009}, which is a single-particle concept. Despite nearly 40 years since the discovery of the IQH state \cite{klitzing1980}, a systematic and comprehensive understanding of invertible topological phases of matter for general symmetry groups $G$ is still lacking.  

An important subset of invertible phases is that of \textit{symmetry-protected topological (SPT) phases} \cite{Chen2013,Senthil2015SPT}. Such phases can be adiabatically connected to the `trivial' gapped insulating phase without closing the bulk energy gap, but only if the symmetry is broken. The difference between invertible phases and SPT phases is that the former may still be nontrivial even if all symmetries are broken; IQH states, for example, have chiral edge modes, characterized by a chiral central charge, that persist even when charge conservation symmetry is broken. Invertible topological phases have the property that a ground state corresponding to an invertible phase possesses an inverse, such that stacking the state and its inverse gives a state that can be adiabatically connected to a trivial product state.
Since invertible topological phases have a unique ground state on any closed spatial manifold, they do not possess topologically nontrivial quasiparticles and therefore a complete classification may be within reach.\footnote{In contrast, for non-invertible topological phases, such as those with anyons in (2+1)D, a complete classification would require a classification of unitary modular tensor (UMTC) categories, which is not believed to be within reach. In this case, one must fix the UMTC describing the fusion and braiding of anyons, and then the remaining symmetry-enriched topological phases can be fully classified systematically \cite{Barkeshli2019, Manjunath2020fqh}}

Over the past several years, a range of techniques have been developed to characterize and classify SPT or invertible phases. Currently, the most comprehensive approach, which is believed to be applicable to all symmetry types and in general dimensions, is to assume that invertible topological phases of matter are described by deformation classes of invertible topological quantum field theories (TQFTs) \cite{Freed:2016rqq}, which are TQFTs whose path integrals have unit magnitude on every closed manifold. Invertible TQFTs, in turn, can be given an abstract classification in terms of bordism theory, although the mathematical results are fully proven only in cases where there is no thermal Hall effect \cite{Kapustin2014,Kapustin:2014dxa,Freed:2016rqq,yonekura2019}. While this approach is believed to be complete, it has two major drawbacks: first, the computations required to carry out the classification for any particular symmetry group $G$ require difficult spectral sequence computations needing significant technical expertise and, as such, have been carried out only in a few cases \cite{Freed:2016rqq,Kapustin:2014dxa,guo2018,guo2020}. Second, this approach is far-removed from the physical properties of the system. This obscures the physical distinction between different invertible phases, and also removes us from the setting of \it topological phases of matter, \rm which rely on the notion of a gapped Hamiltonian acting on a many-body Hilbert space. A more direct approach to topological phases of matter, in terms of operator algebras, is also under development, although the results are so far less comprehensive than the TQFT approach \cite{kapustin2021, sopenko2021, ogata2021, bourne2021}. 

Recently, a complete classification of interacting fermion SPTs was proposed in Ref.~\onlinecite{Wang2020fSPT}, using a theory of fixed-point wavefunctions, building on earlier work developing an approach using group supercohomology \cite{Gu2014Supercoh,Wang2017Supercoh}. The properties of these phases were summarized in a set of data $(n_1, n_2, \nu_3)$, which are related to decorating defects of various codimension with lower-dimensional fermionic states. This data is subject to redundancies and consistency equations, which have been explicitly computed. However, a complete formulation of the group multiplication law associated with stacking fermionic SPT phases has not been presented. A similar characterization has not been available for more general invertible phases with nontrivial chiral central charge, partly because it is more difficult to write down analytically tractable microscopic models for chiral topological phases. One of the main results of this work is to obtain a set of data, redundancies, and consistency equations, which is based on the properties of symmetry defects in invertible phases. The data provide a complete characterization of invertible phases of interacting fermions in (2+1) dimensions, thus extending the results of Ref.~\onlinecite{Wang2020fSPT} to more general chiral central charges, and providing an alternate perspective on the results for $c_- = 0$.\footnote{When the symmetry of the invertible phases is a nontrivial extension of the ``bosonic symmetry'' $G_b$ by the $\mathbb{Z}_2$ fermion parity symmetry, the classification of the invertible phases cannot be obtained from that of $c_-=0$ by stacking with the invertible phases with only the fermion parity symmetry, and therefore the classification depends nontrivially on the chiral central charge. This is discussed in detail in Section~\ref{Sec:Counting} and Section~\ref{Sec:CS}.}

In this paper we carry out our analysis using techniques borrowed from the related field of \textit{symmetry-enriched} topological phases (SETs). Every fermionic system has a $\Z_2^f$ symmetry, corresponding to the conservation of fermion parity. The full fermionic symmetry group, $G_f$, is the symmetry group that acts nontrivially on fermionic operators, and is a group extension of the bosonic symmetry $G_b$ by $\Z_2^f$, characterized by $[\omega_2] \in \mathcal{H}^2(G_b, \Z_2)$. Upon gauging the fermion parity symmetry, we obtain a bosonic topological phase with nontrivial topological order, corresponding to a state in the 16-fold way \cite{kitaev2006}, which is also $G_b$ symmetric. Therefore we can classify $G_f$-symmetric fermionic invertible phases in terms of $G_b$-symmetry enriched bosonic topological phases. 

The above program can be carried out using two main theoretical tools, both of which we pursue in this paper. One of these is through the framework of $G$-crossed braided tensor categories \cite{Barkeshli2019,Barkeshli2020Anomaly,Bulmash2020}, which is the mathematical theory of SET phases. The other is through $Spin(2c_-)_1$ Chern-Simons theory coupled to a background $G_b$ gauge field. 

Our analysis contains several new results. The main result, as we summarize in Section~\ref{Sec:Summary}, is that each (2+1)D invertible phase with fermionic symmetry $G_f$ can be fully characterized by a set of data $(c_-, n_1, n_2, \nu_3)$, subject to certain redundancies and consistency conditions. Importantly, these consistency conditions depend nontrivially on the value of $c_-$. The data $(c_-, n_1, n_2, \nu_3)$ encode the braiding and fusion properties of symmetry defects. 

A second important result is the explicit derivation of ``stacking rules" for invertible phases. Since invertible phases obey an Abelian group structure under stacking, a complete classification theory needs to explain how the $(c_-, n_1, n_2,\nu_3)$ data of two invertible phases $a$ and $b$ are related to the corresponding data of the phase obtained by stacking $a$ and $b$. In previous works, such stacking rules were only proposed in various special cases \cite{Gu2014Supercoh,Cheng2018fSPT,Chen2019freeinteracting}. The most detailed derivations to date have been given in Ref.~\onlinecite{Bhardwaj_2017,brumfiel2018pontrjagin}; Ref.~\onlinecite{Bhardwaj_2017} derives stacking rules for the case $c_- = 0$, $[\omega_2] = 0$ (i.e. $G_f = G_b \times \Z_2$), and $n_1 = 0$, while Ref.~\onlinecite{brumfiel2018pontrjagin} derives the stacking rule for $c_- = 0$, $[\omega_2] = 0$ (i.e. $G_f = G_b \times \Z_2$), and general $n_1$. Here we use the framework of anyon condensation \cite{Bischoff_2019,Jiang2017} and Chern-Simons theory to explicitly derive the stacking rules in more generality than previous works have considered, and we conjecture a formula for the complete stacking rules. 

Central to both these results is the question of whether a $G_b$-crossed BTC (i.e. the SET phase obtained by gauging the fermion parity) can completely and uniquely describe any given invertible phase. In this work we find that different invertible phases may actually correspond to the same $G_b$-crossed BTC. In other words, the $G_b$-crossed theory for SETs may treat two sets of $(n_1,n_2,\nu_3)$ data as equivalent even though they describe distinct invertible phases. These phases are distinguished by taking the additional step of carefully tracking the $G_f$ fluxes in the invertible phases, and seeing how they manifest in the $G_b$-crossed theories after the gauging procedure. By adding this extra data about flux labels, our work extends the usual $G$-crossed BTC framework. 

Finally, this paper contains several results on specific symmetry groups. We obtain the classification of (2+1)D interacting invertible phases whose symmetries are given by the 10-fold way (Section~\ref{Sec:PdTable}) \cite{Altland1997,Ryu_2010}. Thus, for example, we show how the integer quantum Hall states and the $\Z_2$ time-reversal invariant topological insulator in Class AII are described within the generalized $G_b$-crossed framework. We also study in detail the symmetry groups $G_f = \Z_2 \times \Z_2^f$ (Section~\ref{Sec:Z2xZ2fExample}), reproducing the results of Ref.~\onlinecite{gu2014}, and $G_f = U(1)^f$ (Section~\ref{Sec:U(1)symm}), providing a complete set of topological invariants along with their physical interpretation and a stacking analysis.

Interestingly, we find that the allowed choices of the symmetry group $G_f$ can be constrained by anomalies, depending on the value of $c_-$. For example, we show that a system where local fermions carry half-integer isospins under $G_f=SU(2)$ symmetry, while bosons carry integer isospins, must have even $c_-$ (Section~\ref{Sec:HalvingEgs})
: there exists an $\H^3$ obstruction which does not permit a symmetric (2+1)-dimensional invertible phase with odd $c_-$. While this kind of constraint may be expected from a free fermion band theory perspective, our results give an analysis applicable in the strongly interacting case. This type of anomaly, which can also be understood as a nontrivial manifestation of 2-group symmetry (see {\it e.g.} Ref.~\onlinecite{Benini2018}),  has been discussed previously in Refs.~\onlinecite{Kapustin2014anomaly,fidkowski2018surface}. It arises here because although a certain type of symmetry fractionalization ({\it e.g.} the spin-1/2 property when $G_b = SO(3)$) is well-defined for fermions, there may be an obstruction to extending the symmetry fractionalization to the fermion parity fluxes in the $G_b$-crossed theory. We also show that this anomaly is absent in certain situations, if the symmetry permutes the fermion parity fluxes in a suitable manner, which may happen for instance if $SO(3)$ is broken down to a discrete group $\Z_2 \times \Z_2$ (Section~\ref{Sec:HalvingZ2Z2}). 

An interesting mathematical application of this paper is that it suggests an explicit solution to the group of deformation classes of invertible TQFTs in (2+1)D with the appropriate tangential structures, in terms of the data $(c_-, n_1, n_2, \nu_3)$ and its stacking rules. This extends the result proven mathematically in Ref.~\onlinecite{brumfiel2018pontrjagin}, which applies to the case of spin cobordisms that is applicable for $G_f = G_b \times \Z_2^f$ and $c_- = 0$, to the most general possible symmetry group $G_f$ and chiral central charges. 

\subsection{Organization of paper}

The paper is organized as follows. Section~\ref{Sec:Summary} contains an abbreviated summary of our results. 
Section~\ref{Sec:MainGxCalcs} contains a derivation of the data and equations describing $G_b$ SETs obtained from gauging fermion parity using the framework of $G_b$-crossed BTCs. Section~\ref{Sec:Counting} discusses how the usual $G_b$-crossed theory fails to accurately count invertible phases, and provides a resolution involving the specification of $G_b$ flux labels. In Section~\ref{Sec:CS}, we show how to obtain the same results through the Chern-Simons framework. Section~\ref{Sec:stacking} applies both the $G_b$-crossed theory and the Chern-Simons formalism to compute the classification of invertible phases, by deriving explicit stacking rules. Section~\ref{Sec:Examples} discusses various examples demonstrating the use of our theory to classify and characterize invertible phases. Finally, Section~\ref{Sec:Discussion} concludes and discusses future directions. 

The more abstract or computationally involved details have been presented in the appendices. In Appendix~\ref{Sec:Gxreview} we provide a brief review of $G$-crossed braided tensor categories as applicable to this work. In Appendix~\ref{sec:highercup} we summarize some mathematical background for the higher cup product formalism used in this paper. In Appendix~\ref{Sec:RelAnomComps} we compute the 't Hooft anomaly for fermionic invertible phases and use it to derive certain stacking rules. 
In Appendix~\ref{sec:refstate} we discuss a non-anomalous $G_b$ symmetry enriched theory which can serve as a non-anomalous reference state and is useful in obtaining the 't Hooft anomaly in other $G_b$ symmetry-enriched theories. In Appendix \ref{Sec:Psquare} we show that the anomaly for the symmetry that does not permute the anyons can be obtained from the anomaly of the one-form symmetry.

\section{Summary of results} \label{Sec:Summary}

\subsection{Preliminaries}

First we establish some notation. The fermionic symmetry group $G_f$ of an invertible phase always has a $\Z_2^f$ subgroup corresponding to the conservation of fermion parity. In general, $G_f$ is a central extension of a symmetry group $G_b$ by $\Z_2^f$. The group law in $G_f$ is specified by a 2-cocycle $\omega_2 \in Z^2(G_b,\Z_2)$ (this notation is explained in Appendix \ref{sec:highercup}), as follows. Denote a general element in $G_f$ as a pair $({\bf g},a)$, where ${\bf g} \in G_b$ and $a \in \Z_2^f$. Then, the group law in $G_f$ is
\begin{equation}
({\bf g}_1,a_1) ({\bf g}_2,a_2) = ({\bf g}_1{\bf g}_2,a_1+a_2+\omega_2({\bf g}_1,{\bf g}_2))~.
\end{equation}
A nontrivial $[\omega_2]$ class implies that the local fermion transforms as a projective representation of $G_b$, which is still a linear representation of $G_f$.
The fermion carries fractional $G_b$ quantum numbers as specified by the projective representation.
We also define the homomorphism $s_1: G_b \rightarrow \Z_2$. If ${\bf g}$ is unitary, $s_1({\bf g}) = 0$; if ${\bf g}$ is antiunitary, $s_1({\bf g}) = 1$. For instance, if the symmetry is the time-reversal symmetry $\mathbf{T}^2=(-1)^F$ 
with the local fermion transforming as a Kramer's doublet, then the bosonic symmetry $\mathbf{T}^2=1$ is extended by $\mathbb{Z}_2^f$ with nontrivial extension class $[\omega_2]$ characterized by the nontrivial component
$\omega_2(\mathbf{T},\mathbf{T}) = 1$.
 
We will make repeated use of the \textit{cup product} of cochains below: these are also reviewed in Appendix~\ref{sec:highercup}. Let $f_p$ and $g_q$ be $p$ (respectively, $q$) variable functions from $G_b$ to some abelian group (generally $\Z_2$ or $U(1)$). $f_p$ and $g_q$ are called $p$-and $q$-cochains respectively. Then, we define their cup product $f_p \cup g_q$ as follows: 
\begin{equation}
    (f_p \cup g_q)({\bf g}_1,\dots , {\bf g}_{p+q}) = f_p({\bf g}_1,\dots , {\bf g}_{p})g_q({\bf g}_{p+1},\dots , {\bf g}_{p+q}).
\end{equation}
Similarly we can define objects called higher cup products (see Appendix~\ref{sec:highercup}), which are useful in organizing our formulas.

\subsection{Defining equations}

Each invertible phase is described by a set of data $(c_-,n_1,n_2,\nu_3)$, where $2c_- \in \Z$ sets the chiral central charge, and
\begin{align}
\label{dataFinal}
(n_1,n_2,\nu_3) \in C^1(G_b, \Z_2) \times C^2(G_b, \Z_2) \times C^3(G_b, U(1)_T)~,
\end{align}
where $C^k$ denotes $k$-cochains. For a given value of $c_-$ the equations are summarized in Table~\ref{Table:MainEqs}.
Three of them constrain $n_1, n_2$ and $\nu_3$. The rest are equivalence relations on $n_2$ and $\nu_3$. To write the equivalence relations, we define $b_1 \in C^1(G_b,\Z_2), b_2 \in C^2(G_b,\Z_2)$, and $\chi \in Z^1(G_b,\Z_2)$. The full set of equivalence relations is obtained by choosing all possible $b_1,b_2,\chi$. Throughout this paper we also define
\begin{align}
\theta = e^{2\pi i c_-/8}~, 
\end{align}
which equals the topological twist of the fermion parity fluxes in the 16-fold way UMTC for each $c_-$.
The data discussed above appeared in special cases in previous works. For $c_- = 0$, $s_1 = 0$, and $[\omega_2] = 0$, this data also appeared in  Refs.~\onlinecite{Gaiotto:2015zta,Bhardwaj_2017, Cheng2018fSPT, brumfiel2018pontrjagin}. For $c_- = 0$ and general $s_1$, $[\omega_2]$, this data appeared in Ref.~\onlinecite{Wang2020fSPT}. 

For a fixed $c_-$, the data $(n_1, n_2, \nu_3)$ form a torsor over a group extension of the group $\mathcal{H}^3(G_b, U(1)_T)$ and (sub)groups of $\mathcal{H}^1(G_b, \Z_2)$, $\mathcal{H}^2(G_b, \Z_2)$. That is, starting with a given choice of $(c_-,n_1,n_2,\nu_3)$, other possible invertible phases with the same central charge $c_-$ can be obtained by certain actions characterized by some $[\alpha_1] \in \mathcal{H}^1(G_b, \Z_2)$, $[\beta_2] \in \mathcal{H}^2(G_b, \Z_2)$, and $[\gamma_3] \in \mathcal{H}^3(G_b, U(1)_T)$, where $\alpha_1,\beta_2,\gamma_3$ are cocycle representatives. The detailed actions will be discussed in Sec.~\ref{sec: Derivation from Gb crossed}.



Invertible topological phases form an abelian group, under an operation called stacking. Physically, stacking two phases can be thought of as taking a double layer system, with each layer consisting of one of the two phases, and viewing the combined system as a single invertible topological phase where the $G_f$ symmetry acts on both layers simultaneously. The general group multiplication law, which we also refer to as the stacking rule, is also summarized in Table~\ref{Table:MainEqs}.
For trivial $\omega_2$, the stacking rule we derive is exact, and reproduces the result in Ref.~\onlinecite{brumfiel2018pontrjagin}. For nontrivial $\omega_2$, the stacking rules for $n_2$ and $\nu_3$ in Table~\ref{Table:MainEqs} are conjectures that are compatible with our expressions for the 't Hooft anomaly $\mathcal{O}_4$.  
For nontrivial $s_1$, we know the stacking rule for $c_-$ and $n_1$ exactly, and our stacking rule for $n_2$ is again a conjecture; we do not propose a conjecture for the $\nu_3$ stacking in this case. 

Note that in the previous literature, the stacking rule for $\nu_3$ is known only in the case $G_f = G_b \times \Z_2^f$, i.e. $\omega_2 = 0$.  These stacking rules have previously been derived in the special cases $c_- = 0$, $n_1 = 0$, and $[\omega_2] = 0$ in Ref.~\onlinecite{Bhardwaj_2017}, and for $c_- = 0$, $[\omega_2] = 0$ (with general $n_1$) in Ref.~\onlinecite{brumfiel2018pontrjagin}. They were also guessed but not fully derived for $c_- = 0$, $n_1 = 0$, $[\omega_2] = 0$ in Ref.~\onlinecite{Cheng2018fSPT}. 

One interesting consequence of our results is that if an invertible fermionic phase corresponding to a particular choice of $(c_-, n_1, n_2, \nu_3)$ has a nontrivial $[\mathcal{O}_4]$, the system cannot exist in (2+1)D. However, the corresponding invertible fermionic phase can exist at the surface of a (3+1)D bosonic SPT characterized by $[\mathcal{O}_4] \in \mathcal{H}^4(G, U(1))$. This gives an intriguing situation where the surface of a nontrivial (3+1)D bosonic SPT can be symmetry-preserving, gapped, and yet not topologically ordered, at the expense of introducing fermions to the surface. (If fermions are introduced in both the (3+1)D bulk and the (2+1)D surface, we expect that the bosonic SPT becomes a trivial fermionic SPT, and that the (2+1)D system is an example of an anomalous fermionic SPT as described in Ref.~\onlinecite{wang2019spt}). 

As we will describe below and in Section~\ref{Sec:Counting}, there are two equivalent ways of parameterizing the data that classifies invertible phases. 
The description presented above in terms of $(c_-, n_1, n_2, \nu_3)$ follows the notation given in Ref.~\onlinecite{Wang2020fSPT} for $c_- = 0$ i.e. for fSPT phases. There is a second description which is more natural to the $G_b$-crossed braided tensor category approach, which consists of a set of data $(c_-,\tilde{n}_1,\tilde{n}_2,\nu_3,\{L_{\bf g}\})$, and which will be summarized below.  

We note that 
our consistency equations as summarized in Table~\ref{Table:MainEqs} are mostly equivalent to those of Ref.~\onlinecite{Wang2020fSPT} when $c_-=0$, with some differences in the equivalence relations, which are summarized in Section \ref{sec:relation_wanggu}.

\begin{widetext}
\begin{savenotes}
\begin{table*}[t]
    \renewcommand{\arraystretch}{1.0}
    \centering
    \begin{tabular}{c}
		\hline \\
		Data for invertible fermion phases: $(c_-,n_1,n_2,\nu_3) \in \frac{1}{2}\Z\times C^1(G_b, \Z_2) \times C^2(G_b, \Z_2) \times C^3(G_b, U(1)_T)$ \\
		~\\
		\hline \\
\parbox{0.9\textwidth}{
    General equations 
    \begin{align}
	    dn_1 &= 0 \quad (\text{mod } 2) ~,\\
	    dn_2 &= n_1 \cup (\omega_2 + s_1 \cup n_1) + c_- \omega_2 \cup_1 \omega_2 \quad (\text{mod } 2)~,\\
	    d\nu_3&= \mathcal{O}_4[c_-,n_1,n_2] ~,\\ 
	    (n_2,\nu_3) &\simeq (n_2,\nu_3\times db_2) \simeq (n_2 + db_1, \nu_3\times (-1)^{ b_1\cup \omega_2+db_1\cup_1n_2+db_1\cup b_1}) \nonumber\\
	    &\simeq (n_2+ \om, \nu_3 \times (-1)^{\om\cup_1 n_2})~, \label{eq:equivs}
	    \\
	    &\quad \forall b_1 \in C^1(G_b,\Z_2), \forall b_2 \in C^2(G_b,U(1)_T)~. \nonumber 
    \end{align}} \\
    \hline \\
\parbox{0.9\textwidth}{
Formulas for $\mathcal{O}_4$ 
\begin{align}
\mathcal{O}_4[c_-,n_1=0,n_2,s_1=0] &= (-1)^{n_2  \cup (n_2+ \omega_2)}e^{i \pi\frac{c_-}{4} \mathcal{P}(\omega_2)}\label{eqn:O40} ~,\\	
\mathcal{O}_4[c_-=k+1/2,n_1,n_2,s_1=0] &= (-1)^{n_2 \cup n_2} \label{eqn:O4i} ~,\\
\mathcal{O}_4[c_-=k,n_1,n_2,s_1=0] &= (-1)^{{\cal P}(n_2) + n_2 \cup \om + \zeta(n_1,\omega_2) }i^{ n_1 \cup n_1 \cup \om} e^{\frac{i \pi c_-}{4} {\cal P}(\om)} (-1)^{c_- (n_1 \cup \om) \cup_2 (\om \cup_1 \om)}
\label{eqn: O4 c=k}
~, \\
 \mathcal{O}_4[c_-=0,n_1,n_2; s_1 \ne 0] &= (-1)^{{\cal P}(n_2) + n_2 \cup \omega'_2 + \zeta(n_1,\omega_2') }i^{ n_1 \cup n_1 \cup \omega'_2} (-1)^{s_1 \cup n_1 \cup n_2 } ~.
\label{eqn:O4f}
\end{align}
}
\\\hline \\
	\parbox{0.9\textwidth}{Stacking rules (Group multiplication law for invertible phases)
    \begin{alignat}{3}
    & &&(c_-^\text{tot}, n_1^\text{tot}, n_2^\text{tot}, \nu_3^\text{tot}) = (c_-^a, n_1^a, n_2^a, \nu_3^a) \times (c_-^b, n_1^b, n_2^b, \nu_3^b) & &
    \nonumber \\
    & & &c_-^\text{tot} = c_-^a + c_-^b~, &  \\
    & & &n_1^\text{tot} = n_1^a + n_1^b~,  &  \\
    &  &\text{for }s_1 = 0: \qquad& \mkern-17mu
    \begin{cases}
        n_2^\text{tot} =& n_2^a + n_2^b + n_1^a \cup n_1^b  ~,\\
        \nu_3^{\text{tot}}=&\nu_3^a\nu_3^b\,
        (-1)^{(n_1^a \cup n_1^b) \cup_1 (n_2^a + n_2^b) + n_2^a \cup_1 n_2^b + dn_2^a \cup_2 n_2^b + n_1^a \cup (n_1^a \cup_1 n_1^b) \cup n_1^b + c^a (\om \cup_1 \om) \cup_3 (n_1^b \cup \om)} \\
        & \cdot ~ i^{ (n_1^a \cup_1 n_1^b) \cup \om + n_1^a \cup n_1^b \cup n_1^b}~,
    \end{cases} & \\
    & &\text{for }s_1 \ne 0:\qquad &n_2^\text{tot} =n_2^a+n_2^b+n_1^a\cup n_1^b
    + n_1^a\cup (s_1\cup_1 n_1^b)
    +n_1^b\cup (s_1\cup_1 n_1^a)
    +(n_1^a\cup_1 n_1^b)\cup s_1~. &
\end{alignat}
}
\\ 
\hline
\end{tabular}
\caption{Data and equivalences for invertible fermion phases for different values of the chiral central charge $c_-$ with symmetry $G_f$, which is a central extension of $G_b$ by $\Z_2^f$ with the 2-cocycle $\om \in \H^2(G_b, \Z_2)$. The parameter $k$ is an integer.
In the formulas for ${\cal O}_4$, $\zeta = n_1 \cup [(n_1 \cup \om) \cup_2 \om +n_1 \cup \om]$ is the Cartan coboundary \cite{Cartan_coboundary_2020}, and $\omega'_2 = \om + s_1 \cup n_1$.
Here $\mathcal{P}(X) = X \cup X - X \cup_1 dX$ is the Pontryagin square of the 2-cochain $X$ \cite{Whitehead1949}. 
The data $\omega_2,s_1$ are constrained by $c_-$: when $c_-$ is a half integer, $\omega_2=0$, while 
when $c_-\neq 0$, we assume that $s_1= 0$. (One can in principle also consider $s_1 \ne 0$ for $c_- = 4 \mod 8$, which could potentially be interpreted as describing an anomalous invertible state, although we do not consider this here.)
Our stacking rules for $n_2$ and $\nu_3$ are derived from Chern-Simons theory in the case where $s_1 = 0$ and $\omega_2 = 0$, and also in the case $G_f = U(1)^f$. For $\omega_2$ or $s_1$ non-zero, the general stacking rule for $n_2$ is conjectured (an updated version of Ref. \cite{aasen2021characterization} has proven the $n_2$ stacking rules in the case $s_1 = 0$); the general stacking rule for $\nu_3$ has an additional undetermined $3$-cocycle which we have conjectured to vanish. The transformations of $\nu_3$ in the equivalence relations of Eq.~\eqref{eq:equivs} in principle may include certain 3-cocycles if the $\nu_3$ stacking conjecture is incorrect (see Sec. \ref{sec:stackfromO4} for a detailed discussion). 
}

\label{Table:MainEqs}
\end{table*}
\end{savenotes}
\end{widetext}

\subsection{Derivation from $G_b$-crossed BTCs} \label{sec: Derivation from Gb crossed}

Here we will briefly sketch how the classification summarized above arises from the perspective of $G_b$-crossed BTCs. We note that Ref.~\onlinecite{Cheng2018fSPT} gave a classification for the case $c_- = 0$ and $\omega_2 = 0$ using the framework of $G$-crossed BTCs as well. However, there is a crucial conceptual difference between this paper and the approach of Ref.~\onlinecite{Cheng2018fSPT}, which characterized fSPTs as a $G_f$-crossed extension of the super-modular category $\{1,\psi\}$. In our approach, we are characterizing invertible phases using a $G_b$-crossed extension of the bosonic phase (referred to as the bosonic shadow), described by a UMTC $\mathcal{C}$, obtained by gauging fermion parity. This change in perspective is useful to properly account for a nontrivial $[\omega_2]$ and to compute the $\mathcal{O}_4$ obstructions; however, the price to pay is that we will need to keep track of certain additional data and equivalences beyond the $G_b$-crossed extension.

For simplicity let us assume here that the symmetry is unitary, so $s_1 = 0$. We will include $s_1 \neq 0$ in the main text. 

First, starting with the invertible fermionic phase with $G_f$ symmetry, we gauge fermion parity. This gives a bosonic topologically ordered phase with $G_b$ symmetry. The intrinsic topological order is characterized by a unitary modular tensor category $\mathcal{C}(c_-)$. There are 16 distinct possibilities for $\mathcal{C}(c_-)$, which are referred to as the 16-fold way, depending on the value of $2c_- \text{ mod } 16$ \cite{kitaev2006}. Mathematically these are the 16 distinct minimal modular extensions of the super-modular category $\{I,\psi\}$ and are summarized in Table~\ref{Table:16fold}. If $c_-$ is even, the anyons are written as $\{I, e, m,\psi\}$; if $c_-$ is odd, they are written as $\{I,v,\psi,\bar{v}\}$; if $c_-$ is half-integer, then they are written as $\{I, \sigma, \psi\}$. $m$, $v$, $\bar{v} = \psi \times v$, $e = m \times \psi$, and $\sigma$ all have the physical interpretation of being a fermion parity flux, as a full braid with the fermion $\psi$ gives a sign. By convention, if $c_-$ is unspecified, we will denote a fermion parity flux as $m$ and its counterpart as $m \times \psi$.
\begin{center}
    \begin{table*}[t]
        \renewcommand{\arraystretch}{1.5}
		\begin{tabular}{|c|c|c|c|}
			\hline 
			$c_-$ & Anyons& $\{d_a\}$ & Fusion rules\\ \hline
			$2k$& $I,\psi,e,m$& $d_{\psi} = d_e = d_m = 1$ & $e^2 = m^2 = \psi^2 = I, ~e \times m = m \times e = \psi$\\
			\hline
			$2k+1$& $I,\psi,v,\bar{v}$& $d_{\psi} = d_{v} = d_{\bar{v}} = 1$ & $v^2 = \psi, ~v^3 = \bar{v}, ~v^4 = I$\\
			\hline
			$k+1/2$& $I,\psi,\sigma$& $d_{\psi} =1, ~ d_{\sigma} = \sqrt{2}$ & $\psi \times \sigma = \sigma \times \psi = \sigma, ~\sigma \times \sigma = I + \psi$\\
			\hline
		\end{tabular}
		\caption{Properties of the UMTC $\mathcal{C}(c_-)$ described in the '16 fold way'. These are obtained by gauging the fermion parity of an invertible phase with chiral central charge $c_-$, where $k$ are integers. A fermion parity flux always has the topological twist $\theta = e^{i \pi c_-/4}$. $d_a$ denotes the quantum dimension of the anyon $a$. The complete fusion and braiding properties of the anyons are listed in Table~\ref{Table:AnyonFRUn}.} \label{Table:16fold}
	\end{table*}
\end{center}
We note the following constraint between the symmetry $G_f$ and the possible chiral central charge $c_-$.
If $c_- = k + 1/2$, where $k$ is an integer, then we must have $[\omega_2] = 0$, {\it i.e.} $G_f=G_b\times\mathbb{Z}_2^f$. $[\omega_2] \neq 0$ means that $\psi$ carries fractional $G_b$ quantum numbers, which is inconsistent with the possibility of a fermion parity vortex $\sigma$ that can absorb a fermion: $\sigma \times \psi = \sigma$. Therefore in the equations that follow, if $c_- = k+ 1/2$, we will implicitly assume that $\omega_2 = 0$. 
	
In order to classify fermionic invertible topological phases in terms of $G_b$ SETs, we first must specify how the $G_b$ symmetries permute the anyons. However, importantly, there is a constraint, which is that the permutation should keep the fermion $\psi$ invariant. Physically, a symmetry in an invertible fermionic phase cannot permute a fermion into a fermion parity flux. Consequently, we see that not all possible $G_b$ SETs correspond to valid invertible fermionic phases. The permutation action is therefore specified by a group homomorphism
\begin{align}
    \tilde{n}_1 : G_b \rightarrow \text{Aut}_\psi(\mathcal{C}(c_-)) ~,
\end{align}
where $\text{Aut}_\psi(\mathcal{C}(c_-)) \subset \text{Aut}(\mathcal{C}(c_-))$ is the subgroup of braided autoequivalences (also referred to as topological symmetries \cite{Barkeshli2019}) which keep $\psi$ invariant. Note that $\text{Aut}_\psi(\mathcal{C}(c_-)) = \Z_2$ if $c_-$ is integer and is trivial otherwise. 

The choice of $\tilde{n}_1$ determines, up to certain gauge transformations, a set of $U(1)$ phases $\{U_{\bf g}(a,b;c)\}$, which determine how each ${\bf g} \in G_b$ acts on the fusion and splitting spaces of the anyon theory. 

Next, SETs are specified by a set of symmetry fractionalization data \cite{Barkeshli2019}. For the cases of relevance here, this is specified by a set of $U(1)$ phases $\{\eta_{a}({\bf g},{\bf h})\}$, for ${\bf g}, {\bf h} \in G_b$ which determine the fractional $G_b$ quantum numbers carried by the anyons. They are subject to certain consistency equations and gauge transformations. We can always fix $U_{\bf g}$ by fixing a gauge, and fixing a canonical reference state which sets a reference value $\{\eta^{\text{ref}}_{a}({\bf g},{\bf h})\}$. All other symmetry fractionalization classes can then be related to the reference as
\begin{align}
    \eta_{a}({\bf g}, {\bf h}) = \eta^{\text{ref}}_{a}({\bf g},{\bf h}) M_{a,\mathfrak{t}({\bf g}, {\bf h})}~,
\end{align}
where $\mathfrak{t} \in Z^2(G_b, \mathcal{A})$ and where $\mathcal{A}$ is an Abelian group determined by fusion of the Abelian anyons in $\mathcal{C}(c_-)$. $M_{a,x}$ is the phase obtained by a double braid between $a$ and an Abelian anyon $x$. We define $\mathfrak{t}$ as 
\begin{align}
    \mathfrak{t}
    = \begin{cases}
    m^{\omega_2} \times \psi^{\tilde{n}_2}  & \text{ for } c_- = 2k \\
    v^{\omega_2} \times \psi^{\tilde{n}_2} & \text{ for } c_- = 2k+1 \\
    \psi^{\tilde{n}_2} & \text{ for } c_- = k+ 1/2 ~,
    \end{cases}
\end{align}
where $k$ is an integer and $\tilde{n}_2 \in C^2(G_b, \Z_2)$, which we take to be valued in $\tilde{n}_2({\bf g},{\bf h}) \in \{0,1\}$. We choose our reference $\eta$ symbols as follows:
\begin{align}\label{eqn:refetanontrivial}
    \eta_a^{\text{ref}} = 
    \begin{cases}
    M_{a,\psi^{\tilde{n}_1 \cup \tilde{n}_1}}  & \text{ for } c_- = 4k+3 \\
    1 & \text{ otherwise } \\
    \end{cases}~.
\end{align}
These particular reference states are chosen to simplify the formulas for the 't Hooft anomaly $\mathcal{O}_4$  (see Sec. \ref{sec:SF_oddc}). We can show using Chern-Simons theory that these reference states are all non-anomalous (note $G_b$ is unitary in the present discussion). 

The equations above ensure that 
\begin{align}
    \eta_\psi({\bf g}, {\bf h}) = (-1)^{\omega_2({\bf g}, {\bf h})}, 
\end{align}
which corresponds to the statement that the fermion $\psi$ carries fractional $G_b$ quantum numbers, as specified by $\omega_2 \in Z^2(G_b, \Z_2)$. 
	
The requirement that $\mathfrak{t}$ be a $2$-cocycle leads to the equation 
\begin{align}
    d \tilde{n}_2 = \tilde{n}_1 \cup \omega_2 + c_- \omega_2 \cup_1 \omega_2 ~.
\end{align}
Since both $m$ and $e = m \times \psi$ are fermion parity fluxes and both $e$ and $m$ are physically on equal footing, we are free to interchange $m$ and $e$ in the above equation; the same holds true for $v$ and $\bar{v} = v \times \psi$. This leads to the redundancy 
\begin{align}
    \tilde{n}_2 \simeq \tilde{n}_2 + \omega_2~.
\end{align}

The $G_b$-crossed BTC $\mathcal{C}_{G_b}^\times(c_-)$ is a $G_b$-graded fusion category:
\begin{align}
    \mathcal{C}_{G_b}^\times(c_-) = \bigoplus_{{\bf g} \in G_b} \mathcal{C}_{\bf g} ~,
\end{align}
where $\mathcal{C}_{\bf 0} = \mathcal{C}(c_-)$, and the objects of $\mathcal{C}_{\bf g}$ are the topologically distinct ${\bf g}$ defects, $a_{\bf g} \in \mathcal{C}_{\bf g}$. If we write the fusion rules of the defects in the canonical reference state as
\begin{align}
    a_{\bf g} \times b_{\bf h} = \sum_{c_{\bf gh} \in \mathcal{C}_{\bf gh}} N_{a_{\bf g}, b_{\bf h}}^{c_{\bf gh}} c_{\bf gh}~,
\end{align}
then the state in our symmetry fractionalization class of interest has the defect fusion rules 
\begin{align}
    a_{\bf g} \times b_{\bf h} = \mathfrak{t}({\bf g}, {\bf h}) \sum_{c_{\bf gh} \in \mathcal{C}_{\bf gh}} N_{a_{\bf g}, b_{\bf h}}^{c_{\bf gh}} c_{\bf gh} ~.
\end{align}
Thus we can see the physical meaning of $\tilde{n}_2$. Changing $\tilde{n}_2({\bf g}, {\bf h})$ corresponds to changing the fusion outcome of ${\bf g}$ and ${\bf h}$ defects by $\psi$.

The next important ingredient is that the $G_b$-crossed BTC by design keeps track of topologically distinct ${\bf g}$ defects $a_{\bf g}$, for ${\bf g} \in G_b$. However in the invertible fermionic phase, we physically have $G_f$ defects, which we can label as $({\bf g}, 0), ({\bf g}, 1) \in G_f$, for ${\bf g} \in G_b$. When we gauge fermion parity, $({\bf g},0)$ gets mapped to some element in $\mathcal{C}_{\bf g}$ and $({\bf g}, 1)$ gets mapped to another element in $\mathcal{C}_{\bf g}$. Therefore to fully resolve the $G_f$ fluxes, we need additional data corresponding to a preferred element $L_{\bf g} \in \mathcal{C}_{\bf g}$ that specifies which ${\bf g}$ defect corresponds to a $({\bf g}, 0)$ defect.
These preferred fluxes satisfy the fusion rules 
\begin{align}
    L_{\bf g} \times L_{\bf h} = \begin{cases}
    m^{\om({\bf g}, {\bf h})} L_{\bf gh} \\
    m^{\om({\bf g}, {\bf h})} (L_{\bf gh} + \psi \times L_{\bf g h}) ~,
    \end{cases}
\end{align}
depending on whether $L_{\bf g}$ or $L_{\bf h}$ are Abelian or both non-Abelian. Here $m$ is any fermion parity flux. The factor $m^{\om({\bf g}, {\bf h})}$ ensures that $L_{\bf gh}$ corresponds to a $({\bf gh},0)$ defect if $L_{\bf g}$ and $L_{\bf h}$ correspond to a $({\bf g},0)$ and an $({\bf h},0)$ defect respectively. Here we have another redundancy,
\begin{align}
    L_{\bf g} \simeq L_{\bf g} \times \psi ~,
\end{align}
because fusing with the fermion does not physically change the choice $({\bf g}, 0)$.  

The choice of $\tilde{n}_1$ and symmetry fractionalization class, specified by $\omega_2$ and $\tilde{n}_2$, then specifies the 't Hooft anomaly of the $G_b$ SET. In particular, this specifies an element $[\mathcal{O}_4] \in \mathcal{H}^4(G_b, U(1))$, which must vanish for the SET to be a well-defined (2+1)D system. General methods to compute the 't Hooft anomaly were presented in Ref.~\onlinecite{Barkeshli2020Anomaly,Bulmash2020}. In particular, Ref.~\onlinecite{Barkeshli2020Anomaly} provided simple formulas for the relative anomaly between two SETs whose symmetry fractionalization class differ by an element $[\mathfrak{t}] \in \mathcal{H}^2(G_b, \mathcal{A})$. Therefore, 
assuming we choose reference states that are non-anomalous, the relative anomaly formulas can be used to give  explicit expressions for $\mathcal{O}_4$ (see Eq.~\eqref{eqn:O40}-\eqref{eqn:O4f}). We can show that such a non-anomalous reference state always exists for unitary symmetry; see Appendix~\ref{sec:refstate} for an explicit construction. 

When the anomaly vanishes, 
$\mathcal{O}_4$ is cohomologically trivial:
\begin{align}
    \mathcal{O}_4 = d \nu_3~, 
\end{align}
for some $\nu_3 \in C^3(G_b, U(1))$. The data $\nu_3({\bf g}, {\bf h}, {\bf k})$ can be related to certain additional data required to fully specify the $G_b$-crossed BTC. More specifically, if we start with some reference theory with $F$ symbols denoted by $\hat{F}$, and change the symmetry fractionalization class by $[\mathfrak{t}]$, then the defect $F$-symbols of the new theory are given by
\begin{align}
    F^{a_{\bf g}, b_{\bf h}, c_{\bf k}}_{\mathfrak{t}({\bf g},{\bf h})\mathfrak{t}({\bf gh},{\bf k}) d_{\bf ghk}} = \tilde{F}^{a_{\bf g}, b_{\bf h}, c_{\bf k}}_{\mathfrak{t}({\bf g},{\bf h})\mathfrak{t}({\bf gh},{\bf k}) d_{\bf ghk}} (\nu_3)^{-1}({\bf g}, {\bf h}, {\bf k})~,
\end{align}
where $\tilde{F}^{a_{\bf g}, b_{\bf h}, c_{\bf k}}_{\mathfrak{t}({\bf g},{\bf h})\mathfrak{t}({\bf gh},{\bf k}) d_{\bf ghk}}$ is obtained by multiplying the $\hat{F}$ symbol in the reference theory with additional $F$ and $R$ symbols of the Abelian anyons \cite{Barkeshli2020Anomaly}. 
$\nu_3$ can be thought of as a kind of local counterterm that corrects $\tilde{F}$, so that $F$ obeys the pentagon equation exactly, and not just up to a 4-coboundary. Applying the pentagon equation gives the constraint
\begin{equation}
    d\nu_3 = \mathcal{O}_4
\end{equation}
for some 4-cocycle $\mathcal{O}_4$ that can be obtained in purely terms of anyon data. 

Note that $\tilde{n}_2, \nu_3$ are thus both defined relative to the reference states discussed previously, while $c_-, \tilde{n}_1$ are defined absolutely. 

$\tilde{n}_2$ and $\nu_3$ are subject to three kinds of equivalences. The first equivalence has the form 
\begin{align}
\nu_3 \simeq \nu_3 \times db_2~,
\end{align}
where $b_2 \in C^2(G_b,U(1))$, and describes the change in the defect $F$ symbols under (vertex-basis) gauge transformations of the $G_b$-crossed BTC. 

The second equivalence is of the form 
\begin{align}
    (\tilde{n}_2,\nu_3) &\simeq (\tilde{n}_2 + db_1, 
    \nu_3 \times (-1)^{b_1 \cup \om + db_1 \cup_1 n_2 + db_1 \cup b_1}\times \alpha_3')
\end{align}
where $b_1 \in C^1(G_b,\Z_2)$. In the $G_b$-crossed BTC, this equivalence is obtained by relabeling the defects as $a_{\bf g} \rightarrow a_{\bf g} \times \psi^{b_1({\bf g})}$.  $\alpha_3'$ is an undetermined 3-cocycle, see below. 

When $b_1 = \chi$ is a 1-cocycle, the above transformation reduces to
\begin{align}
\nu_3 \simeq \nu_3 \times (-1)^{\chi\cup\omega_2}~. 
\end{align}
Finally, shifting $\tilde{n}_2 \rightarrow \tilde{n}_2 + \om$ also changes $\nu_3$; therefore the full equivalence is
\begin{equation}
  (\tilde{n}_2,\nu_3)\simeq (\tilde{n}_2+ \om, \nu_3 \times (-1)^{\om\cup_1 \tilde{n}_2}\times \alpha_3'')  
\end{equation}
The 3-cocycles $\alpha_3',\alpha_3''$ can be fixed exactly knowing the full stacking rule for $\nu_3$ (see Section \ref{sec:stackfromO4}); if our conjectured stacking rule is exact, we will have $\alpha_3' = \alpha_3'' = 0$. 

We now comment on how to obtain the classification of invertible phases from the above data. We show that the data $(n_1, n_2, \nu_3)$ for a fixed $c_-$ form a torsor over a group extension involving subgroups of $\mathcal{H}^1(G_b, \Z_2)$, $\mathcal{H}^2(G_b, \Z_2)$, $\mathcal{H}^3(G_b, U(1)_T)$.
First, we notice that $\nu_3$ satisfies $d \nu_3 = \mathcal{O}_4$ and therefore if $\nu_3$ is a solution, $\nu_3  \gamma_3$ for any $[\gamma_3] \in \mathcal{H}^3(G_b, U(1)_T)$ is also a solution.
We define the $\mathcal{H}^3$-action by $[\gamma_3] \in \mathcal{H}^3(G_b, U(1)_T)$ on the basic data as follows:
\begin{eqs}
    \nu_3 \rightarrow \nu_3 \gamma_3,
\end{eqs}
where $\gamma_3$ is a cocycle representative. 
Due to the equivalence relation $\nu_3\simeq \nu_3(-1)^{\chi\cup \omega_2}$ for $\chi\in \mathcal{H}^1(G_b,\mathbb{Z}_2)$,
the $\nu_3$ data thus forms a torsor over $\mathcal{H}^3(G_b, U(1)_T)/\Gamma_3$, where $\Gamma_3 \subseteq \mathcal{H}^3(G_b, U(1)_T)$ is generated by cocycles of the form $(-1)^{\chi \cup \omega_2}$.

Next, given some $[\beta_2] \in \mathcal{H}^2(G_b, \Z_2)$ and $\nu_3'$ satisfying
\begin{eqs}
    d \nu_3^\prime = (-1)^{\beta \cup (\beta + \omega_2)},
\end{eqs}
we can define an invertible phase with data $(c_-^\prime=0, n_1^\prime = 0, n_2^\prime=\beta, \nu_3^\prime)$. Accordingly, the basic data transforms under the following action of $\H^2(G_b,\Z_2)$:
\begin{eqs}
    (c_-,n_1, n_2, \nu_3) \rightarrow&   (0,0, \beta, \nu_3^\prime) \times (c_-,n_1, n_2, \nu_3) \\
    &= (c_-, 0 , n_2 + \beta, \nu_3 \nu_3^\prime \times (-1)^{\beta \cup_1 n_2}).
\end{eqs}
(Since some choices of $\beta$ are trivial, the above action is really by a subgroup of $\H^2(G_b,\Z_2)$.) $\nu_3^\prime$ is not uniquely chosen, and can be modified by the $\mathcal{H}^3$-action above. Moreover, if the quantity $\beta \cup \beta + \beta \cup \omega_2$ is cohomologically trivial for $\beta = B_2, B_2'$, it is also trivial for $\beta = B_2 + B_2'$. Hence, this $\mathcal{H}^2$-action on the data is well-defined and compatible with the $\mathcal{H}^3$-action.

Lastly, from $dn_1 = 0$, the allowed choices of $n_1$ lie in $\mathcal{H}^1(G_b, \Z_2)$. Given $[\alpha_1^\prime] \in \mathcal{H}^1(G_b, \Z_2)$ and a consistent solution $(c_-^\prime=0, n_1^\prime =\alpha_1^\prime, n_2^\prime, \nu_3^\prime)$,\footnote{For $\omega_2=0$, we can always choose $n_2^\prime=0$ and $\nu_3^\prime=1$ for any $n_1^\prime \in \mathcal{H}^1(G_b, \Z_2)$. However, for nontrivial $\omega_2$, some $n_1^\prime$ may not be allowed and for allowed $n_1^\prime$, there is no canonical solution for $n_2^\prime, \nu_3^\prime$.} we can define an $\mathcal{H}^1(G_b,\Z_2)$-action on the data $(c_-,n_1, n_2, \nu_3)$ by
\begin{eqs}
    (c_-,n_1, n_2, \nu_3) \rightarrow&   (0,\alpha_1^\prime, n_2^\prime, \nu_3^\prime) \times (c_-,n_1, n_2, \nu_3) \\
    &= (c_-, n_1 + \alpha_1^\prime, n_2^{\text{tot}}, \nu_3^{\text{tot}}),
\end{eqs}
where $n_2^{\text{tot}}$ and $\nu_3^{\text{tot}}$ are defined through the stacking rules given in Table~\ref{Table:MainEqs}. The choices of $n_2^\prime$ and $\nu_3^\prime$ are not unique since there may be multiple solutions related by the $\mathcal{H}^2$ and $\mathcal{H}^3$ actions discussed above. 
The data $(n_1, n_2, n_3)$ for a fixed $c_-$ ultimately forms a torsor over some group formed by a sequence of group extensions: $\mathcal{H}^1(G_b, \Z_2)$ extended by a subgroup of $\mathcal{H}^2(G_b, \Z_2)$, in turn extended by a subgroup of  $\mathcal{H}^3(G_b, U(1)_T)$.


The equivalence between the data $(\tilde{n}_1, \tilde{n}_2, \nu_3, L_{\bf g})$ that arises in the $G_b$-crossed BTC description and the data $(n_1, n_2, \nu_3)$ depends on the value of $c_-$ and is discussed in detail in Section~\ref{Sec:Counting}. 
Finally, the generalization to the case where $s_1 \neq 0$ is explained in Section~\ref{Sec:MainGxCalcs}.

A complementary way of understanding the above results is in terms of the Chern-Simons theory description for the bosonic shadow theory.
The $G_b$ symmetry enriched bosonic shadow theory can be described using the intrinsic symmetries of the Chern-Simons theory \cite{Benini2018}. 
We will discuss this approach in detail in Section~\ref{Sec:CS}, where some of the advantages and shortcomings will also be seen. 
    
\subsection{Relation to Wang-Gu fSPT classification} \label{sec:relation_wanggu}
Here we compare our results to those of Ref. \cite{Wang2020fSPT}, assuming $c_-=0$. We note that our consistency equations as summarized in Table~\ref{Table:MainEqs} are mostly equivalent to those of Ref.~\onlinecite{Wang2020fSPT} when $c_-=0$, with some differences, which are summarized below. Importantly, in Appendix~\ref{sec: equivalence of O4}, we show that the obstruction ${\cal O}_4$ in Eq.~(136) of Ref.~\onlinecite{Wang2020fSPT} is equal to Eq.~\eqref{eqn: O4 c=k} with $c_-=0$ and $s_1=0$, up to a coboundary (see Eq. \eqref{eqn:O4c=4n} in Appendix~\ref{Sec:RelAnomComps} for our derivation). The key observation is that the two formulas happen to choose opposite conventions for branching structures; after reversing our convention, the formulas defining $n_1, n_2$ and $\nu_3$ all agree. An advantage of our expression is that we can analytically show that $\mathcal{O}_4[c_-,n_1,n_2]$ is closed. 
There is one key difference in our formulas with respect to those of Ref.~\cite{Wang2020fSPT}. Certain equivalence relations were given therein, which change $n_2$ but keep the topological phase invariant. Here we show that these equivalences also come with a change in $\nu_3$, as summarized in Table~\ref{Table:MainEqs}.

\begin{center}
	\begin{table}[t]
	    \renewcommand{\arraystretch}{1.2}
		\begin{tabular}{|l|l|l|l|l|}
			\hline
			\multicolumn{5}{|c|}{The interacting tenfold way} \\ \hline
			Cartan & $G_b$ & $G_f$&Free & Interacting\\ \hline
			A & $U(1)$& $U(1)^f$&$\Z$ & $\Z^2$ \\
			AI & $U(1) \rtimes \Z_2^T$& $U(1)^f \rtimes \Z_2^T$& $\Z_1$ & $\Z_1$ \\
			AII & $U(1)\rtimes\Z_2^T$& $U(1)^f\rtimes\Z_4^{Tf}/\Z_2$ & $\Z_2$ & $\Z_2$ \\
			AIII & $U(1)\times \Z_2^T$ & $U(1)^f\times \Z_2^T$  &$\Z_1$ &$\Z_1$ \\
			D & $\Z_1$ &$\Z_2^f$ &$\Z$ &$\Z$\\
			DIII &$\Z_2^T$ & $\Z_4^{Tf}$ & $\Z_2$& $\Z_2$ \\
			BDI &  $\Z_2^T$ & $\Z_2^T\times\Z_2^f$& $\Z_1$& $\Z_1$\\
			C & $SO(3)$& $SU(2)^f$& $2\Z$& $2\Z\times 2\Z$\\
			CI & $SO(3)\times \Z_2^T$& $SU(2)^f \times \Z_4^{Tf}/\Z_2$ &$\Z_1$ &$\Z_1$\\
			CII & $SO(3)\times \Z_2^T$ & $SU(2)^f \times \Z_2^T$& $\Z_1$& $\Z_1$ \\ \hline
		\end{tabular}
		\caption{Classification of invertible (2+1)D fermion phases for the symmetry classes of the 'tenfold way', with free and interacting fermions. } \label{Table:10fold}
	\end{table}
\end{center}

\subsection{Applications}

With our general classification in hand, we then proceed to apply the theory to a number of physically relevant situations. 
One of our main applications (Section~\ref{Sec:PdTable}) is to consider the symmetry groups appropriate to the 10-fold way, using our approach to derive an interacting 10-fold way classification of fermionic topological insulators and superconductors. This is summarized in Table~\ref{Table:10fold}. As has been noted previously (see {\it e.g.} Ref.~\onlinecite{Morimoto2015}), all the free fermion phases in this classification survive in the presence of interactions. Moreover, in class AIII and class C, the interacting classification acquires an extra factor of $\Z$, which comes from the classification of bosonic SPT phases with $G_b = U(1)$ and $G_b = SO(3)$ respectively. Some entries within the periodic table are discussed in separate examples, as we summarize below.

In Section~\ref{Sec:Z2xZ2fExample}, we study the classification and associated invariants for invertible phases with $G_f = \Z_2 \times\Z_2^f$. We also give an explicit mapping between the free and interacting classifications for this symmetry group. We consider $G_b=U(1)^f$ in Section~\ref{Sec:U(1)symm}; here we derive the Hall conductivity using the $G_b$-crossed theory, and give an alternative discussion on gauging fermion parity using $U(1)$ Chern-Simons theory.

We also discuss the classification of invertible phases with $G_f = SU(2)^f$ symmetry in a separate example, Section~\ref{Sec:HalvingEgs}, as it realizes an $\H^3$ obstruction associated with the failure to satisfy the $dn_2$ equation.
For free fermions, this obstruction is simply the statement that we cannot place isospin-1/2 fermions in a band with odd Chern number. In the interacting case, it follows from the fact that if we gauge fermion parity in a system with odd $c_-$ and with the fermion carrying projective representation of half integer isospin under $G_b = SO(3)$, the fermion parity flux $v$ would 
would have to carry isospin ``1/4" or ``3/4", which is ill-defined and ruled out by the $\H^3$ obstruction mentioned above.

Motivated by this example, in Section~\ref{Sec:HalvingZ2Z2} we consider a $c_-=1$ theory with $G_b = \Z_2\times\Z_2$ that has nontrivial $n_1$ and $\om$, so that $G_f = \mathbb{D}_8$. Unlike $SU(2)^f$, this theory admits a solution for $n_2$ as well as $\nu_3$. We believe it is the simplest example of an invertible phase in which $n_1,\om$ are both nontrivial; it is known \cite{Wang2020fSPT} that such examples do not exist when $c_-=0$ and $G_b$ is a finite abelian group.

\begin{center}
	\begin{table*}
	    \renewcommand{\arraystretch}{1.25}
		\begin{tabular}{|l|l|l|l||l|}
			\hline
			\multicolumn{5}{|c|}{$F,R,U,\eta$ symbols of anyons in the 16-fold way} \\ \hline
			$c_-$& $F^{a b c}$ & $R^{a b}$ & $U_{\bf g}(a,b)$ & $\eta_a({\bf g},{\bf h})$ \\ \hline
			 $4k$&1 &$(-1)^{a_e b_m + \frac{c_-}{4}(a_e b_e + a_m b_m)}$ & $  \left( \frac{i^{a_e a_m + b_e b_m}}{i^{[a_e+b_e]_2[a_m+b_m]_2}}(-1)^{a_m b_e} \right)^{\tilde{n}_1({\bf g})}$ & $(-1)^{(a_e+a_m) \tilde{n}_2 + a_e \omega_2 + a_e(1+a_m) s_1 \cup \tilde{n}_1}$ \\ 
			 $4k+ 2$& $(-1)^{a_e b_e c_e + a_m b_m c_m}$ & $\theta^{a_e b_e + a_m b_m}$ & 1 & $(-1)^{(a_e + a_m) \tilde{n}_2 + a_m \om}$ \\
			 $4k+1$& $\theta^{[a]_4([b]_4+[c]_4-[b+c]_4)}$& $\theta^{[a]_4 [b]_4}$ &  $\begin{cases}
			 (-1)^{a \tilde{n}_1({\bf g})} & b \ne 0 \\ 1 & b = 0
			 \end{cases}$ & $e^{i \pi [a]_4 \frac{c_-}{2}[\omega_2]_2} \times (-1)^{[a]_4 \tilde{n}_2}$\\
			 $4k+3$& $\theta^{[a]_4([b]_4+[c]_4-[b+c]_4)}$& $\theta^{[a]_4 [b]_4}$ &  $\begin{cases}
			 (-1)^{a \tilde{n}_1({\bf g})} & b \ne 0 \\ 1 & b = 0
			 \end{cases}$ & $e^{i \pi [a]_4 \frac{c_-}{2}[\omega_2]_2} \times (-1)^{[a]_4 (\tilde{n}_2+\tilde{n}_1 \cup \tilde{n}_1)}$\\
			 $k+1/2$& $F^{\psi\psi\psi} = 1$& $R^{\psi\psi} = -1$ & 1 & $\eta_{\psi} = 1, \eta_{\sigma} = (-1)^{\tilde{n}_2}$ \\ \hline
		\end{tabular} 
	\caption{$G$-crossed data for the anyons in the SETs obtained by gauging the fermion parity in an invertible fermion phase with symmetry group $G_f$. We define $\theta = e^{i 2\pi c_-/8}$, and $[a]_m = a \mod m$; the integer $k$ is a parameter. When $c_-$ is not a multiple of 4, we assume that the symmetry group $G_b$ has no antiunitary operations, i.e. $s_1 = 0$. We have not written the $F$ and $R$ symbols involving $\sigma$ when $c_-=k+1/2$ as they are not required to determine $\tilde{n}_2$ or $\nu_3$. For a fixed $\tilde{n}_1$, the reference states correspond to $\tilde{n}_2=0$, $\omega_2 = 0$ in the above equations.}
	\label{Table:AnyonFRUn}
	\end{table*} 
\end{center}

\section{$G_b$-crossed extensions of the 16-fold way}\label{Sec:MainGxCalcs}

In this section we discuss the $G_b$ symmetry enriched  bosonic ``shadow" theories obtained by gauging the fermion parity symmetry in the fermionic invertible phases,  summarized in Section~\ref{Sec:Summary}.
The discussion uses the formalism of $G_b$-crossed braided tensor category reviewed in Appendix~\ref{Sec:Gxreview}. In Section~\ref{Sec:Counting} we will use the bosonic shadow theories to classify the fermionic invertible phases.

The $G_b$ symmetry-enriched bosonic theories are described by $F,R,U,\eta$ symbols that specify the rules between defects and the anyons \cite{Barkeshli:2014cna}. They are summarized in Table~\ref{Table:AnyonFRUn}. The  theories can suffer from obstructions that prevent their realization in purely (2+1)D; we compute the obstructions using the  relative anomaly formula in Ref.~\onlinecite{Barkeshli2020Anomaly} by choosing a non-anomalous reference state given by $\tilde n_2=0$ (notation defined in Eq.~\eqref{eqn:n2tilde}) and $\omega_2=0$.

\subsection{Topological order}
	
There are 16 different topological orders obtained by gauging the fermion parity symmetry in
invertible fermionic phases with $\mathbb{Z}_2^f$ symmetry 
\cite{kitaev2006}.
The anyon fusion rules were discussed in Section~\ref{Sec:Summary} and summarized in Table~\ref{Table:16fold}. The $F$ and $R$ symbols of the anyons are given in Table~\ref{Table:AnyonFRUn}. Besides the identity particle and the fermion $\psi$, the topological order has either one or two fermion parity fluxes, whose topological twist equals $\theta = e^{i \pi c_-/4}$.

We will label the anyons for integer $c_-$ as follows. For even $c_-$, the anyons obey $\mathbb{Z}_2\times \mathbb{Z}_2$ fusion algebra, and we will label them by $(a_e,a_m)$ with $a_e,a_m=0,1$.  we will sometimes use the vector representation $I = (0,0), \psi = (1,1), e = (1,0), m = (0,1)$ for the $\Z_2\times \Z_2$ anyons.
For odd $c_-$, the anyons obey $\mathbb{Z}_4$ fusion algebra, and we will label the anyons by $a=0,1,2,3$. We will also label them by $I = [0], v = [1], \psi = [2], \bar{v} = [3]$.

When $c_-$ is a half-integer, the $\Z_2$ abelian anyons are $I,\psi$, and there is also one fermion parity flux $\sigma$ of quantum dimension $\sqrt{2}$. If $c_-$ is not specified, we will use the symbol $m$ to refer to a generic fermion parity flux. In that case, the other fermion parity flux (if $c_-$ is an integer) will be referred to as $\psi \times m$.

\subsection{Symmetry action and structure of defects}

\subsubsection{integer $c_-$}

In the usual $G$-crossed theory, we consider all anyon permutations in the group $\text{Aut}(\mathcal{C})$. However, when describing the bosonic shadow of invertible phases, we restrict our analysis to autoequivalences that preserve $\psi$. This is because $\psi$ is a local fermion in the invertible phase, and a global symmetry cannot permute the local fermion and the fermion parity flux. For integer $c_-$, these autoequivalences form a group $\Z_2$. Therefore we define $\tilde{n}_1: G_b \rightarrow \Z_2$ such that the symmetry $G_b$ permutes the fermion parity fluxes if $\tilde{n}_1({\bf g}) = 1$, and does not if $\tilde{n}_1({\bf g}) = 0$. Since the symmetry operations ${\bf g},{\bf h}$ applied separately should permute each anyon in the same way as the operation ${\bf gh}$, we have
\begin{align}
    \tilde{n}_1({\bf g}) + \tilde{n}_1({\bf h}) &= \tilde{n}_1({\bf gh}) \mod 2 \\
    \implies d\tilde{n}_1 &= 0~ \mod 2.
\end{align}
The choice of $\tilde{n}_1$ determines the number of ${\bf g}$-defects as well as their quantum dimension, for each ${\bf g}$ \cite{Cheng2018fSPT}. First we note that the total quantum dimension of $\C_{\bf g}$ equals that of $\C_{\bf 0}$, which is 2. Now if $\tilde{n}_1({\bf g}) = 0$, all 4 anyons are ${\bf g}$-invariant, so from Eq.~\eqref{Eq:numGdefects} we obtain $|\C_{\bf g}| = 4$. The only possibility is that all 4 ${\bf g}$-defects are abelian. Denoting one of them as $I_{\bf g}$, we can in fact write the others as $a_{\bf g} := a \times I_{\bf g}$, where $a \in \C$ (this is proven in Section~X of Ref.~\onlinecite{Barkeshli2019}). The fusion of an anyon $b$ with a defect $a_{\bf g}$ is specified by
\begin{equation}
    a_{\bf g} \times b = b \times a_{\bf g} = (ab)_{\bf g}~.
\end{equation}

If $\tilde{n}_1({\bf g}) = 1$, only 2 anyons ($I,\psi$) are ${\bf g}$-invariant, so from Eq.~\eqref{Eq:numGdefects} we obtain $|\C_{\bf g}| = 2$. In this case, we can show that the two ${\bf g}$-defects must have the same quantum dimension $\sqrt{2}$, i.e. they correspond to Majorana zero modes. Labelling them as $\sigma^+_{\bf g},\sigma^-_{\bf g}$, we can further show that
\begin{align}
    \psi \times \sigma_{\bf g}^{\pm} = \sigma_{\bf g}^{\pm}\times\psi &= \sigma_{\bf g}^{\pm} \\
    m \times \sigma_{\bf g}^{\pm} = \sigma_{\bf g}^{\pm} \times m &= \sigma_{\bf g}^{\mp}~,
\end{align}
where $m$ is a fermion parity flux.

\subsubsection{half-integer $c_-$}

In this case, there are no anyon permutations. As a result, $\C_{\bf g}$ is in bijection with $\C_{\bf 0}$ (see Section~X of Ref.~\onlinecite{Barkeshli2019} for a proof). The ${\bf g}$-defects can therefore be written as $I_{\bf g}, \psi_{\bf g}$ and $\sigma_{\bf g}$. The fusion of $\psi$ with $\sigma_{\bf g}$ is $\psi \times \sigma_{\bf g} = \sigma_{\bf g}\times\psi = \sigma_{\bf g}$.

Note that there is no analog of $\tilde{n}_1$ in this case. This is the first hint that the usual $G$-crossed theory is missing some information about the invertible phase. We will introduce an analog of $\tilde{n}_1$ in this case through additional flux labels in Section~\ref{Sec:Counting}.

\subsection{Symmetry fractionalization}

In this section, we will obtain conditions on the symmetry fractionalization classes by solving Eqs.~\eqref{Eq:FR-Uconsistency},~\eqref{kappaU},~\eqref{etaConsistency},~\eqref{etacocycle}. This will give us two main results: (i) an explicit form for the $\eta$ and $U$ symbols for the anyons, as summarized in Table~\ref{Table:AnyonFRUn}, and (ii) relatedly, a definition of the data $\tilde{n}_2$ as well as a constraint on $\tilde{n}_2$, for each value of $c_-$. We will also see that the cocycle $\omega_2$ determines the fractional quantum numbers of the fermion $\psi$.

\subsubsection{$c_- = 4k$}
We will separately deal with the case where $c_-$ is a multiple of 4, because the symmetry $G_b$ is then allowed to be antiunitary. For all other values of $c_-$, we will find that Eq.~\eqref{Eq:FR-Uconsistency} only admits unitary symmetries. Note, however, that for $c_-=4 \mod 8$ the antiunitary case is anomalous, even though there is no obstruction to solving for the $U$ and $\eta$ symbols of the anyons \cite{barkeshli2019tr}. Therefore our discussion of anti-unitary symmetries will only be physically relevant for purely (2+1)D systems when $c_-=0$.

For $c_-=4k$ we can choose the anyon $F$ symbols to be trivial, while the $R$ symbols are
\begin{equation}
   R^{ab} = (-1)^{a_e b_m + \frac{c_-}{4}(a_e b_e + a_m b_m)}~. 
\end{equation}
Eq.~\eqref{Eq:FR-Uconsistency} gives the following constraint on $U$:
\begin{equation}
\frac{U_{\bf g}(a,b)}{U_{\bf g}(b,a)} = (-1)^{\tilde{n}_1({\bf g}) (a_e b_m + a_m b_e)}.
\end{equation}
We can satisfy this constraint by choosing 
\begin{equation}
U_{\bf g}(a,b) = \frac{\gamma_a({\bf g}) \gamma_b({\bf g})}{\gamma_{a\times b}({\bf g})} 
(-1)^{a_m b_e \tilde{n}_1({\bf g})} ~
\end{equation} 
for arbitrary functions $\gamma_a({\bf g})$. 
It will be useful to define $\gamma_a({\bf g}) = i^{\alpha({\bf g})a_ea_m}$, where $\alpha$ can be an arbitrary homomorphism from $G_b \rightarrow \Z_2$. Let us take $\alpha = \tilde{n}_1$. This is now used to fix the $\eta$ symbols. Substituting into Eq.~\eqref{kappaU}, we see that  
\begin{equation}
\kappa_{ {\bf g}, {\bf h}}(a,b) = (-1)^{ (s_1 \cup \tilde{n}_1 )({\bf g},{\bf h}) (a_e b_m + a_m b_e)}~.
\end{equation} 
Here we have defined the cup product $s_1 \cup \tilde{n}_1({\bf g},{\bf h}) = s_1({\bf g}) \tilde{n}_1({\bf h})$ (see Appendix~\ref{sec:highercup} for the general definition of cup products). Substituting this into Eq.~\eqref{etaConsistency} with $a = e, b=m$ gives
\begin{equation}
\frac{\eta_e({\bf g},{\bf h}) \eta_m({\bf g},{\bf h})}{\eta_{\psi}({\bf g},{\bf h})} = (-1)^{ (s_1 \cup \tilde{n}_1)({\bf g},{\bf h})}~.
\end{equation}
We also have $\eta_a^2 = 1$ for $a = e,m,\psi$. A general parametrization of the solutions is given by 
\begin{equation}
    \eta_a({\bf g},{\bf h}) = (-1)^{a_e a_m s_1({\bf g}) \tilde{n}_1({\bf h})} M_{a,\mathfrak{t}({\bf g},{\bf h})}~.
\end{equation}
where $\mathfrak{t}$ is an anyon.\footnote{This can be seen as follows. A particular solution can be obtained by taking $\mathfrak{t}$ to be trivial above. Any other solution must differ by some $\eta'$ symbols satisfying $\eta'_a \eta'_b = \eta'_{a \times b}$. These can be parameterized in general as $\eta'_a = M_{a,\mathfrak{t}}$, as proven in Sec IIB of \cite{Barkeshli2019}).} Applying Eq.~\eqref{etacocycle} now imposes the following condition on $\mathfrak{t}$:
\begin{equation}
    \mathfrak{t}({\bf g},{\bf h}) \times \mathfrak{t}({\bf gh},{\bf k}) = {^{\bf g}}\mathfrak{t}({\bf h},{\bf k}) \times \mathfrak{t}({\bf g},{\bf hk})~.
\end{equation}
At this point, we make specific choices for the $\eta$ symbols guided by physical considerations. When $[\omega_2]$ is nontrivial, the $\psi$ particle carries fractional quantum numbers under $G_b$, as specified by the cocycle $\omega_2$. This means that we should set
\begin{equation}
\eta_{\psi}({\bf g},{\bf h}) = (-1)^{\omega_2({\bf g},{\bf h})}~.
\end{equation}
Since $M_{\psi,\psi} = 1$, this can be done by defining
\begin{equation}\label{eqn:n2tilde}
    \mathfrak{t} := m^{\omega_2 + s_1 \cup \tilde{n}_1}\times \psi^{\tilde{n}_2}~,
\end{equation}
for some $\tilde{n}_2 \in C^2(G_b,\Z_2)$. Using the condition on $\mathfrak{t}$ we recover the constraint derived in Ref.~\onlinecite{Wang2020fSPT} (up to a 2-coboundary):
\begin{equation}\label{n2Const:c=0}
   d\tilde{n}_2 = \tilde{n}_1\cup (\omega_2 + s_1 \cup \tilde{n}_1) \mod 2~. 
\end{equation}
The explicit expressions for $\eta_m$ and $\eta_e$ are
\begin{align}
\eta_{m} &= (-1)^{\tilde{n}_2} \nonumber \\
\eta_{e} &= (-1)^{(\tilde{n}_2 + \omega_2 + s_1 \cup \tilde{n}_1)}~.
\end{align}
Note that we made a particular choice of $U$ symbols in the above calculation. Thus there are multiple possible equations for $d\tilde{n}_2$ which are physically equivalent, depending on our choice of $\gamma_a({\bf g})$. 

The $\eta$ symbols can be used to find a simple reference theory and its relation to an arbitrary theory (this will be useful in computing anomalies.) Specifically, by setting $\mathfrak{t}$ trivial, we obtain a reference system with 

\begin{align}
     \eta_e^{\text{ref}} &= \eta_m^{\text{ref}} = 1; \quad \eta_{\psi}^{\text{ref}} = (-1)^{s_1 \cup \tilde{n}_1} \\
   \implies \eta_a^{\text{ref}} &= (-1)^{a_e a_m s_1 \cup \tilde{n}_1}~.
\end{align}
By our definition of $\omega_2$ and $\tilde{n}_2$, such a system has 
\begin{equation}
    \omega_2^{\text{ref}} := s_1 \cup \tilde{n}_1, \quad \tilde{n}_2^{\text{ref}} = 0~.
\end{equation}

Note that when $G_b$ is unitary, we can canonically set $\eta_a^{\text{ref}} = 1$, so we simply have $\eta_a({{\bf g},{\bf h}}) = M_{a,\mathfrak{t}({\bf g},{\bf h})}$. We can explicitly check that the $c_-=0$ reference is non-anomalous by computing the defect $F$ symbols. For $c_-=4k$ and $s_1 = 0$, Appendix \ref{sec:refstate} shows that there is a state with trivial symmetry fractionalization class (i.e. all $\eta=1$) which is non-anomalous.

Finally, we are unable to prove that the reference is non-anomalous when both $s_1 \ne 0$ and $\tilde{n}_1 \ne 0$. However, since the reference still has $\eta_e = \eta_m = 1$, we conjecture that this is the case.

In the $G_b$-crossed theory, the symmetry fractionalization classes form a torsor over $[\mathfrak{t}] \in \H^2_{\tilde{n}_1}(G_b,\Z_2 \times \Z_2)$. For future reference, we note that our definition of the $\eta$ symbols can be expressed in a general form as
\begin{equation}
\eta_a({\bf g},{\bf h}) = (-1)^{ (a_e \omega_2 + (a_e + a_m) \tilde{n}_2 + a_e(1+a_m) s_1 \cup \tilde{n}_1)({\bf g},{\bf h})}~. 
\end{equation}   

\subsubsection{$c_- = 4k+2$}
In this case we have
\begin{align}
    F^{a b c} &= (-1)^{a_e b_e c_e + a_m b_m c_m} \\
    R^{a b} &= \theta^{a_e b_e + a_m b_m}~,
\end{align}
with $\theta = e^{i \pi c_-/4}$. First we show that $s_1$ must be trivial. The physical reason is that the statistics $\theta = \pm i$ of the $e$ and $m$ particles will not remain invariant under an antiunitary symmetry operation. Formally, we can obtain the following condition on $U$ from Eq.~\eqref{Eq:FR-Uconsistency}:
\begin{equation}
    \frac{U_{\bf g}(b,a)}{U_{\bf g}(a,b)} = (-1)^{s_1({\bf g}) (a_e b_e + a_m b_m)}~.
\end{equation}
Taking $a = b = m$ gives $1 = (-1)^{s_1({\bf g})}$ for each ${\bf g}$, therefore $s_1 = 0$. We can now satisfy Eq.~\eqref{Eq:FR-Uconsistency} by choosing all $U = 1$. This implies that $\eta_a \eta_b = \eta_{a \times b}$. As in the case $c_-=0$, we define
\begin{align}
    \eta_{\psi}({\bf g},{\bf h}) &= (-1)^{\omega_2({\bf g},{\bf h})}~.
\end{align}
Applying Eq.~\eqref{etacocycle} to $\eta_{\psi}$ enforces that $\omega_2$ is a cocycle. Applying Eq.~\eqref{etacocycle} to $\eta_{m}$ gives
\begin{eqs}
    &\eta_m({\bf g},{\bf h})\eta_m({\bf gh},{\bf k}) \\
    =&~ \eta_m({\bf g},{\bf hk})\eta_m({\bf h},{\bf k})\left( \frac{\eta_e({\bf h},{\bf k})}{\eta_m({\bf h},{\bf k})} \right)^{\tilde{n}_1({\bf g})} \\
    \implies & (-1)^{d\tilde{n}_2({\bf g},{\bf h},{\bf k})} = (-1)^{(\tilde{n}_1 \cup \omega_2)({\bf g},{\bf h},{\bf k})}~.
\end{eqs} 
This implies the constraint
\begin{equation}\label{n2Const:ceven}
    d\tilde{n}_2 = \tilde{n}_1 \cup \omega_2 \mod 2~.
\end{equation}
For each $\tilde{n}_1$, we take the reference to satisfy all $\eta_a^{\text{ref}} = 1$. This reference state is non-amonalous, as can be seen from the computations in Appendix \ref{sec:refstate}.

Then, the symmetry fractionalization classes are specified by $\eta_a({\bf g},{\bf h}) = \eta_a^{\text{ref}}({\bf g},{\bf h}) M_{a,\mathfrak{t}({\bf g},{\bf h})}$, where we define
\begin{equation}
    \mathfrak{t}({\bf g},{\bf h}) = m^{\omega_2({\bf g},{\bf h})} \times \psi^{\tilde{n}_2({\bf g},{\bf h})}~.
\end{equation}
We note that $M_{\psi,\mathfrak{t}} = (-1)^{\omega_2}$, consistent with our definition of $\eta_{\psi}$. This also means that
\begin{align}
    \eta_m({\bf g},{\bf h}) &= (-1)^{(\tilde{n}_2+ \omega_2)({\bf g},{\bf h})} \\
    \eta_e({\bf g},{\bf h}) &= (-1)^{\tilde{n}_2({\bf g},{\bf h})}~.
\end{align} 

\subsubsection{$c_- = 2k+1$} \label{sec:SF_oddc}
\begin{align}
    F^{a b c} &= \theta^{[a]_4 ([b]_4 + [c]_4 - [b+c]_4)} \\
    R^{a b} &= \theta^{[a]_4 [b]_4}~,
\end{align}
where $[\cdot]_4$ denotes reduction mod 4.
The $R$ symbols impose the following constraint on $U$:
\begin{equation}
    \frac{U_{\bf g}(b,a)}{U_{\bf g}(a,b)} = \frac{K^{s_1({\bf g})} R^{a b}  K^{s_1({\bf g})}}{ R^{{^{\bf g}} a {^{\bf g}} b}} = \frac{\theta^{(1-2 s_1({\bf g}))[a]_4 [b]_4}}{\theta^{[4-a]_4 [4-b]_4}}~.
\end{equation}
Setting $a=b=v = [1]$, we see that $1 = \theta^{-1-9} = -1$ if $s_1({\bf g}) = 1$. To avoid this contradiction we will assume $s_1 = 0$ in the rest of this subsection. We can solve Eq.~\eqref{Eq:FR-Uconsistency} with
\begin{equation}
    U_{\bf g}(a,b) =
    \begin{cases}
    (-1)^{a \tilde{n}_1({\bf g})} & b \ne 0 \\1 & b=0~.
    \end{cases}
\end{equation}
Eq.~\eqref{kappaU} then gives 
\begin{equation}
    \kappa_{ {\bf g}, {\bf h}}(a,b) = 1~.
\end{equation} 
In turn, Eq.~\eqref{etaConsistency} gives $\eta_a \eta_b = \eta_{a \times b}$. Now we consider two separate cases. When $c_-$ is of the form $4k+1$, we use a reference solution with $\eta^{\text{ref}} \equiv 1$. From Appendix \ref{sec:refstate}, we can see that this reference state is non-anomalous. A general symmetry fractionalization class can now be described by $\eta_a({\bf g},{\bf h}) = M_{a,\mathfrak{t}({\bf g},{\bf h})}$ for some anyon $\mathfrak{t}({\bf g},{\bf h})$. Since on physical grounds we should have $\eta_{\psi} = (-1)^{\omega_2}$, the most general choice of $\mathfrak{t}$ involves $v$ and $\psi$, in the following manner: 
\begin{equation}
    \mathfrak{t}({\bf g},{\bf h}) = v^{\omega_2({\bf g},{\bf h})} \times \psi^{\tilde{n}_2({\bf g},{\bf h})} = v^{[([\omega_2]_2+2\tilde{n}_2)({\bf g},{\bf h})]_4}~
\end{equation} 
for some $\tilde{n}_2 \in C^2 (G_b,\Z_2)$. Here we define $[x]_k := a \mod k$; the last equality was obtained using $\psi = v^2$. In general for $c_- = 4k+1$ we have
\begin{align}
    \eta_a({\bf g},{\bf h}) &= \theta^{2 [a]_4 [([\omega_2]_2+2\tilde{n}_2)({\bf g},{\bf h})]_4} \nonumber \\ &= e^{i \pi [a]_4 \frac{c_-}{2} [\omega_2({\bf g},{\bf h})]_2} \times (-1)^{[a]_4 \tilde{n}_2({\bf g},{\bf h})}~.
\end{align}
When $c_-$ is of the form $4k+3$, we define $\eta^{\text{ref}}$ as follows:
\begin{equation}
    \eta^{\text{ref}}_a = M_{a,\psi^{\tilde{n}_1 \cup \tilde{n}_1}}~. 
\end{equation}
Indeed, using the stacking rules from Section \ref{Sec:stacking}, we can show that this reference state corresponds to a stack of $c_-$ copies of the reference state for $c_-=1$. This relationship is useful in simplifying our eventual expressions for the anomaly $\mathcal{O}_4$. Moreover, Appendix \ref{sec:refstate} shows that for $c_-=4k+3$, there is a non-anomalous state with all $\eta=1$. Using the relative anomaly formula, Eq. \eqref{Eq:RelAnom}, we can show that our chosen reference has trivial relative anomaly with this state; therefore our reference is also non-anomalous.

Now a general $\eta$ symbol is of the form $\eta_a = \eta_a^{\text{ref}} M_{a, \mathfrak{t}}$, where $\mathfrak{t}$ is defined in the same way as for $c_- = 4k+1$. In this case we have
\begin{align}
\eta_a({\bf g},{\bf h}) &= e^{i \pi [a]_4 \frac{c_-}{2} [\omega_2({\bf g},{\bf h})]_2} \times (-1)^{[a]_4 (\tilde{n}_2 + \tilde{n}_1 \cup \tilde{n}_1)({\bf g},{\bf h})}~.
\end{align}
Finally we constrain $\tilde{n}_2$. Applying Eq.~\eqref{etacocycle} to $\eta_{\psi}$ enforces that $\omega_2$ is a cocycle. Applying Eq.~\eqref{etacocycle} to $\eta_{v}$ gives
\begin{widetext}
\begin{align}
    \eta_v({\bf g},{\bf h})\eta_v({\bf gh},{\bf k}) &=\eta_v({\bf g},{\bf hk})\eta_v({\bf h},{\bf k})\left( \frac{\eta_{\bar{v}}({\bf h},{\bf k})}{\eta_v({\bf h},{\bf k})} \right)^{\tilde{n}_1({\bf g})} \\
    \implies (-1)^{(\tilde{n}_1 \cup \omega_2)({\bf g},{\bf h},{\bf k})} &= e^{i \pi \frac{c_-}{2}([\omega_2({\bf g},{\bf h})]_2 + [\omega_2({\bf gh},{\bf k})]_2 - [\omega_2({\bf h},{\bf k})]_2 - [\omega_2({\bf g},{\bf hk})]_2)}\times (-1)^{d\tilde{n}_2({\bf g},{\bf h},{\bf k})}~.
\end{align} 
\end{widetext}
The first term on the rhs can be simplified. Note that 
\begin{align}
    & e^{i \pi \frac{c_-}{2} 
    ([\omega_2({\bf g},{\bf h})]_2 + [\omega_2({\bf gh},{\bf k})]_2 - [\omega_2({\bf g},{\bf h}) 
    + \omega_2({\bf gh},{\bf k})]_2)} \cr
    & =  
    e^{i \pi \frac{c_-}{2}([\omega_2({\bf h},{\bf k})]_2 + 
    [\omega_2({\bf g},{\bf hk})]_2 - [\omega_2({\bf h},{\bf k}) + \omega_2({\bf g},{\bf hk})]_2)}\cr 
    &\quad \cdot 
    (-1)^{\omega_2({\bf g},{\bf h}) \omega_2({\bf gh},{\bf k})} \cr
    &=  (-1)^{\omega_2({\bf h},{\bf k}) \omega_2({\bf g},{\bf hk})}  ~.
\end{align}
Upon dividing these equations, and using the cocycle condition on $\omega_2$, we see that the term of interest is equal to 
\begin{align*}
 (-1)^{\omega_2({\bf g},{\bf h}) \omega_2({\bf gh},{\bf k}) + \omega_2({\bf h},{\bf k}) \omega_2({\bf g},{\bf hk})} = (-1)^{(\omega_2 \cup_1 \omega_2)({\bf g},{\bf h},{\bf k})}~.
\end{align*}
Here we used the definition of the cup-1 product of two 2-cochains, which was developed in Ref.~\onlinecite{Steenrod1947}. Putting this back in the constraint equation, we obtain
 \begin{equation}\label{n2Const:codd}
 d\tilde{n}_2 = \omega_2 \cup_1 \omega_2 + \tilde{n}_1 \cup \omega_2 \mod 2~.
 \end{equation}
Although our definition of $\eta$ was different for $c_-=4k+1$ and $c_-=4k+3$, the resulting equations for $d\tilde{n}_2$ are the same. For a more abstract derivation of the same constraint, see Appendix~\ref{app:Steenrod}.
	
	\subsubsection{$c_- = k+1/2$}
	
	A system with $c_- = k+1/2$, where $k$ is an integer, must have trivial $[\omega_2]$ and $s_1$. Physically speaking, $s_1$ should be trivial because an antiunitary operation will convert a left-moving edge state into a right-moving edge state, so that $c_- \rightarrow -c_-$ under such an operation. Furthermore, $[\omega_2]$ should be trivial, because if $\psi$ transforms projectively under $G_b$, that would mean that inserting $G_b$ flux into the system induces a fermion parity flux. However, for SETs with $c_- = k + 1/2$ this flux is a nonabelian anyon, and therefore cannot be induced through symmetry fractionalization. The same conclusion can be arrived at through different arguments using Chern-Simons theory; this is done in Section~\ref{sec:CShalfc}. 
 
 Formally, we argue as follows. Assume that ${\bf g} \in G_b$ is antiunitary, so that $s_1({\bf g}) = 1$. Then the $U$ symbols are constrained by the $R$ symbols as follows (Eq.~\eqref{Eq:FR-Uconsistency}):
 \begin{equation}
 \frac{U_{\bf g}(\sigma,\sigma;1)}{U_{\bf g}(\sigma,\sigma;1)} = \frac{(R^{\sigma \sigma}_1)^*}{R^{\sigma \sigma}_1}~.
 \end{equation}
 The lhs of this equation is 1. However, $R^{\sigma \sigma}_1 = (-i)^{\nu}$, where $\nu = 2 c_-$. Therefore the rhs equals $(-1)^{\nu} = -1$, because $\nu$ is odd. This contradiction implies that there is no consistent solution for $U$ when $s_1$ is nontrivial. 
 
 We can set all $U=1$. With this we obtain $\kappa_{ {\bf g}, {\bf h}} = 1$, implying $\eta_a \eta_b = \eta_{a \times b}$. Since $\psi \times \sigma = \sigma$, we find that $\eta_{\psi} = 1$, i.e. $\psi$ must transform as a linear representation of $G_b$. This forces $[\omega_2] = 0$. There is still freedom in choosing $\eta_{\sigma} = (-1)^{\tilde{n}_2}$ for some $\tilde{n}_2 \in C^2(G_b,\Z_2)$. Applying Eq.~\eqref{etacocycle} using this ansatz shows that $d\tilde{n}_2 = 0$.
 
\subsection{Defect obstruction ('t Hooft anomaly)} \label{Sec:DefObs}

Let us compute the obstruction ${\cal O}_4$, which is the anomaly of the $G_b$ symmetry. It is an obstruction to a well-defined $G_b$-crossed braided tensor category \cite{ENO2010,Barkeshli2019}. In the case with trivial permutations, the absolute anomaly can be directly computed from Eq. \eqref{Eq:AbsAnomNoPerm}; in Appendix \ref{Sec:Psquare} we show how this formula can be conveniently represented in terms of Pontryagin squares. 

When the symmetry permutes the anyons, we will first find a non-anomalous reference bosonic theory $\C_{G,\text{ref}}^{\times}$ that has the same $G_b$ symmetry and the same anyon permutation (specified by $\tilde n_1$).
Then we use a relative anomaly formula presented in \cite{Barkeshli2020Anomaly} to compute the anomaly $\mathcal{O}_4^{(0)}$ of the given theory relative to $\C_{G,\text{ref}}^{\times}$. Finally we simplify $\mathcal{O}_4^{(0)}$ to the final result $\mathcal{O}_4 = \mathcal{O}_4^{(0)} \times dX^{-1}$, by subtracting a suitable 4-coboundary $dX$. Since the reference theory chosen is non-anomalous, this procedure gives the absolute anomaly $[\mathcal{O}_4] \in \mathcal{H}^4(G_b, U(1))$. 

Let us discuss the reference theory in more detail.
The discussion can be separated into $\tilde n_1=0$ and $\tilde n_1\neq 0$; in the latter case the $G_b$ symmetry permutes the anyons (for integer $c_-$).
When $\tilde{n}_1 = 0$, there is a reference theory where the symmetry acts trivially, $U=1$ and also $\eta^{\text{ref}} \equiv 1$, and thus the reference theory has trivial anomaly. 
When $\tilde{n}_1 \ne 0$, we will focus on the case of unitary symmetry, $s_1 = 0$.\footnote{
When $\tilde n_1\neq 0$ and $s_1\neq 0$, we expect the reference theory where the fermion parity fluxes do not carry fractional quantum numbers, i.e. $\eta_e = \eta_m=1$, is also non-anomalous, but do not have a proof.
} 
For half integer $c_-$, the symmetry in our reference theory does not permute the anyons, and we can choose a reference where the symmetry acts trivially, with $U=1$ and $\eta=1$, so this theory is non-anomalous.
For each integer $c_-$,
a non-anomalous theory is given in Appendix \ref{sec:refstate}, which has $\omega_2=0$ and corresponds to setting all $\eta=1$ in the $G_b$-crossed language. As stated in the previous section, when $c_-=3 \mod 4$ we use a different reference with $\eta^{\text{ref}}_a = M_{a,\psi^{\tilde{n}_1 \cup \tilde{n}_1}}$. We can show that this theory has trivial relative anomaly with the one constructed in Appendix \ref{sec:refstate}.

Upon fixing a non-anomalous reference, we can compute the relative anomaly between $\C_{G,\text{ref}}^{\times}$ and $\C_{G}^{\times}$ by using Eq.~\eqref{Eq:RelAnom}. 
In the next Section, we reparameterize the data $\tilde{n}_1$ and $\tilde{n}_2$ (along with additional data tracking $G_b$ flux labels) in terms of new data $n_1$ and $n_2$. Combining the results for different $c_-$ then leads to the expressions summarized previously in Section~\ref{Sec:Summary}. The details of the computations can be found in Appendix~\ref{Sec:RelAnomComps}.

We remark that the results in Table \ref{Table:MainEqs} for the anomaly ${\cal O}_4$ use a non-anomalous reference theory where the $G_b$ symmetry has non-trivial fractionalization (\ref{eqn:refetanontrivial}) for $c_-=3$ mod 4 to simplify the formula, instead of the reference theory with trivial symmetry fractionalization. If we instead use the latter reference theory $\eta_a=1$ for all values of $c_-$, the expressions of ${\cal O}_4$ will have extra term $(-1)^{n_1\cup n_1\cup \omega_2}$ for $c_-=3$ mod 4.
This extra term in ${\cal O}_4$ can also be generated by the redefinition $n_2\rightarrow n_2+n_1\cup n_1$ in the formula of ${\cal O}_4$.
As a consistency check, one can perform the time-reversal transformation, which changes $c_-\rightarrow -c_-$ and complex conjugates ${\cal O}_4$. Due to the term $i^{n_1\cup n_1\cup \omega_2}$ in ${\cal O}_4$, 
we find that the anomaly for the theories with chiral central charges $c_-,-c_-$ differ by $(-1)^{n_1\cup n_1\cup\omega_2}$ if the anomaly is evaluated with respect to $\eta_a=1$.
We remark that one can show both reference theories are non-anomalous, since they have vanishing relative anomaly, and the reference theory with trivial symmetry fractionalization $\eta_a=1$ has vanishing anomaly as shown in Appendix \ref{sec:refstate}.

\section{Characterizing invertible fermionic phases using the $G_b$-crossed theory}\label{Sec:Counting}

Up to this point, 
 we have studied the $G_b$-crossed theories as if we were only interested in the SET phases themselves. 
As we will now see, in order to obtain an accurate counting of invertible fermionic phases, we need to introduce some additional data.

\subsection{The need for flux labels}

First we discuss why the data $\tilde{n}_1, \tilde{n}_2,\nu_3$ in the usual $G_b$-crossed BTC theory do not give a complete description of invertible fermion phases. In Section~\ref{Sec:Z2xZ2fExample}, we discuss a detailed example which studies the counting and classification of invertible phases with $G_f = \Z_2\times\Z_2^f$. The example will concretely illustrate the general points made in this section.   

Topological invariants for invertible phases are often interpreted as the symmetry charges associated to $G_b$ flux insertion. For example, in the integer quantum Hall states, the Hall conductivity measures the integer $U(1)$ charge transported upon inserting a quantum of $U(1)$ flux, while in the $\Z_2$ quantum spin Hall insulator, inserting a $2\pi$ flux of the bosonic $U(1)$ symmetry changes the fermion parity (charge under $\Z_2^f$) in the ground state. Now, a complete theory of defects in a ``$G_f$-crossed" theory of topological phases, where the only anyons are the identity and the fermion, has not been fully developed (see Ref.~\onlinecite{Cheng2018fSPT} for a partial theory in the fSPT case). The basic difficulty is that in this case, the anyon category $\C$ is not modular. Nevertheless, we can make several statements about symmetry defects in such theories. In particular, we can define a set of ``pure" $G_b$ fluxes, represented as $a_{({\bf g},0)}$ (recall that a general element of $G_f$ is given by $({\bf g},a)$, where ${\bf g} \in G_b$ and $a \in \Z_2^f$), while the product of a pure $G_b$ flux and a fermion parity flux is given by some $a'_{({\bf g},1)}$. 

In computing topological invariants, we would like to measure the symmetry charge of $a_{({\bf g},0)}$. However, the usual $G_b$-crossed theory does not distinguish between $a_{({\bf g},0)}$ and $a'_{({\bf g},1)}$. Upon gauging $\Z_2^f$, the $({\bf g},0)$ and $({\bf g},1)$ defects form a single ${\bf g}$-defect sector. Hence there is an ambiguity in determining topological invariants in this case: the $G_b$-crossed theory does not specify which defect to consider. This specification must be included as additional data. For each ${\bf g} \in G_b$, we thus define $L_{\bf g}$ as an ${\bf g}$-defect that corresponds to a $({\bf g},0)$ flux in the original $G_f$-crossed theory. This lets us physically distinguish between $L_{\bf g}$ and $L_{\bf g}\times m$, where $m$ is a fermion parity flux. 

\subsection{Assigning $L_{\bf g}$}
We now explain how to assign the label $L_{\bf g}$. For integer $c_-$ the assignment proceeds as follows. Suppose $L_{\bf g},L_{\bf h}$ correspond to $({\bf g},0)$ and $({\bf h},0)$ defects in the invertible phase. Then we know that the fusion product of $L_{\bf g}\times L_{\bf h}$ is a $({\bf gh},\om({\bf g},{\bf h}))$ defect (if $L_{\bf g}$ and $L_{\bf h}$ are both nonabelian, there will be two fusion products). Therefore, the fusion product of $L_{\bf g}\times L_{\bf h} \times m^{\om({\bf g},{\bf h})}$ must be $({\bf gh},0)$ defect, and we can define $L_{\bf gh}$ through the equation
\begin{align}
    &L_{\bf g}\times L_{\bf h} \times m^{\om({\bf g},{\bf h})} = \nonumber \\ &\begin{cases}L_{\bf gh}, & \text{$L_{\bf g}$ or $L_{\bf h}$ is abelian} \\
    L_{\bf gh} \times (1 + \psi), & \text{$L_{\bf g}$ and $L_{\bf h}$ are nonabelian} \end{cases}
\end{align}
This defines $L_{\bf gh}$ up to multiplication by $\psi$. The ambiguity is due to the fact that defects can always be relabelled by fermions without changing the $G_b$-crossed theory.

We thus assign $L_{\bf g}$ arbitrarily on a complete set of generators of $G_b$; then $L_{\bf g}$ is fully determined by the above equation. Different assignments of $L_{\bf g}$ on the generators in general lead to descriptions of physically distinct invertible phases.

Next we consider the case $c_-=k+1/2$. Here the assignment is simple: for each generator of ${\bf g}$, we define $L_{\bf g}$ as either the abelian defect $I_{\bf g}$, or as the nonabelian defect $\sigma_{\bf g}$. This assignment will in fact define a homomorphism $n_1: G_b \rightarrow \Z_2$, as we will see below. Here $L_{\bf gh}$ is simply one of the fusion products of $L_{\bf g}\times L_{\bf h}$.

\subsection{Relation between the representations $(n_1,n_2,\nu_3)$ and $(\tilde{n}_1,\tilde{n}_2,\nu_3,\{L_{\bf g}\})$}

From previous sections, we see that an invertible phase is described completely by the data $(\tilde{n}_1,\tilde{n}_2,\nu_3,\{L_{\bf g}\})$. We can relabel the defects so that $L_{\bf g}$ gets redefined into some ``canonical" form $L_{\bf g}^0$. Then the same information can actually be encoded within a new set of data $(n_1, n_2,\nu'_3,L_{\bf g}^0)$. Since $L_{\bf g}^0$ is canonical, we can drop it and simply use the data $(n_1, n_2,\nu'_3)$ as introduced in Section~\ref{Sec:Summary}. There are multiple reasons for redefining the data in this way: (i) we can more easily compare our results to those of Ref.~\onlinecite{Wang2020fSPT} when $c_-=0$; and (ii) our equations and stacking formulas for different $c_-$ can be expressed in a more compact form. 

The explicit correspondence is as follows. First let $c_-$ be an integer. We define a `canonical' flux labelling $L_{\bf g}^0$, say $L_{\bf g}^0 = I_{\bf g}$ or $\sigma^+_{\bf g}$ depending on whether $\tilde{n}_1({\bf g}) = 0$ or $1$. $L_{\bf g}^0$ is related to $L_{\bf g}$ by
\begin{equation}
    L_{\bf g}^0 := \begin{cases}
    L_{\bf g}\times m^{l_1({\bf g})}\times \psi^{b_1({\bf g})}, & c_-=2k \\
    L_{\bf g}\times v^{[l_1({\bf g})]_2}\times \psi^{b_1({\bf g})}, & c_-=2k+1~. 
    \end{cases}
\end{equation}
Here we define $l_1, b_1 \in C^1(G_b,\Z_2)$; we also use the notation $[a]_2 = a \mod 2$. Then, by relabelling the defects such that $L_{\bf g} \rightarrow L_{\bf g}^0$, we find that
\begin{align}\label{eqn:n2l1}
    n_1 &= \tilde{n}_1; \\
    n_2 &= \begin{cases}
    \tilde{n}_2 + n_1 \cup l_1, & c_-=2k \\
    \tilde{n}_2 + n_1 \cup l_1 + l_1 \cup l_1, & c_-=2k+1~.
    \end{cases}
\end{align}
Thus $n_2$ is defined as a specific instance of $\tilde{n}_2$, for which the flux labels are canonically defined. From this $n_1$ and $n_2$, we determine $\nu_3$ in the usual manner.

We can prove this as follows. Before the relabelling we have the fusion rule
\begin{equation}
    a_{\bf g}\times b_{\bf h} = \mathfrak{t}({\bf g},{\bf h}) \times \sum_c c_{\bf gh}.
\end{equation}
For $c_-$ even, if we relabel $a_{\bf g} \rightarrow a_{\bf g} \times m^{l_1({\bf g})} \times \psi^{b_1({\bf g})}$, we find that
\begin{equation}
    \mathfrak{t} \rightarrow \mathfrak{t} \times m^{dl_1} \times \psi^{db_1 + n_1 \cup l_1}~.
\end{equation}
Up to coboundaries, we see that $\tilde{n}_2 \rightarrow \tilde{n}_2 + n_1 \cup l_1$. We can repeat this calculation for the case with odd $c_-$; here $\tilde{n}_2$ is additionally shifted by a term $l_1 \cup l_1$. This leads to the result quoted above. We note that specifying $L_{\bf g}$ above is equivalent to specifying the relabelling 1-cochain $l_1$.

Now consider $c_-=k+1/2$. Here we define $L_{\bf g}^0 = I_{\bf g}$ for each ${\bf g}$. The relation between $L_{\bf g}$ and $L_{\bf g}^0$ now defines $n_1$:
\begin{equation}
    L_{\bf g} := L_{\bf g}^0\times \sigma^{n_1({\bf g})}~.
\end{equation}
We can check that $n_1({\bf gh}) = n_1({\bf g}) + n_1({\bf h}) \mod 2$. This is equivalent to saying that (i) $L_{\bf gh}$ is abelian if $L_{\bf g}$ and $L_{\bf h}$ are both abelian or nonabelian; and (ii)$L_{\bf gh}$ is nonabelian if exactly one of $L_{\bf g}$ and $L_{\bf h}$ is nonabelian. This proves that $n_1$ is a homomorphism, in agreement with the definition when $c_-=k$. We then define
\begin{equation}
    n_2 := \tilde{n}_2~,
\end{equation}
and determine $\nu_3$ as before in terms of $n_2$. 

This redefinition has the following interesting consequence. In the bosonic shadow theory, the pair $(\omega_2,\tilde{n}_2)$ is classified by a torsor over $\H^2_{\tilde{n}_1}(G_b,\A)$ where $\A$ is the group of abelian anyons in the bosonic shadow; the coboundary equivalences that define $\H^2_{\tilde{n}_1}(G_b,\A)$ arise because in the bosonic shadow, relabeling symmetry defects by fusing them with Abelian anyons are considered trivial operations. However, once we choose our data so that the flux labels are canonical, we no longer allow relabelings of defects that shift them by a fermion parity flux, because such shifts would change the flux labels away from the canonical choices. Therefore the equivalence on $n_2$ is reduced to arise from only those relabelings of symmetry defects corresponding to fusion with the fermion $\psi$. As such, the different choices of $n_2$ form a torsor over $\H^2(G_b,\Z_2)$.

\subsection{Summary and a general counting procedure}
We now summarize the details given in previous sections by stating a general procedure to count the number of invertible phases with arbitrary symmetry group $G_f$. The steps are as follows, and lead to the results in Table~\ref{Table:MainEqs}:

\begin{enumerate}
    \item For each choice of $c_-$, pick an $n_1 \in \H^1(G_b,\Z_2)$. Then compute $\H^2_{n_1}(G_b,\Z_2)$. Find the allowed choices of $n_2$ in this case: the distinct classes of $[n_2]$ will form a $\H^2(G_b,\Z_2)$ torsor, if unobstructed.
    \item In this representation, we do not need to assign flux labels. If $c_-$ is an integer, the use of $n_1$ and $n_2$ implies that the flux labels are canonically chosen. If $c_-$ is a half-integer, $L_{\bf g}$ is fixed by the choice of $n_1$. 
    \item Using the above choice of $n_2$, obtain a solution for $\nu_3$. The remaining solutions can be obtained by shifting $\nu_3$ by cocycle representatives of $\H^3(G_b,U(1)_T)$.  
    \item The final step is to count redundancies in $n_1,n_2, \nu_3$. The various redundancies are discussed below in Section~\ref{Sec:Equivs}. To summarize that section, first we show that there is always a redundancy $n_2 \simeq n_2 + \omega_2$ for integer $c_-$, which arises from relabelling the fermion parity fluxes. There are also redundancies in $\nu_3$ corresponding to a fixed $n_2$. These take the form $\nu_3 \simeq \nu_3 \times (-1)^{\chi \cup \omega_2}$, where $\chi \in \H^1(G_b,\Z_2)$. They arise due to the freedom in relabelling defects by fermions. After incorporating these redundancies, we obtain the final count of invertible phases. 
\end{enumerate}

\subsection{Equivalences}\label{Sec:Equivs}

In this section we will discuss various redundancies in the counting of invertible phases. The redundancies in $n_2$ and $\nu_3$ were discussed for (2+1)D fSPT phases in Ref.~\onlinecite{Wang2020fSPT}. We will see that similar redundancies hold for all integer values of $c_-$ (but not for half-integer $c_-$). 

The redundancies in $n_2$ can be understood in terms of a relabelling of anyons or symmetry defects by $\psi$; this is considered an equivalence because in the invertible theory, $\psi$ is a local fermion that should not change the topological character of the simple objects. The redundancies in $n_2$ also transform $\nu_3$.

\subsubsection{Equivalence from relabelling fermion parity fluxes}

In the analysis below we can use either the representation $(\tilde{n}_1,\tilde{n}_2,\nu_3,\{L_{\bf g}\})$ or the representation $(n_1,n_2,\nu_3)$. 
Let $c_-$ be an integer. The first result, proven below, is that under a relabelling of the fermion parity fluxes we should take $\tilde{n}_2 \rightarrow \tilde{n}_2 + \om$ together with a change in $\nu_3$. This result holds for both unitary and antiunitary symmetries. If we switch to the variables $(n_1,n_2,\nu_3)$ the same equivalence takes $n_2 \rightarrow n_2 + \om$.

The relabelling equivalence can be proven in a 2-step process. We start with some $G$-crossed theory with known $F,R,U,\eta$ symbols. In the first step, the fermion parity fluxes are relabelled, thus giving a new theory with data $\tilde{F}, \tilde{R}, \tilde{U},\tilde{\eta}$. To prove the equivalence between these descriptions, we perform gauge transformations that send $\tilde{F}, \tilde{R}, \tilde{U}$ back to the original $F,R,U$. In order to send $\tilde{\eta}$ back to $\eta$, we will need to redefine $\tilde{n}_2$ as $\tilde{n}_2 + \omega_2$.

First we consider $c_-=0$. Here we have the following anyon data for a general $G_b$-crossed theory (Table~\ref{Table:AnyonFRUn}):
\begin{align}
    F^{a b c} &= 1 \nonumber \\
    R^{a b} &= (-1)^{a_e b_m} \nonumber \\
    U_{\bf g}(a,b;c) &= ((-1)^{a_m b_e}i^{a_e a_m + b_e b_m - c_e c_m})^{\tilde{n}_1({\bf g})} \nonumber \\
    \eta_a &= (-1)^{a_e(n_2 + \om + s_1 \cup \tilde{n}_1) + a_m \tilde{n}_2 + a_e a_m s_1 \cup \tilde{n}_1}~.
\end{align}

Step 1: Upon relabelling $e \leftrightarrow m$ we get 
\begin{align}
    \tilde{F}^{a b c} &= 1 \nonumber \\
    \tilde{R}^{a b} &= (-1)^{a_m b_e} \nonumber \\
    \tilde{U}_{\bf g}(a,b;c) &= ((-1)^{a_e b_m}i^{a_e a_m + b_e b_m - c_e c_m})^{\tilde{n}_1({\bf g})} \nonumber \\
    \tilde{\eta}_a &= (-1)^{a_m(\tilde{n}_2 + \om + s_1 \cup \tilde{n}_1) + a_e \tilde{n}_2 + a_e a_m s_1 \cup \tilde{n}_1}~.
\end{align}
Step 2: We perform a vertex basis gauge transformation (see Eq.~\eqref{Eq:VBGT}) given by $\Gamma^{a b}_{a \times b} = i^{a_m b_e-a_e b_m}$. This takes 
\begin{eqs}
    \tilde{F}^{a b c}_d &\rightarrow \frac{\Gamma^{a b}_e \Gamma^{e c}_{d}}{\Gamma^{a f}_d \Gamma^{b c}_f}\tilde{F}^{a b c} = 1 = F^{a b c} \nonumber \\
    \tilde{R}^{a b} &\rightarrow \tilde{R}^{a b} \frac{\Gamma^{b a}}{\Gamma^{a b}}= (-1)^{a_e b_m} = R^{a b} \nonumber \\
    \tilde{U}_{\bf g}(a,b) &\rightarrow \frac{\Gamma^{{^{\bf g}}a ^{\bf g}b}}{(\Gamma^{a b})^{1-2s_1({\bf g})}}\tilde{U}_{\bf g}(a,b) \nonumber \\ &= (-1)^{(a_m b_e + a_e b_m)(\tilde{n}_1+s_1)({\bf g})} \tilde{U}_{\bf g}(a,b)  \nonumber \\ &=(-1)^{(a_m b_e + a_e b_m) s_1({\bf g})} U_{\bf g}(a,b)\nonumber \\
    \tilde{\eta} &\rightarrow \tilde{\eta}~.
\end{eqs}
These vertex basis gauge transformations do not completely correct $\tilde{U}$ if $s_1 \ne 0$. The only other gauge transformation that can do so is called a symmetry action gauge transformation; it takes
\begin{equation}
    \tilde{U}_{\bf g}(a,b) \rightarrow \frac{\gamma_a \gamma_b}{\gamma_{a\times b}}({\bf g})\tilde{U}_{\bf g}(a,b).
\end{equation}
If we choose $\gamma_m = \gamma_{\psi} = 1, \gamma_e = (-1)^{s_1}$, we can check that the $U$ symbols are fully corrected; moreover, the $\eta$ symbols change as follows (Eq.~\eqref{Eq:SAGT}):
\begin{align}
    \tilde{\eta}_a \rightarrow (-1)^{(a_e + a_m)s_1 \cup \tilde{n}_1} \tilde{\eta}_a.
\end{align}
The original relabelling shifted $\tilde{n}_2$ by $\om + s_1 \tilde{n}_1$. The above gauge transformation $\gamma$ only shifts it back by $\tilde{n}_1 s_1$, which is coboundary equivalent to $s_1 \tilde{n}_1$. This is not sufficient to fully restore the $\eta$ symbols. The only way to do so is to additionally redefine $\tilde{n}_2 \rightarrow \tilde{n}_2 + \om$ by hand. We can see that this corrects the $\eta$ symbols fully. Since we then obtain the same data as in the original theory, we conclude that a relabelling of $e$ and $m$ particles is equivalent to a shift of $\tilde{n}_2 \rightarrow \tilde{n}_2 + \om$, along with suitable gauge transformations.

We have verified in Appendix \ref{sec:gtappendix} that when $s_1 = 0$, additional gauge transformations cannot produce any further equivalences on $n_2$. However, when $s_1 \ne 0$ we do formally obtain another equivalence, namely $\tilde{n}_2 \rightarrow \tilde{n}_2 + \om + s_1 \tilde{n}_1$.  The gauge transformations associated to this equivalence all require a particular choice $\Gamma^{\psi \psi} = -1$. On the other hand, in the transformation $\tilde{n}_2 \rightarrow \tilde{n}_2 + \om$ above we used a gauge transformation with $\Gamma^{\psi \psi} = +1$. Ref. \cite{bulmashSymmFrac} noted that allowing $\Gamma^{\psi \psi} = -1$ leads to several inconsistencies in the classification of fermionic SET phases. Following Ref. \cite{bulmashSymmFrac}, we thus impose $\Gamma^{\psi \psi} = +1$ on our allowed gauge transformations; as noted there, it is still an open question to find a more physically motivated argument for this condition.

Next we consider $c_- \ne 0$. As before, we find that a relabelling of the fermion parity fluxes changes the $G_b$-crossed data; moreover the $\eta$ symbols cannot be restored to their original values by gauge transformations. Now the $\eta$ symbols can equivalently be expressed in terms of the cocycle $\mathfrak{t}$: 
\begin{equation}
\mathfrak{t}({\bf g},{\bf h}) = m^{\omega_2({\bf g},{\bf h})} \times \psi^{\tilde{n}_2({\bf g},{\bf h})} ~.
\end{equation}
(We denote a fermion parity flux as $m$ or $\psi \times m$ whenever $c_-$ is unspecified.) When we shift $\tilde{n}_2 \rightarrow \tilde{n}_2 + \omega_2$, the cocycle is correspondingly modified as
\begin{align}
\mathfrak{t}({\bf g},{\bf h}) &\rightarrow (\psi m)^{\omega_2({\bf g},{\bf h})} \times \psi^{\tilde{n}_2({\bf g},{\bf h})}  \\
&= m^{\omega_2({\bf g},{\bf h})} \times \psi^{(\tilde{n}_2+\omega_2)({\bf g},{\bf h})} ~.
\end{align}
Therefore when $c_-$ is an integer, shifting $\tilde{n}_2 \rightarrow \tilde{n}_2 + \omega_2$ is equivalent to relabelling the fermion parity fluxes. We can check that the $F,R,U$ symbols for the anyons remain invariant up to gauge transformations when such a relabelling is performed. 
Note that this equivalence does not arise when $c_-$ is a half-integer, because $\omega_2$ must be trivial in that case.

We note that for any change in $\tilde{n}_2$, the expression for $\mathcal{O}_4$ changes, and as a result $\nu_3$ must also change. For the above equivalence on $\tilde{n}_2$, the change in $\mathcal{O}_4$ can be computed directly; integrating this gives an expression for the change in $\nu_3$, up to a 3-cocycle $\alpha''$:
\begin{equation}\label{eq:equivn2om2}
    (\tilde{n}_2,\nu_3) \simeq (\tilde{n}_2 + \om, \nu_3 \times (-1)^{\om \cup_1 \tilde{n}_2}\times \alpha'')
\end{equation}
If our conjectured stacking rules are correct, we will have $\alpha''=0$ (see the argument in Section \ref{sec:stackfromO4}) 

\subsubsection{Equivalences from relabelling symmetry defects}\label{sec:equivs-defects}

We can obtain additional equivalences as follows: we relabel the symmetry defects with fermions, compute the change in the defect $F,R,U,\eta$ symbols, and then apply gauge transformations which shift the data back to their original form, up to some redefinition of $\nu_3$. In this section we use a result from Ref.~\onlinecite{aasen2021torsorial}, which considers general relabellings of defects in a $G$-crossed BTC by anyons. Consider a $G_b$-crossed BTC with some $\tilde{n}_1,\tilde{n}_2$ and $\nu_3$. 
Let us relabel each defect as $a_{\bf g} \rightarrow a_{\bf g}\times \psi^{\chi({\bf g})}$, where $\chi \in Z^1(G_b,\Z_2)$. Then, the
results of Ref.~\onlinecite{aasen2021torsorial}, specialized to our case, imply that the defect $F$ symbols transform by a 3-cocycle $\xi \in Z^3(G_b,U(1))$, defined as follows: 
\begin{align}
    \xi({\bf g},{\bf h},{\bf k}) &:= \eta_{\psi^{\chi({\bf k})}}({\bf g},{\bf h}) U_{\bf g}(\psi^{\chi({\bf h})},\psi^{\chi({\bf k})}) \\
    &= (-1)^{\chi({\bf g}) \om({\bf h},{\bf k})} \times 1= (-1)^{(\chi \cup \om)({\bf g},{\bf h},{\bf k})}~.
\end{align}
This result is true for all integer values of $c_-$, and agrees with the result derived for $c_-=0$ in Ref.~\onlinecite{Wang2020fSPT} through a different construction. We note that this idea of relabelling defects using elements of $\H^1(G,\A)$, where $\A$ is a group of abelian anyons, has also been studied in other recent works (see e.g. Appendix G of Ref.~\onlinecite{Manjunath2020fqh}). 

 More generally, consider a relabelling of the form
\begin{equation}
    a_{\bf g} \rightarrow a_{\bf g} \times \psi^{b_1({\bf g})}~.
\end{equation}
For a general 1-cochain $b_1$, this changes the fusion rules. In particular, the symmetry fractionalization cocycle transforms as follows:
\begin{equation}
    \mathfrak{t}({\bf g},{\bf h}) \rightarrow \mathfrak{t}({\bf g},{\bf h}) \times \psi^{db_1({\bf g},{\bf h})}~. 
\end{equation}
We can absorb the extra piece into the definition of $\tilde{n}_2$ by redefining it as follows:
\begin{equation}
    \tilde{n}_2 \rightarrow \tilde{n}_2 + db_1~.
\end{equation}
The relabelling also changes $\nu_3$: the shift in $\nu_3$ can be computed from the formula for $\mathcal{O}_4$. We have
\begin{align}\label{eq:equivn2db2}
    \frac{\mathcal{O}_4^{\text{new}}}{\mathcal{O}_4^{\text{old}}} = (-1)^{db_1 \cup \om + db_1 \cup \tilde{n}_2 + \tilde{n}_2 \cup db_1 + db_1 \cup db_1}~,
\end{align}
and therefore we must have
\begin{equation}
    \nu_3 \rightarrow \nu_3 \times (-1)^{b_1 \cup \om + db_1 \cup_1 \tilde{n}_2 + b_1 \cup d b_1 }\times \alpha_3'~,
\end{equation}
where $\alpha_3'$ is an undetermined 3-cocycle. 

In Section \ref{Sec:stacking} we discuss the stacking rules for invertible fermionic phases. The stacking rule for $\nu_3$ also has an undetermined 3-cocycle which we call $\alpha_3$. In Section \ref{sec:stackfromO4} we show that $\alpha_3'$ can be written exactly in terms of $\alpha_3$: in particular, if $\alpha_3 =0$ then $\alpha_3' = 0$ as well. We conjecture that $\alpha_3 ,\alpha_3'$ indeed vanish.

To summarize, the most general defect relabeling by fermions induces the following transformation on the data (assuming the conjecture $\alpha_3' = 0$):
\begin{align}
    \tilde{n}_2 &\rightarrow \tilde{n}_2 + db_1 \nonumber \\
    \nu_3 &\rightarrow \nu_3 \times (-1)^{b_1 \cup \om  + db_1 \cup_1 \tilde{n}_2 + b_1 \cup db_1}~.
\end{align}
The same equivalences apply to the data $(n_1, n_2, \nu_3)$ without tildes, as written in Table \ref{Table:MainEqs}. This is because the above arguments go through without any changes if we assume that the flux labels are canonically defined.

What do these redundancies physically mean? As pointed out in Ref.~\onlinecite{Jiang2017}, the cocycle $\nu_3 = (-1)^{\omega_2 \cup \chi}$ describes a symmetry enriched toric code in which there is no symmetry fractionalization, but the $m$ particles transform as \textit{linear} representations of $G_b$, and have nontrivial integer quantum numbers. Using this, we can heuristically understand the relabelling equivalence as follows. The action of ${\bf g}$ on an $m$ particle in the original system consists of braiding $m$ around a ``bare" ${\bf g}$-defect $I_{\bf g}$. If we redefine $I_{\bf g}$ as $I_{\bf g} \times \psi^{\chi({\bf g})}$, then the symmetry action on an $m$ particle in the redefined system consists of braiding the $m$ particle around the bound state of $I_{\bf g}$ and the anyon $\psi^{\chi({\bf g})}$. The extra braiding phase $M_{m,\psi^{\chi({\bf g})}} = (-1)^{\chi({\bf g})}$ indicates that the $m$ particle now carries an additional linear representation under $G_b$, as specified by $\chi$, relative to the original system. This shifts $\nu_3$ by an SPT 3-cocycle, for the following reason. When we insert $G_b$ flux in the relabelled system through the operations ${\bf g},{\bf h}$, we induce the anyon $ m^{\omega_2({\bf g},{\bf h})}$ (up to a possible fermion). This anyon transforms as a linear representation under a third symmetry operation given by ${\bf k}$. Therefore, a $G_b$ flux inserted through ${\bf g},{\bf h}$ has integer $G_b$ charge, which is measured by the 3-cocycle
\begin{equation}
M_{\mathfrak{t}({\bf g},{\bf h}) \psi^{\chi({\bf k})}} = (-1)^{\omega_2({\bf g},{\bf h}) \chi({\bf k})}~.
\end{equation}     

Since relabelling defects by fermions is considered a trivial modification, such cocycles should also be modded out of the classification. However, as we have argued previously, relabelling defects in the $G_b$-crossed theory by fermion parity fluxes has no physical analog in the fSPT picture, and should not be considered as an equivalence between fSPT phases.

\section{$\text{Spin}(2c_-)_1$ Chern-Simons theory approach} \label{Sec:CS}

In this section, we will recover the results of the $G_b$-crossed approach using Chern-Simons theory. As in Sections \ref{Sec:MainGxCalcs} and \ref{Sec:Counting}, we will primarily be interested in constructing ``bosonic shadow" theories for $G_b$ SET phases. We can also directly construct actions for invertible phases, which are useful in studying the stacking rules: we discuss this in Section~\ref{Sec:stacking}. Furthermore, there are potential applications of realizing these phases using Chern-Simons theory coupled to matter fields, but we will not discuss them here.

Let us begin with the invertible fermionic phases with only the fermion parity symmetry. The phases are classified by the chiral central charge $c_-$, and they can be described by
the Chern-Simons theory $SO(L)_1$ with $L=2c_-$, which depends on the spin structure.\footnote{When $L=1$, we define the theory by the duality $SO(1)_1\leftrightarrow SO(17)_1\times (E_8)_{-1}$ where $(E_8)_1$ is the invertible bosonic TQFT that will not participate in the discussion. Similarly, when $L=0$ we define the theory by the duality
\begin{equation}
    SO(0)_1\leftrightarrow SO(16)_1\times (E_8)_{-1}~.
\end{equation}
}
The local fermion particle is described by the Wilson line in the vector representation. The theory is invertible \cite{Freed:2004yc}, with effective action given by the gravitational Chern-Simons term $-2c_- \text{CS}_\text{grav}$, with
\begin{equation}
\text{CS}_\text{grav}=\frac{1}{192\pi} \text{Tr }\left(\omega d\omega +\frac{2}{3}\omega^3\right)~,
\end{equation}
where $\omega$ is the spin connection of the spacetime manifold (not to be confused with the $SO(L)$ gauge field). This effective action generates a $\mathbb{Z}$ classification of the fermionic invertible phases.
The fermion parity symmetry can be identified with the magnetic 0-form symmetry of the Chern-Simons theory \cite{Cordova:2017vab}, generated by $(-1)^{\oint w_2^{SO}}$ where $w_2^{SO}$ is the second Stiefel-Whitney class of the $SO(L)$ bundle.\footnote{Physically, it can be understood as the total GNO charge \cite{Goddard:1976qe} mod 2.}

The fermionic invertible phase with symmetry $G_b\times \mathbb{Z}_2^f$ or an extension of $G_b$ by $\mathbb{Z}_2^f$ corresponds to coupling the $SO(L)_1$ Chern-Simons theory to background $G_b$ gauge field and the spin structure. 

\subsection{Relation between chiral central charge and symmetry extension}\label{sec:CS_w2defn}

When the symmetry is $G_b\times \mathbb{Z}_2^f$, the gravitational Chern-Simons term is well-defined for any integer coefficient and gives a well-defined fermionic invertible phase (where $G$ acts trivially).
Thus we can focus on classifying the phases with $c_-=0$: the phases with nonzero $c_-$ are obtained by stacking with additional fermionic invertible phase with only fermion parity symmetry ({\it e.g.} a $p+ip$ superconductor).

On the other hand, when the symmetry is a nontrivial extension of $G_b$ by $\mathbb{Z}_2^f$ specified by $[\omega_2] \in \H^2(G_b,\Z_2)$, the fermionic invertible phases with only fermion parity symmetry is no longer well-defined: there are $G_f$ transformations that are not in $\mathbb{Z}_2^f$ but can compose into the nontrivial element of $\mathbb{Z}_2^f$. For example, if the fermion has spin 1/2, a spatial rotation by the angle $2\pi$ is equivalent to multiplication by $(-1)^F$ and not $1$. In such cases, it is not consistent for a system to only transform under $\mathbb{Z}_2^f$ but not other elements of $G_f$. In particular, the $p+ip$ topological superconductor does not admit such a symmetry extension.

In fact, we will show later in the section that fermionic invertible phases with $[\omega_2] \ne 0$ must have integer chiral central charge, and can never have half integer chiral central charge. The same result is obtained in the $G_b$-crossed approach by studying the constraints on symmetry fractionalization when $c_-$ is a half integer.

These results can be understood more formally. When the symmetry is a nontrivial extension, the total symmetry group is no longer $Spin_\text{Lorentz}\times G_b$ but\footnote{Here we assume the symmetry $G_b$ is unitary for simplicity. For antiunitary symmetry, the $Spin$ Lorentz symmetry is replaced by $Pin^\pm$ and there is additional $\mathbb{Z}_2$ quotient to identify the time-reversal element in $Pin^\pm,G_b$.} 
\begin{equation}
 {Spin_\text{Lorentz}\times G_f\over \mathbb{Z}_2}~.  
\end{equation}
Now the spin structure is replaced by a ``spin-$G_f$ structure", given by a $\mathbb{Z}_2$ one-cochain $\rho$ satisfying
\begin{equation}
   d \rho = w_2(TM)+(A)^*\omega_2~.
\end{equation}
Here 
$A$ is the background gauge field for $G_b$ symmetry, and $w_2(TM)$ is the second Stiefel-Whitney class of the tangent bundle, and $\omega_2\in \H^2(G_b,\mathbb{Z}_2)$ specifies the extension $G_f$. When $\omega_2=0$, $\rho$ is independent of the details of $G_b$. Since the allowed gravitational CS terms depend only on the definition of $\rho$, they can also be defined independent of the details of $G_b$ for each value of $c_-$. On the other hand, when $[\omega_2] \ne 0$, $\rho$ depends on $\omega_2$, and in turn the gravitational Chern-Simons term is only well defined for specific values of $c_-$: see the examples below. 
We remark that since the phases do not depend on the usual spin structure, they are classified by the cobordism group with spin-$G_f$ structure instead of the spin cobordism group \cite{Kapustin:2014tfa,Freed:2016rqq,Yonekura:2018ufj}.

For example, when $G_b=U(1)$ and $G_f$ is its double covering, such a symmetry is present in systems that satisfy the spin/charge relation with respect to the global $U(1)$ symmetry: local particles with odd charge are fermion, and with even charge are bosons. 
The systems require a $spin^c$ structure, where $Spin^c(d)=\left(Spin(d)_\text{Lorentz}\times U(1)\right)/\mathbb{Z}_2$.
The effective actions for the invertible phases with $spin^c$ structure $A$ are fully given by integer linear combinations of the terms (see {\it e.g.} Ref.~\onlinecite{Seiberg:2016rsg,Hsin:2016blu})
\begin{align}
 & I[A]=\frac{1}{4\pi}AdA+2\text{CS}_\text{grav},\nonumber \\   
 & 16\text{CS}_\text{grav}~.
\end{align}
The latter is the effective action for the bosonic $\overline{(E_8)_1}$ invertible topological order. The above two effective actions have chiral central charge $c_-=-1$ and $c_-=-8$ respectively, and thus the invertible phases they generate always have integer chiral central charge. For this symmetry, the fermionic invertible phases do not allow half integer chiral central charges.

Another example is the free fermion invertible phase. Consider free massive Majorana fermions transformed under some representation $\mathbf{r}$ of flavor symmetry $G_f$, where the center $\mathbb{Z}_2^f\subset G_f$ flips the sign of all fermions. 
Massive free fermions describe invertible phases \cite{Hastings_2008}. 
If we turn on background $A$ for $G_f$ symmetry, the fermion has action
\begin{equation}
  {\cal L}=  i\bar \psi \slashed{D}_{A}\psi-m\bar\psi\psi~,
\end{equation}
where $D_A \equiv\partial-iA  -\frac{i}{2}\omega$ is the covariant derivative.
Integrating out the fermions generates different gravitational Chern-Simons terms for positive and negative mass. The difference is equal to $(\text{dim }\mathbf{r})\text{CS}_\text{grav}$, where $\text{dim }\mathbf{r}$ is the dimension of the representation (it is also equal to the number of Majorana fermions in the multiplet).\footnote{
The fermion partition function is given by the eta invariant coupled to background $A$ for the $G_f$ symmetry \cite{RevModPhys.88.035001}, and by Atiyah-Patodi-Singer index theorem \cite{Atiyah:1975jf} it can be expressed as a combination of gravitational Chern-Simons term and a part that depends on the $A$ background gauge field.
}
Thus the chiral central charge of the fermionic invertible phase of $G_f$ symmetry is given by 
\begin{equation}
    c_-=-\frac{1}{2}\text{dim }\mathbf{r}~.
\end{equation}
If the group $G_f$ is a nontrivial extension of $G_b$ by $\mathbb{Z}_2$, i.e. $G_f$ is a nontrivial double covering of $G_b$, then the dimension of the representation that transforms under the $\mathbb{Z}_2$ center must be even, $\text{dim }\mathbf{r}\in 2\mathbb{Z}$, and therefore the chiral central charge of the fermionic invertible phase is an integer instead of a half integer.
Such a relation between integer $c_-$ and nontrivial symmetry extension $[\omega_2]$ holds in general for fermionic invertible phases.

\subsection{Classifying invertible phases using ``bosonic shadow" $Spin(2c_-)_1$ Chern-Simons theory}\label{sec:CS_idea}

To obtain the full classification of fermionic invertible phases, and in particular when the symmetry is a nontrivial extension of $G_b$ by $\mathbb{Z}_2^f$, it is easier to work with the bosonic theory obtained by gauging the fermion parity symmetry \footnote{More precisely, summing over spin structures. The difference is explained in \cite{Hsin:2019gvb}.} in the symmetry-enriched $SO(2c_-)_1$ Chern-Simons theory. This theory is called the ``bosonic shadow" for the fermionic invertible phase \cite{Bhardwaj_2017}. Since the fermion parity symmetry is the magnetic 0-form symmetry in the $SO(2c_-)_1$ Chern-Simons theory, gauging this symmetry produces the double-covering $Spin(2c_-)_1$ Chern-Simons theory. The latter theory is enriched by the the dual $\mathbb{Z}_2$ one-form symmetry corresponding to the fermion $\psi$. 

The existence of an additional one-form symmetry means that we couple the Chern-Simons theory to the background gauge field not just for the $G_b$ 0-form symmetry, but also a two-form gauge field $B_2$ for this one-form symmetry. If we gauge this one-form symmetry, {\it i.e.} sum over the background gauge field $B_2$, we recover the original fermionic theory. The gauging produces a fermionic theory from a bosonic theory because the one-form symmetry is generated by $\psi$ and hence has a particular anomaly \cite{Gaiotto:2015zta}. In the rest of this section we will discuss the classification of fermionic invertible phases using the bosonic shadow theories.

To relate this to the classification using $G_b$-crossed BTCs discussed in previous sections, we note that turning on the background for the 0-form symmetry $G_b$ is equivalent to inserting $G_b$ symmetry defects. The properties of the anyons and symmetry defects partially specify the invertible phase, but we need to add some extra information in the form of flux labels (see Section~\ref{Sec:Counting}). Now shifting the definition of flux labels in the $G_b$-crossed theory is related to a shift in spin structure in the fermionic invertible phase.
We note that the bosonic theory has a ``dynamical spin structure" $\rho$ which is summed over in the path integral, whose Wilson line is the fermion line that generates the dual $\mathbb{Z}_2$ one-form symmetry. The one-form symmetry has background $B_2$ that couples as
\begin{equation}
    \pi\int B_2\cup \rho~.
\end{equation}
Thus shifting the spin structure by $A_b^*n_1$ for $n_1\in \H^1(G_b,\mathbb{Z}_2)$, $\rho\rightarrow \rho+A_b^*n_1$, induces the effective action
\begin{equation}
    \pi\int_{3d} B_2\cup A_b^*n_1,\quad n_1\in \H^1(G_b,\mathbb{Z}_2)~,
\end{equation}
which describes a symmetry protected topological phase with $\mathbb{Z}_2$ one-form symmetry and $G_b$ 0-form symmetry.

Thus, the $G_b$-crossed data for anyons and defects, together with flux labels, can be equivalently encoded by coupling the $Spin(2c_-)_1$ theory to background gauge field for 0-form symmetry $G_b$ and $\mathbb{Z}_2$ one-form symmetry.

We remark that the classification for the fermionic SPT phases with $c_-=0$ using this approach (where $Spin(0)_1$ is equivalent to the untwisted $\mathbb{Z}_2$ gauge theory of the toric code) is discussed in Refs.~\onlinecite{Bhardwaj_2017,Kapustin:2017jrc}.\footnote{
Note the duality $Spin(16)_{1}\times (E_8)_{-1}\leftrightarrow (\mathbb{Z}_2)_0$ where the right hand side is the untwisted $\mathbb{Z}_2$ gauge theory  \cite{Cordova:2017vab}.}

\subsection{Summary of the intrinsic symmetry in $Spin(2c_-)_1$ Chern-Simons theory}\label{sec:CS_symmetry}

Here we summarize some properties of the $Spin(2c_-)_1$ Chern-Simons theory. See {\it e.g.} Appendix C of Ref.~\onlinecite{Seiberg:2016rsg}, and Ref.~\onlinecite{Cordova:2017vab,Benini2018,Hsin:2020nts} for a more detailed discussion of the global symmetry and anomaly of the theory.
The theory has chiral central charge $c_-$, which can be an integer or half integer. These results are also summarized in Tables \ref{Table:MainEqs} and \ref{Table:16fold}.

For every $c_-$, the theory has a fermion denoted by $\psi$ that generates $\mathbb{Z}_2$ one-form symmetry. Gauging the one-form symmetry changes the theory into $SO(2c_-)_1$, which is a fermionic invertible phase with $(-1)^F$ symmetry dual to $2c_-$ copies of the p+ip topological superconductor.
We will construct $Spin(2c_-)_1$ theories enriched with $G_b$ symmetry such that gauging the one-form symmetry produces the fermionic theory with $G_f$ symmetry.

\subsubsection{For $c_-$ a half integer}

For $c$ a half integer, the theory has three anyons, denoted by $I,\psi,\sigma$ of spin $0,\frac{1}{2},\frac{2c_-}{16}$ mod 1. They obey Ising fusion rules (see Table~\ref{Table:16fold}) and are Wilson lines in the trivial, vector and spinor representation of $Spin(2c_-)$.
The theory has $\mathbb{Z}_2$ one-form symmetry generated by $\psi$, with the anomaly described by an SPT phase in one dimension higher \cite{Gaiotto:2014kfa,Hsin:2018vcg}:
\begin{equation}\label{eqn:anomchalfinteger}
    \pi\int_{4d} {\cal P}(B_2)=\pi\int_{4d} Sq^2(B_2)~,
\end{equation}
where $B_2$ is the background two-form gauge field for the $\mathbb{Z}_2$ one-form symmetry, and it satisfies $d B_2=0$. ${\cal P}(B_2)=B_2\cup B_2-B_2\cup_1d B_2$ is the Pontryagin square of $B_2$ \cite{Whitehead1949},~\footnote{For a review of Pontryagin square, see Appendix B,C of \cite{Benini2018}.} and for $d B_2=0$ mod 2 we have $\pi\int {\cal P}(B_2)=\pi\int Sq^2(B_2)=\pi\int B_2\cup B_2$. 
The background modifies the gauge bundle into $SO(2c_-)$ bundle with second Stiefel-Whitney class equal to $B_2$. That is, the transitions for triple overlapped coordinate patches do not close, up to a sign fixed by $B_2$. The Wilson line in the spinor representation (corresponding to $\sigma$), that transforms under the one-form symmetry, is attached with a ``Wilson surface" $\int B_2$ to be gauge invariant under one-form gauge transformation\footnote{Note $\psi$ generates the one-form symmetry, but it does not transform under the one-form symmetry.}: this is the generalization of a Wilson line for a 2-form gauge field.
The Wilson lines for $I,\psi$ are not attached with a Wilson surface; they are in representations of $SO(2c_-)$. \footnote{
In gauge theory with gauge group $K$ and center $C$, coupling the theory to background gauge field $B_2$ for the center one-form symmetry $C$ inserts a one-form symmetry defect at the Poincar\'e dual of $B_2$, such that the Wilson line transforms under the one-form symmetry braids with the defect. This implies that the transition function of the $K$ field is no longer closed on the triple overlap of the coordinate patches, and only closed up to an element in $C$ as specified by $B_2$, and thus the bundle around the one-form symmetry defect becomes a $K/C$ bundle.
}
    
We note that there are no anyon-permuting symmetries that act on $Spin(2c_-)_1$ for half integer $c_-$. 
    
\subsubsection{For $c_-$ an even integer}    
For $c_-$ an even integer, the theory has $\mathbb{Z}_2\times\mathbb{Z}_2$ anyons, $I,e,m,em=\psi$ of spins $0,\frac{2c_-}{16}, \frac{2c_-}{16} ,\frac{1}{2}$, where $e,m$ are permuted by a $\mathbb{Z}_2$ 0-form symmetry.
They are Wilson lines in the trivial, spinor, cospinor and vector representation of $Spin(2c_-)$. The $\mathbb{Z}_2$ 0-form symmetry is the electric-magnetic duality symmetry of $Spin(2c_-)$; it is also referred to as a charge conjugation symmetry. \footnote{When $c_-=4$ mod 8, the theory is the three fermion theory, and it has enlarged $S_3$ symmetry that permutes the fermions $e,m,em$. Since the $S_3$ symmetry does not leave the fermion line invariant, it is sufficient to focus on $\mathbb{Z}_2$ $e-m$ duality 0-form symmetry.}
The Chern-Simons theory has $\mathbb{Z}_2\times\mathbb{Z}_2$ one-form symmetry generated by (say) the $\psi,m$ lines, with the anomaly described by the topological spins of the symmetry lines \cite{Kapustin:2014gua,Gaiotto:2014kfa,Hsin:2018vcg}. Explicitly, the anomaly is described by the SPT phase in 3+1d with effective action
\begin{equation}\label{eqn:anomceven}
    \pi\int_{4d} {\cal P}(B_2^{(1)})+\pi\int_{4d} B_2^{(1)}\cup B_2^{(2)}+2\pi\frac{2c_-}{16}\int_{4d}{\cal P}(B_2^{(2)})~,
\end{equation}
where $B_2^{(1)},B_2^{(2)}$ are the background two-form gauge fields for the $\mathbb{Z}_2\times\mathbb{Z}_2$ one-form symmetry. We take $B_2^{(1)}$ to be the background for the one-form symmetry generated by $\psi$; it evaluates to the cochain $n_2$, while $B_2^{(2)}$ is generated by $m$ and evaluates to $\omega_2$. 
${\cal P}(B_2^{(2)})=B_2^{(2)}\cup B_2^{(2)}-B_2^{(2)}\cup_1 dB_2^{(2)}$ is the Pontryagin square of $B_2^{(2)}$ and it is a well-defined $\mathbb{Z}_4$ class. The background turns the gauge bundle into a $Spin(2c_-)/(\mathbb{Z}_2\times \mathbb{Z}_2)$ bundle with second Stiefel-Whitney class (valued in $\mathbb{Z}_2\times\mathbb{Z}_2$) equalling the $\mathbb{Z}_2$-valued two-forms $B_2^{(1)},B_2^{(2)}$.

\paragraph{Charge conjugation 0-form symmetry}
    
    If we also turn on a background $B_1$ for the $\mathbb{Z}_2$ charge conjugation 0-form symmetry, where $d B_1=0$, the bundle has first Stiefel-Whitney class given by $B_1$, and the $\mathbb{Z}_4$ gauge field $B_2$ can be expressed as $\mathbb{Z}_2\times\mathbb{Z}_2$ gauge fields $B_2^{(1)},B_2^{(2)}$ that satisfies \cite{Hsin:2019fhf,Hsin:2020nts}
\begin{equation}\label{eqn:w2bundleevenc}
    d B_2^{(1)}=B_1\cup B_2^{(2)},\quad d B_2^{(2)}=0~.
\end{equation}
The gauge fields $B_1, B_2^{(1)},B_2^{(2)}$ evaluate to $n_1,n_2,\omega_2$ respectively on 2-simplices. The backgrounds $(B_2^{(1)},B_2^{(2)})$ describe a twisted $\mathbb{Z}_2\times\mathbb{Z}_2$ two-cocycle under the action of the $\mathbb{Z}_2$ 0-form symmetry with background $B_1$. Moreover, the gauge transformations of this twisted theory encode the various equivalences among the data noted in the $G_b$-crossed formalism. For instance, under the 0-form symmetry transformation $B_1\rightarrow B_1+d \lambda$ by 0-form $\lambda=0,1$,  $B_2^{(1)}\rightarrow B_2^{(1)}+B_2^{(2)}\cup \lambda$, and for the global transformation $\lambda=1$ this produces the shift $B_2^{(1)}\rightarrow B_2^{(1)}+B_2^{(2)}$. In the $G_b$-crossed language, this gauge transformation implements an $e-m$ relabelling which gives the equivalence $n_2 \simeq n_2 + \omega_2$.

The charge conjugation symmetry does not have anomaly by itself, and it does not have intrinsic mixed anomaly with the one-form symmetry.\footnote{
The intrinsic mixed anomaly between one-form symmetry and permutation symmetry is described by projective representation of the permutation symmetry. Gauging the one-form symmetry of $B_2^{(1)}$ gives an $SO(2c_-)$ gauge theory which also has charge conjugation symmetry, and the possible mixed anomaly with $B_2^{(2)}$ is a coboundary $ B_1^2\cup B_2^{(2)}=d\left(B_2^{(1)}\cup B_1\right)$.
}
The anomaly of the full symmetry (i.e. the $\mathcal{O}_4$ obstruction) is given by the Pontryagin square for the twisted cocycle. 
When the anomaly is expressed in terms of $(B_2^{(1)},B_2^{(2)},B_1)$, the expression (\ref{eqn:anomceven}) has the following correction from ``Cartan coboundaries".\footnote{
The modified cocycle condition $dB_2^{(1)}=B_1\cup B_2^{(2)}$ for even $c_-$ implies that ${\cal P}(B_2^{(1)})$ is no longer well-defined: $d{\cal P}(B_2^{(1)})=dB_2^{(1)}\cup_1 dB_2^{(1)}=(B_1\cup B_2^{(2)})\cup_1 (B_1\cup B_2^{(2)})=Sq^2(B_1\cup B_2^{(2)})$ mod 2. Similarly, $B_2^{(1)}\cup B_2^{(2)}$ is not well-defined, $d\left(B_2^{(1)}\cup B_2^{(2)}\right)=B_1 \cup B_2^{(2)}\cup B_2^{(2)}$. The two problems compensate each other using the Cartan formula (\ref{eqn:Cartan}).
} 

\begin{align}\label{eqn:bulktwistedevenc}
        &\pi\int_{4d} {\cal P}(B_2^{(1)})+\pi\int_{4d} B_2^{(1)}\cup B_2^{(2)}+
        2\pi\frac{2c_-}{16}\int_{4d}{\cal P}(B_2^{(2)})\cr
        &+
        \pi\int \zeta(B_1,B_2^{(2)})
        +c_-\pi\int (B_1\cup B_2^{(2)})\cup_2 Sq^1B_2^{(2)}\cr
        &+\frac{\pi}{2}\int \tilde B_1 \cup \tilde B_1 \cup \tilde B_2^{(2)}~,
\end{align}
where $\tilde B_1,\tilde B_2^{(2)}$ are the canonical lift of $B_1,B_2^{(2)}$ to $\mathbb{Z}_4$ cochains that take value in $0,1$ (we note $\frac{\pi}{2}\int B_1 \cup B_1 \cup \tilde B_2^{(2)}=\frac{\pi}{2}\int (\tilde B_1\cup d \tilde B_2^{(1)}$). In the second line of the above expression, the second term is absent for even $c_-$, and the first term $\zeta(B_1,B_2^{(2)})$ satisfies $\zeta(0,B_2^{(2)})=0=\zeta(B_1,0)$ is the Cartan coboundary in the following Cartan formula for Steenrod squares:
\begin{align}\label{eqn:Cartan}
    & d \zeta(B_1,B_2^{(2)})=Sq^2(B_1\cup B_2^{(2)})\nonumber \\ &-\left(B_1\cup Sq^2(B_2^{(2)})+Sq^1(B_1)\cup Sq^1(B_2^{(2)})\right) \mod 2~.
\end{align}
For an explicit expression, see {\it e.g.} Ref.~\onlinecite{Cartan_coboundary_2020} or Appendix~\ref{Sec:RelAnomComps}.
A similar Cartan coboundary (for a different Cartan formula) is discussed in Appendix F of Ref.~\onlinecite{Kapustin:2017jrc}.
The correction from the last three terms in (\ref{eqn:bulktwistedevenc}) ensures the action is closed.
We note that the equation for $\zeta$ can have two solutions, but they differ only by a coboundary $B_1^2\cup B_2^{(2)}=d(B_1\cup B_2^{(1)})$, where we used $dB_2^{(1)}=B_1\cup B_2^{(2)}$. Since the coboundary does not change the anomaly, the solution for $\zeta$ is unique. 
In Appendix~\ref{Sec:RelAnomComps} we also derive the above anomaly using the relative anomaly formula in Ref.~\onlinecite{Barkeshli2020Anomaly}.

\subsubsection{For $c_-$ an odd integer}

 For $c_-$ an odd integer, the theory $Spin(2c_-)_1$ has $\mathbb{Z}_4$ anyons $I,v,v^2,v^3=v^{-1}$ with spin $0,\frac{2c_-}{16},\frac{1}{2},\frac{2c_-}{16}$ mod 1. They are Wilson lines in the trivial, spinor, vector, and cospinor representation of $Spin(2c_-)$. The anyons $v,v^{-1}$ are permuted by a $\mathbb{Z}_2$ 0-form symmetry (the charge conjugation symmetry of $Spin(2c_-)$).
Thus the theory has a $\mathbb{Z}_4$ one-form symmetry generated by (say) the $v$ line, with the anomaly described by the SPT phase
    \begin{equation}\label{eqn:anomcodd}
        2\pi\frac{2c_-}{16}\int_{4d} {\cal P}(Y_2)~,
    \end{equation}
    where $Y_2$ is the background two-form gauge field for the $\mathbb{Z}_4$ one-form symmetry. (Evaluating $Y_2$ on 2-simplices gives the cochain $\omega_2 + 2n_2 \mod 4$.)
    The background turns the gauge bundle into a $Spin(2c_-)/\mathbb{Z}_4$ bundle with second Stiefel-Whitney class equal to $Y_2$. ${\cal P}(Y_2)=Y_2\cup Y_2-Y_2\cup_1 d Y_2$ is the (generalized) Pontryagin square of $Y_2$, and for $\mathbb{Z}_4$ valued $Y_2$ it is a well-defined $\mathbb{Z}_8$ class.
    
    We can also rewrite the anomaly by expressing the $\mathbb{Z}_4$ valued background two-from as a pair of $\mathbb{Z}_2$ valued two-forms $B_2^{(1)},B_2^{(2)}$ (which evaluate to $(n_2,\omega_2)$) satisfying
    \begin{equation}
       d B_2^{(1)}=B_2^{(2)}\cup_1 B_2^{(2)},\quad d B_2^{(2)}=0.
    \end{equation}
    The right hand side of the first equation equals $d \tilde B_2^{(2)}/2$ mod 2, where here and in the following discussion in this section we use tilde to denote the lift from $\mathbb{Z}_2$ to $\{0,1\}$ in $\mathbb{Z}_4$. It is the image of the Bockstein homomorphism for the short exact sequence $1\rightarrow\mathbb{Z}_2\rightarrow \mathbb{Z}_4\rightarrow \mathbb{Z}_2\rightarrow 1$.
    The pair $(B_2^{(1)},B_2^{(2)})$ gives a $\mathbb{Z}_4$ cocycle $Y_2=2\tilde B_2^{(1)}+B_2^{(2)}$ that satisfies $d Y_2=0$ mod 4.

    In this representation, the anomaly in Eq.~\eqref{eqn:anomcodd} equals
    \begin{equation}\label{eqn:anomcoddp}
        \pi\int_{4d} {\cal P}(B_2^{(1)})+\pi\int_{4d} B_2^{(1)}\cup B_2^{(2)}+2\pi\frac{2c_-}{16}\int_{4d}{\cal P}(B_2^{(2)})~.
    \end{equation}
    
\paragraph{Charge conjugation 0-form symmetry}    
    
     If we also turn on background $B_1$ for the $\mathbb{Z}_2$ charge conjugation 0-form symmetry,  where $d B_1=0$, the bundle has first Stiefel-Whitney class given by $B_1$. The $\mathbb{Z}_4$ gauge field $Y_2$ is now a twisted cocycle due to the action of charge conjugation symmetry. It can be expressed as $\mathbb{Z}_2\times\mathbb{Z}_2$ gauge fields $B_2^{(1)},B_2^{(2)}$ that satisfy \cite{Hsin:2019fhf,Hsin:2020nts}
\begin{equation}\label{eqn:w2bundleoddc}
    d B_2^{(1)}= B_1\cup B_2^{(2)}+B_2^{(2)}\cup_1 B_2^{(2)},\quad d B_2^{(2)}=0~.
\end{equation}
The backgrounds $(B_2^{(1)},B_2^{(2)})$ describe a twisted $\mathbb{Z}_4$ two-cocycle under the action of the $\mathbb{Z}_2$ 0-form symmetry with background $B_1$.
The anomaly can be expressed as a Pontryagin square for this twisted cocycle \eqref{eqn:anomcoddp}. When the anomaly is expressed in terms of $(B_2^{(1)},B_2^{(2)},B_1)$, the expression (\ref{eqn:anomcoddp}) has additional corrections as in (\ref{eqn:bulktwistedevenc}).\footnote{
We note that when $c_-$ is odd, the solution of $\zeta$ is not unique, since $\zeta$ can change by a nontrivial cocycle $B_1 \cup B_1 \cup B_2^{(2)}$ and satisfy the same equation, where $dB_2^{(1)}=B_1\cup B_2^{(2)}+Sq^1B_2^{(2)}$ for odd $c_-$, and thus $B_1 \cup B_1 \cup B_2^{(2)}$ is no longer exact.
 We will fix this ambiguity using the relative anomaly formula \cite{Barkeshli2020Anomaly} in Appendix~\ref{sec:coddanomaly}, which uses the $G$-crossed BTC approach.
}
\begin{align}\label{eqn:bulktwistedevencp}
            &\pi\int_{4d} {\cal P}(B_2^{(1)})+\pi\int_{4d} B_2^{(1)}\cup B_2^{(2)}+
            2\pi\frac{2c_-}{16}\int_{4d}{\cal P}(B_2^{(2)})\cr
            &+
            \pi\int \zeta(B_1,B_2^{(2)})
            +c_-\pi\int (B_1B_2^{(2)})\cup_2 Sq^1B_2^{(2)}\cr
    &        +\frac{\pi}{2}\int \tilde B_1 \cup \tilde B_1 \cup \tilde B_2^{(2)}
            ~.
\end{align}
One can verify that the bulk action is closed.\footnote{
Explicitly, acting with coboundary on the bulk action gives
\begin{align}
    &\pi d B_2^{(1)}\cup_1 d B_2^{(1)}
    +\pi d B_2^{(1)}\cup B_2^{(2)}+\pi d \zeta(B_1,B_2^{(2)})\cr
    &+\pi c_- \left((B_1B_2^{(2)})\cup_1 Sq^1B_2^{(2)}+
    Sq^1B_2^{(2)}\cup_1(B_1B_2^{(2)})
    \right)\cr
    &+\pi B_1 \cup B_1 \cup {d B_2^{(2)}\over 2}
    +\pi c_-({d B_2^{(2)}\over 2}\cup B_2^{(2)}-{d B_2^{(2)}\over 2}\cup_1 {d B_2^{(2)}\over 2})~. \nonumber
\end{align}
Using the relation $d B_2^{(1)}=B_1B_2^{(2)}+Sq^1B_2^{(2)}$ and ${d B_2^{(2)}\over 2}=Sq^1B_2^{(2)}$, we can show the above expression vanishes mod $2\pi$.
}

By combining the results from the preceding subsections, it is straightforward to obtain the results in Table~\ref{Table:MainEqs}. The rest of this section will show the explicit calculations involved.  

\subsection{Fermionic invertible phases with half integer $c_-$}
\label{sec:CShalfc}

\subsubsection{Bosonic shadow}

The theory $Spin(2c_-)_1$ can be enriched by $G_b$ symmetry, giving Chern-Simons theory
\begin{equation}
    {Spin(2c_-)_1\times (\tilde G)_\nu\over \mathbb{Z}_2}~.
\end{equation}
Denote the $G_b$ background gauge field by $A_b$. It couples to the $Spin(2c_-)_1$ bundle by modifying it into an $SO(2c_-)$ bundle, characterized by the second Stiefel-Whitney class
\begin{equation}
    w_2^{SO}=A_b^*n_2,\quad n_2\in \H^2(G_b,\mathbb{Z}_2).
\end{equation}
$n_2$ also determines the extension
\begin{equation}
    1\rightarrow \mathbb{Z}_2\rightarrow \tilde G\rightarrow G_b=\tilde G/\mathbb{Z}_2\rightarrow 1~.
\end{equation}
The Wilson line in the spinor representation of $Spin(2c_-)$ carries projective representation of $G_b$ symmetry as specified by $(-1)^{n_2}$ \cite{Benini2018}.

For instance, we can construct the bosonic shadow as a $Spin(2c_-)_1$ Chern-Simons matter theory with massive scalars $\sigma$, which do not condense and transform in the spinor representation of $Spin(2c_-)$, and in the projective representation of $G_b$ with projective factor $(-1)^{n_2}$. The Chern-Simons term at level 1 for $SO(2c_-)$ bundles is not well-defined and is anomalous (see \eqref{eqn:anomchalfinteger}). This anomaly can be characterized by the bulk term 
\begin{equation}
    \pi\int_{4d} {\cal P}(w_2^{SO})=\pi\int_{4d} A_b^* {\cal P}(n_2)~.
\end{equation}
Thus for the theory to be well-defined, the anomaly can be compensated by demanding the $G_b$ gauge field Chern-Simons term $\nu_3\in C^3(BG_b,U(1))$ to satisfy
\begin{equation}
    d \nu_3=(-1)^{{\cal P}(n_2)}=(-1)^{Sq^2(n_2)}~.
\end{equation}

In addition, we can turn on a background $B_2$ for the $\mathbb{Z}_2$ one-form symmetry generated by $\psi$. This is done by shifting $w_2^{SO}=B_2+A_b^*n_2$ and adding the following SPT phase effective action to the bosonic shadow
\begin{equation}\label{eqn:1-form0-form}
    \pi\int_{3d} B_2\cup A_b^*n_1,\quad n_1\in \H^1(BG_b,\mathbb{Z}_2)~.
\end{equation}

\subsubsection{Fermionic invertible phase}
Suppose we gauge the one-form symmetry by replacing $w_2^{SO}$ with $w_2^{SO}+B_2$ for dynamical $\mathbb{Z}_2$ two-form gauge field $B_2$ that satisfies $d B_2=0$. Then the theory has an anomaly
\begin{equation}
    \pi\int_{4d} B_2\cup B_2=\pi\int_{4d} B_2 \cup w_2(TM) ~,
\end{equation}
where $w_2(TM)$ is the second Stiefel-Whitney class of the tangent bundle. Here we used the Wu formula \cite{milnor1974characteristic}. The anomaly
 is cancelled by choosing a spin structure $\rho_1$, which couples as $\pi\int_{3d}B_2\cup \rho_1$, and obeys
 \begin{equation}
  d\rho_1=w_2(TM)~.   
 \end{equation}
 In this case, $\rho$ is independent of $G_b$. Thus gauging the one-form symmetry gives a fermionic invertible phase with direct product $G_b\times \Z_2^f$ symmetry: the `bosonic" symmetry is not extended by $\mathbb{Z}_2^f$. 

We remark that the mixed one-form/0-form symmetry SPT phase in the bosonic shadow (Eq.~\eqref{eqn:1-form0-form}) has the following consequence for the fermionic invertible phase. After gauging the one-form symmetry we find the spin structure is given by $\rho_1+A_b^*n_1$ instead of just $\rho_1$. For instance, when $G_b=\mathbb{Z}_2$, we have a $\mathbb{Z}_8$ classification for fermionic SPT phases. The effective action for the root phase of this classification can be obtained from $SO(2c_-)_1$ with half integer $c_-$ by shifting the spin structure as above, and stacking another $SO(2c_-)_{-1}$ (see {\it e.g.} Ref.~\onlinecite{Cordova:2017vab}).

For comparison, in the $G_b$-crossed theory, $n_1$ determines whether the $G_b$ fluxes in the invertible phase are abelian or nonabelian. Thus, a shift in the spin structure determines whether a given $G_b$ flux is associated to a Majorana zero mode or not.

\subsubsection{Absence of symmetry extension by fermion parity}

In Section~\ref{sec:CS_w2defn} we had stated that when $c_-$ is a half integer, we must have $[\omega_2]$ trivial. The proof is straightforward: $\omega_2$ enters the theory through the one-form symmetry with background $B_2^{(2)}$. Since $B_2^{(2)}$ is absent in the bosonic shadow $Spin(2c_-)_1$ for half integer $c_-$, there is no nontrivial $\omega_2$ for the corresponding fermionic invertible phases. In other words, the symmetry in the fermionic invertible phase must be a direct product $\mathbb{Z}_2^f\times G_b$.

\subsection{Fermionic invertible phases with even integer $c_-$ and $n_1 = 0$}

\subsubsection{Bosonic shadow}

For even $c_-$ there are two cases: $G_b$ symmetry either permutes or does not permute the two anyons $e,m$ in the spinor representation. The permutation is encoded by $n_1\in \H^1(BG_b,\mathbb{Z}_2)$. Let us begin with the case that the symmetry does not permute the anyons.

The $G_b$ enriched theory is described by the Chern-Simons theory
\begin{equation}
    {Spin(2c_-)_1\times (\tilde G)_\nu\over \mathbb{Z}_2\times\mathbb{Z}_2}~.
\end{equation}
The $G_b$ gauge field $A_b$ couples to the theory by modifying the $Spin(2c_-)$ bundle into a $Spin(2c_-)/(\mathbb{Z}_2\times\mathbb{Z}_2)$ bundle, with Stiefel-Whitney classes $w_2^{(1)},w_2^{(2)}$ specified by $A_b$ as follows:
\begin{equation}
    w_2^{(1)}=A_b^*n_2,\, w_2^{(2)}=A_b^*\omega_2,\, 
\end{equation}
with $n_2,\omega_2\in \H^2(BG_b,\mathbb{Z}_2)$. 
The Wilson lines that transforms under the $\mathbb{Z}_2\times\mathbb{Z}_2$ one-form symmetry carry projective representation of $G_b$ symmetry specified by $n_2,\omega_2$ \cite{Benini2018}. For instance, the
Wilson lines in the spinor representation of $Spin(2c_-)$ (corresponding to the $e$ and $m$ anyons) carry a projective representation of $G_b$ symmetry as specified by $(-1)^{n_2}$.

We can construct the bosonic shadow as $Spin(2c_-)_1$ Chern-Simons matter theory with two types of massive scalars, say $m$ and $em = \psi$ (that do not condense). The scalar $m$ transforms in the spinor (cospinor) representation of $Spin(2c_-)$ for $c_-=0$ mod 4 ($c_-=2$ mod 4), and the projective representation of $G_b$ with projective factor $(-1)^{n_2}$. The scalar $\psi$ transforms in the vector representation of $Spin(2c_-)$, and the projective representation of $G_b$ with projective factor $\omega_2$.

The Chern-Simons term at level one for $Spin(2c_-)/(\mathbb{Z}_2\times\mathbb{Z}_2)$ gauge field is anomalous, due to Eq.~\eqref{eqn:anomceven}. The anomaly is characterized by the bulk term
\begin{widetext}
\begin{align}
    \pi\int_{4d} {\cal P}(w_2^{(1)})+\pi\int_{4d} w_2^{(1)}\cup w_2^{(2)}+\frac{2\pi (2c)}{16}\int_{4d} {\cal P}(w_2^{(2)})
    &=
    \pi\int_{4d} A_b^*{\cal P}(n_2)+\pi\int_{4d} A_b^*n_2\cup \omega_2+
    \frac{2\pi (2c_-)}{16}\int_{4d} A_b^* {\cal P}(\omega_2)~,
\end{align}
\end{widetext}

where ${\cal P}(x)=x\cup x-x\cup_1 d x$ is the Pontryagin square of the two-cocycle $x$.
The anomaly can be compensated by demanding $\nu_3\in C^3(BG_b,U(1))$ to satisfy
\begin{equation}
  d \nu_3=(-1)^{{\cal P}(n_2)+n_2\cup \omega_2} i^{(c/2){\cal P}(\omega_2)}~.
\end{equation}
The above equation agrees with that obtained from the $G_b$-crossed theory, as computed in Appendix~\ref{Sec:RelAnomComps}.

We could consider an additional mixed 1-form/0-form  SPT phase with effective action
\begin{equation}
    \pi\int_{3d}B_2^{(1)}\cup A_b^*n_1',\quad n_1'\in \H^1(BG,\mathbb{Z}_2)~.
\end{equation}
However, such an SPT phase is redundant, and it can be removed by a global one-form transformation of $B_2^{(2)}$ due to the mixed anomaly involving both $B_2^{(1)}$ and $B_2^{(2)}$. Thus $n_1'$ does not contribute to additional data in the classification and we will ignore it in the following.

\subsubsection{Fermionic invertible phase}

Let us gauge the $\mathbb{Z}_2$ one-form symmetry generated by $\psi$, by replacing $w_2^{(1)}$ with $w_2^{(1)}+B_2$ for the dynamical $\mathbb{Z}_2$ two-form gauge field $B_2$. The theory now has anomaly
\begin{align}
    &\pi\int_{4d} B_2\cup  B_2+\pi\int_{4d} B_2 \cup w_2^{(2)}\cr &=\pi\int_{4d}  B_2\cup  w_2(TM)+\pi\int_{4d} B_2\cup  A_b^*\omega_2~.
\end{align}
The anomaly described by the above bulk term can be cancelled by choosing a spin structure $\rho_1$ that couples to $B_2$ as $\pi\int_{3d} B_2\cup \rho_1$, with
\begin{equation}\label{eqn:extendevencnotperm}
    d\rho_1= w_2(TM)+A_b^*\omega_2~.
\end{equation}
The resulting fermionic invertible phase has $G_f$ symmetry, where $G_f$ is the extension of $G_b$ by $(-1)^F$ specified by $\omega_2\in \H^2(BG_b,\mathbb{Z}_2)$,
\begin{equation}
    1\rightarrow \mathbb{Z}_2=\mathbb{Z}_2^f\rightarrow G_f \rightarrow G_b\rightarrow 1~.
\end{equation}
To see this, we note that Eq.~\eqref{eqn:extendevencnotperm} describes the gauge field for the extension of $SO_\text{Lorentz}\times G_b$ by $\mathbb{Z}_2$ (whose background is $\rho_1$), given by
\begin{equation}
    1\rightarrow\mathbb{Z}_2\rightarrow \frac{Spin_\text{Lorentz}\times G_f}{\mathbb{Z}_2}\rightarrow SO_\text{Lorentz}\times G_b\rightarrow 1~.
\end{equation}
Here the $\mathbb{Z}_2$ quotient identifies $(-1)^F\subset Spin_\text{Lorentz}$ with $\mathbb{Z}_2\subset G_f$, and thus the symmetry in the fermionic invertible phase can be interpreted as $G_b$ 0-form symmetry extended by $(-1)^F$.

\subsection{Fermionic invertible phases with odd integer $c_-$ and $n_1 = 0$}

\subsubsection{Bosonic shadow}

Here the $G_b$ enriched theory is described by the Chern-Simons theory
\begin{equation}
    {Spin(2c_-)_1\times (\tilde G)_\nu\over \mathbb{Z}_4}~.
\end{equation}
The $G_b$ gauge field $A_b$ couples to the theory by modifying the $Spin(2c_-)$ bundle into a $Spin(2c_-)/\mathbb{Z}_4$ bundle, with Stiefel-Whitney classes $w_2^{PSO}$ specified by $A_b$ as
\begin{equation}
    w_2^{PSO}=A_b^*\eta_2,\quad \eta_2\in \H^2(BG_b,\mathbb{Z}_4)~.
\end{equation}
We can describe $\eta_2\in \H^2(BG_b,\mathbb{Z}_4)$ by a pair $n_2,\omega_2$ with $\omega_2\in \H^2(BG_b,\mathbb{Z}_2)$ and
\begin{equation}
   d n_2=\omega_2\cup_1\omega_2~.
\end{equation}
Explicitly, $ \eta_2 = 2\tilde n_2-\tilde\omega_2$ mod 4,
where tilde denotes a lift from $\mathbb{Z}_2$ to $\mathbb{Z}_4$. 
This corresponds to
\begin{equation}
    B_2^{(1)}=A_b^*n_2,\quad B_2^{(2)}=A_b^*\omega_2~.
\end{equation}

We can construct the bosonic shadow as $Spin(2c_-)_1$ Chern-Simons matter theory with massive scalars $v,\bar{v}$ (that do not condense) transforming in the spinor  representation of $Spin(2c_-)$ and in the projective representation of $G_b$ with projective factor $(-1)^{\eta_2}$.

The Chern-Simons term at level one for $Spin(2c_-)/\mathbb{Z}_4$ gauge field is anomalous, see (\ref{eqn:anomcodd}), as characterized by the bulk term
\begin{equation}
    {2\pi (2c_-)\over 16}\int_{4d} {\cal P}(w_2^{PSO})=
\frac{2\pi (2c_-)}{16}\int_{4d} A_b^* {\cal P}(\eta_2)~.
\end{equation}

The anomaly described by the above bulk term can be compensated by modifying the $G_b$ Chern-Simons term $\nu_3\in C^3(BG_b,U(1))$ to satisfy
\begin{equation}
   d \nu_3=e^{(\pi i c/4){\cal P}(\eta_2)}~.
\end{equation}
The above equation agrees with that obtained from the $G_b$-crossed theory, as computed in Appendix~\ref{Sec:RelAnomComps}.

We could consider stacking an additional mixed 1-form/0-form SPT phase with effective action
\begin{equation}
    \pi\int_{3d}Y_2\cup A_b^*n_1',\quad n_1'\in \H^1(BG,\mathbb{Z}_2)~.
\end{equation}
However, such SPT phase is redundant, and it can be removed by a global one-form transformation of $B_2^{(2)}$ due to the mixed anomaly involving both $B_2^{(1)}$ and $B_2^{(2)}$. Thus $n_1'$ does not contribute to additional data in the classification and we will ignore it in the following.

\subsubsection{Fermionic invertible phase}

Let us gauge the $\mathbb{Z}_2$ one-form symmetry corresponding to $\psi$, by replacing $w_2^{PSO}$ with $w_2^{PSO}+2B_2^{(1)}$. Here $2B_2^{(1)}$ is the image of $B_2^{(1)}$ under the inclusion map $\mathbb{Z}_2\rightarrow \mathbb{Z}_4$, for the dynamical $\mathbb{Z}_2$ two-form gauge field $B_2^{(1)}$. The theory has anomaly described by the bulk term
\begin{align}
&\frac{2\pi (2c_-)}{16}\int_{4d} {\cal P}(2B_2^{(1)}+A_b^*\eta_2)-\frac{2\pi (2c)}{16}\int_{4d} {\cal P}(A_b^*\eta_2)\cr
&=\pi \int_{4d} {\cal P}(B_2^{(1)})+\pi \int_{4d} B_2^{(1)} \cup A_b^*\eta_2\cr&=\pi\int_{4d} B_2^{(1)}\cup w_2(TM) +\pi\int_{4d} B_2^{(1)}\cup A_b^*\eta_2~.
\end{align}
The anomaly described by the above bulk term can be cancelled by choosing spin structure $\rho_1$ that couples to $B_2^{(1)}$ as $\pi\int_{3d} B_2^{(1)}\cup \rho_1$, with
\begin{equation}
    d\rho_1= w_2(TM)+A_b^*\eta_2\text{ mod }2~.
\end{equation}

As in the case with $c_-$ even, we conclude that gauging the one-form symmetry produces a fermionic invertible phase with $G_f$ symmetry, where $G_f$ is the extension of $G_b$ by $(-1)^F$ specified by $\omega_2$.

\subsection{Fermionic invertible phases with even $c_-$ and general $n_1$}

\subsubsection{Bosonic shadow}

If the symmetry permutes the anyons, then the $G_b$ enriched theory can be described by the Chern-Simons theory
\begin{equation}
    {Pin^\pm(2c)_1\times (\tilde G)_\omega\over \mathbb{Z}_2\times \mathbb{Z}_2\times\mathbb{Z}_2}~.
\end{equation}
Let us explain the quotient. One $\mathbb{Z}_2$ quotient denotes gauging a $\mathbb{Z}_2$ quantum one-form symmetry that identifies the first Stiefel-Whitney class $w_1$ of the $Pin^\pm$ bundle with $A_b^*n_1$ for $G_b$ gauge field $A_b$. The other $\mathbb{Z}_2\times\mathbb{Z}_2$ quotient modifies the $Pin^\pm$ gauge field into $Pin^\pm/(\mathbb{Z}_2\times\mathbb{Z}_2)$, with Stiefel-Whitney class $w_2^{(1)},w_2^{(2)}$, which satisfies \eqref{eqn:w2bundleevenc} \cite{Hsin:2019fhf,Hsin:2020nts}:
\begin{align}
    &d w_2^{(1)}=w_1\cup w_2^{(2)},\cr & 
 w_2^{(1)}=A_b^*n_2,\, w_2^{(2)}=A_b^*\omega_2,\quad 
 w_1=A_b^* n_1\cr
    &n_2\in C^2(BG,\mathbb{Z}_2),\quad \omega_2\in \H^2(BG,\mathbb{Z}_2),\quad 
    n_1\in \H^1(BG,\mathbb{Z}_2)\cr
&d n_2=n_1\cup \omega_2~
\end{align}
for $Pin^+/(\mathbb{Z}_2\times\mathbb{Z}_2)$, and similar for $Pin^-/(\mathbb{Z}_2\times\mathbb{Z}_2)$ with $w_2^{(1)}$ replaced by $w_2^{(1)}+w_1^2$. Without loss of generality we can focus on $Pin^+$. 

We can construct the bosonic shadow as $Spin(2c_-)_1$ Chern-Simons matter theory with massive scalars $e,m$ (that do not condense). They transform in the spinor (cospinor) representation of $Spin(2c_-)$ for $c=0$ mod 4 ($c=2$ mod 4). The theory also has a fermion $\psi$ in the vector representation of $Spin(2c_-)$. $(e,m)$ together also transform in the representation of $\tilde G$ which is the twisted projective representation of $G_b$ specified by $(n_2,\omega_2)$.

The Chern-Simons term for the $Pin^+/(\mathbb{Z}_2\times\mathbb{Z}_2)$ bundle at level one is not well-defined and has an anomaly, characterized by the bulk term (\ref{eqn:bulktwistedevenc}):
\begin{widetext}
\begin{align}
    &\pi\int_{4d}{\cal P}( w_2^{(1)})+\pi\int_{4d} w_2^{(1)}\cup w_2^{(2)}+\int \zeta(\tilde w_1^{Pin},\tilde w_2^{(2)})+c_-\pi\int (w_1^{Pin} w_2^{(2)})\cup_2 Sq^1w_2^{(2)}\cr
&\quad     +{\pi\over 2}\int (w_1^{Pin})^2\cup w_2^{(2)}
+2\pi \frac{2c_-}{16}\int_{4d} {\cal P}(w_2^{(2)})  \cr
&    =
    \pi\int_{4d} A_b^*({\cal P}(n_2))+\pi\int_{4d} A_b^* n_2\cup \omega_2
    +\pi\int A_b^*\zeta(n_1,\omega_2)+
    +c_-\pi\int A_b^*(n_1\omega_2)\cup_2 Sq^1\omega_2\cr
    &\quad +{\pi\over 2}\int A_b^*(\tilde n_1)^2\cup \tilde \omega_2
    +2\pi \frac{2c_-}{16}\int_{4d} A_b^*{\cal P}(\omega_2)    ~.
\end{align}  
\end{widetext}
Note that $(n_1)^2\cup \omega_2=n_1 d n_2$ mod 2.

The anomaly can be compensated by demanding $\nu_3\in C^3(BG_b,U(1))$ to satisfy 
\begin{align}
    d\nu_3
    =& ~ (-1)^{{\cal P}(n_2)+n_2\cup \omega_2}\cr
    &\cdot (-1)^{\zeta(n_1,\omega_2)+c_-(n_1\omega_2)\cup_2Sq^1(\omega_2)}\cr
    &\cdot i^{\tilde n_1^2\cup\tilde \omega_2}i^{{c_-\over 2}{\cal P}(\omega_2)}~.
\end{align}
The above equation agrees with that obtained from the $G_b$-crossed theory, as computed in Appendix~\ref{Sec:RelAnomComps}.

\subsubsection{Fermionic invertible phase}

Let us gauge the $\mathbb{Z}_2$ one-form symmetry of $\psi$ by replacing $w_2^{(1)}$ with $w_2^{(1)}+B_2$ for some dynamical $\mathbb{Z}_2$ two-form gauge field $B_2$ that satisfies $dB_2=0$. The theory now has an anomaly described by the bulk term
\begin{align}
    &\pi\int_{4d} B_2 \cup B_2+\pi\int_{4d} B_2\cup  A_b^*\omega_2\cr
    &=\pi\int_{4d} B_2\cup w_2(TM) +\pi\int_{4d} B_2 \cup A_b^*\omega_2~,
\end{align}
which can be compensated by choosing a spin structure $\rho_1$ that couples to $B_2$ as $\pi\int_{3d} B_2\cup \rho_1$, with
\begin{equation}
    d\rho_1=w_2(TM)+A_b^*\omega_2~.
\end{equation}
Thus the resulting theory after gauging the one-form symmetry becomes a fermionic invertible phase with $G_f$ symmetry given by the extension of $G_b$ by $(-1)^F$ as specified by $\omega_2\in \H^2(BG_b,\mathbb{Z}_2)$.

\subsection{Fermionic invertible phases with odd $c_-$ and general $n_1$}

\subsubsection{Bosonic shadow}

If the symmetry permutes the anyons, then the $G_b$ symmetry enriched theory can be described by
\begin{equation}
    {Pin^\pm(2c)_1\times (\tilde G)_\nu\over \mathbb{Z}_2\times \mathbb{Z}_4}~.
\end{equation}
Let us explain the quotient. 
One $\mathbb{Z}_2$ quotient denotes gauging a $\mathbb{Z}_2$ quantum one-form symmetry that identifies the first Stiefel-Whitney class $w_1$ of the $Pin^\pm$ bundle with $A_b^*n_1$ for $G$ gauge field $A_b$. The other $\mathbb{Z}_4$ quotient modifies the $Pin^\pm$ bundle into $Pin^\pm/\mathbb{Z}_4$ bundle, with the quotient described by two $\mathbb{Z}_2$ Stiefel-Whitney class $w_2^{(1)},w_2^{(2)}$  that satisfy \cite{Hsin:2019fhf,Hsin:2020nts}
\begin{align}
    &d w_2^{(1)}=w_1\cup w_2^{(2)}+w_2^{(2)}\cup_1 w_2^{(1)},\nonumber \\
    & w_2^{(1)}=A_b^* n_2,\quad w_2^{(2)}=A_b^*\omega_2,
    \quad w_1=A_b^* n_1 ; \cr
    &n_2\in C^2(BG,\mathbb{Z}_2),
    ~\omega_2\in \H^2(BG,\mathbb{Z}_2),
    ~n_1\in \H^1(BG,\mathbb{Z}_2)\cr
    &  d n_2= n_1\cup \omega_2+\omega_2\cup_1 \omega_2~
\end{align}
for $Pin^+/\mathbb{Z}_4$, and similarly for $Pin^-/\mathbb{Z}_4$ with $w_2^{(1)}$ replaced by $w_2^{(1)}+w_1^2$.
Without loss of generality we can focus on $Pin^+$. In the above, we used $w_2^{(2)}\cup_1 w_2^{(2)}=\text{Bock}(w_2^{(2)})$, the Bockstein homomorphism for $1\rightarrow\mathbb{Z}_2\rightarrow\mathbb{Z}_4\rightarrow\mathbb{Z}_2\rightarrow 1$.

The resulting bosonic shadow is a $Spin(2c_-)_1$ Chern-Simons matter theory with massive scalars $(I,v,\psi=v^2,\bar{v})$ (that do not condense) transforming in the spinor (cospinor) representation of $Spin(2c_-)$ for $v,\bar{v}$, and vector representation of $Spin(2c_-)$ for $\psi$. $v,\bar{v}$ also transform under the twisted projective representation of $G_b$ specified by $(n_2,\omega_2)$.

The Chern-Simons term for the $Pin^+/\mathbb{Z}_4$ bundle at level one is not well-defined and has an anomaly, characterized by the bulk term (\ref{eqn:anomcoddp}):
\begin{widetext}
\begin{align}
    &\pi\int_{4d}{\cal P}( w_2^{(1)})+\pi\int_{4d} w_2^{(1)}\cup w_2^{(2)}
+
+\pi\int \zeta(w_1^{Pin},w_2^{(2)})+c_-\pi\int (w_1^{Pin}w_2^{(2)})\cup_2 Sq^1w_2^{(2)}\cr
&\quad+{\pi\over 2}\int (\tilde w_1^{Pin})^2\cup \tilde w_2^{(2)}
+2\pi \frac{2c_-}{16}\int_{4d} {\cal P}(w_2^{(2)})    \cr
&    =
    \pi\int_{4d} A_b^*({\cal P}(n_2))+\pi\int_{4d} A_b^* n_2\cup \omega_2
    +
    \pi\int A_b^*\zeta(n_1,\omega_2)
    +c_-\pi\int A_b^* (n_1\omega_2)\cup_2Sq^1\omega_2\cr
    &\quad +{\pi\over 2}\int A_b*(\tilde n_1)^2\cup \tilde \omega_2
    +2\pi \frac{2c_-}{16}\int_{4d} A_b^*{\cal P}(\omega_2)    ~.
\end{align}  
\end{widetext}

It can be compensated by demanding $\nu_3\in C^3(BG_b,U(1))$ to satisfy
\begin{align}
    d\nu_3=&(-1)^{{\cal P}(n_2)+n_2\cup \omega_2+\zeta(n_1,\omega_2)+c_-(n_1\omega_2)\cup_2Sq^1(\omega_2)}\cr
    &\cdot 
    i^{\tilde n_1^2\cup \tilde \omega_2}e^{(\pi ic/4){\cal P}(\omega_2)}~.
\end{align}
The above equation agrees with that obtained from the $G_b$-crossed theory, as computed in Appendix~\ref{Sec:RelAnomComps}.

\subsubsection{Fermionic invertible phase}

We can gauge the $\mathbb{Z}_2$ one-form symmetry of $\psi$ by replacing $w_2^{(1)}$ with $w_2^{(1)}+B_2$ for some dynamical $\mathbb{Z}_2$ two-form gauge field $B_2$ that satisfies $d B_2=0$. The theory now has an anomaly
\begin{align}\label{eqn:fermionSPT}
    &\pi\int_{4d} B_2\cup  B_2+\pi\int_{4d} B_2\cup  A_b^*\omega_2\cr
    &=\pi\int_{4d} B_2\cup  w_2(TM)+\pi\int_{4d} B_2\cup  A_b^*\omega_2~,
\end{align}
which can be compensated by choosing a spin structure $\rho_1$ that couples to $B_2$ as $\pi\int_{3d} B_2\cup \rho_1$, with
\begin{equation}\label{eqn:fermionSPTrho}
    d\rho_1=w_2(TM)+A_b^*\omega_2~.
\end{equation}
Thus resulting theory after gauging the one-form symmetry becomes a fermionic invertible phase with $G_f$ symmetry given by the extension of $G_b$ by $(-1)^F$ as specified by $\omega_2\in \H^2(BG_b,\mathbb{Z}_2)$,
\begin{equation}
    1\rightarrow \mathbb{Z}_2^f\rightarrow G_f\rightarrow G_b\rightarrow 1~.
\end{equation}

\subsection{Equivalence relation from gauge transformation}
\label{sec:gaugetransform}

The equivalence relation discussed in Section~\ref{Sec:Equivs} can be derived using the gauge transformation of the background fields for the intrinsic symmetries.
Let us illustrate this using the theories with integer $c_-$.

\paragraph{Gauge transformation of $B_2^{(1)}$}

Consider the background gauge transformation $B_2^{(1)}\rightarrow B_2^{(1)}+d\lambda$, for $B_2^{(1)}=A_b^*n_2$ this is
\begin{equation}
    n_2\rightarrow n_2+d\chi_1
\end{equation}
for $\mathbb{Z}_2$ one-cochain $\chi_1$, where $\lambda_1=A_b^*\chi_1$.
Due to the anomaly between the one-form symmetry for background $B_2^{(1)}$ and $G_b$ symmetry,
 one finds $\nu_3$ changes as
\begin{equation}
    \nu_3\rightarrow \nu_3 (-1)^{\chi_1\cup \omega_2} (-1)^{d\chi_1\cup_1 n_2+d\chi_1\cup \chi_1}~.
\end{equation}
If we take $\chi_1$ which is closed, $d\chi_1=0$, $n_2$ does not transform, and thus we find the equivalence relation
\begin{equation}
    \nu_3\sim \nu_3(-1)^{\chi_1\cup \omega_2},\quad d\chi_1=0\text{ mod }2~.
\end{equation}

\paragraph{Gauge transformation of $B_2^{(2)}$ }

Consider the gauge transformation $B_2^{(2)}\rightarrow B_2^{(2)}+d\lambda_1'$, for $B_2^{(2)}=A_b^*\omega_2$ this is
\begin{equation}
    \omega_2\rightarrow\omega_2+d\xi_1~
\end{equation}
for $\mathbb{Z}_2$ one-cochain $\xi_1$, where $\lambda_1'=A_b^*\xi_1$.
Then due to the equation
\begin{equation}
    dB_2^{(1)}=B_1\cup B_2^{(2)}+c_- Sq^1B_2^{(2)}~,
\end{equation}
one finds $B_2^{(1)}$ also transforms, and $B_2^{(1)}=A_b^*n_2$ changes as
\begin{equation}
    n_2\rightarrow n_2+n_1\cup \xi_1+c_-\left(\xi_1\cup\xi_1+\xi_1\cup_1d\xi_1+\omega_2\cup_2 d\xi_1\right)~.
\end{equation}
For $d\xi=0$ we find 
\begin{equation}\label{eqn:n2xi}
    n_2\sim n_2+ n_1\cup \xi_1+c_- \xi_1\cup \xi_1~.
\end{equation}

On the other hand, this does not lead to an equivalence relation in the fermionic theory after gauging the one-form symmetry by promoting $B_2^{(1)}$ to be dynamical. The theory has dual $\mathbb{Z}_2^f$ symmetry with background $\rho$ that couples as
\begin{equation}
    \pi\int B_2^{(1)}\cup \rho~.
\end{equation}
Thus shifting $B_2^{(1)}$ due to the background gauge transformation actually changes the effective action by
\begin{equation}
    \pi\int A_b^*\left(n_2+ n_1\cup \xi_1+c_- \xi_1\cup \xi_1\right)\cup\rho~.
\end{equation}
Thus (\ref{eqn:n2xi}) is not an equivalence relation in the fermionic theory; it reproduces the relabelling in (\ref{eqn:n2l1}).
On the other hand, the background gauge transformation of $B_2^{(1)}$ only changes it by an exact cocycle and it gives an equivalence relation in the fermionic theory.

\subsection{Interpretation of the obstructions to the constraint equations}\label{sec:obstructions}
The data that describes the bosonic shadow theory enriched by $G_b$ symmetry needs to obey the constraints 
\begin{align}
    dn_2 &= n_1\cup \omega_2+c_-Sq^1\omega_2+n_1\cup s_1\cup n_1\quad  (\text{for integer }c_-)\cr
    d\nu_3&={\cal O}_4(n_1,n_2,\omega_2)~.
\end{align}
When the first equation cannot be satisfied for integer $c_-$,  we cannot simply fix the backgrounds for the intrinsic symmetries to be  $B_2^{(1)}=A_b^*n_2,B_2^{(2)}=A_b^*\omega_2,B_1=A_b^*n_1$ for integer $c_-$, where $A_b$ is the $G_b$ background gauge field. 
Instead, we need to turn on general $B_2^{(1)}$ such that
\begin{equation}\label{eqn:H3twogroup}
    dB_2^{(1)}=A_b^*\Theta_3,\quad \Theta_3=n_1\cup \omega_2+c_-Sq^1\omega_2+n_1\cup s_1\cup n_1~,
\end{equation}
where we cannot express $B_2^{(1)}$ as $A_b^*n_2$ for some $n_2$.
Thus the theory is enriched by the two-group symmetry that extends the $G_b$ symmetry by $\mathbb{Z}_2$ one-form symmetry generated by the fermion line, where the two-group is specified by the Postnikov class $\Theta$ given in the above equation \cite{kapustin2013,Cordova:2018cvg,Benini2018}.
If we gauge the $\mathbb{Z}_2$ one-form symmetry in the bosonic shadow theory by summing over the background $B_2^{(1)}$, this produces an ``anomalous SPT phase" with $G_f$ symmetry, which is an
almost trivial theory with effective action that depends on the extension of the $G_f$ background gauge field to the bulk \cite{Tachikawa:2017gyf,Hsin:2020nts}. After gauging the one-form symmetry in the bosonic shadow theory, the theory has dual $\mathbb{Z}_2^f$ symmetry, whose background $\rho$ couples as $\pi\int B_2^{(1)}\cup \rho$. Then (\ref{eqn:H3twogroup}) contributes the bulk dependence
\begin{align}
   & S_{n_2}=\pi\int_{4d} dB_2^{(1)}\cup \rho=
    \pi\int_{4d} (A_b^*\Theta)\cup \rho
    \cr 
    &=\pi\int_{4d} A_b^*\left(n_1\cup \omega_2+c_-Sq^1\omega_2+n_1\cup s_1\cup n_1\right)\cup \rho~.\cr
\end{align}
This contribution to the anomaly is only present for integer $c_-$, and is absent for half integer $c_-$, where the constraint equation is instead $dn_2=0$ and it can always be satisfied.

Similarly, if the constraint equation for $\nu_3$  cannot be satisfied for any $\nu_3$, then the theory after gauging the one-form symmetry generated by the fermion line has additional bulk dependence: $S_\text{total}=S_{n_2}+S_{\nu_3}$ with
\begin{equation}
    S_{\nu_3}=\int A_b^*\phi_4,\quad {\cal O}_4\equiv e^{i\phi_4}~.
\end{equation}

Since the bosonic shadow theory becomes an invertible theory with $G_f$ symmetry after gauging the one-form symmetry generated by the fermion line, the total anomaly $S_\text{tot}$ for the $G_f$ symmetry must belong to the trivial class, {\it i.e.} it can be expressed as $\int_{4d} d\alpha_3$ where $\alpha_3$ is a classical action that depends on the $G_f$ background gauge field. For instance, when $\omega_2=0$, the symmetry is $G_f=G_b\times\mathbb{Z}_2^f$, and the anomaly ${\cal O}_4=e^{i\phi_4}$ can be expressed as a boundary term for the $G_f=G_b\times\mathbb{Z}_2^f$ background gauge fields $(A_b,\rho)$:
\begin{align}
 \int_{4d} &A_b^*\phi_4
 =  \pi\int_{4d} A_b^* n_2\cup A_b^* n_2\cr
 &=\pi\int_{4d} A_b^*n_2\cup w_2(TM)\cr &=\pi\int_{4d}d(A_b^*n_2\cup \rho)]=0\text{ mod }2\pi~,
\end{align}
where we used the Wu formula, $\rho$ is the spin structure ({\it i.e.} the background gauge field for $\mathbb{Z}_2^f$ symmetry) that satisfies $d\rho=w_2(TM)$, and we used $dn_2=0$.
Another case is when $n_1=0$ while $\omega_2\neq 0$, where the background for the $G_f$ symmetry satisfies
$d\rho=A_b^*\omega_2+w_2(TM)$ being the trivial class.
The anomaly belongs to the trivial class on the bulk manifolds:
\begin{align}
    &\int_{4d}A_b^*\phi_4\cr 
    &=\pi\int_{4d} A_b^* {\cal P}(n_2)+\pi\int_{4d} A_b^*n_2\cup \omega_2+\frac{\pi c_-}{4}\int {\cal P}(A_b^*\omega_2)\cr 
    &=\pi\int_{4d} A_b^* n_2\cup w_2(TM)
     +\pi\int_{4d} A_b^*n_2\cup w_2(TM)\cr
     &\qquad +{\pi c_-\over 4}\int_{4d} \frac{dA'}{2\pi}\frac{dA'}{2\pi}~,
\end{align}
where in the first line we used the Wu formula, and we used the property that $w_2(TM)$ admits an integral lift $dA'/2\pi$ on orientable four-manifolds. Thus the anomaly can be cancelled by a Chern-Simons term on the boundary and belongs to the trivial class on bulk manifolds.

We remark that usually anomaly is defined up to adding local counterterms for the background gauge field. Here, when the obstructions are nontrivial, the anomaly in the theory after gauging the one-form symmetry is nontrivial only in the sense that we cannot cancel the bulk dependence by a local counterterm $A_b^*\nu_3$ on the boundary that only depends on the $G_b$ background gauge field, but the anomaly can be cancelled by a local counterterm $\alpha_3$ on the boundary that depends on $G_f$ background gauge field. Such a boundary theory is an example of an ``anomalous" SPT phase with $G_f$ symmetry \cite{PhysRevLett.123.207003}.

\section{Stacking rules and overall classification} \label{Sec:stacking}
	The computations in previous sections have given a set of defining equations for invertible fermion phases, as well as a set of equivalence relations among the data. From this we now wish to obtain a general \textit{stacking rule} for invertible fermion phases. Specifically, given two invertible phases $a$ and $b$ defined by the data $(c_-^i, \tilde{n}_1^i, \tilde{n}_2^i, \nu_3^i, \{L^i_{\bf g}\})$ for $i = a,b$, suppose we stack $a$ and $b$ to obtain a phase whose data we will indicate using the symbol $'\text{tot}'$. In a Hamiltonian formulation, the stacking operation is defined by imposing the following relation on the Hilbert spaces $\H_a, \H_b, \H_{\text{tot}}$ and the Hamiltonians $H_a, H_b, H_{\text{tot}}$ of $a,b$ and $'{\text{tot}}'$:
	\begin{align}
	    \H_{\text{tot}} &= \H_a \otimes \H_b \\
	    H_{\text{tot}} &= H_a \otimes \mathds{1}_b + \mathds{1}_a \otimes H_b~.
	\end{align}
	The goal of this section is to express the data $(c_-^{\text{tot}}, \tilde{n}_1^{\text{tot}}, \tilde{n}_2^{\text{tot}}, \nu_3^{\text{tot}},\{L^{\text{tot}}_{\bf g}\})$ in terms of the corresponding data of $a$ and $b$. The results, referred to as ``stacking rules", are summarized in Table~\ref{Table:MainEqs}. When $G_b$ is unitary, the stacking rules for $c_-$ and $\tilde{n}_1$ can be obtained straightforwardly from either the $G_b$-crossed theory or from Chern-Simons theory. Using the latter, the complete stacking rules can be derived exactly when $G_f = G_b \times \Z_2^f$ and $G_f = U(1)^f$. For a general unitary $G_b$, we make conjectures for the $n_2$ and $\nu_3$ stacking rules which are compatible with our formulas for $\mathcal{O}_4$. To obtain the stacking rule for $\nu_3$ we follow the algebraic approach described in Refs. \cite{Gu2014Supercoh,Cheng2018fSPT}: we consider the quantity $\frac{\mathcal{O}_4^{\text{tot}}}{O_4^a O_4^b}$, and show that it is equal to a coboundary $db_3$. Therefore, up to a possible 3-cocycle we can write
	\begin{equation}
	    \nu_3^{\text{tot}} = \nu_3^a \nu_3^b \times b_3~.
	\end{equation}
	Although we cannot prove an exact stacking rule for $n_2$, this process gives us an answer that reproduces exact results when $\om = 0$. Deriving the exact stacking rules for $n_2$ and $\nu_3$ within the $G_b$-crossed theory requires a more complete theory of anyon condensation in $G$-crossed BTCs, which we do not develop here. 
	
	When $s_1 \ne 0$, we can obtain the stacking rule for $n_2$ up to a possible 2-cocycle, by integrating the equation for $dn_2$. Since we do not have an exact stacking rule for $n_2$, we do not attempt to calculate the stacking rule for $\nu_3$ in this case.

\subsection{Stacking rules from $G_b$-crossed theory}

In this section, we will present stacking rules for $c_-,n_1$ and $\om$ within the $G_b$-crossed theory. We will assume the symmetry is unitary, $s_1=0$.
The corresponding rules for $n_2,\nu_3$ are not derived exactly, since this requires a complete theory of anyon condensation in $G$-crossed BTCs, which we do not develop here. Note that in this section we will use the representation $n_1,n_2,\nu_3$, assuming that the flux labels are canonically chosen and denoted as $L_{\bf g}^0$. 

We first define the notion of stacking two MTCs $\C^a,\C^b$. The anyons in $\C^a,\C^b$ are $\{I^a, \psi^a, m^a, (\psi^a m^a)\}$ and $\{I^b,\psi^b,m^b,(\psi^b m^b)\}$ respectively. (If $c_-$ is a half-integer, $m$ and $\psi m$ are identified.) We consider the Deligne product $\C^a \boxtimes \C^b$, which is defined so that anyons in $\C^a$ are assumed to braid trivially with those in $\C^b$, and vice versa. Then we condense the bound state $\psi^a \psi^b$. Upon confining all anyons that braid nontrivially with $\psi^a \psi^b$, we obtain the ``stacked" topological order $\C^{\text{tot}}$.  
The stacking of two $G_b$-crossed theories $\C_{G_b}^{\times, a},\C_{G_b}^{\times, b}$ is defined similarly: we write
\begin{equation}
    \C_{G_b}^{\times, {\text{tot}}} := \frac{\C_{G_b}^{\times, a}\boxtimes \C_{G_b}^{\times, b}}{ \psi^a \psi^b \sim 1}~.
\end{equation}

\subsubsection{Stacking rule for $c_-$}

The chiral central charge follows the rule $c_-^{{\text{tot}}} = c_-^a + c_-^b \mod 8$. This can be seen by computing the topological spin of a fermion parity flux in $\C^{\text{tot}}$. If $m^a$ and $m^b$ denote fermion parity fluxes in $\C^a$ and $\C^b$, then $m^a m^b$ is a fermion parity flux in $\C^a\boxtimes\C^b$. It is deconfined, since $\psi^a\psi^b$ braids trivially with it: $M_{m^a m^b,\psi^a \psi^b} = M_{m^a,\psi^a} M_{m^b,\psi^b} = (-1)^2 = 1$. Its topological twist moreover remains unchanged after condensing $\psi^a\psi^b$. Thus we have
\begin{align}
    e^{i \pi c_-^{\text{tot}}/4} &:= \theta_{m^{\text{tot}}} = \theta_{m^a m^b} = \theta_{m^a} \theta_{m^b} M_{m^a m^b} \nonumber \\ &= e^{i \pi (c_-^a + c_-^b)/4}~.
\end{align}
 
\subsubsection{Stacking rule for $n_1$}

\begin{figure}
    \centering
    \includegraphics[width=0.25\textwidth]{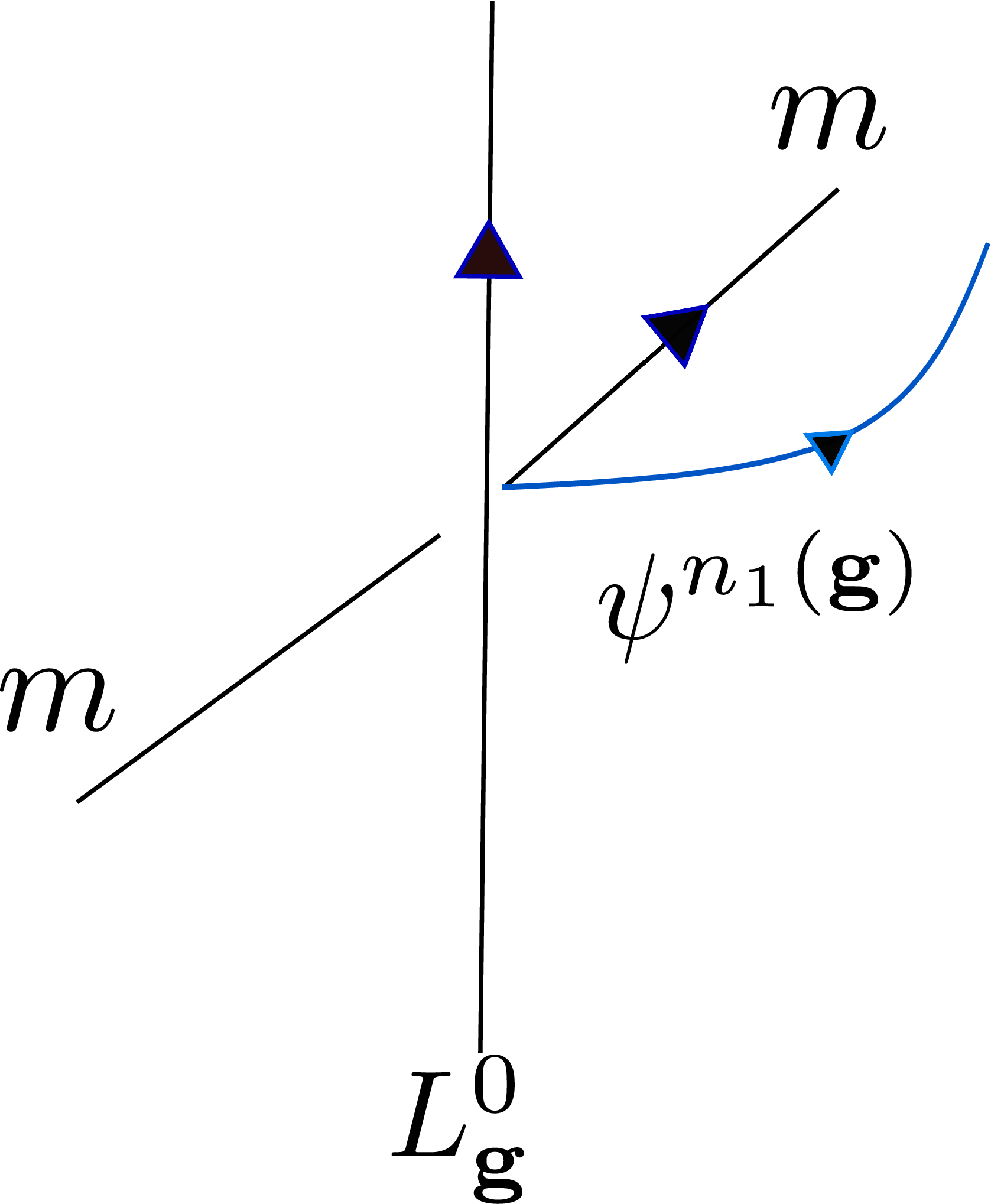}
    \caption{A fermion parity flux $m$ is permuted by ${\bf g}$ action.}
    \label{fig:fpermute}
\end{figure}

We next prove the result $n_1^{\text{tot}} = n_1^a + n_1^b$. Recall that the definition of $n_1$ implies two things: 1) if $n_1({\bf g}) = 1$, then $L_{\bf g}^0$ is nonabelian; and 2) given a fermion parity flux $m$, the action of ${\bf g}$ takes $m \rightarrow m \times \psi^{n_1({\bf g})}$. This is shown diagramatically in Fig.~\ref{fig:fpermute}, where the symmetry action corresponds to moving $m$ under $L_{\bf g}^0$. Note that these two statements hold for any value of $c_-$.

\begin{figure}
    \centering
    \includegraphics[width=0.3\textwidth]{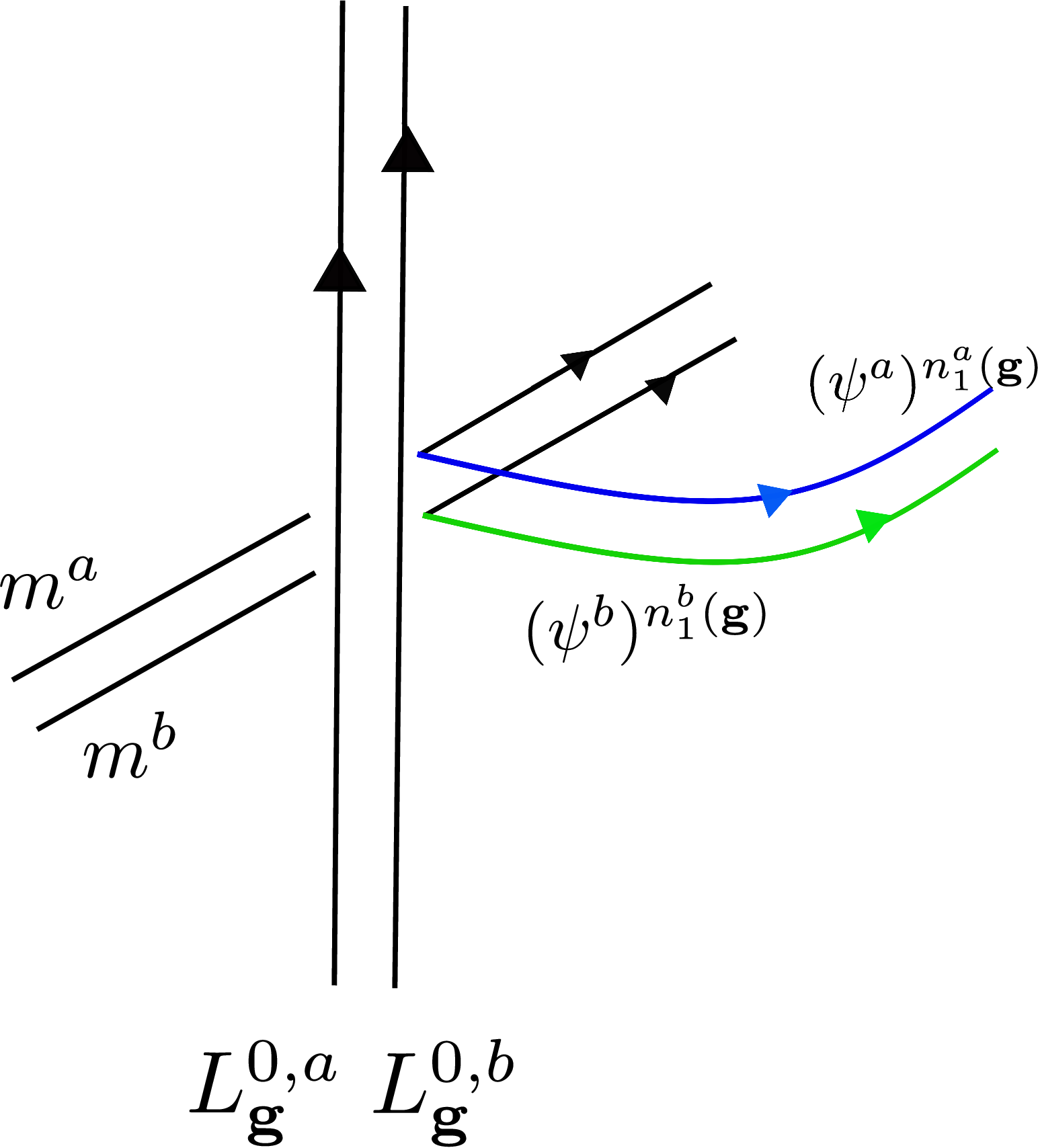}
    \caption{Stacking rule for $n_1$. Upon condensing, we identify $\psi^{\text{tot}} \sim \psi^a \sim \psi^b$.}
    \label{fig:n1stack}
\end{figure}

Upon stacking $a$ and $b$, we have the situation shown in Fig.~\ref{fig:n1stack}. Prior to the condensation, the defect $L_{\bf g}^{0,\text{tot}}$ can be expressed as a pair $(L_{\bf g}^{0,a}, L_{\bf g}^{0,b})$, while the anyon $m^{\text{tot}}$ can be expressed as $m^a m^b$. The process of passing $m^{\text{tot}}$ under $L_{\bf g}^{0,\text{tot}}$ can be broken up into independent processes occurring in $a$ and $b$.  As a result, we see that
\begin{align}
    m^am^b &\rightarrow (m^a \times (\psi^a)^{n_1^a({\bf g})})(m^b \times (\psi^b)^{n_1^b({\bf g})}) \nonumber \\ &= (m^a m^b) \times ((\psi^a)^{n_1^a({\bf g})} (\psi^b)^{n_1^b({\bf g})})~.
\end{align}
After condensation, we have the identity $ \psi^a \sim \psi^b \sim \psi^{\text{tot}}$ and $m^a m^b \rightarrow m^{\text{tot}}$. The above equation then becomes
\begin{align}
   m^{\text{tot}} &\rightarrow m^{\text{tot}} \times (\psi^{\text{tot}})^{(n_1^a + n_1^b)({\bf g})}~.
\end{align}

\subsubsection{Stacking rule for $\omega_2$}

\begin{figure}
    \centering
    \includegraphics[width=0.5\textwidth]{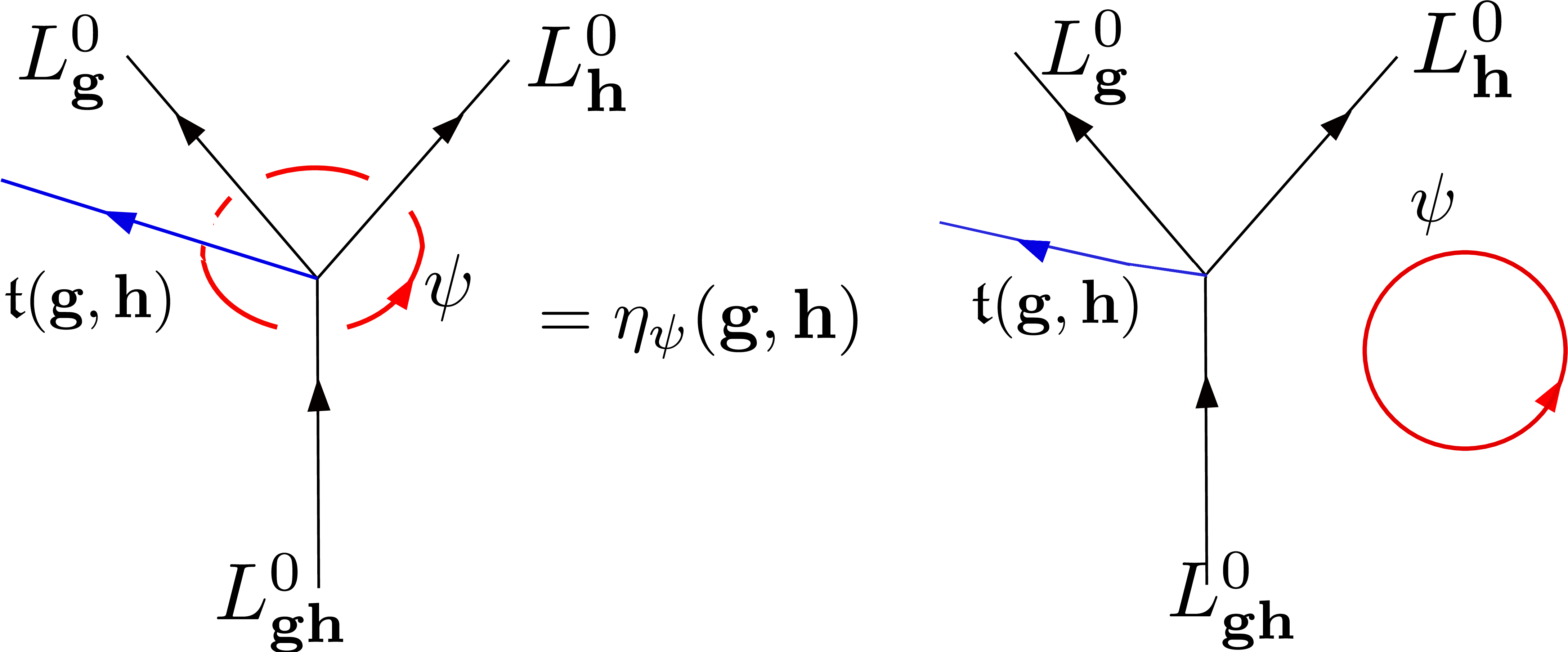}
    \caption{Computation of $\eta_{\psi}$ by performing a full braid of $\psi$ around an $L_{\bf g}^0,L_{\bf h}^0,L_{\bf gh}^0$ trijunction.}
    \label{fig:eta_psi}
\end{figure}

Below we show that if the stacking of $G_b$-crossed theories is to be meaningfully related to the stacking of invertible fermionic phases, then we must have $\omega_2^a = \omega_2^b = \omega_2$. Although this is intuitively obvious if we consider stacking the original $G_f$-crossed theories, it is not so obvious in the $G_b$-crossed theory, where the complete set of symmetry fractionalization classes contains different choices of $[\omega_2]$. We argue as follows. First we have the definitions $\eta_{\psi^a}({\bf g},{\bf h}) = \eta_{\psi^b}({\bf g},{\bf h}) =(-1)^{\omega_2({\bf g},{\bf h})}$. Diagramatically, these $\eta$ symbols are calculated by braiding $\psi^a$ (respectively, $\psi^b$) around a junction formed by the fusion of a ${\bf g}$ and an ${\bf h}$ defect into a ${\bf gh}$ defect (see Fig.~\ref{fig:eta_psi}). The red line can be removed from the junction at the cost of a phase factor $\eta_{\psi}({\bf g},{\bf h})$. When the symmetry is unitary, $\eta_{\psi}({\bf g},{\bf h})$ can be set equal to the braiding phase $M_{\psi,\mathfrak{t}({\bf g},{\bf h})}$ between $\psi$ and an anyon $\mathfrak{t}({\bf g},{\bf h})$, defined as
\begin{equation}
    \mathfrak{t}({\bf g},{\bf h}) = m^{\omega_2({\bf g},{\bf h})} \times \psi^{n_2({\bf g},{\bf h})}~.
\end{equation}
Since $M_{\psi,\psi} = 1$, there will be no contribution to $\eta_{\psi}$ from $n_2$.

\begin{figure}
    \centering
    \includegraphics[width=0.3\textwidth]{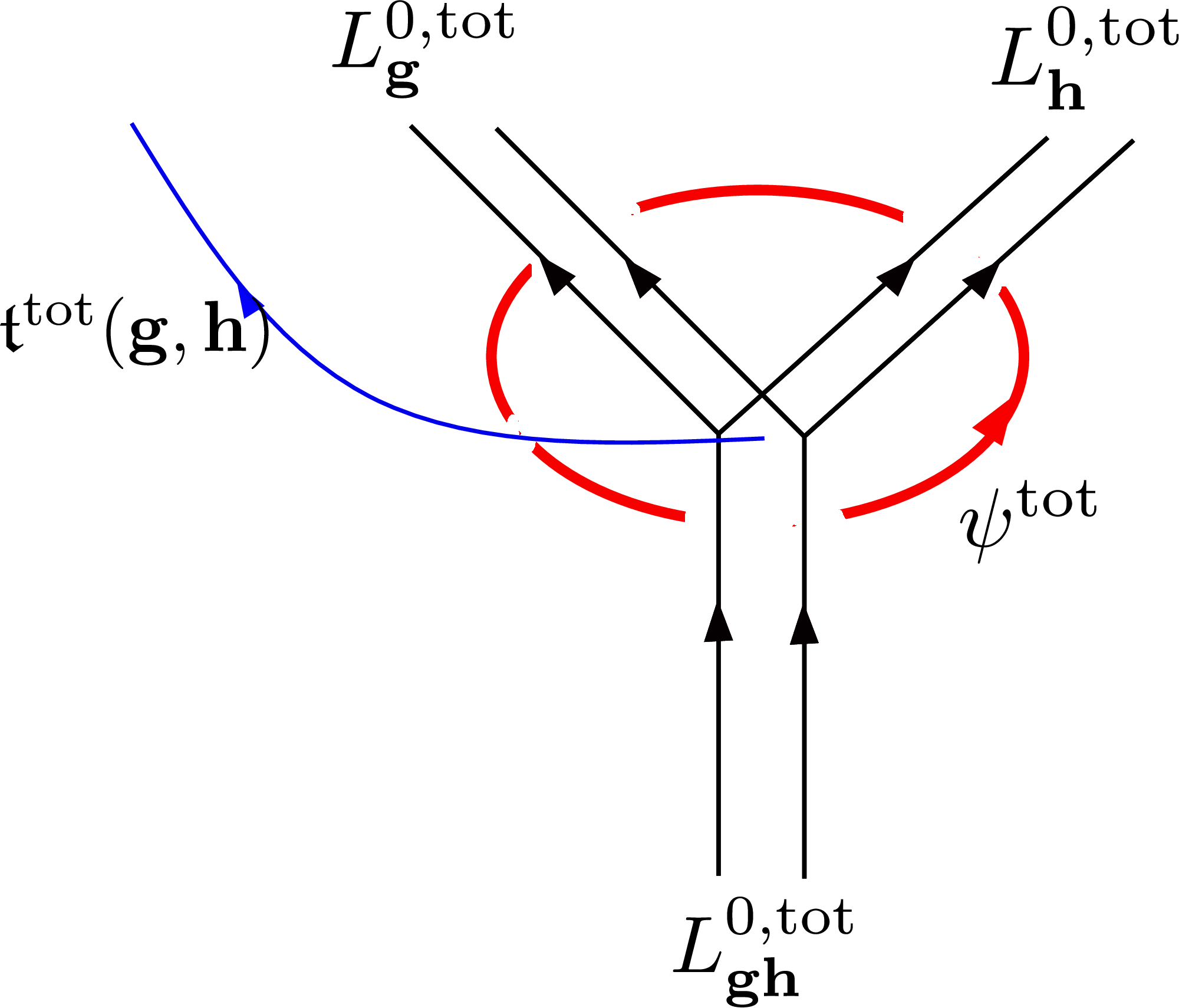}
    \caption{The computation of $\eta_{\psi}$ in the stacked phase splits into independent braiding processes within $a$ and $b$. This is because there are no additional crossings generated by the interaction of $a$ and $b$ lines.}
    \label{fig:w_2stack}
\end{figure}

To interpret this stacking of $G_b$-crossed theories as a stacking procedure for invertible phases, we need to be able to condense $\psi^a\psi^b$ without breaking symmetry. For this we must have $\eta_{\psi^a\psi^b} = 1$, i.e. $\psi^a\psi^b$ should not have any fractional $G_b$ quantum numbers \cite{Bischoff_2019}. The symbol $\eta_{\psi^a\psi^b}$ can be measured by braiding $\psi^a\psi^b$ around a double junction obtained by stacking $a$ and $b$. Since we are working prior to the $\psi^a\psi^b$ condensation, the only braiding factors come from the crossing of $\psi^a$ with $(m^a)^{\omega_2^a}$ and of $\psi^b$ with $(m^b)^{\omega_2^b}$. These braiding phases are given by $\eta_{\psi^a}$ and $\eta_{\psi^b}$ respectively. As a result, we can write
\begin{equation}
    \eta_{\psi^a\psi^b} = \eta_{\psi^a}\eta_{\psi^b} = (-1)^{\omega_2^a+\omega_2^b}~.
\end{equation}
For the lhs to be 1, we must have $\omega_2^a = \omega_2^b = \omega_2$. 

Now let us show that $\om^{\text{tot}} = \om$ for the stacking of invertible phases. We have $\psi^{\text{tot}} \sim \psi^a \sim \psi^b$. The value of $\eta_{\psi^{\text{tot}}}$ is measured by braiding $\psi^{\text{tot}}$ around a stacked junction, as depicted in Fig.~\ref{fig:w_2stack}. Here $\psi^{\text{tot}}$ crosses the anyon $\mathfrak{t}^{\text{tot}}$.
The statement $\om^{\text{tot}} = \om$ is equivalent to saying that $\mathfrak{t}^{\text{tot}}({\bf g},{\bf h})$ is a fermion parity flux if and only if $\mathfrak{t}^a({\bf g},{\bf h})$ and $\mathfrak{t}^a({\bf g},{\bf h})$ are fermion parity fluxes. The 'if' part is clear: if $\mathfrak{t}^a({\bf g},{\bf h})$ and $\mathfrak{t}^a({\bf g},{\bf h})$ are fermion parity fluxes, their bound state is also a fermion parity flux, and determines $\mathfrak{t}^{\text{tot}}({\bf g},{\bf h})$ up to a fermion. For the other direction, we argue that if $\mathfrak{t}^{\text{tot}}$ is a fermion parity flux, then both $\mathfrak{t}^a({\bf g},{\bf h})$ and $\mathfrak{t}^a({\bf g},{\bf h})$ are fermion parity fluxes. Indeed, the only deconfined bound states in the stacked phase are $m^am^b, (m^a\psi^a) m^b$, and $\psi^a$ (up to equivalences). The only bound state which gives a fermion parity flux is $m^a m^b$. 

\subsubsection{Conjectured stacking rule for $n_2$}\label{Sec:n2stack}

In the $G_b$-crossed theory, the basic data are given by $\tilde{n}_1,\tilde{n}_2,\nu_3$ and the flux labels $L_{\bf g}$. For integer $c_-$ one passes to $n_2$ by relabelling the defects such that $L_{\bf g} \rightarrow L_{\bf g}^0$, and then computing the change in $\tilde{n}_2$. Thus in order to obtain the stacking rule for $n_2$ from the $G_b$-crossed theory, we must first obtain the stacking rules for symmetry defects $L_{\bf g}$ and for the symmetry fractionalization parameter $\tilde{n}_2$. The former step requires a more complete theory of anyon condensation in $G$-crossed BTCs than is currently available, and so we do not pursue this here. 

An exact stacking rule for $n_2$ can be independently obtained from Chern-Simons theory when $s_1 = 0$ and $G_f = G_b \times \Z_2^f$ (Section~\ref{sec:stackingGbxZ2f}): it has the form
\begin{equation}\label{n2stack}
    n_2^{\text{tot}} = n_2^a + n_2^b + n_1^a \cup n_1^b.
\end{equation}
We conjecture that this rule should also hold when $\om \ne 0$. In support of this conjecture when $\om \ne 0$, we find that $\frac{\mathcal{O}_4[n_1^{\text{tot}},n_2^{\text{tot}}]}{\mathcal{O}_4[n_1^a,n_2^a]\mathcal{O}_4[n_1^a,n_2^a]}$ is a coboundary even when $[\om]$ is nontrivial. Thus Eq.~\eqref{n2stack} gives a meaningful guess for the stacking rule for $\nu_3$, as we will discuss below.

\subsection{Stacking rule for $\nu_3$ from the $\mathcal{O}_4$ anomaly}\label{sec:stackfromO4}

Here we obtain an explicit expression for the stacking rule for $\nu_3$, by integrating the equation for $d\nu_3$. We will assume $s_1 = 0$, and also take the $n_2$ stacking rule to be given by Eq.~\eqref{n2stack}. This approach determines $\nu_3^{\text{tot}}$ up to a 3-cocycle; we conjecture that this 3-cocycle should be trivial, which we verify in the case $\om = 0$ in Section~\ref{sec:stackingGbxZ2f} using the CS formulation. However we do not have a general direct proof when $\om \ne 0$. 

Note that for half integer $c_-$ the symmetry is $G_b\times\mathbb{Z}_2^f$, which is covered by the analysis of Section~\ref{sec:stackingGbxZ2f}. Therefore here we will focus on the case with integer $c_-$.
The condition satisfied by $\nu_3$ is
\begin{align}
    d \nu_3 &= (-1)^{ {\cal P}(n_2) + n_2 \cup \om +\zeta (n_1, \om) + c_- (n_1 \cup \om) \cup_2 (\om \cup_1 \om)}\cr
    &\quad \;(-1)^{\frac{c_-}{4} {\cal P}(\om)}
    i^{  n_1 \cup n_1 \cup \om  }~,
\end{align}
with the Pontryagin square ${\cal P}(X_2) \equiv X_2 \cup X_2 - X_2 \cup_1 d X_2$ and the Cartan coboundary $\zeta(n_1, \om) = n_1 \cup [(n_1 \cup \om) \cup_2 \om + n_1 \cup \om]$. We can first verify that the obstruction is closed by expanding out the various terms.
To compute the stacking rule for $\nu_3$, we use the property that ${\cal O}_4(n_1^{\text{tot}},n_2^{\text{tot}},\omega_2^{\text{tot}},c_-^{\text{tot}})$ differs from ${\cal O}_4(n_1^a,n_2^a,\omega_2^a,c_-^a){\cal O}_4(n_1^b,n_2^b,\omega_2^b,c_-^b)$ by a coboundary.
The details of the computation are in Appendix~\ref{sec:stackingnu3}. The answer, up to a possible 3-cocycle, is\footnote{
We thank David Aasen for pointing out a term involving $\cup_3$ which was inadvertently dropped in an earlier version of this paper.
}
\begin{align}
    \nu_3^{\text{tot}}=&\nu_3^a\nu_3^b\, i^{ \frac{1}{2} n_1^a \cup n_1^b \cup n_1^b + n_1^a \cup (n_1^a \cup_1 n_1^b) \cup n_1^b}i^{ (n_1^a \cup_1 n_1^b) \cup \om}\cr
    &\times (-1)^{(n_1^a \cup n_1^b) \cup_1 (n_2^a + n_2^b) + n_2^a \cup_1 n_2^b + dn_2^a \cup_2 n_2^b}\cr 
    &\times (-1)^{c^a_-(\omega_2\cup_1\omega_2)\cup_3(n_1^b\cup \omega_2)}~.
\end{align}

In the special case $\omega_2^a=\omega_2^b=\omega_2^{\text{tot}}=0$,  the stacking rule for $\nu_3$ becomes
\begin{align}\label{eqn:stackingnu3p}
    \nu_3^{\text{tot}}&=\nu_3^a\nu_3^b\, i^{ n_1^a \cup n_1^b \cup n_1^b + n_1^a \cup (n_1^a \cup_1 n_1^b) \cup n_1^b}\cr &(-1)^{(n_1^a \cup n_1^b) \cup_1 (n_2^a + n_2^b) + n_2^a \cup_1 n_2^b }~.
\end{align}
This reproduces the stacking rule of the fermionic invertible phases with $G_b\times\mathbb{Z}_2^f$ symmetry in Ref.~\onlinecite{brumfiel2018pontrjagin}, which is exact, and is derived using the Chern-Simons theory in Section~\ref{sec:stackingGbxZ2f}.

\subsubsection{Application: equivalences on the data}

For general $\om$, the above stacking rule is undetermined up to a possible 3-cocycle $\alpha_3(\om, D^a, D^b)$, where the data $D^i = (c^i, n_1^i, n_2^i,\nu_3^i)$ for $i = a,b$ defines the invertible phase $i$. We conjecture that $\alpha_3 = 0$. 

If $\alpha_3 \ne 0$, the equivalences on $(n_2, \nu_3)$ stated in Eq.~\eqref{eq:equivs} of Table \ref{Table:MainEqs} will also need to be modified. First consider the equivalence (see Eq.~\eqref{eq:equivn2db2})
\begin{equation}
    (n_2,\nu_3) \simeq (n_2 + db_1, \nu_3 \times (-1)^{b_1 \cup \om + db_1 \cup_1 n_2 + db_1 \cup b_1} \times \alpha_3')
\end{equation}
where $\alpha_3'$ is an undetermined 3-cocycle. This equivalence can be thought of as coming from stacking a trivial phase with the data $D'_1 = (0,0,db_1,(-1)^{db_1 \cup b_1 + b_1 \cup \om})$ onto an invertible phase with data $D'_2 = (c_-,n_1,n_2,\nu_3)$. In this case, the $\nu_3$ data of the stacked phase equals
\begin{equation}
    \nu_3^{\text{tot}} = \nu_3 \times (-1)^{b_1 \cup \om + db_1 \cup_1 n_2 + db_1 \cup b_1} \times \alpha_3(\om, D'_1, D'_2)
\end{equation}
and therefore we can express $\alpha'_3$ in terms of $\alpha_3$:
\begin{equation}
    \alpha'_3 \equiv \alpha_3(\om, D'_1, D'_2).
\end{equation}

We can similarly analyze the equivalence (from Eq.~\eqref{eq:equivn2om2})
\begin{equation}
    (n_2,\nu_3) \simeq (n_2 + \om, \nu_3 \times (-1)^{\om \cup_1 n_2} \times \alpha''_3)
\end{equation}
and determine $\alpha''_3$ in terms of $\alpha_3$. Here we can interpret the equivalence in terms of stacking two phases with data $D''_1 = (0,0,\om,1)$ and $D''_2 = (c_-,n_1,n_2,\nu_3)$. Then we find that
\begin{equation}
    \alpha''_3 \equiv \alpha_3(\om,D''_1,D''_2).
\end{equation}
In particular, if $\alpha_3 = 0$ as we conjecture, then we must have $\alpha'_3 = \alpha''_3 = 0$.
\subsection{Stacking rules when $s_1 \neq 0$}

We will discuss the stacking rule of $n_2$ for $s_1\neq 0$ and general $n_1$ using the constraint equations for $dn_2,d\nu_3$. For simplicity, we will focus on the stacking rule of $n_2$, while leaving the stacking rule of $\nu_3$ to future work.
The constraint equation for $dn_2$ is
\begin{equation}
    dn_2=n_1\cup \omega_2+n_1\cup s_1\cup n_1~.
\end{equation}
This implies that
\begin{equation}
    dn_2^{\text{tot}} = dn_2^a + dn_2^b + n_1^a \cup s_1 \cup n_1^b + n_1^b \cup s_1 \cup n_1^a.
\end{equation}
Thus we find
\begin{align}
    n_2^\text{tot}
    &=n_2^a+n_2^b+n_1^a\cup n_1^b+
     (n_1^a \cup_1 s_1) \cup n_1^b\cr 
     &\;\;+ (n_1^b \cup_1 s_1) \cup n_1^a
     + s_1 \cup (n_1^a \cup_1 n_1^b)+x~.
\end{align}
where $x$ is a $\mathbb{Z}_2$ two-cocycle that vanishes for $s_1=0$.
We propose the minimal stacking rule with $x=0$:
\begin{align}
    n_2^\text{tot}&=n_2^a+n_2^b+n_1^a\cup n_1^b+
     (n_1^a \cup_1 s_1) \cup n_1^b \cr 
    &\quad
     + (n_1^b \cup_1 s_1) \cup n_1^a+ s_1 \cup (n_1^a \cup_1 n_1^b)~.
\end{align}

\subsection{Stacking rules from Chern-Simons theory when $G_f = G_b \times \Z_2^f$}\label{sec:stackingGbxZ2f}

While the bosonic shadow theories are useful for relating the Chern-Simons approach to the $G_b$-crossed BTC, we can directly study the invertible fermion phase using Chern-Simons theory. In this section, we assume the symmetry is $G_b\times \mathbb{Z}_2^f$.  
We can couple the Chern-Simons theory to background $G_b$ gauge field $A_b$ via the magnetic $\mathbb{Z}_2$ 0-form symmetry: we turn on background for the magnetic 0-form symmetry given by the $\mathbb{Z}_2$ one-form $A_b^*n_1$, where $n_1\in \H^1(G_b,\mathbb{Z}_2)$.
Equivalently, we shift the spin structure by $A_b^*n_1$.
We can also add a local action $\int_{3d}A_b^*\phi_3$ for the background $G_b$ gauge field, which depends on the bulk extension by
\begin{equation}
    \int_{4d} A_b^*d \phi_3,\quad \phi_3\in C^3(BG,\mathbb{R}/2\pi\mathbb{Z})~.
\end{equation}
In bosonic theory, we require the dependence to be trivial to have a well-defined local action; on spin manifolds, $w_2(TM)$ is trivial, and it can instead satisfies 
\begin{align}
A_b^* d\phi_3 &=\pi w_2(TM)\cup A_b^* n_2\cr
&=\pi A_b^*Sq^2(n_2),\quad n_2\in \H^2(BG,\mathbb{Z}_2)~,
\end{align}
where we used the Wu formula $Sq^2(n_2)=n_2\cup w_2$ mod 2 on orientable four-manifolds \cite{milnor1974characteristic}.
For instance, when the symmetry of the fermionic phase is $\mathbb{Z}_{2n}\times\mathbb{Z}_2^f$, $\phi_3$ is given by (possibly spin) Chern-Simons terms of the $\mathbb{Z}_{2n}$ background gauge field, and as demonstrated in Ref.~\onlinecite{Cordova:2017vab} this reproduces the $\mathbb{Z}_{4n}\times \mathbb{Z}_2$ classification for even $n$ and $\mathbb{Z}_{8n}$ classification for odd $n$ \cite{PhysRevB.95.195147}.
The cochain $\phi_3$ can be exponetiated to $\nu_3=e^{i\phi_3}$.

More explicitly, the effective action for the fermionic invertible phase can be expressed as follows.
We would like to understand the effective action for the fermionic invertible phase that is not expressed as $\phi_3$.
Following Ref.~\onlinecite{Cordova:2017vab}, we denote the effective action produced by shifting the spin structure with $A_b^*n_1$ in $SO(2c_-)_1$ equals $2c_- f(A_b^*n_1)$, where $f$ is defined by
\begin{equation}
    e^{iL f(B_1)}=\frac{Z_{SO(L)_1}[0]}{Z_{SO(L)_1}[B_1]}~,
\end{equation}
where $Z_{SO(L)_1}[B_1]$ is the partition function of $SO(L)_1$ Chern-Simons theory coupled to background $B_1=A_b^*n_1$ by the $\mathbb{Z}_2$ magnetic 0-form symmetry.
Equivalently, since the magnetic 0-form symmetry in $SO(L)_1$ is identified with the fermion parity \cite{Cordova:2017vab}, $e^{iLf[B_1]}$ is the change of the partition function for $SO(L)_1\leftrightarrow -L\text{CS}_\text{grav}$ under shifting the spin structure by $B_1=A_b^*n_1$, and the exponent is proportional to $L$. The effective action $f(B_1)$ is the effective action for the fermionic SPT phase with $\mathbb{Z}_2\times\mathbb{Z}_2^f$ unitary symmetry that belongs to the unit class of $\mathbb{Z}_8$ classification. We note that $A_b^*\phi_3$ in general cannot be expressed using $f$, since the later is at most $\mathbb{Z}_8$ valued.
We remark that although $f(A_b^*n_1)$ cannot be expressed as a special case of $A_b^*\phi_3$, on the other hand, $2f(A_b^*n_1)$ is equivalent to a special case $A_b^*\phi_3$ \cite{Cordova:2017vab}.
This is because for $c_-=1$, the gravitational Chern-Simons term can be expressed as products of $U(1)_1$ Chern-Simons theory, and shifting the spin structure by $\mathbb{Z}_2$ gauge field $B_1=\pi A_b^*n_1$ (we normalize $n_1=0,1$) in $U(1)_1$ produces a Chern-Simons term for the background $\mathbb{Z}_2$ gauge field $A_b^*n_1$:
\begin{equation}
    \frac{1}{4\pi} ada+\frac{1}{2\pi}a d B_1\leftrightarrow \frac{1}{4\pi}a'da' -\frac{1}{4\pi}B_1dB_1,\quad a'=a+B_1~,
\end{equation}
where the Chern-Simons term of $B_1=\pi A_b^*n_1$ can be expressed as integral of an auxiliary term in the bulk of the form $Sq^2(n_2)$ with $n_2=dn_1/2=Sq^1(n_1)$:
\begin{align}
    &\frac{1}{4\pi}\int_{3d} dB_1dB_1 = \pi\int_{4d} \frac{dB_1}{2\pi}\frac{dB_1}{2\pi}\cr 
    &=\pi\int Sq^2\left(\frac{dB_1}{2\pi}\right)=\pi\int A_b^* Sq^2(dn_1/2)~.
\end{align}

Thus, it is sufficient to parametrize the effective action 
of the fermionic invertible phase with $G_b\times \mathbb{Z}_2^f$ symmetry by $(n_1,n_2,\phi_3)$ as
\begin{align}\label{eqn:fspteffectiveaction}
    S(n_1,n_2,\phi_3;A_b)&=\int A_b^*\phi_3+ f(A_b^*n_1)-2c_-\text{CS}_\text{grav},\cr d \phi_3&=\pi Sq^2(n_2)~,
\end{align}
where $A_b$ is the background $G_b$ gauge field. This agrees with the classification in Ref.~\onlinecite{Bhardwaj_2017}, since $f$ is the effective action of the ``root" fermionic SPT phase with unitary $\mathbb{Z}_2\times\mathbb{Z}_2^f$ symmetry. 

Let us consider the stacking of two fermionic invertible phases with $G_b\times\mathbb{Z}_2^f$ symmetry \cite{brumfiel2018pontrjagin}, {\it i.e.} how the data $(n_1,n_2,\phi_3)$ composes under addition of the effective action:
\begin{align}
    &S(n_1^a,n_2^a,\phi_3^a;A_b)+S(n_1^b,n_2^b,\phi_3^b;A_b)\cr 
    &\quad =
    S(n_1^{\text{tot}},n_2^{\text{tot}},\phi_3^{\text{tot}};A_b)~.
\end{align}
From the gravitational Chern-Simons term  $-2c_-\text{CS}_\text{grav}$ we find $c_-^\text{tot}=c_-^a+c_-^b$.
We note that\footnote{
 We use the property that $f(B_1+B_1')-f(B_1)-f(B_1')$ gives ``half of the mixed Chern-Simons term of $B_1,B_1'$'' \cite{Cordova:2017vab}.
}
\begin{align}\label{eqn:f}
    f&(A_b^*n_1^a)+f(A_b^*n_1^b)=f(A_b^*(n_1^a+n_1^b))+A_b^*\xi(\tilde n_1^a,\tilde n_1^b),\cr 
    d \xi
     &=\pi (n_1^a)^2\cup (n_1^b)^2\cr
     &=Sq^2(n_1^a\cup n_1^b)
    +d(n_1^a\cup (n_1^a\cup_1\eta_b^1 )\cup n_1^b)~,
\end{align} 
where $\tilde n_1^a,\tilde n_1^b$ are the canonical integral one-cochain lifts of $n_1^a,n_1^b$ that takes value in $0,1$.
Explicitly, $\xi=(\pi/2)\tilde n_1^a\cup (\tilde n_1^b)^1$.
We used the identity 
$(n_1^a\cup n_1^b+n_1^b\cup n_1^a)=d(n_1^a\cup_1 n_1^b)$.
In Eq.~\eqref{eqn:f}, the term $f(A_b^*(n_1^a+n_1^b)=f(A_b^*n_1^\text{tot})$ gives the stacking rule $n_1^\text{tot}=n_1^a+n_1^b$.
In the last line of (\ref{eqn:f}), the first term $Sq^2(n_1^a\cup n_1^b)$ together with $Sq^2(n_2^a)+Sq^2(n_2^b)$ contribute to $Sq^2(n_2^\text{tot})$.
Thus by comparing with (\ref{eqn:fspteffectiveaction}), we find the stacking rule 
\begin{equation}
 n_1^{\text{tot}}=n_1^a+n_1^b,\quad 
 n_2^{\text{tot}}=n_2^a+n_2^b+n_1^a\cup n_1^b~.
\end{equation}
The stacking rule of $\phi_3$ is 
derived by writing the bulk terms other than $Sq^2(n_2^{\text{tot}})$ as a boundary cochain \footnote{
Explicitly,
\begin{align}
    &Sq^2(n_2^a+n_2^b+n_1^a\cup n_1^b)-Sq^2(n_2^a)\cr
    &\quad -Sq^2(n_2^b)-Sq^2(n_1^a\cup n_1^b)\cr
    &\quad =d ( n_1^a\cup_1 n_2^b+ (n_1^a+n_2^b)\cup_1 (n_1^a\cup n_1^b))~.
\end{align}
}. In the end we find
$\phi_3^{\text{tot}}$ is related to $\phi_3^a,\phi_3^b$ by
\begin{align}
    \phi_3^{\text{tot}}=&\phi_3^a+\phi_3^b+\pi (n_2^a\cup_1 n_2^b
    +(n_2^a+n_2^b)\cup_1(n_1^a\cup n_1^b)\cr
    &+
    n_1^a\cup (n_1^a\cup_1n_1^b)\cup n_1^b )+\frac{\pi}{2}\tilde n_1^a\cup (\tilde n_1^b)^2.
\end{align}
Thus we reproduce the stacking rules for $G_b\times\mathbb{Z}_2^f$ symmetry fermionic invertible phases in Ref.~\onlinecite{brumfiel2018pontrjagin}.

\subsection{Stacking rules for $G_f = U(1)^f$ using CS theory with background $Spin^c=(U(1)\times\mathbb{Z}_2^f)/\mathbb{Z}_2$ structures}

Let us use an example to illustrate the stacking rule when the symmetry of the fermionic invertible phase is an extension of $G_b$  by $\mathbb{Z}_2^f$ (strictly speaking such phases are not fermionic, in the sense that they do not have neutral fermionic particles that are singlet under $G_b$).
We consider the symmetry to be $Spin^c=(U(1)\times\mathbb{Z}_2^f)\mathbb{Z}_2$.
If we turn on the background gauge field $A$, the fermionic invertible phase has effective action
\begin{equation}
    k_1\left(\frac{1}{4\pi}AdA+2\text{CS}_\text{grav}\right)+16 k_2 \text{CS}_\text{grav}~
\end{equation}
for background  $Spin^c$ connection $A$, where
$(k_1,k_2)\in\mathbb{Z}\times\mathbb{Z}$ gives the stacking rule. In particular, although the gravitational part is not well-defined by itself for general $k_1$, the stacking rule for the part that depends on $A$ and the gravitational part can be treated separately: when adding the effective action for different classes $(k_1,k_2),(k_1',k_2')$, we add the gravitational part and the ``non-gravitational" part that depends only on $A$ separately. This behavior of the stacking rule is similar to the case when the symmetry is a direct product $G_b\times\mathbb{Z}_2^f$. As we will argue in the following, this separation of the gravitational and non-gravitational stacking rule continues to hold when the symmetry is a general extension of any $G_b$ by $\mathbb{Z}_2^f$. 

We remark that the separation of the stacking rule for gravitational and non-gravitational parts is also related to the absence of topological thermo-electric coupling in unitary theories: there is no topological term that mixes the background gauge field and background gravity. For a recent discussion of absence of topological thermo-electric transport, see Ref.~\onlinecite{PhysRevB.103.235101}.

\section{Examples}\label{Sec:Examples}
We now illustrate our classification approach with several examples. To conveniently account for the various equivalences involved, we use the following strategy: (i) find the allowed choices of $n_1 \in \H^1(G_b,\Z_2)$. (ii) For each $n_1$, find a representative set of allowed choices of $n_2$. These form a torsor over $\H^2(G_b,\Z_2)$ modulo the possible equivalence $n_2 \simeq n_2 + \om$. (iii) With the representative $n_2$ fixed, find all possible $\nu_3$, accounting for all equivalences that preserve $n_2$. 
\subsection{$G_f = \Z_2\times\Z_2^f$}\label{Sec:Z2xZ2fExample}
In this example, we will use the $G_b$-crossed theory to classify invertible phases with $G_f = \Z_2\times\Z_2^f$. It is well-known from previous work that the classification of fSPTs with this symmetry should be $\Z_8$ \cite{Gu2014interaction,Ryu2012interaction,Qi_2013}. From cobordism theory, we also expect that the full classification of invertible phases is $\Z\times\Z_8$, with 8 phases for each $c_-$ \cite{Kapustin:2014dxa}. Although a naive count of $G_b$-crossed theories does not agree with these results, we recover this classification by keeping track of flux labels. Finally, we show how the free and interacting classifications can be explicitly related in this case. We use this correspondence to illuminate the stacking rules.

\begin{table*}[t]
    \renewcommand{\arraystretch}{1.4}
    \centering
    \begin{tabular}{|c|c|c|c|}
        \hline
        Free fermion parameter& Interacting fermion parameter & $G_b$-crossed data with flux labels & Fully gauged theory \\ 
        $(\nu^{\uparrow},-\nu^{\downarrow})$ & $(n_1,n_2,\nu_3)$ & $(\tilde{n}_1,\tilde{n}_2,\nu_3,L_{\bf h})$ &  \\
        \hline
        $(0,0)$& $(0,0,1)$ & $(0,0,1,I_{\bf h})$& $\Z_2 \text{t.c.}^{\uparrow} \times \Z_2 \text{t.c.}^{\downarrow}$ \\
        \hline
        $(1,-1)$& $(1,0,1)$ & $(1,0,1,\sigma_{\bf h}^+)$ & $\text{Ising}^{\uparrow} \times \overline{\text{Ising}}^{\downarrow}$ \\
        \hline
        $(2,-2)$& $(0,1,-i)$ & $(0,1,-i,I_{\bf h})$& $U(1)^{\uparrow}_4 \times U(1)^{\downarrow}_{-4}$ \\
        \hline
        $(3,-3)$& $(1,1,-i)$ & $(1,0,-1,\sigma^-_{\bf h})$& $\text{Ising}^{(3),\uparrow} \times \text{Ising}^{(-3),\downarrow}$ \\
        \hline
        $(4,-4)$& $(0,0,-1)$ & $(0,0,-1,I_{\bf h})$& $\text{D.S. }(+i)^{\uparrow} \times \text{D.S.}(-i)^{\downarrow}$ \\
        \hline
        $(5,-5)$& $(1,0,-1)$ & $(1,0,-1,\sigma^+_{\bf h})$& $\text{Ising}^{(5),\uparrow} \times \text{Ising}^{(-5),\downarrow}$ \\
        \hline
        $(6,-6)$& $(0,1,i)$ & $(0,1,-i,m_{\bf h})$ & $U(1)_{-4}^{\uparrow}\times U(1)_4^{\downarrow}$ \\
        \hline
        $(7,-7)$& $(1,1,i)$ & $(1,0,1,\sigma_{\bf h}^-)$ & $\text{Ising}^{(7),\uparrow} \times \text{Ising}^{(-7),\downarrow}$ \\
        \hline
    \end{tabular}
    \caption{Comparison of data for $c_-=0$ phases with $G_f = \Z_2\times\Z_2^f$ in terms of (1) a free fermion construction, (2) the $(n_1,n_2,\nu_3)$ representation used in Ref.~\onlinecite{Wang2020fSPT}, (3) the $(\tilde{n}_1,\tilde{n}_2,\nu_3,L_{\bf g})$ representation coming from the $G_b$-crossed theory, and 4) the theory obtained by fully gauging $G_f$. Here ``t.c." means ``toric code" while ``D.S.$(\pm i)$" means ``double semion theory with semion statistics $\pm i$". All arguments of $n_1,n_2,\tilde{n}_1,\tilde{n}_2$, and $\nu_3$ are set equal to ${\bf h}$.}
    \label{Table:Z2SPTdata}
\end{table*}

\subsubsection{Classification when $c_-=0$}
Let ${\bf h}$ be the generator of the unitary $\Z_2$ symmetry. We begin by naively counting the number of $G_b$-crossed extensions of the toric code UMTC. This is done by counting all possible inequivalent triples $(\tilde{n}_1,\tilde{n}_2,\nu_3)$. First consider the case where $\tilde{n}_1({\bf h}) = 1$. After setting $\omega_2 = 0$, we have $[(\omega_2,\tilde{n}_2)] \in \H^2_{\tilde{n}_1}(\Z_2,\Z_2\times\Z_2) = \Z_1$. Although there are 2 choices for the function $\tilde{n}_2$, given by $\tilde{n}_2({\bf h},{\bf h}) = 0$ or 1, these choices are coboundary-equivalent. (In particular, these two choices correspond to the symmetry fractionalization cocycles with $\mathfrak{t}({\bf h},{\bf h}) = I$ or $\psi$; they are related by the coboundary $d\mathfrak{b}$, where $\mathfrak{b}({\bf h}) = m$.)
Therefore we will fix $\tilde{n}_2 = 0$. We also have two defect classes, which are inequivalent under relabellings; these correspond to $\nu_3({\bf h},{\bf h},{\bf h}) = \pm 1$. In total there are two $G_b$-crossed theories with $\tilde{n}_1 \ne 0$.

Next, we look at theories with $\tilde{n}_1 = 0$. In this case there are two symmetry fractionalization classes, with $\tilde{n}_2({\bf h},{\bf h}) = 0,1$; they are classified by the subgroup of $\H^2(G_b,\Z_2\times\Z_2)$ with $\omega_2=0$, and are inequivalent. For each symmetry fractionalization class, we have two defect classes (i.e. 2 choices of $\nu_3$). However, the two defect classes associated to $\tilde{n}_2({\bf h},{\bf h}) = 1$ (i.e. the ones with $\nu_3({\bf h},{\bf h},{\bf h}) = \pm i$) are equivalent under a relabelling of defects (this was shown previously in Refs.~\onlinecite{Lu2016,Barkeshli2019}). Therefore we only have 3 SET phases in this case, giving a total of 5 $G_b$-crossed theories with $c_-=0$. Indeed, by examining the third column of Table~\ref{Table:Z2SPTdata}, we can verify that there are only 5 distinct triples $(\tilde{n}_1,\tilde{n}_2,\nu_3)$. 

Below we show that these 5 theories in fact contain a total of 8 topological invariants which take the values $e^{2\pi i k/8}$, for $k = 0,1, \dots , 7$. Thus the same $G_b$-crossed BTC might correspond to two different fSPT phases. In order to fully specify the fSPT, we need to define a flux label $L_{\bf h}$, corresponding to the $({\bf h},0)$ defect in the ungauged $G_f$-crossed theory. 

Consider the fSPT phases with $\tilde{n}_1 \ne 0$. Here we can define 
\begin{align}
    \tilde{n}_2({\bf h},{\bf h}) &= 0 \\
    \nu_3({\bf h},{\bf h},{\bf h}) &= (-1)^q~,
\end{align}
 where $q \in \Z_2$. In this case, the $G_b$-crossed theory has two ${\bf h}$-defects $\sigma^+_{\bf h}$ and $\sigma^-_{\bf h}$. For each $q$, the $G_b$-crossed theory has two invariants, denoted as $\mathcal{I}(\sigma^+_{\bf h})$ and $\mathcal{I}(\sigma^-_{\bf h})$. The explicit form of the invariants is
\begin{equation}
\mathcal{I}(\sigma^{\pm}_{\bf h}) := \theta_{\sigma^{\pm}_{\bf h}}^2 \eta_{\sigma^{\pm}_{\bf h}} ({\bf h},{\bf h})~.
\end{equation} 

We can solve for the $G$-crossed data in a gauge with all $\eta = 1$, and with $\theta_{\sigma^{\pm}_{\bf h}} = i^{\pm [q]_2} e^{\pm i \frac{\pi}{8}}$, where $q \in \Z_2$ (see Section~X.I of Ref.~\onlinecite{Barkeshli2019}). This implies that
\begin{align}
\mathcal{I}(\sigma^{\pm}_{\bf h}) &= (-1)^q e^{\pm i \frac{2\pi}{8}}~.
\end{align}
By taking $q = 0,1$ and $L_{\bf h} = \sigma^+_{\bf h},\sigma^-_{\bf h}$, we see that the invariants can take 4 values $e^{i \pi/4}, e^{i 3\pi/4}, e^{i 5\pi/4}, e^{i 7\pi/4}$. If we fix $\tilde{n}_2$ and $\nu_3$, we cannot interchange these invariants by relabelling defects with fermions. This means that the 4 invariants really describe distinct phases. 

When $\tilde{n}_1 = 0$, we have 4 abelian ${\bf h}$-defects $I_{\bf h},e_{\bf h},m_{\bf h},\psi_{\bf h}$. The symmetry fractionalization is specified by $\tilde{n}_2({\bf h},{\bf h}) \in \{0,1\}$. When $\tilde{n}_2 = 0$, we find that
\begin{align}
    \nu_3({\bf h},{\bf h},{\bf h}) &= (-1)^q \\
    \mathcal{I}(a_{\bf h}) &= (-1)^q, \quad \forall a_{\bf h}~.
\end{align}
We thus have two fSPT invariants taking values $\pm 1$. When $\tilde{n}_2({\bf h},{\bf h}) = 1$, we find that
\begin{align}
    \nu_3({\bf h},{\bf h},{\bf h}) &= i(-1)^q \\
    \mathcal{I}(a_{\bf h}) &= (-i)(-1)^{q + a_e + a_m}~.
\end{align}
Once again, there are only two invariants, which take the values $\pm i$. The theory with $\nu_3 = \pm i,L_{\bf h} = I_{\bf h}$ is the same $G_b$-crossed theory as the one with $\nu_3 = \mp i, L_{\bf h} = m_{\bf h}$. Thus if we were just counting SET phases, we would need to treat the two defect classes with $\tilde{n}_2({\bf h},{\bf h}) = 1$ as equivalent. However, relabelling defects by fermion parity fluxes is not a trivial operation in the $G_b$-crossed theory, and therefore the two invariants taking the values $\pm i$ correspond to distinct fSPT phases.

Note that we always have $\mathcal{I}(L_{\bf h})=\mathcal{I}(\psi\times L_{\bf h})$. This reflects the fact that it is physically equivalent to choose $I_{\bf h}$ or $\psi_{\bf h}$ as the $\Z_2$ flux, since defects can always be relabelled by attaching fermions. 

To summarize, we count the number of different values of $\mathcal{I}(L_{\bf h})$ by varying over the distinct choices of $(n_1,\tilde{n}_2,\nu_3,L_{\bf h})$. When $n_1 = 0$, there are 4 values, $\pm 1$ and $\pm i$. Considering the cases $n_1=0$ and $n_1 \ne 0$ together, we obtain 8 different invariants given by $e^{2\pi i k/8}$ for $k = 0,1, \dots , 7$. Even without considering formal stacking rules, this suggests that the group corresponding to the fSPT classification should be $\Z_8$.

\subsubsection{Relating the $G_b$-crossed data to the $(n_1, n_2, \nu_3)$ parameterization}

We summarize the result of the classification of fSPTs using the $(n_1, n_2, \nu_3)$ parameterization \cite{Wang2020fSPT} for this example in the second column of Table~\ref{Table:Z2SPTdata}. We have $|\H^1(G_b,\Z_2)| = |\H^2(G_b,\Z_2)| = |\H^3(G_b,U(1))| = 2$. The final classification has 8 fSPTs, obtained by considering all possible values of $n_1, n_2$ and $\nu_3$ which satisfy the defining equations. In this case there are no obstructions, so we indeed have 8 phases where each choice of $n_1$ gives 4 fSPTs. 

We can straightworwardly relate the $G_b$-crossed data $(\tilde{n}_1,\tilde{n}_2,\nu_3,L_{\bf g})$ to the $(n_1, n_2, \nu_3)$ parameterization. For $c_-=0$ we have $n_1 := \tilde{n}_1$. Next, we fix a canonical definintion $L_{\bf h}^0$ of the flux labels, as follows: if $n_1({\bf h}) =0, L_{\bf h}^0 = I_{\bf h}$, while if $n_1({\bf h}) =1, L_{\bf h}^0 = \sigma^+_{\bf h}$. Then we use the result from Section~\ref{Sec:Counting} that if $L_{\bf h} = L_{\bf h}^0 \times m^{l_1({\bf h})}$, then
\begin{equation}
    n_2({\bf h},{\bf h}) = \tilde{n}_2({\bf h},{\bf h}) + l_1({\bf h}) n_1({\bf h})~.
\end{equation}
The resulting formula for $\nu_3$ in terms of $n_2$ is
\begin{align}
   \nu_3({\bf h},{\bf h},{\bf h}) &= (-1)^q i^{[n_2({\bf h},{\bf h})]_2}~.
\end{align}
This gives us the desired relations. A similar redefinition is possible for each integer $c_-$, although we will not demonstrate it explicitly. 
\subsubsection{Classification when $c_- \ne 0$}
When $c_-$ is an arbitrary integer, the analysis is almost identical to the $c_- = 0$ case. We present only the results. The data of the usual $G_b$-crossed theory are given by
\begin{align}
   n_1({\bf h}) &\in \{0,1\} \\
   \tilde{n}_2({\bf h},{\bf h}) &\in \{0,1\} \\ 
   \nu_3({\bf h},{\bf h},{\bf h}) &= (-1)^q i^{[n_2({\bf h},{\bf h})]_2}~.
\end{align}
For each integer $c_-$, we can compute 8 invariants which take the values $e^{2\pi i k/8}$, where $k \in \Z_8$. We can go through the steps discussed previously and form a table similar to Table~\ref{Table:Z2SPTdata} for each $c_-$.

The case $c_-= k+1/2$ with integer $k$ is slightly different. Here we have 3 ${\bf h}$-defects $I_{\bf h}, \psi_{\bf h}, \sigma_{\bf h}$, where $\sigma_{\bf h}$ is nonabelian. For each choice of $[\tilde{n}_2]$ and $[\nu_3]$, we can evaluate $\mathcal{I}(L_{\bf h})$. We get
\begin{align}
    \mathcal{I}(I_{\bf h}) &= \mathcal{I}(\psi_{\bf h}) = (-1)^q i^{-[\tilde{n}_2({\bf h},{\bf h})]}\\
    \mathcal{I}(\sigma_{\bf h}) &= (-1)^q i^{[\tilde{n}_2({\bf h},{\bf h})]}e^{i 2\pi/8}~.
\end{align}
Therefore, we have 8 invariants of the form $e^{i 2\pi k/8}$, where $k = 0,1,\dots , 7$. The 4 invariants with $k$ even are obtained by assigning $L_{\bf h} = I_{\bf h}$ or $\psi_{\bf h}$, while the 4 invariants with $k$ odd are obtained by assigning $L_{\bf h} = \sigma_{\bf h}$. In this case we define $n_1$ as trivial or nontrivial according as $L_{\bf h}$ is abelian or nonabelian. We then set $n_2 = \tilde{n}_2$ and evaluate $\nu_3$ in the usual manner.
\subsubsection{Relation to free fermion constructions}
\begin{figure}
    	\begin{center}
	\begin{tikzpicture}[scale=0.7,x  = {(1cm,0cm)},
	y  = {(0.9cm,0.5cm)},
	z  = {(0cm,1cm)}]
	
	\begin{scope}[canvas is xy plane at z=0,decoration={
		markings,
		mark=at position 0.5 with {\arrow{>}}}]
	\draw[ultra thick] (5,0)--(5,5)--(0,5)--(0,0);
	\fill[yellow!70] (0,0)--(5,0)--(5,5)--(0,5)--cycle;
		\draw[ultra thick,postaction={decorate}] (5,0)--(0,0);
	
	\end{scope}
	\begin{scope}[canvas is xy plane at z=1.5,decoration={
		markings,
		mark=at position 0.5 with {\arrow{>}}}]
	\draw[ultra thick] (5,0)--(5,5)--(0,5)--(0,0);
	
	\fill[cyan!40] (0,0)--(5,0)--(5,5)--(0,5)--cycle;
	\draw[ultra thick,postaction={decorate}] (0,0)--(5,0);
	
	\end{scope}
	\node at (6,0,0) {\Large $-\nu, \downarrow$};
	\node at (6,0,3) {\Large $\nu,\uparrow$};
	
	\end{tikzpicture}
\end{center}
    \caption{Stacking $p+ip$ superconductors with spin-up and spin-down fermions to obtain an fSPT with $G_f = \Z_2\times\Z_2^f$.}
    \label{fig:FFstack}
\end{figure}
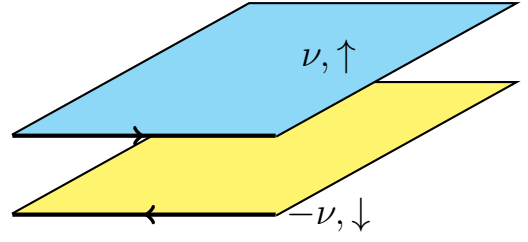
In this section we describe a free fermion construction of the $\Z\times\Z_8$ classification. Our purpose is twofold: (i) to derive an explicit correspondence between the free and interacting classifications, and (ii) to obtain the stacking rules for the interacting case using the free fermion picture, in which the stacking is particularly simple. 

The $\Z_8$ classification of fSPT phases was obtained in Ref.~\onlinecite{Gu2014interaction} by gauging the full $\Z_2\times\Z_2^f$ symmetry in a model of free fermions. We now review this argument. Consider spin-up and spin-down fermions which transform under the unitary $\Z_2$ symmetry as follows:
\begin{align}
    U_{\bf h} c^{\dagger}_{\uparrow} U_{\bf h}^{\dagger} &= -c^{\dagger}_{\uparrow} \\
    U_{\bf h} c^{\dagger}_{\downarrow} U_{\bf h}^{\dagger} &= +c^{\dagger}_{\downarrow}~.
\end{align}
Both types of fermions transform by a minus sign under the $\Z_2^f$ symmetry operator, denoted as $P^f$:
\begin{equation}
    P^f  c^{\dagger}_{\uparrow (\downarrow)} (P^f)^{\dagger} = -c^{\dagger}_{\uparrow (\downarrow)}~.
\end{equation}

Now consider placing the $\uparrow$ fermions in a $p+ip$ superconductor state with Chern number $\nu^{\uparrow}$, and the $\downarrow$ fermions in a $p+ip$ superconductor state with Chern number $\nu^{\downarrow} = -\nu^{\uparrow}$. The resulting stack, denoted as $(\nu^{\uparrow},\nu^{\downarrow})$, has zero net chirality, and is a realization of a free fermion SPT phase with $\Z_2\times\Z_2^f$ symmetry (see Fig.~\ref{fig:FFstack}). 

Ref.~\onlinecite{Gu2014interaction} showed that the quantity $\nu^{\uparrow} \mod 8$ remains a topological invariant of the fSPT in the presence of interactions. Following their argument, we first note that the $\Z_2$ subgroup of $G_f$ generated by ${\bf h}$ acts only on the $\uparrow$ fermions, while the $\Z_2$ subgroup generated by ${\bf h} (-1)^F$ acts only on the $\downarrow$ fermions. Therefore the full symmetry can be gauged by gauging these subgroups separately in each layer. For example, gauging the $(0,0)$ state results in two copies of the $\Z_2$ toric code, while gauging the $(1,-1)$ state gives the topological order $\text{Ising}\times \overline{\text{Ising}}$. The gauged theories for each $\nu^{\uparrow}$ are listed in the third column of Table~\ref{Table:Z2SPTdata}. (In this context, the $G_b$-crossed BTC is an intermediate object obtained by gauging just the $\Z_2^f$ symmetry in both layers.)

The states $(\nu^{\uparrow},-\nu^{\uparrow})$ and $(\nu^{\uparrow}+8,-(\nu^{\uparrow}+8))$ result in the same gauged theory: this can be seen by constructing an explicit isomorphism. On the other hand, the gauged theories corresponding to $(\nu^{\uparrow}+k,-\nu^{\uparrow}-k)$ are all different for $k=0,1,\dots , 7$. However, this statement needs some clarification. Upon gauging the $(1,-1)$ state as indicated, we obtain $\text{Ising}\times \overline{\text{Ising}}$. On the other hand, upon gauging the state $(-1,1) \equiv (7,-7)$, we obtain $\overline{\text{Ising}}\times \text{Ising}$, which is the same topological order. Indeed, if we only gauge $\Z_2^f$, the $G_b$-crossed theories for the $(1,-1)$ and $(-1,1)$ states are the same: they are both described by the data $\tilde{n}_1 \ne 0, [\tilde{n}_2] = 0$, and $\nu_3 = 1$.

To distinguish between the $(1,-1)$ and $(-1,1)$ states we must keep track of which layer transforms nontrivially under the unitary $\Z_2$ symmetry. Thus, the more accurate description of the topological order obtained by gauging the $(1,-1)$ state is $\text{Ising}^{\uparrow}\times \overline{\text{Ising}}^{\downarrow}$. This is different from $\overline{\text{Ising}}^{\uparrow}\times \text{Ising}^{\downarrow}$, which is obtained by gauging the $(-1,1)$ state, because the nonabelian anyon corresponding to the $\uparrow$ layer is different in the two cases. Thus there are indeed 8 different fSPTs with this symmetry group. In a similar way, we can check that the free fermion construction gives 8 different invertible phases associated to each $c_-$. The need to keep track of $\uparrow$ and $\downarrow$ fermions is closely related to the requirement of specifying flux labels in the $G_b$-crossed theory. 

The $G_b$-crossed invariants allow us to easily relate the free fermion data $(\nu^{\uparrow},-\nu^{\downarrow})$ to the corresponding $G_b$-crossed data $(\tilde{n}_1,\tilde{n}_2,\nu_3,L_{\bf h})$. From the $G_b$-crossed theory we find that
\begin{equation}\label{Eq:kinvt}
    \mathcal{I}(L_{\bf h}) = e^{i 2\pi \left(\frac{[n_1({\bf h})]_2}{8} - \frac{[n_2({\bf h},{\bf h})]_2}{4} + \frac{q}{2}\right)}~.
\end{equation}
Now the $\Z_8$ topological invariant of the $(\nu^{\uparrow},-\nu^{\downarrow})$ state should measure the $\Z_2$ charge of the ${\bf h}$ flux in the $\uparrow$ layer, which transforms nontrivially under ${\bf h}$. This is given by $e^{2\pi i \nu^{\uparrow}/8}$. By setting $\mathcal{I}(L_{\bf h}) = e^{2\pi i \nu/8}$, we can straightforwardly obtain the desired relations:
\begin{align}
    c_- &= \frac{\nu^{\uparrow}+\nu^{\downarrow}}{2} \\
    n_1({\bf h}) &= \nu^{\uparrow} \mod 2 \\
    n_2({\bf h},{\bf h}) &= \frac{\nu^{\uparrow}(\nu^{\uparrow}-1)}{2} \mod 2 \\
    \nu_3({\bf h},{\bf h},{\bf h}) &= (-i)^{\lfloor \frac{\nu^{\uparrow}}{2} \rfloor \mod 4}~.
\end{align}

\subsubsection{Stacking rules}

The above results allow us to explicitly verify the stacking rules that were discussed in Section~\ref{Sec:stacking}. In the free fermion picture, we have a state $a$ with parameters $(\nu^{\uparrow, a},\nu^{\downarrow, a})$ and $(\nu^{\uparrow, b},\nu^{\downarrow, b})$. Stacking them gives a state with 
\begin{equation}
  (\nu^{\uparrow, \text{tot}},\nu^{\downarrow, \text{tot}}) = (\nu^{\uparrow, a}+\nu^{\uparrow, b},\nu^{\downarrow, a}+\nu^{\downarrow, b}).  
\end{equation}
We can easily verify that the stacking rule for $n_1$ is additive, because $n_1^{\text{tot}}({\bf h}) = \nu^{\uparrow,\text{tot}}$. Moreover, the definition of $n_2$ in terms of free fermion parameters implies that
\begin{align}
    n_2^{\text{tot}}({\bf h},{\bf h}) &= \frac{(\nu^a+\nu^b)(\nu^a+\nu^b-1)}{2} \\
    &= \frac{\nu^a(\nu^a-1)}{2} + \frac{\nu^b(\nu^b-1)}{2} + \nu^a \nu^b~.
\end{align}
This is consistent with the following stacking rule:
\begin{equation}
    n_2^{\text{tot}} = n_2^a + n_2^b + n_1^a \cup n_1^b \mod 2~.
\end{equation}

We can also check the stacking rule for $\nu_3$, but will not perform the calculation here as it is not very illuminating.

\subsection{Invertible fermion phases with $G_f = U(1)^f$} \label{Sec:U(1)symm}
In this example, we study $G_b = U(1)$. Formally we have two possibilities for $G_f$: either $G_f = U(1)\times\Z_2^f$ or $G_f = U(1)^f$. The latter case is more physically relevant as it defines the symmetry of the integer quantum Hall states and many other fermionic systems. We will study this example as it provides insight into the difference between ``$G_b$ quantum numbers" and ``$G_f$ quantum numbers". After working out the mathematical classification using $G$-crossed BTCs data, we will perform the same calculation using $K$-matrix Chern-Simons theory, and explicitly show how to gauge the fermion parity.

\subsubsection{Analysis using $G$-crossed theory}
In this case, $G_f = U(1)^f$ is a central extension of $G_b = U(1)$ by $\Z_2^f$. We represent the group $U(1)$ as $\R/\Z$. A representative cocycle $\omega_2$ is
\begin{equation}
\omega_2({\bf g},{\bf h}) = \begin{cases}
1,& {\bf g} + {\bf h} \ge 1\\
0,&{\bf g} + {\bf h} < 1
\end{cases}~.
\end{equation}

	Since $\H^1(U(1),\Z_2) = \Z_1$, we must have $n_1 = 0$ for each $c_-$. Since $\omega_2$ is nontrivial, only integer values of $c_-$ are allowed. The defining equation for $n_2$ becomes
	\begin{equation}
	dn_2 = c_- \omega_2 \cup_1 \omega_2 \mod 2~.
	\end{equation}
	We can show that $\omega_2 \cup_1 \omega_2$ is in fact the zero function. We have
	\begin{align}
	\omega_2 \cup_1 \omega_2({\bf g},{\bf h},{\bf k}) &= \omega_2({\bf g},{\bf h}) \omega_2({\bf gh},{\bf k}) + \omega_2({\bf g},{\bf hk}) \omega_2({\bf h},{\bf k})~. 
	\end{align}
	The first term is nonzero if and only if
	\begin{align}
	{\bf g}+{\bf h} &\ge 1 \\
	\text{and} \quad [{\bf g}+{\bf h}] + {\bf k} &\ge 1 \\
	\implies {\bf g}+{\bf h}+ {\bf k} &\ge 2~.
	\end{align}
	This implies the following relations:
	\begin{align}
	{\bf g}+[{\bf h}+ {\bf k}] &\ge 2-1 = 1 \\
	\text{and} \quad{\bf h}+ {\bf k} \ge 2 - {\bf g} &> 1~. 
	\end{align} 
	Therefore when the first term in the expression for $\omega_2 \cup_1 \omega_2$ is 1, the second term is also 1 (and vice versa). We conclude that $\omega_2 \cup_1 \omega_2 \equiv 0$, and therefore $dn_2 = 0$ for each $c$.
	
	Since $\H^2(G_b,\Z_2) \cong \Z_2$, there are two choices for $n_2$: we can choose $n_2 = 0$ or $n_2 = \omega_2$. These two choices are equivalent. Therefore we fix $n_2 = 0$, and obtain
	\begin{equation}
	\mathfrak{t}({\bf g},{\bf h}) = m^{\omega_2({\bf g},{\bf h})}~,
	\end{equation}
	where $m$ is a fermion parity flux.
	
	Under the above assumptions, the obstruction equation reads
	\begin{equation}
	\mathcal{O}_4({\bf g},{\bf h},{\bf k},{\bf l}) = R^{\mathfrak{t}({\bf g},{\bf h})\mathfrak{t}({\bf k},{\bf l})} = e^{2\pi i \frac{c_-}{8} [\omega_2]_2 \cup [\omega_2]_2({\bf g},{\bf h},{\bf k},{\bf l})}~.
	\end{equation}
	(One also needs to evaluate the $F$ symbols in the anomaly expression, Eq.~\eqref{Eq:AbsAnomNoPerm}, but that contribution is trivial because its exponent is of the form $\omega_2 \cup_1 (\omega_2 \cup_1 \omega_2)$.) This equation can be solved by setting
	\begin{equation}
	\nu_3({\bf g},{\bf h},{\bf k}) = e^{2\pi i \left(\frac{c_-}{8} + n\right){\bf g}({\bf h}+{\bf k}-[{\bf h}+{\bf k}])}~.
	\end{equation}
	where $n \in \Z$ parametrizes the $\H^3(U(1),U(1))$ freedom in choosing $\nu_3$. 
	
	If we consider \textit{bosonic} integer quantum Hall states, we have a similar equation with $c_-=0$. In that case, setting $n=1$ corresponds to $\sigma_H = 2$ \cite{Manjunath2020fqh}. Therefore naively it would seem that the fermionic $c_-=1$ IQH states are described by $\sigma_H = 2n + 1/4$. However, this is incorrect, for the following reason. 
	
	The fermion has unit charge under $G_f$. Now a $\pi$ flux of $G_f$ corresponds to a $2\pi$ flux of $G_b$. Also the charge of the fermion under $G_b$ is measured by braiding it around a $2\pi$ flux of $G_b$, which is the same as a $\pi$ flux of $G_f$. Therefore the fermion has charge 1/2 under $G_b$. 
	
	The Hall conductivity $\sigma_H$ is defined as the $G_f$ charge of a $2\pi$ flux of $G_f$. $\nu_3$ is however measuring a different quantity $\sigma_H^b$, namely the $G_b$ charge of a $2\pi$ flux of $G_b$. We have $\sigma_H^b = \frac{1}{4}\sigma_H$, because 
	\begin{align}
	\sigma_H &= \text{$G_f$ charge of a $2\pi$ flux of $G_f$} \nonumber \\
	&= 2\times\text{$G_f$ charge of a $2\pi$ flux of $G_b$} \nonumber\\
	&= 4\times \text{$G_b$ charge of a $2\pi$ flux of $G_b$} \nonumber\\
	&= 4 \sigma_H^b~.
	\end{align} 
	Therefore we really have $\sigma_H^b = 2n+c_-/4$, which implies $\sigma_H^f = c_-+8n$.
	
	\subsubsection{Invariants}
	We now formally compute $\sigma_H$ from the $G_b$-crossed theory, and address certain mathematical subtleties that arise in the case of $U(1)^f$ symmetry. In Ref.~\onlinecite{Manjunath2020fqh}, invariants for bosonic SET phases with $U(1)$ symmetry were obtained as follows. Fix a defect $a_{\bf g}$, where ${\bf g}^k = {\bf 0}$ for some integer $k$. Then evaluate
	\begin{equation}
	    \mathcal{I}_k(a_{\bf g}) := \theta_{a_{\bf g}}^k \prod\limits_{j=1}^{k-1} \eta_{a_{\bf g}}({\bf g},{\bf g}^j)~.
	\end{equation}
	This quantity is invariant under gauge transformations, for each $k$ and each $a_{\bf g}$. It is however not clear what values of $k$ and $a_{\bf g}$ to choose. 
	
	One resolution is as follows. If we choose a gauge in which $I_{\bf g}$ braids trivially with each anyon for each ${\bf g}$, then we find that $\mathcal{I}_k(I_{\bf g}) = e^{\frac{2\pi i}{k} \left(\frac{c_-}{8}+n\right)} = e^{2\pi i\frac{\sigma_H^b}{k}}$. This invariant does give the value of $\sigma_H^b \mod k$. Thus one solution is to simply choose the gauge in which the chosen ${\bf g}$-defect braids trivially with all the anyons, and then evaluate the invariant for diferent values of $k$. Once $k > \sigma_H^b$, the values of $\sigma_H^b \mod k$ will stabilize, so this can be taken as the Hall conductivity. 
	
	Now we consider what happens if this particular gauge choice is not made. Since there are no anyon permutations, we always have four abelian defects, namely $a_{\bf g} = a \times I_{\bf g}$ for each anyon $a$. In Ref.~\onlinecite{Manjunath2020fqh}, it was also shown that when $a^k = I$, we have
	\begin{equation}
	    \frac{\mathcal{I}_k(a_{\bf g})}{\mathcal{I}_k(I_{\bf g})} = M_{a,m} \theta_a^k~.
	\end{equation}
	If we pick $a = \psi$, we must take $k$ even. The rhs then becomes $M_{\psi,m}\theta_{\psi}^2 = -1$. Thus, in a gauge where we define $\mathcal{I}_k(I_{\bf g}) = e^{i 2\pi \sigma_H^b/k}$, we would have $\mathcal{I}_k(\psi_{\bf g}) = -e^{i 2\pi \sigma_H^b/k} = e^{i 2 \pi \frac{\sigma_H^b + k/2}{k}}$. This invariant gives $\sigma_H^b + k/2 \mod k$ and not $\sigma_H^b \mod k$. Which of the two corresponds to the true Hall conductivity?  
	
	We argue that the second case is a pathology and should not be used to define the Hall conductivity. Suppose $k > \sigma_H^b$. Then, in the usual gauge, the value of $\mathcal{I}_k(I_{\bf g})$ will remain \textit{fixed} at $\sigma_H^b \mod k$, as $k$ is increased. However, the value of $\mathcal{I}_k(\psi_{\bf g})$ will continuously increase with $m$; it is physically impossible for the Hall conductivity to behave in this manner. For the same reason, we do not consider the invariants $\mathcal{I}_k(m_{\bf g})$, which are also gauge invariant quantities within the $G$-crossed BTC, but do not give a fixed value of $\sigma_H^b$ as $k$ is increased. 
	
	In general, the requirement that each $I_{\bf g}$ should braid trivially with each anyon $a$ appears to be more restrictive than is needed for our purpose. Since we are interested in what happens at large values of $k$ (or small values of ${\bf g}$), it is sufficient to demand that the braiding data of $I_{\bf g}$ be continuously connected to the identity as ${\bf g} \rightarrow {\bf 0}$ \cite{Manjunath2020fqh}. This will also ensure that the pathological cases mentioned above are ruled out. 
	
	\subsubsection{Analysis using $K$-matrix Chern-Simons theory}
	Here we will use $U(1)$ Chern-Simons theory to corroborate the results from the previous section. To make things clearer, let us consider $G_b = \Z_n$ with $n$ even, which implies that $G_f = \Z_{2n}^f$. The calculations are identical for $G_f = U(1)^f$, but it is easier to see the main idea by taking a discrete group instead. By taking $n \rightarrow \infty$ we can recover the result for $U(1)^f$.
	
	If we couple an fSPT to a $G_f \cong \Z_{2n}$ gauge field $A$ (which takes values in $\frac{2\pi}{2n}\Z$), we have the following general CS action for the invertible state: 
	\begin{equation}
	\mathcal{L} = -\frac{1}{4\pi} a^I \wedge K_{IJ} da^J + \frac{t_I}{2\pi} a^I \wedge dA~.
	\end{equation}
	Any element $w \in \Z_{2n}$ can be written as a pair $(w_1,w_2)$, where $w_1 \in \Z_2$ and $0 \le w < n$. Specifically, we can write $w_1 = \floor{\frac{w}{n}}$ and $w_2 = w \mod n$. In the same way, we can split the $\Z_{2n}$ gauge field $A$ into a gauge field for its $\Z_2^f$ subgroup, given by $\frac{\floor{2A}}{2}$, and the $\Z_n$ coset component $\frac{\{2A\}}{2}$. Here $\floor{x}$ is the greatest integer less than or equal to the real number $x$, and $\{x\} = x-\floor{x}$. The $\Z_2^f$ gauge field $\frac{\floor{2A}}{2}$ takes values in $\{0,\pi\}$, while $\frac{\{2A\}}{2}$ takes values in $\{0, \pi/n, 2\pi/n, \dots , (n-1)\pi/n\}$, and is not an independent $\Z_n$ gauge field. However, $\{2A\}$ is a well-defined $\Z_n$ gauge field. 
	
	For fSPT phases, we have $K = \begin{pmatrix}
	1 & 0 \\ 0 & -1 \end{pmatrix}$. The above Lagrangian can then be reexpressed as
	\begin{align}\label{Eq:U(1)Kmat}
	\mathcal{L} &= -\frac{1}{4\pi} (a^1 \wedge da^1 - a^2 \wedge da^2) \nonumber \\ &+ \frac{1}{2\pi}(t_1 a^1 + t_2 a^2) \wedge d\left(\frac{\floor{2A}}{2} + \frac{\{2A\}}{2}\right)~.
	\end{align}

	The $\Z_2^f$ subgroup is gauged by introducing the additional term
	\begin{equation}
	\mathcal{L}_1 = -\frac{1}{\pi} \left(\frac{\floor{2A}}{2}\right) \wedge dB~.
	\end{equation}
	Integrating out $\frac{\floor{2A}}{2}$ leads to the constraint $t_1 a^1 + t_2 a^2 = 2B$. Since we want the flux of each $a^i$ to have unit charge under $U(1)^f$, we set $t_1 = t_2 = 1$. The constraint is solved by setting $a^1 = B+C, a^2 = B-C$. Substituting this into the expression for $\mathcal{L}' = \mathcal{L} + \mathcal{L}_1$ gives 
	
	\begin{equation}
	\mathcal{L}' = -\frac{1}{2\pi} ( B \wedge dC + C \wedge dB) + \frac{1}{2\pi} B \wedge \{2A\}~,
	\end{equation} 
	which corresponds to an SET coupled to a $\Z_n$ gauge field $\{2A\}$, with 
	\begin{equation}
	K' = \begin{pmatrix} 0 & 2 \\ 2 & 0
	\end{pmatrix}, t' = \begin{pmatrix}
	1 \\ 0 \end{pmatrix}~.
	\end{equation}
	Crucially, the gauged theory couples the topological order to the well-defined $\Z_n$ gauge field $\{2A\}$, which is valued in $\frac{2\pi}{n}\Z$. The given value of $t'$ corresponds to the symmetry fractionalization class in which $2\pi$ flux insertion induces the anyon $e = (1,0)$. Note that the fermion has a half charge under $\{2A\}$: we have $e^{2\pi i Q_{\psi}^{\{2A\}}} = M_{e,\psi} = -1$, thus $Q_{\psi}^{\{2A\}} = 1/2$. The Hall conductivity $\sigma_H^b$ is constrained to be an even integer, since in such a system $e^{i \pi \sigma^b_H} = \theta_e = 1$.
	
	By relabelling $B$ and $C$, we can show that the SET with symmetry fractionalization anyon $m$ is identical to the SET with symmetry fractionalization anyon $e$. Furthermore, we cannot obtain $t' = \psi$ by starting from Eq.~\eqref{Eq:U(1)Kmat}. Therefore there is only one SET phase obtained by gauging the fSPT with a nontrivial $[\omega_2]$ class. These classification results hold when we replace $G_b = \Z_n$ by $G_b = U(1)$, for example by taking the limit $n \rightarrow \infty$.
	
	Now we generalize this argument to nonzero integers $c_- =k$. We again consider $G_b = \Z_n$, with $\omega_2$ nontrivial. We can repeat the above steps starting with $K = \mathbf{1}_{k \times k}$ and $t = (1,1,\dots , 1)^T$. Upon gauging the fermion parity we obtain the Lagrangian
	\begin{equation}
	\mathcal{L}' =  \sum_I \left(-\frac{1}{4\pi}a^I da^I + \frac{1}{2\pi} a^I d \frac{\{2A\}}{2}\right) ~,
	\end{equation}
	together with the constraint
	\begin{equation}
	a^1 + a^2 + \dots + a^{k} = 2B~.
	\end{equation}
	Upon eliminating $a^1$ and expressing the result in terms of the vector $(B,a^2, \dots , a^{k})$, we get 
	
	\begin{equation}
	K' = \begin{pmatrix}
	4 & -2 & -2 & \dots & -2 & -2 \\
	-2 & 2 & 1 & \dots & 1 & 1 \\
	-2 & 1 & 2 & \dots & \dots &1 \\
	\vdots&.&.&.&.&\vdots \\
	\vdots&.&.&\dots&2&1 \\
	-2&1&1&1&\dots&2 \end{pmatrix}, t' = \begin{pmatrix}
	1 \\0 \\ 0\\\vdots \\ \\0
	\end{pmatrix}~.
	\end{equation}
	First consider the case with $c_-$ odd. Here we can show that the anyon associated to $t'$ is of order 4, thus $t' = v$. This means that (i) the fermion has 1/2 charge under $\{2A\}$, and (ii) the Hall conductivity is obtained through the relation $e^{i \pi \sigma_H^b} = \theta_v = e^{i \pi c_-/4}$. Thus we obtain $\sigma^b_H = c_-/4$. 
	
	Next, consider the case with $c_-$ even. Now we have $t' = m$. This means that (i) the fermion has 1/2 charge under $\{2A\}$, and (ii) the Hall conductivity is obtained through the relation $e^{i \pi \sigma_H^b} = \theta_m = e^{i \pi c_-/4}$. Thus we once again obtain $\sigma_H^b = c_-/4 \mod 2$.

	\subsection{Obstruction to ``halving" a fermion for odd $c_-$}\label{Sec:HalvingEgs}
	In this section we will give examples of a specific $\H^3$ obstruction that can be realized in systems with odd $c_-$. This obstruction arises entirely from the definition of $\omega_2$, i.e. it prohibits the fermion from having certain fractional $G_b$ quantum numbers. This example, with $G_b = SO(3)$, demonstrates that isospin-1/2 fermions cannot form a symmetric invertible phase with odd $c_-$. In Section~\ref{Sec:HalvingZ2Z2}, we consider $G_b = \Z_2\times\Z_2$ with nontrivial $[\om]$ and show how this anomaly may sometimes be cancelled by a suitable definition of $n_1$. 
	\subsubsection{$G_b = SO(3)$}\label{Sec:SO(3)example}
	Let us now consider isospin-1/2 fermions with the symmetry $G_b = SO(3)$, for which we have two possible extensions $G_f$, classified by $\H^2(SO(3),\Z_2) \cong \Z_2$. The nontrivial extension where $G_f = SU(2)^f$ corresponds to the fermion having isospin 1/2. When $c_-$ is odd, this would imply that a $v$ particle in such a system should have ``isospin 1/4" upon gauging the fermion parity. As we now argue, this is mathematically prohibited even for strongly interacting fermions. 

For an isospin-1/2 fermion, the single-particle $SO(3)$ rotation operators by an angle $\theta \in [0,2\pi)$ about the axis $\hat{n}$ are given by
  \begin{equation}
  U(\theta; \hat{n}) = e^{i \frac{\theta}{2} \hat{n} \cdot \vec{\sigma}}~.
  \end{equation}
Let the group element corresponding to a $\pi$ rotation about the $x$ axis be denoted $\tilde{X}$; similarly, define $\tilde{Y}$ and $\tilde{Z}$. The group elements $1, \tilde{X},\tilde{Y},\tilde{Z}$ form a $\Z_2 \times \Z_2$ subgroup of $SO(3)$; for our purposes we can work entirely in this subgroup. A spin-1/2 projective representation of these rotations is given by
\begin{equation}
U(\tilde{X}) = i X, U(\tilde{Y}) = i Y, U(\tilde{Z}) = i Z~,
\end{equation}
where $X,Y,Z$ are the $2\times2$ Pauli matrices. For this representation we compute $\omega_2(\tilde{X},\tilde{Y})$ through the relation

\begin{align}
U(\tilde{X}) U(\tilde{Y}) &= (-1)^{\omega_2(\tilde{X},\tilde{Y})} U(\tilde{X}\tilde{Y}) \\
\implies (i X) (i Y) &= (-1)^{\omega_2(\tilde{X},\tilde{Y})} (iZ) \\
\implies \omega_2(\tilde{X},\tilde{Y})  &= 1~.
\end{align}
We can similarly show that
\begin{align}
 \omega_2(\tilde{X},\tilde{X}) =  \omega_2(\tilde{Y},\tilde{Y}) =  \omega_2(\tilde{Z},\tilde{Z}) &= 1 \nonumber \\
 \omega_2(\tilde{X},\tilde{Y}) =  \omega_2(\tilde{Y},\tilde{Z}) =  \omega_2(\tilde{Z},\tilde{X}) &= 1 \nonumber \\
 \omega_2(\tilde{Z},\tilde{Y}) =  \omega_2(\tilde{X},\tilde{Z}) =  \omega_2(\tilde{Y},\tilde{X}) &= 0~.
\end{align}
The $\Z_2$ invariant in $\{0,1\}$ which characterizes the cohomology class of a 2-cocycle $f_2 \in Z^2(SO(3),\Z_2)$ is given by
\begin{equation}
I_2[f_2] = f_2(\tilde{X},\tilde{X}) + f_2(\tilde{Y},\tilde{Y}) + f_2(\tilde{Z},\tilde{Z}) \mod 2~.
\end{equation} 
Taking $f_2 = \omega_2$, we obtain $I_2[\omega_2] = 1$. This confirms that $\omega_2$ is a nontrivial cocycle.

Now we also have the result $\H^3(SO(3),\Z_2) \cong \Z_2$. Given a cocycle $f_3 \in Z^3(SO(3),\Z_2)$, the invariant which characterizes $[f_3]$ is given by
\begin{equation}
I_3[f_3] = f_3(\tilde{X},\tilde{Y},\tilde{Y}) + f_3(\tilde{Y},\tilde{X},\tilde{Y}) + f_3(\tilde{Y},\tilde{Y},\tilde{X})~.
\end{equation} 
(One can replace $\tilde{X},\tilde{Y}$ in this formula by any two distinct elements in the set $\{\tilde{X},\tilde{Y},\tilde{Z}\}$). Now we have
\begin{align}
&(\omega_2\cup_1 \omega_2)({\bf g},{\bf h},{\bf k}) \nonumber \\ &= \omega_2({\bf g},{\bf h})\omega_2({\bf gh},{\bf k}) + \omega_2({\bf h},{\bf k})\omega_2({\bf g},{\bf hk})~.
\end{align}
Using this, we can verify that
\begin{align}
I_3[\omega_2\cup_1 \omega_2]  &= (\omega_2\cup_1 \omega_2)(\tilde{X},\tilde{Y},\tilde{Y}) \nonumber \\ &+ (\omega_2\cup_1 \omega_2)(\tilde{Y},\tilde{X},\tilde{Y}) + (\omega_2\cup_1 \omega_2)(\tilde{Y},\tilde{Y},\tilde{X}) \nonumber \\
&= 0 + 1 + 0 = 1 \mod 2~.
\end{align}
Therefore $\omega_2 \cup_1 \omega_2$ belongs to the nontrivial class of $\H^3(SO(3),\Z_2)$. If $c_-$ is odd, this means that the equation for $n_2$, which is $dn_2 = \omega_2 \cup_1 \omega_2$, has no solution, because the lhs must be cohomologically trivial by definition, while we just showed that the rhs is not. Physically, this obstruction points to the impossibility of putting the $v$ particle in an ``isospin $1/4$" representation of $SU(2)^f$.

Let us provide another explanation using the effective action of the invertible phase. 
The effective action is expressed in terms of an $SU(2)$-spin background gauge field for a $(SU(2)\times \text{Lorentz})/\mathbb{Z}_2$ bundle. The quotient implies that fermion particles carry half-integer spin under $SU(2)$ symmetry, while boson particles carry integer spin. This is an analogue of the spin/charge relation.
The background gauge field can be expressed as an $SO(3)=SU(2)/\mathbb{Z}_2$ background gauge field with fixed $\mathbb{Z}_2=\pi_1(SO(3))$ magnetic flux given by the cocycle representing the second Stiefel Whitney class $w_2$.
The ``root" effective action is generated by the level one Chern-Simons term for the $SU(2)$ background gauge field subject to the $\mathbb{Z}_2$ quotient, with an additional correction from a gravitational Chern-Simons term.
The $\mathbb{Z}_2$ quotient on the $SU(2)_1$ Chern-Simons term leads to an anomaly described by the bulk effective action
\begin{equation}\label{eqn:SU2-spineffectiveaction}
    2\pi\frac{1}{4}\int {\cal P}(w_2)~,
\end{equation}
which can be obtained from the fractional instanton number of $SU(2)/\mathbb{Z}_2$ bundle in the bulk, or the property that the generator of the $\mathbb{Z}_2$ one-form symmetry in $SU(2)_1$ Chern-Simons theory has spin $1/4$ \cite{Hsin:2018vcg}.
Using the mathematical identities on orientable manifolds \cite{Wu:1954_3,milnor1974characteristic}
\begin{equation}
    {\cal P}(w_2)=p_1+2w_4\text{ mod }4,\quad w_4+w_2^2=0\text{ mod }2
\end{equation}
we find ${\cal P}(w_2)=-p_1$ mod 4, and thus the bulk term (\ref{eqn:SU2-spineffectiveaction}) can be written as
\begin{equation}
    -\frac{1}{48\pi}\int \text{Tr }R\wedge R
\end{equation}
which can be cancelled by a boundary gravitational Chern-Simons term that contributes $c_-=-2$. Thus the chiral central charge of such invertible phases must be even.

\subsection{Cancellation of the ``halving" anomaly with $G_b = \Z_2\times \Z_2$} \label{Sec:HalvingZ2Z2}

Above we saw that there is an $\H^3$ obstruction due only to $\om$ with $G_f = SU(2)^f$. Here we ask if such an obstruction can be cancelled in a separate system that admits $n_1 \ne 0$. Indeed, we find that when $G_f = \mathbb{D}_8^f$ and $c_-$ is odd, the data with $n_1 = 0$ are obstructed as above. However, for a particular choice of $n_1 \ne 0$, the remaining data $n_2$ and $\nu_3$ can both be defined consistently. 

We will denote the elements of $G_b = \Z_2 \times \Z_2$ by ${\bf g_i} = (x_i,y_i) \in \{(0,0),(0,1), (1,0),(1,1)\}$. Let the generators of $G_b$ be given by $\bx = (1,0), \by = (0,1)$. Define $\lambda_x,\lambda_y \in \H^1(G_b,\Z_2)$ so that $\lambda_x({\bf g}_i) = x_i, \lambda_y({\bf g}_i) = y_i$.  
We take $\omega_2=\lambda_x\cup \lambda_y$,
\begin{equation}
\omega_2({\bf g_1}, {\bf g_2}) := x_1 y_2 \mod 2~.
\end{equation}
Physically, this means that the symmetry operators $U_{\mathbf{a}}$ and $U_{\mathbf{b}}$ do not commute when acting on the fermionic Hilbert space:
\begin{equation}
    U_{\mathbf{x}} U_{\mathbf{y}}=U_{\mathbf{y}}U_{\mathbf{x}} (-1)^F~.
\end{equation}
The symmetry $G_f$ is isomorphic to the dihedral group of order 8, $\mathbb{D}_8$. This group is nonabelian,
with $\mathbb{Z}_2^f$ center that corresponds to an order two rotation.

First assume $n_1 = 0$, so that $n_2$ must satisfy
\begin{equation}
dn_2 = \omega_2 \cup_1 \omega_2~.
\end{equation}
With the above choice of $\omega_2$, we obtain
\begin{eqs}
&(\omega_2 \cup_1 \omega_2)({\bf g_1}, {\bf g_2},{\bf g_3}) \\
=& x_1 y_2 (x_1+x_2) y_3 + x_2 y_3 x_1 (y_2 + y_3) \\
=& x_1 (x_2+y_2) y_3~.
\end{eqs}
Note that $\H^3(\Z_2\times\Z_2,\Z_2) \cong \Z_2^4$. The cohomology invariants for a general 3-cocycle $f({\bf g_1},{\bf g_2},{\bf g_3}) \in Z^3(\Z_2\times \Z_2,\Z_2)$ are given by
\begin{eqs}
\mathcal{I}_1[f] &= f(\bx,\bx,\bx) \\
\mathcal{I}_2[f] &= f(\by,\by,\by) \\
\mathcal{I}_3[f] &= f(\bx,\bx,\by) + f(\bx,\by,\bx) + f(\by,\bx,\bx)  \\
\mathcal{I}_4[f] &= f(\bx,\by,\by) + f(\by,\bx,\by) + f(\by,\by,\bx)~,
\end{eqs}
where $\bx = (1,0), \by = (0,1)$. Applying these formulas with $f = \omega_2 \cup_1 \omega_2$, we see that $\mathcal{I}_1 = \mathcal{I}_2 = 0, \mathcal{I}_3 = \mathcal{I}_4 = 1$. Therefore $\omega_2 \cup_1 \omega_2$ is a nontrivial 3-cocycle, and no solution for $n_2$ exists. 

On the other hand, by choosing a suitable $n_1$, this anomaly can be cancelled. Define 
$n_1 = \lambda_x + \lambda_y$.
Then we have
\begin{align}
dn_2 &= n_1 \cup \omega_2 + \omega_2 \cup_1 \omega_2 \\
\implies dn_2({\bf g_1},{\bf g_2},{\bf g_3}) &= (x_1+y_1) x_2 y_3 + x_1(x_2+y_2) y_3 \nonumber \\
&= (y_1 x_2 + x_1 y_2) y_3~.
\end{align}
We can check that the invariants associated to $dn_2$ are $\mathcal{I}_1 = \mathcal{I}_2 = \mathcal{I}_3 = \mathcal{I}_4 = 0$. Indeed, this equation can be solved by setting $n_2({\bf g_1},{\bf g_2}) = x_1 y_1 y_2$, which can also be expressed as
$n_2 = (\lambda_x \cup_1 \lambda_y) \cup \lambda_y$.
The other solutions differ from it by two-cocycles $\lambda_x^2,\lambda_y^2$ and $\lambda_x^2+\lambda_y^2$; moreover, adding the term $\om = \lambda_x\cup \lambda_y$ to $n_2$ does not change it in any physical way due to the equivalence relation. 
Therefore we obtain an $\H^3$ obstruction that prohibits a certain choice of $[\omega_2]$ with $n_1 = 0$, but this obstruction no longer exists with $n_1 = \lambda_x + \lambda_y$. Such a situation was not possible with the continuous $SO(3)$ symmetry in the previous example, because in that case $n_1$ was forced to be the zero function.

\subsubsection{Cancellation of the ${\cal O}_4$ anomaly for $G_b = \ZZ_2 \times \ZZ_2$}\label{sec:Z2xZ2}
We continue on this example of $G_b = \ZZ_2 \times \ZZ_2$. Given $n_1 = \lambda_x + \lambda_y$, $\om = \lambda_x \cup \lambda_y$, and $n_2 = (\lambda_x \cup_1 \lambda_y) \cup \lambda_y$ or $n_2 = (\lambda_x \cup_1 \lambda_y) \cup \lambda_y + \lambda_x \cup \lambda_x + \lambda_y \cup \lambda_y$, we study whether the following cocycle $\Theta_4 \in \H^4(\ZZ_2 \times \ZZ_2, \RR/\ZZ)$ is trivial ($\mathcal{O}_4 \equiv e^{2 \pi i \Theta}$):
\begin{eqs}
     \Theta_4 =& \frac{1}{2}({\cal P}(n_2) + n_2 \cup \om + \zeta(n_1,\omega_2)) \\
     & +\frac{1}{2} (n_1 \cup \om) \cup_2 (\om \cup_1 \om)\\
     & + \frac{1}{4} n_1 \cup n_1 \cup \om + \frac{c_-}{8} {\cal P}(\om),
\end{eqs}
for $c_- = 1,3,5,7~(\text{mod }8)$. $\Theta_4$ has a complicated closed form, so we compute the topological invariants for this $\Theta_4 \in \H^4(\mathbb Z_2 \times \mathbb Z_2,\mathbb R/\mathbb Z)$ to identify the cohomology class, which completely distinguish cocycles in this cohomology group \cite{WL15,T17,CET21}. Explicitly, the invariants of the cohomology group $\H^4(\ZZ_{2} \times \ZZ_{2}, \RR/\ZZ) \cong \ZZ_{2}^2$ are given by:
\begin{eqs}
    \mathcal I_1 =&\sum_{k=0}^{1} i_{(1,0)} \Theta_4 \boldsymbol{(}(0,1),(0,k),(0,1)\boldsymbol{)}, \nonumber\\
    \mathcal I_2 =& \sum_{k=0}^{1} i_{(0,1)} \Theta_4 \boldsymbol{(}(1,0),(k,0),(1,0)\boldsymbol{)}. \nonumber
\end{eqs}
where $i_{g} \Theta_4$ is the slant product of $g$ and $\Theta_4$ and is defined as
\begin{eqs}
  i_{g}\Theta_4(x,y,z)&=\Theta_4(g,x,y,z)-\Theta_4(x,g,y,z) \\ &+\Theta_4(x,y,g,z)-\Theta_4(x,y,z,g)~.
\end{eqs}
The quantities $\mathcal I_1$ and $\mathcal I_2$ above are fixed under adding any coboundary to $\Theta_4$. We use Mathematica to compute the invariants numerically and obtain Table~\ref{Table: c I1 I2}. We find that for all odd $c_-$, a solution always exists. This is the smallest group we know of for which all data $(c_-, n_1, n_2, \om, \nu_3)$ are all non-vanishing.

We remark that for the case with $c_-=1$, {\it i.e.} $U(1)_4$ with $\mathbb{Z}_2\times\mathbb{Z}_2$ symmetry, the anomaly is also studied in Ref.~\onlinecite{Bulmash2020} for different symmetry fractionalization classes, by constructing a state sum model for the bulk SPT phase that describes the anomaly.
The
generators of the $\mathbb{Z}_2\times\mathbb{Z}_2$ symmetry are denoted in Ref.~\onlinecite{Bulmash2020} by $\mathbf{X}= U_{{\bf x}},\mathbf{Z}=U_{{\bf x}}U_{{\bf y}}$; $\mathbf{X}$ permutes the anyons, while $\mathbf{Z}$ does not. 
The absence of anomaly found here is consistent with Ref.~\onlinecite{Bulmash2020}, where
our example corresponds to the symmetry fractionalization class described by $(1,-1,1)$ in Table II of Ref.~\onlinecite{Bulmash2020}. 

\begin{table}[ht]
\renewcommand{\arraystretch}{1.5}
\begin{tabular}{|c|c|c|}
    \hline
    $c_-$ & $n_2  = (\lambda_x \cup_1 \lambda_y) \cup \lambda_y$ & $n_2 = (\lambda_x \cup_1 \lambda_y) \cup \lambda_y + \lambda_x^2 + \lambda_y^2$  \\ 
    \hline
    $1$ & $\mathcal I_1=\frac{1}{2}$, $\mathcal I_2=\frac{1}{2}$ & $\mathcal I_1=0$, $\mathcal I_2=0$ \\
    \hline
    $3$ & $\mathcal I_1=0$, $\mathcal I_2=0$ & $\mathcal I_1=\frac{1}{2}$, $\mathcal I_2=\frac{1}{2}$ \\
    \hline
    $5$ & $\mathcal I_1=\frac{1}{2}$, $\mathcal I_2=\frac{1}{2}$ & $\mathcal I_1=0$, $\mathcal I_2=0$ \\
    \hline
    $7$ & $\mathcal I_1=0$, $\mathcal I_2=0$ & $\mathcal I_1=\frac{1}{2}$, $\mathcal I_2=\frac{1}{2}$ \\
    \hline
\end{tabular}
\caption{\label{Table: c I1 I2}The invariants for the $\Theta_4 \in \H^4(\ZZ_2 \times \ZZ_2, \RR/\ZZ)$. For all values of odd $c_-$, there always exists a choice of $n_2$ such that the $\mathcal O_4$ anomaly is trivial.}
\end{table}

\subsection{Periodic table of topological insulators and superconductors}\label{Sec:PdTable}

The classification of topological insulators and topological superconductors of free fermions with different choices of onsite symmetry classes (known as the 'tenfold way') was a milestone result in the study of invertible topological phases of matter \cite{Altland1997,kitaev2009,Ryu_2010} (see Ref.~\onlinecite{Chiu2016review} for a review). Here we use the results stated in Section~\ref{Sec:Summary} to find the analogous classification for interacting fermion phases. The results are given in Table~\ref{Table:10fold}. \footnote{A similar table was presented in a talk by Q.R. Wang; see URL: https://cmsa.fas.harvard.edu/7-22-2020-quantum-matter-seminar/} We verify the previously known result (see {\it e.g.} Ref.~\onlinecite{Morimoto2015}) that every free fermion invertible phase in (2+1) dimensions survives when interactions are added. Moreover, in class A and class C, the interacting classification gets an additional factor isomorphic to $\Z$. As we will show in the following, the symmetry classes AI, AIII, BDI, CI and CII have trivial classification in our approach, while the classes A, AII, D, DIII and C can have nontrivial classification.

\subsubsection{Symmetry classes with trivial classification}
There are 5 symmetry classes (AI, AIII, BDI, CI and CII) for which the interacting classification is trivial. We briefly discuss how to obtain this result below. Note that each of these systems has an antiunitary time-reversal symmetry, with $s_1({\bf T}) = 1$. If the system is to be non-anomalous, we must set $c_-=0 \mod 8$. Additionally, we must set $n_1 = 0$. For these 5 classes, we have $\H^1(G_b,\Z_2) \cong \Z_2$. The nontrivial choice of $n_1$ has $n_1({\bf T}) = 1$. In each of these cases, we obtain $dn_2({\bf T},{\bf T},{\bf T}) = n_1({\bf T})s_1({\bf T})n_1({\bf T}) = 1$. This equation has no solution, implying that $n_2$ is obstructed. Therefore we will assume $n_1 = 0$ and $dn_2 = 0$ for the rest of this section.

\textbf{Class AI}: We have $G_b = U(1)\rtimes\Z_2^T$. Denote a general group element as ${\bf g_i} = (z_i, {\bf T_i})$ where $z_i \in \R/\Z, {\bf T_i} \in \Z_2^T$. Then the group law is
\begin{equation}
(z_1,{\bf T_1}) (z_2,{\bf T_2}) = (z_1+ (-1)^{{\bf T_1}} z_2,{\bf T_1 T_2})~.
\end{equation}
The $U(1)$ subgroup extends $\Z_2^f$ nontrivially: we can write
\begin{equation}
    \omega_2({\bf g_1},{\bf g_2}) = z_1 + (-1)^{{\bf T_1}} z_2 - [z_1 + (-1)^{{\bf T_1}} z_2]~. 
\end{equation}
We also have $\H^1(G_b,\Z_2) \cong \Z_2,\H^2(G_b,\Z_2) \cong \Z_2^2,\H^3(G_b,U(1)_T) \cong \Z_2$. A general 2-cocycle $f_2$ is defined as follows:
\begin{equation}
f_2({\bf g_1},{\bf g_2}) = a(z_1 + (-1)^{{\bf T_1}} z_2 - [z_1 + (-1)^{{\bf T_1}} z_2]) + b {\bf T_1} {\bf T_2}~,
\end{equation} 
where ${\bf g_i} = (z_i,{\bf T_i})$ and $a,b \in \Z_2$.

Taking $n_1=0$, there are two inequivalent choices for $[n_2]$: either $n_2 = 0$ or 
\begin{equation}
n_2({\bf g_1},{\bf g_2}) = {\bf T_1} {\bf T_2}~.
\end{equation}  
In the latter case, the obstruction class $[\mathcal{O}_4]$ is nontrivial, so there is no solution for $\nu_3$. In the former case, there are two solutions for $\nu_3$ given by $[\nu_3] \in \H^3(G_b,U(1)_T) \cong \Z_2$. However, the nontrivial cocycle can be written as
\begin{equation}
    \nu_3({\bf g_1},{\bf g_2},{\bf g_3}) = (-1)^{(z_1 + (-1)^{{\bf T_1}} z_2 - [z_1 + (-1)^{{\bf T_1}} z_2]) {\bf T}_3}~.
\end{equation}
This cocycle has the form $\nu_3 = (-1)^{\omega_2 \cup n_1}$, and therefore it can be trivialized by relabelling defects. We conclude that there is no nontrivial interacting TI in class AI.	
	
\textbf{Class AIII}: We have $G_b = U(1)\times\Z_2^T$. The $U(1)$ subgroup extends $\Z_2^f$ nontrivially. We also have $\H^2(G_b,\Z_2) \cong \Z_2^2,\H^3(G_b,U(1)_T) \cong \Z_1$. 

Let ${\bf g_i} = (z_i, {\bf T_i})$ where $z_i \in U(1), {\bf T_i} \in \Z_2^T$. Taking $n_1 = 0$, there are two inequivalent choices for $[n_2]$: either $n_2 = 0$ or 
\begin{equation}
n_2({\bf g_1},{\bf g_2}) = {\bf T_1} {\bf T_2}~.
\end{equation}  
In the latter case, the obstruction class $[\mathcal{O}_4]$ is nontrivial, so there is no solution for $\nu_3$. In the former case, we have one allowed solution (the trivial insulator) which can be summarized by the data $n_1 = 0, n_2 = 0$ and $\nu_3 = 1$. 
	
\textbf{Class BDI}: We have $G_f = \Z_2^f\times\Z_2^T$. This example was studied previously in Ref.~\onlinecite{Wang2020fSPT}. We have $\H^1(G_b,\Z_2) \cong \Z_2,\H^2(G_b,\Z_2) \cong \Z_2,\H^3(G_b,U(1)_T) \cong \Z_1$. The nontrivial choices of $n_1$ and $n_2$ are both obstructed, while there are no nontrivial $\H^3$ elements.

\textbf{Class CI and CII}: In both cases we have $G_b = \Z_2^T\times SO(3)$, and a general element can be denoted as ${\bf g_i} = (r_i, {\bf T_i})$ where $r_i \in SO(3), {\bf T_i} \in \Z_2^T$.  We also have $\H^1(G_b,\Z_2) \cong \Z_2,\H^2(G_b,\Z_2) \cong \Z_2^2,\H^3(G_b,U(1)_T) \cong \Z_2$. The nontrivial $\H^1$ element is associated to the time-reversal symmetry. The nontrivial $\H^2$ elements are independently associated to the $\Z_2^T$ and $SO(3)$ symmetries. The difference in the two classes co    mes from the definition of $\omega_2$. In class CI, both the $\Z_2^T$ and $SO(3)$ subgroups nontrivially extend $\Z_2^f$. In class CII, it is only the $SO(3)$ subgroup that nontrivially extends $\Z_2^f$.

The nontrivial choice of $n_1$ is obstructed for the reason discussed previously. After accounting for the equivalence $n_2 \simeq n_2 + \omega_2$, we have one nontrivial choice of $n_2$, corresponding to $n_2({\bf g_1},{\bf g_2}) = {\bf T_1} {\bf T_2}.$ However, $\nu_3$ is obstructed in this case. Thus $n_2$ must be trivial. Finally, it turns out that the two distinct $\H^3$ classes defining $\nu_3$ are physically equivalent: in either case, the nontrivial choice of $\nu_3$ can be put the form $\nu_3 = (-1)^{\omega_2 \cup n_1}$. Thus in either case, we obtain a trivial classification.

\subsubsection{Symmetry classes with nontrivial classification} 

\textbf{Class A}:
Class A systems have $G_b = U(1)$. Here a nontrivial chiral central charge is allowed, but we must have $c_- \in \Z$. This is because $\omega_2$ is nontrivial: a $2\pi$ rotation of $G_b$ equals $(-1)^F$. As we showed in Section~\ref{Sec:U(1)symm}, each integer value of $c_-$ is allowed. We have $\H^1(G_b,\Z_2) \cong \Z_1, \H^2(G_b,\Z_2) \cong \Z_2,\H^3(G_b,U(1)) \cong \Z$. The two choices for $n_2$ are equivalent, and so for a given $c_-$ we obtain a $\Z$ classification of IQH states having $\sigma_H = c_- + 8n$ ($\sigma_H$ being measured with respect to $U(1)^f$). The overall classification is parametrized by two integers $(c_-, n = \frac{\sigma_H-c_-}{8})$ which we can independently choose, thus the full classification is $\Z^2$. Note the difference between the free and interacting classifications: in the former case, the only invariant is $c_- \in \Z$. Although IQH states of free fermions have a quantized Hall conductivity due to the $U(1)^f$ symmetry, we have $\sigma_H \propto c_-$, so the two numbers are not independent. Moreover, when this symmetry is broken, $\sigma_H$ becomes undefined, however the state remains nontrivial because $c_-$ is still well defined. 

\textbf{Class AII}:
We next consider Class AII, which predicts the $\Z_2$ quantum spin Hall state for free fermions. The bosonic symmetry of the QSH insulator is $G_b = U(1)_b\rtimes \Z_2^T$. The full symmetry is specified by the conditions $U^b_{2\pi} = (-1)^F$ (i.e. a $2\pi$ rotation by the $U(1)_b$ symmetry gives a sign change on the single-particle     state) and ${\bf T}^2 = (-1)^F$. These conditions are incorporated into the definition of $\omega_2$, as follows. We find that $\H^2(G_b,\Z_2) \cong \Z_2 \times \Z_2$. For a general cocycle $f_2$ we can define the invariants $f_2(\pi,\pi) \in \{0,1\}$ (here $\pi$ refers to a $\pi$ rotation by $U(1)_b$) and $f_2({\bf T},{\bf T}) \in \{0,1\}$. With these definitions, we see that $\omega_2$ is defined by the invariants $\omega_2(\pi,\pi) = \omega_2({\bf T},{\bf T}) = 1$.

 Since the system has an antiunitary symmetry, we set $c_- = 0$. In the corresponding symmetry-enriched toric code, having $\omega_2(\pi,\pi) = 1$ means that a $2\pi$ flux of $U(1)_b$ induces a fermion parity flux, so that the fermion has a half charge under $U(1)_b$. Having $\omega_2({\bf T},{\bf T}) = 1$ means that a fermion carries a local Kramers degeneracy \cite{Barkeshli2020Anomaly}. 

Next we discuss the classification data. We have $\H^1(G_b,\Z_2) \cong \Z_2, \H^2(G_b,\Z_2) \cong \Z_2^2, \H^3(G_b,U(1)_T) \cong \Z_2$. The nontrivial choice of $n_1$ (corresponding to $n_1({\bf T}) = 1$) is obstructed on account of the $n_1 \cup s_1 \cup n_1$ term. Thus we set $n_1 = 0$. 

Just as for $\omega_2$, the possible choices of $n_2$ are specified by the invariant quantities $n_2(\pi,\pi) \in \{0,1\}$ and $n_2({\bf T},{\bf T}) \in \{0,1\}$. Having $n_2(\pi,\pi) = 1$ implies that inserting $2\pi$ flux of $U(1)_b$ (or equivalently, $\pi$ flux of $U(1)^f$) changes the fermion parity of the ground state by 1. In the bosonic shadow, this means that an $m$ particle has a half charge under $U(1)_b$. Having $n_2({\bf T},{\bf T})$ implies that inserting a double crosscap (a pair of time-reversal symmetry defects) changes the fermion parity of the ground state by 1. In the bosonic shadow, this means that an $m$ particle carries a local Kramers degeneracy.

After accounting for the equivalence $n_2 \simeq n_2 + \omega_2$, we obtain two nontrivial choices for $n_2$, namely $n_2(\pi,\pi) = n_2({\bf T},{\bf T}) = 0$ and $n_2(\pi,\pi) = 0, n_2({\bf T},{\bf T}) = 1$. We now argue that the second choice corresponds to the QSH insulator. Since the $U(1)_b$ symmetry nontrivially extends fermion parity, a $\pi$ flux of the $U(1)^f$ symmetry induces an $m$ particle. This particle also carries a local Kramers degeneracy, because $n_2({\bf T},{\bf T}) = 1$. This property is the signature of the bosonic shadow of the QSH insulator. There is an equivalent choice $n_2(\pi,\pi) = 1, n_2({\bf T},{\bf T}) = 0$. This also has a direct interpretation within the QSH insulator: inserting a $\pi$ flux of $U(1)^f$ changes the fermion parity of the ground state. 

Finally, we consider $\H^3(G_b,U(1)_T) \cong \Z_2$. This factor is trivialized, because a cocycle in the nontrivial $\H^3$ class can be written in the form $(-1)^{\omega_2 \cup n_1}$. Therefore, the final classification is $\Z_2$, the same as for free fermions.

\textbf{Class D} The systems in class D correspond to chiral $p+ip$ superconductors characterized by an integer number $\nu$ of chiral Majorana edge states (or equivalently, by their chiral central charge $c_- = \nu/2$). There is no external symmetry, so the full classification is given by $\Z$, and the interacting classification is identical to the free classification.

\textbf{Class DIII}: Next we look at systems in class DIII, for which $G_f = \Z_4^{Tf}$. The free fermion classification predicts a nontrivial state corresponding to ``time-reversal invariant topological superconductors" with ${\bf T}^2 = (-1)^F$. This system is characterized by a Kramers doublet of Majorana bound states at its edge \cite{Qi2009TRS-TSC,Zhang2013MKP}. The ${\bf T}$ operation applied in the neighbourhood an MBS changes its fermion parity. This system was studied in detail previously in Ref.~\onlinecite{Wang2020fSPT}, and we present the results here for completeness. Although $n_1, s_1$ and $\omega_2$ are all nontrivial, the combination $\omega_2 + s_1 \cup n_1$ is coboundary equivalent to the zero function. Therefore $n_2$ is unobstructed, and we can set $dn_2 = 0$. Now we have $\H^2(G_b,\Z_2) \cong \Z_2$, but the two choices of $n_2$ are equivalent under $n_2 \simeq n_2 + \omega_2$. Finally, we have $\H^3(G_b,U(1)_T) \cong \Z_1$. Therefore the final classification is $\Z_2$, and is set by the choice of $n_1$. The fact that the nontrivial insulator has $n_1({\bf T}) = 1$, so that ${\bf T}$-defects are nonabelian, is consistent with our physical expectation, namely that the edge of the system hosts Majorana zero modes that also transform as a Kramers pair.  

\textbf{Class C}: As shown in Section~\ref{Sec:SO(3)example},
spin-1/2 fermionic systems with $SO(3)$ symmetry must have even chiral central charge. We must also set $n_1 = 0$, while the only choices for $n_2$ are $n_2 = 0$ and $n_2 = \omega_2 \simeq 0$. The different choices of $\nu_3$ are parametrized by \textit{even} integers. This is because the spin Hall conductance $\sigma_S$, which is the defining response property of bosonic SPTs with $SO(3)$ symmetry, is quantized to even integer multiples of the elementary unit of spin Hall conductance \cite{Liu2013}. The full classification of spin-1/2 fSPTs with $G_b = SO(3)$ is therefore given by $(c_-, \sigma_S) \in 2\Z\times 2\Z$, where both numbers are even integers. In the free fermion case, there is only one $2\Z$ factor coming from $c_-$. Therefore the free and interacting classifications are different.

\section{Discussion}\label{Sec:Discussion}

In our work we first gauged the fermion parity, and then studied the resulting $G_b$-crossed theory. It is reasonable to consider directly studying the $G_f$-crossed theory for the invertible phase without gauging the fermion parity. We briefly sketch that analysis below. Recall that each element of $G_f$ is given by a pair $({\bf g},a)$ where ${\bf g} \in G_b,a \in \Z_2^f$. The group law for $G_f$ is
\begin{equation}
    ({\bf g_1},a_1) ({\bf g_2},a_2) = ({\bf g_1g_2},a_1+a_2+\omega_2({\bf g_1},{\bf g_2}))~.
\end{equation}
The anyons (i.e. the $({\bf 0},0)$ defects) consist of the identity particle and the fermion $\psi$. The nature of the $({\bf 0},1)$ defects depends on the value of $c_-$. If $c_-$ is an integer, there are two abelian defects $I_{({\bf 0},1)},\psi_{({\bf 0},1)}$. If $c_-$ is a half-integer, there is only one non-Abelian defect, given by $\sigma_{({\bf 0},1)}$. These defects are the same as the fermion parity fluxes in the $G_b$-crossed theory.

Note that there is no anyon permutation in this formulation. The function $n_1$ instead appears as a choice on whether the $({\bf g},0)$ defect is abelian or not. In fact, the label $L_{\bf g}$ in the $G_b$-crossed theory is given to the descendant of the $({\bf g},0)$ defect in the $G_f$-crossed theory.

The symmetry fractionalization can be expressed in terms of the defect fusion rules as follows:
\begin{align}
    &b_{({\bf g_1},a_1)} \times c_{({\bf g_2},a_2)} = \psi^{n_2({\bf g_1},{\bf g_2})}\sum_{e}N_{a b}^e e_{({\bf g_1g_2},\omega_2({\bf g_1},{\bf g_2}))} \\
    &= \psi^{n_2({\bf g_1},{\bf g_2})}\times (I_{({\bf 0},1)})^{\omega_2({\bf g_1},{\bf g_2})}\sum_{e}N_{a b}^e e_{({\bf g_1g_2},0)}~.
\end{align}
This fusion rule appears very similar to what was obtained in the $G_b$-crossed theory: we simply identify $I_{({\bf 0},1)}$ with a fermion parity flux. Note that $\omega_2$ is no longer a variable symmetry fractionalization parameter: the only such parameter is $n_2$. 

Since in this approach we directly study the invertible phase, we can in principle obtain an accurate counting of phases without needing to add any extra information. However, in order to compute the $\H^4$ anomaly, we need to gauge fermion parity and study the anomalies of the resulting bosonic theory, which is ultimately the approach of this paper.

There are several open issues that have arisen from our work which deserve further study. 
When $G_b$ is unitary, and $[\om] \ne [0]$, our stacking rule for $n_2$ and $\nu_3$ are determined only up to possible 2- and 3-cocycles which we have conjectured to vanish. When $G_b$ is anti-unitary, the stacking rule for $n_2$ is also only known up to a 2-cocycle which we have conjectured to vanish; and we do not compute the stacking rule for $\nu_3$. To fully derive the stacking rules, it would be useful to understand how to obtain the full $G$-crossed data of a theory after anyon condensation, which we leave for future work.

Furthermore, we used a relative anomaly formula to compare the desired anomaly of our system to that of some reference state. We have not explicitly proven that the reference is non-anomalous when $c_- = 0$, $n_1, s_1 \neq 0$, although we have verified this in some nontrivial examples and believe this to be the case generally. 

If we consider antiunitary symmetries, $s_1 \ne 0$, relabelling the $e$ and $m$ particles leads to an equivalence $n_2 \simeq n_2 + \om$ (with an associated equivalence on $\nu_3$). Formally, it is also possible to obtain an equivalence $n_2 \simeq n_2 + \om + s_1 \cup n_1$. For purely bosonic $G_b$-enriched SET phases, both equivalences hold. However, in order to obtain the second equivalence in the $G_b$-crossed theory, one needs to perform a certain gauge transformation $\Gamma^{\psi \psi}=-1$ on the data that is forbidden (see Appendix \ref{sec:gtappendix}). On the other hand, the first equivalence can be obtained without requiring such gauge transformations, therefore we expect it to hold equally for fermionic phases. The argument that $\Gamma^{\psi \psi}=-1$ is forbidden is based on observing several inconsistencies when it is allowed; however it would be illuminating to find a more physically motivated argument that this is a forbidden transformation.

Finally, we have explicitly shown the existence of a $c_-=1$ state which has nontrivial $n_1$ and $\om$, and is also non-anomalous. Naively, we might expect that Majorana zero modes localized at symmetry defects and fermions with fractional $G_b$ quantum numbers should not coexist. This is because a Majorana zero mode can absorb the fermion, leading to the non-conservation of the $G_b$ quantum number. However, the above example appears to evade this naive reasoning. The resolution is that the $G_b$ quantum number is simply not well defined for the symmetry defects in this case, just as the fermion parity is not well defined for the vortices of a $p+ip$ superconductor. More precisely, the fractional quantum number of $\psi$ is in this case defined by the property that the generators ${\bf x},{\bf y}$ of $G_b$ anticommute on fermions, due to the relation $\om({\bf x},{\bf y})-\om({\bf y},{\bf x}) = 1$. But since ${\bf x},{\bf y}$ both permute the symmetry defects, it is not meaningful to define such an (anti)commutator for the defects in a gauge-invariant manner using the $G_b$-crossed data.

\section*{Acknowledgements}

We thank Anton Kapustin, Parsa Bonderson, Daniel Bulmash, and Meng Cheng for discussions. Y.-A. C thanks Nathanan Tantivasadakarn for advice on numerical calculations of the obstruction, and Dave Aasen for conceptual discussions. MB is supported by NSF CAREER (DMR- 1753240) and JQI-PFC-UMD. Y.-A. C is supported by the JQI fellowship at the University of Maryland. The work of P.-S.\ H.\ is supported by the U.S. Department of Energy, Office of Science, Office of High Energy Physics, under Award Number DE-SC0011632, and by the Simons Foundation through the Simons Investigator Award.  

\it Note added: \rm As this work was being completed, we also became aware of parallel independent work that overlaps with our results \cite{aasen2021characterization}, which was posted to the arxiv simultaneously with v1 of this paper.
A second paper, Ref.~\cite{ning2021enforced}, also was posted to the arxiv shortly after our paper which overlaps with some of our results. After this paper was posted to the arxiv, we learned that the stacking rules for fermionic SPTs ($c_- = 0$) were also independently derived using the fixed point wave function method \cite{Wang_Gu_unpublished} and agree with our Table \ref{Table:MainEqs}.  

We thank Qing-Rui Wang and David Aasen for pointing out some errors in our previous presentation of the $\nu_3$ stacking rules. An updated version of Ref.~\cite{aasen2021characterization} proves the stacking rules for $n_2$ in the case of unitary symmetries, and their results agree with our (corrected) stacking rules.
	
\addtocontents{toc}{\protect\setcounter{tocdepth}{1}}
\appendix

\section{Brief review of $G$-crossed BTCs}\label{Sec:Gxreview}
In this section we review some results from the $G$-crossed BTC theory developed in Ref.~\onlinecite{Barkeshli2019,Barkeshli2020Anomaly}. We will only discuss those results that are relevant for this work. In particular, we will not require a full theory of the fusion and braiding of symmetry defects. Strictly speaking, the theory in this section applies to bosonic SET phases. The modifications that are required to describe invertible fermion phases are discussed in Section~\ref{Sec:Counting}.
	
\subsection{Overview}

The $G$-crossed BTC is a mathematical object that takes as input two pieces of data: (i) the data of a topological order $\C$ whose objects are anyons, and which is represented as a unitary modular tensor category (UMTC); and (ii) a global symmetry $G$. The UMTC is specifed by a consistent set of $F$ and $R$ symbols for the anyons. The action of the symmetry on the anyons is specified by certain additional data, including a set of $U$ symbols and a set of $\eta$ symbols for each anyon (to be defined below). Thus the full symmetry action on the anyons is specified by a consistent set of $\{F,R,U,\eta\}$ symbols. 

Apart from anyons, the $G$-crossed BTC also contains a set of objects called \textit{symmetry defects}, which are also associated to a consistent set of $\{F,R,U,\eta\}$ symbols. All the properties of the symmetry-enriched topological phase, including the classification and a full set of topological invariants, can in principle be deduced from the above data. If there is no consistent specification of $\{F,R,U,\eta\}$, we say that there is an obstruction to extending the topological order by the global symmetry, i.e. we cannot consistently define an SET phase in 2+1 dimensions. There are different types of obstructions (also referred to as anomalies), as we will see below. 

In this work, define three data elements that specify the bosonic shadow of an invertible fermion phase, referred to as $\tilde{n}_1, \tilde{n}_2$, and $\nu_3$. This notation is adapted from \cite{Wang2020fSPT}. It will be useful to keep in mind how these data arise in the $G_b$-crossed BTC: $\tilde{n}_1$ specifies how the symmetry permutes the anyons, $\tilde{n}_2$ specifies the symmetry fractionalization on the anyons, and $\nu_3$ specifies the $\{F,R,U,\eta\}$ symbols of the symmetry defects.

\subsection{UMTC notation}

Here we briefly review the notation that we use to describe UMTCs. For a more comprehensive review of the notation, see e.g. \cite{Barkeshli2019}. The topologically 
nontrivial quasiparticles of a (2+1)D topologically ordered state are equivalently referred to
as anyons, topological charges, and quasiparticles. In the category theory terminology, they correspond
to isomorphism classes of simple objects of the UMTC. 

A UMTC $\mathcal{C}$ contains splitting spaces $V_{c}^{ab}$, and their dual fusion spaces, $V_{ab}^c$,
where $a,b,c \in \mathcal{C}$ are the anyons. These spaces have dimension 
$\text{dim } V_{c}^{ab} = \text{dim } V_{ab}^c = N_{ab}^c$, where $N_{ab}^c$ are referred
to as the fusion rules. They are depicted graphically as: 
\begin{equation}
\left( d_{c} / d_{a}d_{b} \right) ^{1/4}
\pspicture[shift=-0.6](-0.1,-0.2)(1.5,-1.2)
  \small
  \psset{linewidth=0.9pt,linecolor=black,arrowscale=1.5,arrowinset=0.15}
  \psline{-<}(0.7,0)(0.7,-0.35)
  \psline(0.7,0)(0.7,-0.55)
  \psline(0.7,-0.55) (0.25,-1)
  \psline{-<}(0.7,-0.55)(0.35,-0.9)
  \psline(0.7,-0.55) (1.15,-1)	
  \psline{-<}(0.7,-0.55)(1.05,-0.9)
  \rput[tl]{0}(0.4,0){$c$}
  \rput[br]{0}(1.4,-0.95){$b$}
  \rput[bl]{0}(0,-0.95){$a$}
 \scriptsize
  \rput[bl]{0}(0.85,-0.5){$\mu$}
  \endpspicture
=\left\langle a,b;c,\mu \right| \in
V_{ab}^{c} ~,
\label{eq:bra}
\end{equation}
\begin{equation}
\left( d_{c} / d_{a}d_{b}\right) ^{1/4}
\pspicture[shift=-0.65](-0.1,-0.2)(1.5,1.2)
  \small
  \psset{linewidth=0.9pt,linecolor=black,arrowscale=1.5,arrowinset=0.15}
  \psline{->}(0.7,0)(0.7,0.45)
  \psline(0.7,0)(0.7,0.55)
  \psline(0.7,0.55) (0.25,1)
  \psline{->}(0.7,0.55)(0.3,0.95)
  \psline(0.7,0.55) (1.15,1)	
  \psline{->}(0.7,0.55)(1.1,0.95)
  \rput[bl]{0}(0.4,0){$c$}
  \rput[br]{0}(1.4,0.8){$b$}
  \rput[bl]{0}(0,0.8){$a$}
 \scriptsize
  \rput[bl]{0}(0.85,0.35){$\mu$}
  \endpspicture
=\left| a,b;c,\mu \right\rangle \in
V_{c}^{ab}~,
\label{eq:ket}
\end{equation}
where $\mu=1,\ldots ,N_{ab}^{c}$, $d_i$ is the quantum dimension of anyon $i$ ($i=a,b,c$), 
and the factors $\left(\frac{d_c}{d_a d_b}\right)^{1/4}$ are a normalization convention for the diagrams. 

We denote $\bar{a}$ as the topological charge conjugate of $a$, for which
$N_{a \bar{a}}^1 = 1$, i.e.
\begin{align}
a \times \bar{a} = 1 +\cdots~.
\end{align}
Here $1$ refers to the identity particle, i.e. the vacuum topological sector, which physically describes all 
local, topologically trivial excitations. 

The $F$ symbols are defined as the following basis transformation between the splitting
spaces of $4$ anyons:
\begin{equation}
  \pspicture[shift=-1.0](0,-0.45)(1.8,1.8)
  \small
  \psset{linewidth=0.9pt,linecolor=black,arrowscale=1.5,arrowinset=0.15}
  \psline(0.2,1.5)(1,0.5)
  \psline(1,0.5)(1,0)
  \psline(1.8,1.5) (1,0.5)
  \psline(0.6,1) (1,1.5)
   \psline{->}(0.6,1)(0.3,1.375)
   \psline{->}(0.6,1)(0.9,1.375)
   \psline{->}(1,0.5)(1.7,1.375)
   \psline{->}(1,0.5)(0.7,0.875)
   \psline{->}(1,0)(1,0.375)
   \rput[bl]{0}(0.05,1.6){$a$}
   \rput[bl]{0}(0.95,1.6){$b$}
   \rput[bl]{0}(1.75,1.6){${c}$}
   \rput[bl]{0}(0.5,0.5){$e$}
   \rput[bl]{0}(0.9,-0.3){$d$}
 \scriptsize
   \rput[bl]{0}(0.3,0.8){$\alpha$}
   \rput[bl]{0}(0.7,0.25){$\beta$}
  \endpspicture
= \sum_{f,\mu,\nu} \left[F_d^{abc}\right]_{(e,\alpha,\beta)(f,\mu,\nu)}
 \pspicture[shift=-1.0](0,-0.45)(1.8,1.8)
  \small
  \psset{linewidth=0.9pt,linecolor=black,arrowscale=1.5,arrowinset=0.15}
  \psline(0.2,1.5)(1,0.5)
  \psline(1,0.5)(1,0)
  \psline(1.8,1.5) (1,0.5)
  \psline(1.4,1) (1,1.5)
   \psline{->}(0.6,1)(0.3,1.375)
   \psline{->}(1.4,1)(1.1,1.375)
   \psline{->}(1,0.5)(1.7,1.375)
   \psline{->}(1,0.5)(1.3,0.875)
   \psline{->}(1,0)(1,0.375)
   \rput[bl]{0}(0.05,1.6){$a$}
   \rput[bl]{0}(0.95,1.6){$b$}
   \rput[bl]{0}(1.75,1.6){${c}$}
   \rput[bl]{0}(1.25,0.45){$f$}
   \rput[bl]{0}(0.9,-0.3){$d$}
 \scriptsize
   \rput[bl]{0}(1.5,0.8){$\mu$}
   \rput[bl]{0}(0.7,0.25){$\nu$}
  \endpspicture
.
\end{equation}
To describe topological phases, these are required to be unitary transformations, i.e.
\begin{widetext}
\begin{eqnarray}
\left[ \left( F_{d}^{abc}\right) ^{-1}\right] _{\left( f,\mu
,\nu \right) \left( e,\alpha ,\beta \right) }
= \left[ \left( F_{d}^{abc}\right) ^{\dagger }\right]
_{\left( f,\mu ,\nu \right) \left( e,\alpha ,\beta \right) } 
= \left[ F_{d}^{abc}\right] _{\left( e,\alpha ,\beta \right) \left( f,\mu
,\nu \right) }^{\ast }
~.
\end{eqnarray}
\end{widetext}

The $R$ symbols define the braiding properties of the anyons, and are defined via the the following
diagram:
\begin{equation}
\pspicture[shift=-0.65](-0.1,-0.2)(1.5,1.2)
  \small
  \psset{linewidth=0.9pt,linecolor=black,arrowscale=1.5,arrowinset=0.15}
  \psline{->}(0.7,0)(0.7,0.43)
  \psline(0.7,0)(0.7,0.5)
 \psarc(0.8,0.6732051){0.2}{120}{240}
 \psarc(0.6,0.6732051){0.2}{-60}{35}
  \psline (0.6134,0.896410)(0.267,1.09641)
  \psline{->}(0.6134,0.896410)(0.35359,1.04641)
  \psline(0.7,0.846410) (1.1330,1.096410)	
  \psline{->}(0.7,0.846410)(1.04641,1.04641)
  \rput[bl]{0}(0.4,0){$c$}
  \rput[br]{0}(1.35,0.85){$b$}
  \rput[bl]{0}(0.05,0.85){$a$}
 \scriptsize
  \rput[bl]{0}(0.82,0.35){$\mu$}
  \endpspicture
=\sum\limits_{\nu }\left[ R_{c}^{ab}\right] _{\mu \nu}
\pspicture[shift=-0.65](-0.1,-0.2)(1.5,1.2)
  \small
  \psset{linewidth=0.9pt,linecolor=black,arrowscale=1.5,arrowinset=0.15}
  \psline{->}(0.7,0)(0.7,0.45)
  \psline(0.7,0)(0.7,0.55)
  \psline(0.7,0.55) (0.25,1)
  \psline{->}(0.7,0.55)(0.3,0.95)
  \psline(0.7,0.55) (1.15,1)	
  \psline{->}(0.7,0.55)(1.1,0.95)
  \rput[bl]{0}(0.4,0){$c$}
  \rput[br]{0}(1.4,0.8){$b$}
  \rput[bl]{0}(0,0.8){$a$}
 \scriptsize
  \rput[bl]{0}(0.82,0.37){$\nu$}
  \endpspicture
  .
\end{equation}

Under a basis transformation, $\Gamma^{ab}_c : V^{ab}_c \rightarrow V^{ab}_c$, the $F$ and $R$ symbols change:
\begin{align}
  F^{abc}_d &\rightarrow \tilde{F}^{abc}_d = \Gamma^{ab}_e \Gamma^{ec}_d F^{abc}_d [\Gamma^{bc}_f]^\dagger [\Gamma^{af}_d]^\dagger
  \nonumber \\
  R^{ab}_c & \rightarrow \tilde{R}^{ab}_c = \Gamma^{ba}_c R^{ab}_c [\Gamma^{ab}_c]^\dagger ~.
  \end{align}
  These basis transformations are referred to as vertex basis gauge transformations. Physical quantities correspond to gauge-invariant combinations
  of the data. 
  
The topological twist $\theta_a = e^{2\pi i h_a}$, with $h_a$ the topological spin, is defined
via the diagram:
\begin{equation}
\theta _{a}=\theta _{\bar{a}}
=\sum\limits_{c,\mu } \frac{d_{c}}{d_{a}}\left[ R_{c}^{aa}\right] _{\mu \mu }
= \frac{1}{d_{a}}
\pspicture[shift=-0.5](-1.3,-0.6)(1.3,0.6)
\small
  \psset{linewidth=0.9pt,linecolor=black,arrowscale=1.5,arrowinset=0.15}
  \psarc[linewidth=0.9pt,linecolor=black] (0.7071,0.0){0.5}{-135}{135}
  \psarc[linewidth=0.9pt,linecolor=black] (-0.7071,0.0){0.5}{45}{315}
  \psline(-0.3536,0.3536)(0.3536,-0.3536)
  \psline[border=2.3pt](-0.3536,-0.3536)(0.3536,0.3536)
  \psline[border=2.3pt]{->}(-0.3536,-0.3536)(0.0,0.0)
  \rput[bl]{0}(-0.2,-0.5){$a$}
  \endpspicture
~.
\end{equation}

A quantity that we make extensive use of is the double braid, which is
a phase if either $a$ or $b$ is an Abelian anyon:
\begin{equation}
  \pspicture[shift=-0.6](0.0,-0.05)(1.1,1.45)
  \small
  \psarc[linewidth=0.9pt,linecolor=black,border=0pt] (0.8,0.7){0.4}{120}{225}
  \psarc[linewidth=0.9pt,linecolor=black,arrows=<-,arrowscale=1.4,
    arrowinset=0.15] (0.8,0.7){0.4}{165}{225}
  \psarc[linewidth=0.9pt,linecolor=black,border=0pt] (0.4,0.7){0.4}{-60}{45}
  \psarc[linewidth=0.9pt,linecolor=black,arrows=->,arrowscale=1.4,
    arrowinset=0.15] (0.4,0.7){0.4}{-60}{15}
  \psarc[linewidth=0.9pt,linecolor=black,border=0pt]
(0.8,1.39282){0.4}{180}{225}
  \psarc[linewidth=0.9pt,linecolor=black,border=0pt]
(0.4,1.39282){0.4}{-60}{0}
  \psarc[linewidth=0.9pt,linecolor=black,border=0pt]
(0.8,0.00718){0.4}{120}{180}
  \psarc[linewidth=0.9pt,linecolor=black,border=0pt]
(0.4,0.00718){0.4}{0}{45}
  \rput[bl]{0}(0.1,1.2){$a$}
  \rput[br]{0}(1.06,1.2){$b$}
  \endpspicture
= M_{ab}
\pspicture[shift=-0.6](-0.2,-0.45)(1.0,1.1)
  \small
  \psset{linewidth=0.9pt,linecolor=black,arrowscale=1.5,arrowinset=0.15}
  \psline(0.3,-0.4)(0.3,1)
  \psline{->}(0.3,-0.4)(0.3,0.50)
  \psline(0.7,-0.4)(0.7,1)
  \psline{->}(0.7,-0.4)(0.7,0.50)
  \rput[br]{0}(0.96,0.8){$b$}
  \rput[bl]{0}(0,0.8){$a$}
  \endpspicture
~.
\end{equation}
	
When $a$ or $b$ is abelian, we have $M_{a b} = R^{a b} R^{b a}$.
\subsection{Topological symmetry}

An important property of a UMTC $\mathcal{C}$ is the group of ``topological symmetries,'' which are related
to ``braided auto-equivalences'' in the mathematical literature. They are associated with the symmetries of the
emergent TQFT described by $\mathcal{C}$, irrespective of any microscopic global symmetries of a quantum
system in which the TQFT emerges as the long wavelength description. 

The topological symmetries consist of the invertible maps
\begin{align}
\varphi: \mathcal{C} \rightarrow \mathcal{C} ~.
\end{align}
The different $\varphi$, modulo equivalences known as natural isomorphisms, form a group, which we denote
as Aut$(\mathcal{C})$.

For example, consider the $\Z_2$ toric code, which has 4 anyons $1,e,m,\psi$ forming a group $\Z_2\times\Z_2$ under fusion. This topological order has a $\Z_2$ topological symmetry which permutes $e$ and $m$ (this is an example of a nontrivial $\varphi$).   
	
\subsection{Global symmetry action on anyons}

Let us now suppose that we are interested in a system with a global symmetry group $G$. For example, we may be interested
in a given microscopic Hamiltonian that has a global symmetry group $G$, whose ground state preserves $G$, and whose 
anyonic excitations are algebraically described by $\mathcal{C}$. 

To allow for antiunitary symmetries, we associate the grading $s_1({\bf g})$ by defining $s_1({\bf g}) =1$ when ${\bf g}$ is antiunitary, and $s_1({\bf g}) = 0$ otherwise.

The global symmetry acts on the topological quasiparticles 
and the topological state space through the action of a group homomorphism
\begin{align}
[\rho] : G \rightarrow \text{Aut}(\mathcal{C})~ . 
\end{align}
We use the notation $[\rho_{\bf g}] \in \text{Aut}(\mathcal{C})$ for a specific element ${\bf g} \in G$. The square
brackets indicate the equivalence class of symmetry maps related by natural isomorphisms, which we define below. $\rho_{\bf g}$ is thus a
representative symmetry map of the equivalence class $[\rho_{\bf g}]$. We use the notation
\begin{align}
\,^{\bf g}a \equiv \rho_{\bf g}(a)~. 
\end{align}

In the example of the $\Z_2$ toric code, $[\rho]$ specifies how the $e-m$ permutation is realized by the elements of $G$. If $\rho({\bf g}) = 1$ for some ${\bf g} \in G$, we have $\,^{\bf g}e = m, \,^{\bf g}m = e$, and $^{\bf g}\psi = \psi$. In the context of invertible phases with integer chiral central charge, we denote $\rho = \tilde{n}_1$. 
	
\subsection{$\eta$ and $U$ symbols for the anyons}
To fully specify how the symmetry acts on the anyons, we further need to define (i) a representation of the symmetry action on the fusion/splitting spaces, and (ii) the projective phases associated with the symmetry action on a local region surrounding an anyon. These properties are encoded in the $U$ and $\eta$ symbols.

The permutation $\rho_{\bf g}$ has an action on the fusion/splitting spaces:
\begin{align}
\rho_{\bf g} : V_{ab}^c \rightarrow V_{\,^{\bf g}a \,^{\bf g}b}^{\,^{\bf g}c} ~.   
\end{align}
This map is unitary if $s_1({\bf g}) = 0$ and anti-unitary if $s_1({\bf g}) = 1$. We write this as
\begin{eqs}
&\rho_{\bf g} |a,b;c, \mu\rangle \cr
=&\sum_{\nu} [U_{\bf g}(\,^{\bf g}a ,
\,^{\bf g}b ; \,^{\bf g}c )]_{\mu\nu} K^{s_1({\bf g})} |\,^{\bf g} a, \,^{\bf g} b; \,^{\bf g}c,\nu\rangle~,
\end{eqs}
where $U_{\bf g}(\,^{\bf g}a , \,^{\bf g}b ; \,^{\bf g}c ) $ is a $N_{ab}^c \times N_{ab}^c$ matrix, and 
$K$ denotes complex conjugation.

Under the map $\rho_{\bf g}$, the $F$ and $R$ symbols transform as well:
\begin{widetext}
	\begin{align} \label{Eq:FR-Uconsistency}
	\rho_{\bf g}[ F^{abc}_{def}] &= U_{\bf g}(\,^{\bf g}a, \,^{\bf g}b; \,^{\bf g}e) U_{\bf g}(\,^{\bf g}e, \,^{\bf g}c; \,^{\bf g}d) F^{\,^{\bf g}a \,^{\bf g}b \,^{\bf g}c }_{\,^{\bf g}d \,^{\bf g}e \,^{\bf g}f} 
	U^{-1}_{\bf g}(\,^{\bf g}b, \,^{\bf g}c; \,^{\bf g}f) U^{-1}_{\bf g}(\,^{\bf g}a, \,^{\bf g}f; \,^{\bf g}d) = K^{s_1({\bf g})} F^{abc}_{def} K^{s_1({\bf g})}
	\nonumber \\
	\rho_{\bf g} [R^{ab}_c] &= U_{\bf g}(\,^{\bf g}b, \,^{\bf g}a; \,^{\bf g}c)  R^{\,^{\bf g}a \,^{\bf g}b}_{\,^{\bf g}c} U_{\bf g}(\,^{\bf g}a, \,^{\bf g}b; \,^{\bf g}c)^{-1} = K^{s_1({\bf g})} R^{ab}_c K^{s_1({\bf g})}~,
	\end{align}
\end{widetext}
where we have suppressed the additional indices that appear when $N_{ab}^c > 1$. We remark that if the $F$ and $R$ symbols are invariant under an overall permutation of their arguments by arbitrary ${\bf g}$, it is often possible to set all $U = 1$. This gives a substantial simplification to the consistency equations. 

Importantly, we have
\begin{align}
\kappa_{{\bf g}, {\bf h}} \circ \rho_{\bf g} \circ \rho_{\bf h} = \rho_{\bf g h} ~,
\end{align}
where the action of $\kappa_{ {\bf g}, {\bf h}}$ on the fusion / splitting spaces is defined as
\begin{align}
\kappa_{ {\bf g}, {\bf h}} ( |a, b;c,\mu \rangle) = \sum_\nu [\kappa_{ {\bf g}, {\bf h}} ( a, b;c )]_{\mu\nu} | a, b;c,\nu \rangle~.
\end{align}
The above definitions imply that
\begin{widetext}
	\begin{align}
	\label{kappaU}
	\kappa_{ {\bf g}, {\bf h}} ( a, b;c ) = U_{\bf g}(a,b;c)^{-1} K^{s_1({\bf g})} U_{\bf h}( \,^{\bar{\bf g}}a, \,^{\bar{\bf g}}b; \,^{\bar{\bf g}}c  )^{-1} K^{s_1({\bf g})} U_{\bf gh}(a,b;c )~,
	\end{align}
\end{widetext}
where $\bar{\bf g} \equiv {\bf g}^{-1}$. Additionally, $\kappa_{ {\bf g}, {\bf h}}$ is a natural isomorphism, which means that by definition,
\begin{align}
[\kappa_{ {\bf g}, {\bf h}} (a,b;c)]_{\mu \nu} = \delta_{\mu \nu} \frac{\beta_a({\bf g}, {\bf h}) \beta_b({\bf g}, {\bf h})}{\beta_c({\bf g}, {\bf h}) }~,
\end{align}
where $\beta_a({\bf g}, {\bf h})$ are $U$ dependent phases. 

The data $\eta_a({\bf g}, {\bf h})$ characterize the difference in phase obtained when acting ``locally'' on an anyon
$a$ by ${\bf g}$ and ${\bf h}$ separately, as compared with ${\bf gh}$. In other words, $\eta_a$ specifies the projective representation of $G$ associated to $a$. There are two important consistency conditions for $U$ and $\eta$, which we
will use repeatedly. The first one is
\begin{align}
\label{etaConsistency}
\frac{\eta_a({\bf g}, {\bf h}) \eta_b ({\bf g}, {\bf h})}{\eta_c({\bf g}, {\bf h})} = \kappa_{ {\bf g}, {\bf h}}(a, b;c) ~,
\end{align}
with $\kappa$ defined in terms of $U$ as in Eq.~\eqref{kappaU}. ($\eta$ and $\beta$ have a slightly different interpretation, see Sec. IV B of \cite{Barkeshli2019} for a discussion; but they are related to $\kappa_{{\bf g},{\bf h}}$ in the same way.) The other one is
\begin{equation}
\eta_a({\bf g},{\bf h})\eta_a({\bf gh},{\bf k})=\eta_a({\bf g},{\bf hk})\eta^{s_1({\bf g})}_{\rho_{\bf g}^{-1}(a)}({\bf h},{\bf k})~.
\label{etacocycle}
\end{equation}

\subsection{Symmetry localization and fractionalization}

In order to fully specify the symmetry action on the anyons, we must obtain a consistent solution to the $U$ and $\eta$ symbols using Eqs.~\eqref{Eq:FR-Uconsistency},~\eqref{kappaU},~\eqref{etaConsistency},~\eqref{etacocycle}. If there is no consistent solution to Eq.~\eqref{Eq:FR-Uconsistency}, we say there is a \textit{symmetry localization obstruction} characterized by an element $[\mathfrak{O}] \in \H^3_{[\rho]}(G,\A)$, where $\A$ is the group of abelian anyons under fusion. In that case, we cannot define the $\eta$ symbols at all, i.e. there is no consistent way to ``localize" the symmetry action on each anyon. The $\H^3$ obstruction can alternatively be viewed as a manifestation of 2-group symmetry \cite{Benini2018}. 

Even if we can solve for Eq.~\eqref{Eq:FR-Uconsistency}, there may be further obstructions to consistently defining $\eta$. Physically these arise from the fact that if we have a given symmetry fractionalization on $\psi$, it may not be possible to extend this to the entirety of $\C$ \cite{fidkowski2018surface}. In our context, these obstructions define constraints on the data $n_2$. In fact, $n_2$ is directly defined in terms of the $\eta$ symbols (see Table~\ref{Table:AnyonFRUn}). Assuming there are no such obstructions, the distinct choices of $\eta$ define the \textit{symmetry fractionalization classes} of the SET phase. Mathematically, it is found \cite{Barkeshli2019} that they are classified by a torsor over the group $\H^2_{[\rho]}(G,\A)$. This means that if we can find a reference solution $\eta^{\text{ref}}$, there exist $|\H^2_{[\rho]}(G,\A)|$ distinct solutions, given by $\eta_a({\bf g},{\bf h}) = M_{a, \mathfrak{t}({\bf g},{\bf h})} \eta_a^{\text{ref}}({\bf g},{\bf h})$, where $\mathfrak{t}({\bf g},{\bf h})$ is an anyon and $[\mathfrak{t}] \in \H^2_{[\rho]}(G,\A)$.
	
\subsection{Symmetry defects}

Another way to understand the classification of symmetry fractionalization is in terms of the properties
of symmetry defects. A symmetry defect consists of a defect line in space, labeled by a group element ${\bf g} \in G$,
which we sometimes refer to as a branch cut, and which can terminate at a point. In the (2+1)D
space-time, the symmetry defect is thus associated with a two-dimensional branch sheet. A given branch cut line associated
with ${\bf g}$ can have topologically distinct endpoints, which thus give rise to topologically distinct types of ${\bf g}$ defects;
a particular topological class of ${\bf g}$ defect is thus labeled as $a_{\bf g}$. An anyon $x$ crossing the ${\bf g}$ defect branch cut
is transformed into its permuted counterpart, $\,^{\bf g}x$. 	

First, we note that the defects can be organized into a $G$-graded fusion category,
\begin{align}
\mathcal{C}_G = \bigoplus_{{\bf g}\in G} \mathcal{C}_{\bf g}~,
\end{align}
where the simple objects of $\mathcal{C}_{\bf g}$ are the topologically distinct set of ${\bf g}$ defects. This means that we have a notion of fusion for defects, just as for anyons.

To determine the quantum dimensions of the defects, we use the following important formula. By considering states on a torus with a ${\bf g}$ defect wrapping one of the cycles, one can show that
\begin{align}\label{Eq:numGdefects}
|\mathcal{C}_{\bf g}| = |\mathcal{C}_{{\bf 0}}^{\bf g}|~,
\end{align}
where $|\mathcal{C}_{\bf g}|$ is the number of topologically distinct ${\bf g}$ defects, and $|\mathcal{C}_{{\bf 0}}^{\bf g}|$ is the number
of ${\bf g}$ invariant anyons.

Now the symmetry fractionalization classes discussed in the previous section can also be defined in terms of the different classes of fusion rules of the defects. Fusion of the defects respects the group multiplication law associated with their branch cuts, so that an allowed fusion rule takes the form
\begin{align}
  \label{defectFusion}
a_{\bf g} \times b_{\bf h} = \sum_{c_{\bf g h}} N_{ab}^c c_{{\bf gh}} ~.
\end{align}
(If there is an $\H^3$ obstruction, such a fusion rule will not exist.) Given this, we can consider the modified rule
\begin{align}
\label{newFusion}
a_{\bf g} \times b_{\bf h} = \mathfrak{t}({\bf g}, {\bf h}) \sum_{c} N_{ab}^c c_{{\bf gh}} ~.
\end{align}
This new rule must respect associativity:
\begin{align}
    (a_{\bf g} \times b_{\bf h}) \times c_{\bf k} = a_{\bf g} \times (b_{\bf h} \times c_{\bf k}) ~. 
\end{align}
For the new fusion rules (\ref{newFusion}) to be associative, we are thus led to the constraint
\begin{align}
\label{2cocycle}
    \mathfrak{t}({\bf g}, {\bf h}) \mathfrak{t}({\bf gh}, {\bf k}) = \,^{\bf g} \mathfrak{t}({\bf h}, {\bf k}) \mathfrak{t}({\bf g}, {\bf hk}) ~. 
\end{align}
From this 2-cocycle condition we infer that the distinct choices of $[\mathfrak{t}]$ are classified by $\H^2_{[\rho]}(G,\A)$.

\subsection{The $\H^4$ anomaly and its computation}
A complete description of symmetry defects involves the specification of $\{F,R,U,\eta\}$ for all defects, subject to several consistency conditions. We will not describe this here; a comprehensive treatment is given in Ref.~\onlinecite{Barkeshli2019}. We only note that the defect $F$ symbols need to satisfy an internal consistency condition known as the pentagon equation, which takes the abbreviated form $FF = \sum FFF$. Even if the symmetry fractionalization class is well defined, we find that instead of the pentagon equation, we have
\begin{equation}
    FF \mathcal{O}_4({\bf g},{\bf h},{\bf k},{\bf l}) = \sum FFF~,
\end{equation}
where $\mathcal{O}_4({\bf g},{\bf h},{\bf k},{\bf l})$ is a cocycle representative of the group $\H^4(G,U(1)_T)$. $\mathcal{O}_4$ is called the \textit{defect obstruction}. If it is a nontrivial cocycle, it signifies that the SET with the given symmetry fractionalization class cannot be realized in (2+1)D; instead, it must exist on the surface of an SPT state in (3+1) dimensions with cohomology class $[\mathcal{O}_4]$.

The function $\nu_3$ arising in the study of invertible phases is closely related to the defect $F$ symbols in the corresponding $G_b$-crossed BTC. $\nu_3$ fixes a subset of the $F$ symbols, and the remaining ones are in turn fixed by the pentagon equation. We can in fact write the pentagon equation as a condition on $\nu_3$:
\begin{equation}
    d\nu_3 = \mathcal{O}_4~.
\end{equation}
Computing the 't Hooft anomaly is often the most difficult technical task in studying a topological phase. We will use an idea developed in Ref.~\onlinecite{Barkeshli2020Anomaly}: rather than compute the anomaly from first principles, we break the problem down into two steps. 
First, we find an SET $\C^{\times}_{G,\text{ref}}$ in a symmetry fractionalization class for which the anomaly is easily computable. Then we find the \textit{relative} anomaly $\mathcal{O}_r({\bf g},{\bf h},{\bf k},{\bf l})$ between $\C^{\times}_{G,\text{ref}}$ and the desired SET $\C^{\times}_{G}$, whose symmetry fractionalization class is related to that of $\C^{\times}_{G,\text{ref}}$ by the anyon $\mathfrak{t}({\bf g},{\bf h})$. A formula for $\mathcal{O}_r$ was derived in Ref.~\onlinecite{Barkeshli2020Anomaly} (see below). When $G_b$ is unitary, we can always find a reference $\C^{\times}_{G,\text{ref}}$ which is non-anomalous. Thus in this case, the relative anomaly between the two phases is equal to the \textit{absolute} anomaly of $\C^{\times}_{G}$. 

In terms of the data $\{F,R,U,\eta^{\text{ref}}\}$ of $\C^{\times}_{G,\text{ref}}$, the relative anomaly is given by the formula
\begin{widetext}
   \begin{align} \label{Eq:RelAnom}
\mathcal{O}_r({\bf g},{\bf h},{\bf k},{\bf l}) &= R^{{^{\bf gh}} \mathfrak{t}({\bf k},{\bf l}) \mathfrak{t}({\bf g},{\bf h})} \eta_{{^{\bf gh}} \mathfrak{t}({\bf k},{\bf l})}({\bf g},{\bf h})\frac{U_{\bf g}({^{\bf g}} \mathfrak{t}({\bf h},{\bf kl}), {^{\bf gh}} \mathfrak{t}({\bf k}.{\bf l}))}{U_{\bf g}({^{\bf g}} \mathfrak{t}({\bf hk},{\bf l}), {^{\bf g}} \mathfrak{t}({\bf h},{\bf k}))} \nonumber \\
& \times \frac{F^{\mathfrak{t}({\bf ghk},{\bf l}) \mathfrak{t}({\bf gh},{\bf k}) \mathfrak{t}({\bf g},{\bf h})}}{F^{\mathfrak{t}({\bf ghk},{\bf l}) \mathfrak{t}({\bf g},{\bf hk}) {^{\bf g}}\mathfrak{t}({\bf h},{\bf k})}} \frac{F^{\mathfrak{t}({\bf g},{\bf hkl}) {^{\bf g}}\mathfrak{t}({\bf hk},{\bf l}) {^{\bf g}}\mathfrak{t}({\bf h},{\bf k})}}{F^{\mathfrak{t}({\bf g},{\bf hkl}) {^{\bf g}}\mathfrak{t}({\bf h},{\bf kl}) {^{\bf gh}}\mathfrak{t}({\bf k},{\bf l})}} \frac{F^{\mathfrak{t}({\bf gh},{\bf kl}) \mathfrak{t}({\bf g},{\bf h}) {^{\bf gh}}\mathfrak{t}({\bf k},{\bf l})}}{F^{\mathfrak{t}({\bf gh},{\bf kl}) {^{\bf gh}}\mathfrak{t}({\bf k},{\bf l}) \mathfrak{t}({\bf g},{\bf h})}}~.
\end{align}

\end{widetext}
We will sometimes abbreviate this as ``$\mathcal{O}_r = R \eta \frac{U}{U} \frac{FFF}{FFF}$". When the symmetry is non-permuting (i.e. $[\rho]$ is trivial), Ref.~\onlinecite{Barkeshli2019} gives a formula for the \textit{absolute} anomaly of $\C_{G}^{\times}$ in terms of the $F$ and $R$ symbols of $\C_{G}^{\times}$:
\begin{widetext}
\begin{equation}\label{Eq:AbsAnomNoPerm}
\mathcal{O}_4({\bf g},{\bf h},{\bf k},{\bf l}) = \frac{F^{\mathfrak{t}({\bf h},{\bf k})\mathfrak{t}({\bf g},{\bf hk})\mathfrak{t}({\bf ghk},{\bf l})}}{F^{\mathfrak{t}({\bf g},{\bf h})\mathfrak{t}({\bf gh},{\bf k})\mathfrak{t}({\bf ghk},{\bf l})}}
\frac{F^{\mathfrak{t}({\bf k},{\bf l})\mathfrak{t}({\bf h},{\bf kl})\mathfrak{t}({\bf g},{\bf hkl})}}{F^{\mathfrak{t}({\bf h},{\bf k})\mathfrak{t}({\bf hk},{\bf l})\mathfrak{t}({\bf g},{\bf hkl})}}
\frac{F^{\mathfrak{t}({\bf g},{\bf h})\mathfrak{t}({\bf k},{\bf l})\mathfrak{t}({\bf gh},{\bf kl})}}{F^{\mathfrak{t}({\bf k},{\bf l})\mathfrak{t}({\bf g},{\bf h})\mathfrak{t}({\bf gh},{\bf kl})}}R^{\mathfrak{t}({\bf g},{\bf h})\mathfrak{t}({\bf k},{\bf l})}~.
\end{equation}
\end{widetext}
For convenience, we will sometimes abbreviate this as $``\mathcal{O}_4 = R\frac{FFF}{FFF}"$. The relative anomaly formula reduces to the absolute anomaly formula in the non-permuting case, because we can always set $U=1$ for the anyons and also find a reference state with $\eta = 1$.

Assuming the anomaly is trivial, the distinct choices of $\nu_3$ for a given $[\mathfrak{t}]$ are related by elements of $\H^3(G,U(1)_T)$.
\subsection{Gauge transformations}
The full $G$-crossed BTC has a consistent set of $F,R,U,\eta$ symbols for anyons as well as symmetry defects. The data are subject to two kinds of gauge transformations, referred to as vertex basis and symmetry action gauge transformations. Let $a_{\bf g},b_{\bf h},c_{\bf k},d_{\bf ghk},e_{\bf gh},f_{\bf hk}$ be topological charges (the group labels are suppressed below). We assume that $N_{a b}^e \le 1$, i.e. there are no fusion multiplicities. The vertex basis transformations generalize those defined for UMTCs, and are defined as follows (here $s_1({\bf g}) = 1$ is understood to act by complex conjugation):
\begin{align}
    [F^{a b c}_d]_{e,f} & \rightarrow \frac{\Gamma^{a b}_e \Gamma^{e c}_f}{\Gamma^{a f}_d \Gamma^{b c}_f}[F^{a b c}_d]_{e,f} \nonumber \\
    R^{a b}_e & \rightarrow \frac{\Gamma^{b {^{\bar{\bf h}}} a}_e}{\Gamma^{a b}_e}R^{a b}_e \nonumber \\
    U_{\bf k}(a,b;e) & \rightarrow \frac{\Gamma^{{^{\bar{\bf k}}}a {^{\bar{\bf k}}} b}_{{^{\bar{\bf k}}}e}}{(\Gamma^{a b}_e)^{s_1({\bf k})}} U_{\bf k}(a,b;e) \nonumber \\
    \eta_c({\bf g},{\bf h}) & \rightarrow \eta_c({\bf g},{\bf h})~. \label{Eq:VBGT}
\end{align}
The symmetry action gauge transformations act as follows:
\begin{align}
[F^{a b c}_d]_{e,f} &\rightarrow [F^{a b c}_d]_{e,f} \nonumber \\
R^{a b}_e & \rightarrow \gamma_a({\bf h}) R^{a b}_e \nonumber \\
U_{\bf g}(a,b;c) &\rightarrow \frac{\gamma_{a}({\bf g}) \gamma_b({\bf g})}{ \gamma_c({\bf g}) } U_{\bf g}(a,b;c)
\nonumber \\
\eta_c({\bf g}, {\bf h}) & \rightarrow \frac{\gamma_c({\bf g h}) }{(\gamma_{\,^{\bf g} c}({\bf g}))^{s_1({\bf g})} \gamma_c({\bf h}) } \eta_c({\bf g}, {\bf h})~. \label{Eq:SAGT}
\end{align}
We have $\Gamma^{a 0}_a = \Gamma^{0 b}_b=1$ and $\gamma_0({\bf h}) = \gamma_a({\bf 0}) = 1$. Imoprtantly, for fermionic systems we additionally impose the constraints \cite{bulmashSymmFrac}
\begin{align}
    \gamma_{\psi}({\bf g}) &= 1 \\
    \Gamma^{\psi \psi} &= 1.
\end{align}
Our main application of these gauge transformations will be in proving equivalences between two different sets of data describing the bosonic shadows of invertible phases (Section~\ref{Sec:Counting}). 
\subsection{Equivalences}
Above we saw that the $G$-crossed BTC is specified by the three properties $[\rho], [\mathfrak{t}], [\nu_3]$. However, two different sets of data might describe the same SET phase. This is because, in addition to the gauge transformations in the theory, two sets of data might also become equivalent under a relabelling of the anyons and defects. As a result, for a given $[\rho]$ the true count of SET phases is generally less than the naive estimate of $|\H^2_{\rho}(G,\A)\times \H^3(G,U(1)_T)|$, even without considering any obstructions. Two systems are treated as physically equivalent if their data can be related by relabelling a subset of the anyons and defects with fermions, and then performing suitable gauge transformations. We will describe in detail how these equivalences affect the counting of invertible phases in Section~\ref{Sec:Counting}.

 \begin{widetext}
	\section{Higher cup products and their properties}
	\label{sec:highercup}

\subsection{Definition of ordinary and higher cup product}
In this section we will assume some background on the definition of group cochains and cohomology groups; this is reviewed, for example, in Ref.~\onlinecite{Chen2013}. Here we only give a very brief overview. An $m$-cochain $f_m \in C^m(G,A)$ is an $m$-variable function from a group $G$ to an abelian group $A$ (which we will often take to be $U(1), \Z_2$ or $\Z$). We can define a map $d: C^m(G,A) \rightarrow C^{m+1}(G,A)$ satisfying $d \circ d = 0$; such a map is called a differential. The cochain $f_m$ is an \textit{$m$-cocycle} if $df_m = 0$; the set of all $m$-cocycles is denoted $Z^m(G,A)$. $f_m$ is an \textit{$m$-coboundary} if it can be written as $f_m = db_{m-1}$ for some $b_{m-1} \in C^{m-1}(G,A)$. The set of all $m$-coboundaries is denoted $B^m(G,A)$. Then, the $m$th cohomology group of $G$ with coefficients in $A$, denoted $\H^m(G,A)$, is defined as
\begin{equation}
    \H^m(G,A) := \frac{Z^m(G,A)}{B^m(G,A)}.
\end{equation}

We now discuss cup products. Let $G$ be a group, and $A$ be a coefficient module. (In this work we often use $A = \Z_2,U(1)$.) Let $f_m \in C^m(G,A), g_n \in C^n(G,A)$ be $m$- and $n$-cochains taking values in $A$. Then, the cup product $\cup: C^m(G,A)\times C^n(G,A)\rightarrow  C^{m+n}(G,A)$ (sometimes referred to as the cup-0 or $\cup_0$ product) of $f_m$ and $g_n$ is defined as follows:
\begin{equation}
    (f_m \cup f_n)({\bf g}_1,{\bf g}_2,\dots , {\bf g}_{m+n}) := f_m({\bf g}_1,\dots , {\bf g}_m) g_{n}({\bf g}_{m+1},\dots , {\bf g}_{m+n})~.
\end{equation}

We can also consider the cochains on simplices, where we label a $p$-simplex by its vertices $(v_0,v_1,\cdots v_p)$. It is related to the previous notation by assigning to the edge connecting the vertices $v_{m},v_{m+1}$ the group element $g_{m+1}$. We will use the two notation interchangeably, i.e., $\mathcal{O}_4({\bf{g}},{\bf{h}},{\bf{k}},{\bf{l}}) = \mathcal{O}_4(01234)$ where group elements ${\bf{g}},{\bf{h}},{\bf{k}}$, ${\bf{l}}$ live on edges $\lr{01}$, $\lr{12}$, $\lr{23}$, $\lr{34}$.

The cup-1 product is a binary operation $\cup_1: C^m(G,A)\times C^n(G,A)\rightarrow  C^{m+n-1}(G,A)$ on cochains, and is defined as follows \cite{Steenrod1947}:
\begin{equation}
    (f_m \cup_1 g_n)({\bf g}_1,\dots , {\bf g}_{m+n-1}) := \sum\limits_{j=0}^{m-1} (-1)^{(m-j)(n+1)} f_m({\bf g}_1,\dots , {\bf g}_j, {\bf g}_{j+n},{\bf g}_{m+n-1}) g_n({\bf g}_j,\dots , {\bf g}_{j+n-1})~.
\end{equation}

Let us look at a few examples explicitly, taking $A=\Z_2$:
\begin{align}
   & m=n=1 & (f_1 \cup_1 g_1)({\bf g}_1) &= f_1({\bf g}_1) g_1({\bf g}_1) \\
   & m=n=2 & (f_2 \cup_1 g_2)({\bf g}_1,{\bf g}_2,{\bf g}_3) &= f_2({\bf g}_1{\bf g}_2,{\bf g}_3) g_2({\bf g}_1,{\bf g}_2) + f_2({\bf g}_1,{\bf g}_2{\bf g}_3) g_2({\bf g}_2,{\bf g}_3) \\
   & m=2,n=3 & (f_2 \cup_1 g_3)({\bf g}_1,{\bf g}_2,{\bf g}_3,{\bf g}_4) &= f_2({\bf g}_1{\bf g}_2{\bf g}_3,{\bf g}_4) g_3({\bf g}_1,{\bf g}_2,{\bf g}_3) + f_2({\bf g}_1,{\bf g}_2{\bf g}_3{\bf g}_4) g_3({\bf g}_2,{\bf g}_3,{\bf g}_4) \\
   & m=3,n=2 & (f_3 \cup_1 g_2)({\bf g}_1,{\bf g}_2,{\bf g}_3,{\bf g}_4) &= f_3({\bf g}_1{\bf g}_2,{\bf g}_3,{\bf g}_4) g_2({\bf g}_1,{\bf g}_2) + f_3({\bf g}_1,{\bf g}_2{\bf g}_3,{\bf g}_4) g_2({\bf g}_2,{\bf g}_3) \nonumber \\ & & &+ f_3({\bf g}_1,{\bf g}_2,{\bf g}_3{\bf g}_4) g_2({\bf g}_3,{\bf g}_4)~.
\end{align}

For $i \ge 0$, the cup-$i$ or $\cup_i$ product is a binary operation on cochains, $\cup_i: C^m(G,A)\times C^n(G,A)\rightarrow  C^{m+n-i}(G,A)$. This operation gives zero if $m<i$ or $n<i$. Note that while the $\cup_0$ product is associative, the higher cup products are not. The expressions for higher cup products are more involved (see e.g. Ref.~\onlinecite{tata2020}), and usually only arise indirectly through the identities discussed below. One useful special case is that for $n$-cochains $f_n,g_n$, $f_n \cup_n g_n$ is also an $n$-cochain with the simple form
\begin{equation}
    (f_n \cup_n g_n)({\bf g}_1,\dots , {\bf g}_n) = f_n({\bf g}_1,\dots , {\bf g}_n) g_n({\bf g}_1,\dots , {\bf g}_n)~.  
\end{equation}

\subsection{Cup product identities}

\begin{enumerate}
    \item Leibnitz rule:
\begin{equation}
    d(f_m \cup f_n) = df_m \cup f_n + (-1)^{m} f_m \cup df_n.
\end{equation}
A consequence of this property is that the cup product is a cohomology operation: given that $f_m,f_n$ are cocycles, $f_m \cup f_n$ is also a cocycle.

The higher cup products satisfy the following Leibnitz rule:
\begin{equation}
    d(f_m \cup_i f_n) = df_m \cup_i f_n + (-1)^m f_m \cup_i df_n + (-1)^{m+n-i}f_m \cup_{i-1} f_n + (-1)^{mn+m+n}f_n \cup_{i-1} f_m~.
\end{equation}
As a result, the $\cup_i$ product is not a cohomology operation for $i \ge 1$. 

\item The Hirsch identity \cite{Hirsch:1955}: for cochains $a,b,c$ of degree $p,q,r$, respectively, we have
\begin{equation}\label{eqn:Hirsch}
    (a_p \cup b_q) \cup_1 c_r = (-1)^p a_p \cup (b_q \cup_1 c_r) + (-1)^{qr}(a_p \cup_1 c_r) \cup b_q~.
\end{equation}
While working in $\Z_2$ coefficients we can neglect the minus signs in the above expressions. 

For $\mathbb{Z}_2$ cocycles $x,y,z$ and working in $\mathbb{Z}_2$ coefficient, we also have
\begin{equation}\label{eqn:Hirschp}
    x\cup_1 (y\cup z) =
    (y\cup z)\cup_1 x+ d(x\cup_2(y
    \cup z))=(y\cup_1 x)\cup z+y\cup (z\cup_1 x)+d(x\cup_2(y
    \cup z))\quad \text{mod }2~.
\end{equation}

\item A useful identity is
\begin{equation}\label{eqn:dXidentity}
    X_2 \cup_1 X_2 + X_2 \cup_2 dX_2 = \frac{1}{2}( d[X_2]_2 - [dX_2]_2)~,
\end{equation}
where $[Y]_2$ denotes the mod 2 reduction for $Y$, and $X_2$ is a $\mathbb{Z}_2$ 2-cochain.
To prove it, we note that 
\begin{align}
    d[X_2]_2-[dX_2]_2&=
    d(X_2\cup_2 X_2)-dX_2\cup_3 dX_2=
    2X_2\cup_1 X_2 + dX_2\cup_2 X_2+X_2\cup_2 dX_2-dX_2\cup_3 dX_2\cr
    &=2(X_2\cup_1 X_2 +X_2\cup_2 dX_2)~.
\end{align}

In particular, when $X_2$ is closed $dX_2=0$ mod 2, we have
\begin{equation}
    X_2\cup_1 X_2=Sq^1X_2=d[X_2]/2~.
\end{equation}

\item Another identity is
$B_2 \cup_2 (B_2\cup_1 B_2) = 0$ mod 2 for $\mathbb{Z}_2$ two-cocycle $B_2$:
\begin{eqs}
    B_2 \cup_2 (B_2\cup_1 B_2)(1234) =& [B_2 (123) + B_2(134)] [B_2(124) B_2 (234) + B_2(134) B_2(123)] \\
    =& [B_2 (123) + B_2(134)] B_2(134) B_2(123) +  [B_2 (124) + B_2(234)] B_2(124) B_2 (234) \\
    =& 0\quad \text{mod }2~.
\end{eqs}
\end{enumerate}

\subsection{Steenrod squares}\label{app:Steenrod}
Take $A=\Z_2$. While the $\cup_i$ product on general cochains is not a cohomology operation, the $i$th Steenrod square, denoted as $Sq^i$, is indeed such an operation. $Sq^i: \H^m(G,\Z_2) \rightarrow \H^{m+i}(G,\Z_2)$ is defined as an additive homomorphism with the following properties:
\begin{enumerate}
    \item (Naturality) For any $f: G \rightarrow G'$, we have $f^*(Sq^i(x)) = Sq^i(f^*(x))$; 
    \item $Sq^0$ is the identity homomorphism 
    \item $Sq^n(f_n) = f_n \cup f_n$, where $f_n$ is a cocycle 
    \item $Sq^i(f_n)=0$ for $i>n$; 
    \item (Cartan formula) $Sq^i(f_m \cup g_n) = \sum\limits_{u+v=i} Sq^u(f_m) Sq^v(g_n)$.
\end{enumerate}
We also note two useful properties of $Sq^1$: (i) for a 2-cocycle $f_2$, $Sq^1(f_2) = f_2 \cup_1 f_2$; and (ii) $Sq^1$ is the Bockstein homomorphism associated to the short exact sequence
\begin{equation}
    1 \rightarrow \Z_2 \rightarrow \Z_4 \rightarrow \Z_2 \rightarrow 1~.
\end{equation}
These properties can be used to understand a particular instance of the '$\H^3$ anomaly'. In the main text we derived through direct calculation that when $c_-$ is odd, and $n_1=0$, the equation for $n_2$ reads
\begin{equation}
    dn_2 = \omega_2 \cup_1 \omega_2~.
\end{equation}
This equation can be understood more abstractly in terms of $Sq^1$. The group $\A = \Z_4$ can be expressed via the short exact sequence
\begin{equation}
    1 \rightarrow  \Z_2^{\psi}  \xrightarrow{\times 2} \Z_4 \xrightarrow{\mod 2} \Z_2^v \rightarrow 1~,
\end{equation}
where $\psi$ generates the normal subgroup, denoted as $\Z_2^{\psi}$, and $v$ generates the second $\Z_2^m$ factor. This short exact sequence induces the following long exact sequence:
\begin{equation}
    \dots \rightarrow \overbrace{\H^2(G_b,\Z_2^{\psi})}^{n'_2} \xrightarrow{\times 2} \overbrace{\H^2(G_b,\Z_4)}^{[\omega_2]_2 + 2n'_2} \xrightarrow{\mod 2} \overbrace{\H^2(G_b,\Z_2^v)}^{\omega_2} \xrightarrow{d} \H^3(G_b,\Z_2^{\psi}) \xrightarrow{\times 2} \H^3(G_b,\Z_4) \rightarrow \dots ~.
\end{equation}
The overbrace above each cohomology group gives the meaning of the cocycle representatives within that group. The homomorphism $d$ which connects $\H^2(G_b,\Z_2^{\psi})$ to $\H^3(G_b,\Z_2^v)$ is the Bockstein homomorphism for the above short exact sequence. It is equal to the first Steenrod square $Sq^1: \H^2(G_b,\Z_2) \rightarrow \H^3(G_b,\Z_2)$, which for the 2-cocycle $\omega_2$ is the same as the cup-1 product:
\begin{equation}
    Sq^1(\omega_2) = \omega_2 \cup_1 \omega_2~.
\end{equation} 
Hence, $\omega_2$ is mapped to $Sq^1(\omega_2)$ by $d$, and then to an element of $\H^3(G_b,\Z_4)$ by the subsequent $\times 2$ map. Physically, this defines a ``symmetry localization anomaly" $\mathfrak{O} = 2Sq^1(\omega_2)$ valued in $Z^3(G_b,\Z_4)$. If $[\mathfrak{O}]$ is nontrivial in $\H^3(G_b,\Z_2)$, there is an obstruction to symmetry fractionalization. 
 
If on the other hand $Sq^1(\omega_2)$ is trivial, this differential can be cancelled by suitably defining $n_2$: we set
\begin{equation}
    dn_2 = \omega_2 \cup_1 \omega_2~,
\end{equation}
as we derived earlier through a more elementary argument. In this case, the physically distinct choices of $n_2$ differ by cocycles $n'_2 \in \H^2(G_b,\Z_2^{\psi})$.

\section{Anomaly computations}\label{Sec:RelAnomComps}

In this section we perform the $\H^4$ anomaly calculations required to complete Section~\ref{Sec:DefObs} and Table~\ref{Table:MainEqs}. We will work in the representation $(n_1,n_2,\nu_3)$. For each choice of $c_-$ we will consider the cases $n_1 = 0$ and $n_1 \ne 0$ separately. In the former case we will directly compute the absolute anomaly from Eq.~\eqref{Eq:AbsAnomNoPerm}, which is reproduced here:
 
\begin{equation}
    \mathcal{O}_4({\bf g},{\bf h},{\bf k},{\bf l}) = \frac{F^{\mathfrak{t}({\bf h},{\bf k})\mathfrak{t}({\bf g},{\bf hk})\mathfrak{t}({\bf ghk},{\bf l})}}{F^{\mathfrak{t}({\bf g},{\bf h})\mathfrak{t}({\bf gh},{\bf k})\mathfrak{t}({\bf ghk},{\bf l})}}
    \frac{F^{\mathfrak{t}({\bf k},{\bf l})\mathfrak{t}({\bf h},{\bf kl})\mathfrak{t}({\bf g},{\bf hkl})}}{F^{\mathfrak{t}({\bf h},{\bf k})\mathfrak{t}({\bf hk},{\bf l})\mathfrak{t}({\bf g},{\bf hkl})}}
    \frac{F^{\mathfrak{t}({\bf g},{\bf h})\mathfrak{t}({\bf k},{\bf l})\mathfrak{t}({\bf gh},{\bf kl})}}{F^{\mathfrak{t}({\bf k},{\bf l})\mathfrak{t}({\bf g},{\bf h})\mathfrak{t}({\bf gh},{\bf kl})}}R^{\mathfrak{t}({\bf g},{\bf h})\mathfrak{t}({\bf k},{\bf l})}~.
\end{equation}
 
 In the latter case we will use Eq.~\eqref{Eq:RelAnom}, also written below, to compute the relative anomaly $\mathcal{O}_r$ between the desired SET $\C_G^{\times}$ and a reference SET $\C_{G,\text{ref}}^{\times}$ which has trivial symmetry fractionalization, and is non-anomalous. In terms of the data of $\C_{G,\text{ref}}^{\times}$, the relative anomaly formula is
 
\begin{align}
    \mathcal{O}_r({\bf g},{\bf h},{\bf k},{\bf l}) &= R^{{^{\bf gh}} \mathfrak{t}({\bf k},{\bf l}) \mathfrak{t}({\bf g},{\bf h})} \eta_{{^{\bf gh}} \mathfrak{t}({\bf k},{\bf l})}({\bf g},{\bf h})\frac{U_{\bf g}({^{\bf g}} \mathfrak{t}({\bf h},{\bf kl}), {^{\bf gh}} \mathfrak{t}({\bf k}.{\bf l}))}{U_{\bf g}({^{\bf g}} \mathfrak{t}({\bf hk},{\bf l}), {^{\bf g}} \mathfrak{t}({\bf h},{\bf k}))} \nonumber \\
    & \times \frac{F^{\mathfrak{t}({\bf ghk},{\bf l}) \mathfrak{t}({\bf gh},{\bf k}) \mathfrak{t}({\bf g},{\bf h})}}{F^{\mathfrak{t}({\bf ghk},{\bf l}) \mathfrak{t}({\bf g},{\bf hk}) {^{\bf g}}\mathfrak{t}({\bf h},{\bf k})}} \frac{F^{\mathfrak{t}({\bf g},{\bf hkl}) {^{\bf g}}\mathfrak{t}({\bf hk},{\bf l}) {^{\bf g}}\mathfrak{t}({\bf h},{\bf k})}}{F^{\mathfrak{t}({\bf g},{\bf hkl}) {^{\bf g}}\mathfrak{t}({\bf h},{\bf kl}) {^{\bf gh}}\mathfrak{t}({\bf k},{\bf l})}} \frac{F^{\mathfrak{t}({\bf gh},{\bf kl}) \mathfrak{t}({\bf g},{\bf h}) {^{\bf gh}}\mathfrak{t}({\bf k},{\bf l})}}{F^{\mathfrak{t}({\bf gh},{\bf kl}) {^{\bf gh}}\mathfrak{t}({\bf k},{\bf l}) \mathfrak{t}({\bf g},{\bf h})}}~.
\end{align}
In the context of invertible fermion phases, we refer to the expression obtained by directly plugging in the anyon data into the relative anomaly formula as $\mathcal{O}_4^{(0)}$. We will refer to each part in the anomaly ${\cal O}^{(0)}_4$ as the $R$ part, the $\eta$ part, the $U/U$ part, and the $FFF/FFF$ part. The computation consists of plugging in the relevant data from Table~\ref{Table:AnyonFRUn} to obtain $\mathcal{O}_4^{(0)}$ and then obtaining a simpler expression $\mathcal{O}_4 := \mathcal{O}_4^{(0)} \times dX^{-1}$ by adding suitable 4-coboundaries. Note that we need to plug in the data of the non-anomalous reference, which is obtained from Table~\ref{Table:AnyonFRUn} by setting $n_2 = \om = 0$. The $F,R,U$ symbols for the reference can be read off directly from Table~\ref{Table:AnyonFRUn}. We remark that it is not necessary to track the 4-coboundary $dX$ in many standard applications. However it is useful to know $X$ in order to obtain an explicit parameterization of the defect $F$ symbols.

We will discuss the following cases separately: $c_-=4k$ (unitary), $c_-=0$ (antiunitary), $c_-=4k+2$, $c_-=2k+1$ and $c_-=k+1/2$. For each case we will first consider $n_1=0$ and then $n_1\neq 0$.

In the following discussion it is convenient to define 4-cochain valued functions $f,g$:
\begin{eqs}\label{eqn:fandg}
    &f(a_1, b_2, c_3) \equiv (a_1 \cup c_3) \cup_2 b_2 + a_1 \cup (c_3 \cup_2 b_2),\quad 
    f(a_1, b_2, c_3)(01234)=
     a(01) b(014) c(1234)  \quad \text{mod } 2 \\
    &g(a_2, b_2, c_2) \equiv (a_2 \cup c_2) \cup_2 b_2 + a_2 \cup (c_2 \cup_2 b_2),\quad 
    g(a_2, b_2, c_2) (01234)= a_2(012) [b(014)+b(124)] c_2(234) 
    ~ 
    \quad \text{mod } 2~, 
\end{eqs}
where the arguments of $f,g$ are cochains with degree given by the subscripts. 
The 4-cochain valued function $f$ satisfies the following properties: for $a_1=n_1,b_2=\omega_2$ that are closed and $c_3=dX$ for integer two-cochain $X$,
\begin{eqs}\label{eqn:f(dX)}
    f(n_1, \om ,dX) &= (n_1 \cup dX) \cup_2 \om + n_1 \cup (dX \cup_2 \om) \\
    &= [dn_1 \cup X - d(n_1 \cup X)] \cup_2 \om + n_1 \cup [d (X \cup_2 \om) - X \cup_2 d \om
    -X \cup_1 \om - \om \cup_1 X] \\
    &\sim  (dn_1 \cup X)\cup_2 \om -(n_1 \cup X) \cup_2 d \om- (n_1 \cup X) \cup_1 \om - \om \cup_1 (n_1 \cup X) \\
    &~~~~ +dn_1 \cup (X \cup_2 \om)- n_1\cup (X \cup_2 d \om) - n_1 \cup (X \cup_1 \om) - n_1 \cup ( \om \cup_1 X)  \\
    &= (n_1 \cup_1 \om) \cup X + \om \cup_1 (n_1 \cup X)
    +n_1 \cup (\om \cup_1 X) \quad \text{mod }2 ~,
\end{eqs}
where $\sim$ means equal up to a coboundary.
We also denote $ \theta \equiv e^{i \pi c_-/4}$ as before.

\subsection{$c_-=4k$}

In this case we have, in vector notation, the following relations:
\begin{align}
    \mathfrak{t}({\bf g},{\bf h}) &= (n_2({\bf g},{\bf h}),(n_2 + \omega_2)({\bf g},{\bf h})) \\
    {^{\bf g}}\mathfrak{t}({\bf h},{\bf k}) &= (n_2({\bf h},{\bf k}) + n_1({\bf g})\omega_2({\bf h},{\bf k}),(n_2 + \omega_2)({\bf h},{\bf k}) + n_1({\bf g})\omega_2({\bf h},{\bf k})) \label{Eq:PermOfTgh}~.
\end{align}
\subsubsection{$n_1 = 0$}
Since the $F$ symbols can be set to 1, we simply have $\mathcal{O}_4 = R$ (Eq.~\eqref{Eq:AbsAnomNoPerm}):
\begin{align}
    \mathcal{O}_4({\bf g},{\bf h},{\bf k},{\bf l})[c_-=4k,n_1=0,n_2] &= R^{\mathfrak{t}({\bf g},{\bf h}) \mathfrak{t}({\bf k},{\bf l})} \\
    &= (-1)^{\mathfrak{t}({\bf g},{\bf h})_e \mathfrak{t}({\bf k},{\bf l})_m + \frac{c_-}{4}(\mathfrak{t}({\bf g},{\bf h})_e \mathfrak{t}({\bf k},{\bf l})_e + \mathfrak{t}({\bf g},{\bf h})_m \mathfrak{t}({\bf k},{\bf l})_m)} \\
    &= (-1)^{n_2 \cup (n_2 + \omega_2) + \frac{c_-}{4}(n_2 \cup n_2 + (n_2 + \omega_2) \cup (n_2 + \omega_2))} \\
    &= (-1)^{n_2 \cup (n_2 + \omega_2) + \frac{c_-}{4} (n_2 \cup \omega_2 + \omega_2 \cup n_2 + \omega_2 \cup \omega_2)} \\
    &= (-1)^{n_2 \cup (n_2 + \omega_2) + \frac{c_-}{4} \omega_2\cup \omega_2} \label{Eq:O4_c0_t}~,
\end{align}
When $dn_2 = 0$, two of the terms proportional to $c_-$ are coboundary terms and hence are dropped in the last line. The $c_- = 0$ result agrees with Ref.~\onlinecite{Wang2020fSPT}.

\subsubsection{$n_1 \ne 0$}
Here we will assume that $s_1 = 0$, i.e. the symmetry is unitary. (The antiunitary case cannot be handled using the relative anomaly formula, because there is no obvious choice of non-anomalous reference theory.) Then we can use the relation $dn_2 = n_1 \cup \omega_2$ to simplify Eq.~\eqref{Eq:PermOfTgh}. The $F$ and $\eta$ symbols in the expression for the relative anomaly $\mathcal{O}_4$ are trivial, so the anomaly ${\cal O}_4$ only contains the $R$ part and the ``$U/U$" part. 

The $R$ part is given by
\begin{align}
R^{{^{\bf gh}} \mathfrak{t}({\bf k},{\bf l}) \mathfrak{t}({\bf g},{\bf h})} &= (-1)^{[{^{\bf gh}} \mathfrak{t}({\bf k},{\bf l})]_e [\mathfrak{t}({\bf g},{\bf h})]_m + \frac{c_-}{4}([{^{\bf gh}} \mathfrak{t}({\bf k},{\bf l})]_e [\mathfrak{t}({\bf g},{\bf h})]_e + [{^{\bf gh}} \mathfrak{t}({\bf k},{\bf l})]_m [\mathfrak{t}({\bf g},{\bf h})]_m) }\\
&= (-1)^{(n_2 \cup n_2 + \omega_2 \cup n_2)({\bf g},{\bf h},{\bf k},{\bf l}) + dn_2({\bf gh}, {\bf k},{\bf l}) (n_2+\omega_2)({\bf g},{\bf h})} \nonumber \\
& \times (-1)^{\frac{c_-}{4} ((n_2 \cup \omega_2 + \omega_2 \cup n_2 + \omega_2 \cup \omega_2 )({\bf g},{\bf h},{\bf k},{\bf l}) + \omega_2({\bf g},{\bf h}) dn_2({\bf gh},{\bf k},{\bf l}) ) }~.
\end{align}  
Next we compute the $U/U$ part. We have $U_{\bf g}(a,b) = \left((-1)^{a_m b_e} \frac{\gamma_a\gamma_b}{\gamma_{a\times b}}\right)^{n_1({\bf g})}$ with $\gamma_a = i^{a_ea_m}$. The $(-1)^{a_m b_e n_1({\bf g})}$ term contributes
\begin{align}
   (-1)^{n_1({\bf g})\{n_2({\bf h},{\bf kl}) ((n_2+\omega_2)({\bf k},{\bf l}) + dn_2({\bf h},{\bf k},{\bf l})) - n_2({\bf hk},{\bf l})(n_2+\omega_2)({\bf h},{\bf k})\}}~.  
\end{align}
The $\gamma$ factors together give $\left(\frac{\gamma_a\gamma_b}{\gamma_c\gamma_d}\right)^{n_1({\bf g})}$, with $a = {^{\bf g}}\mathfrak{t}({\bf h},{\bf kl}), b = {^{\bf gh}}\mathfrak{t}({\bf k},{\bf l}), c = {^{\bf g}}\mathfrak{t}({\bf hk},{\bf l}), d = {^{\bf g}}\mathfrak{t}({\bf h},{\bf k})$. The factor $\gamma_{a b}/\gamma_{c d}$ cancels out, since $a b = c d$. Furthermore, we observe that $\gamma_{a} = \gamma_{{^{\bf g}} a}$, therefore we can ignore the group action in the definition of $a,b,c,d$. Using the definition $\gamma_{
\mathfrak{t}({\bf g},{\bf h})} = i^{n_2({\bf g},{\bf h})[1 + \omega_2({\bf g},{\bf h})]_2} := i^{f({\bf g},{\bf h})}$, we see that the $\gamma$ factors now contribute
\begin{align}
     & i^{n_1({\bf g}) (f({\bf h},{\bf kl}) + f({\bf k},{\bf l}) - f({\bf hk},{\bf l}) - f({\bf h},{\bf k}))} \nonumber \\
     &\sim (-1)^{n_1({\bf g}) n_1({\bf h}) n_2({\bf k},{\bf l})(1+\omega_2({\bf k},{\bf l}))}~,
\end{align}
where $\sim$ in the last line means equal up to a coboundary, which can be absorbed into a redefinition of $\nu_3$. Explicitly, the coboundary is
$db$, where $b = i^{n_1 \cup f}$. Combining terms, we obtain

\begin{align}
\frac{U_{\bf g}({^{\bf g}} \mathfrak{t}({\bf h},{\bf kl}), {^{\bf gh}} \mathfrak{t}({\bf k},{\bf l}))}{U_{\bf g}({^{\bf g}} \mathfrak{t}({\bf hk},{\bf l}), {^{\bf g}} \mathfrak{t}({\bf h},{\bf k}))}  &= 
(-1)^{n_1({\bf g})\{n_2({\bf h},{\bf kl}) ((n_2+\omega_2)({\bf k},{\bf l}) + dn_2({\bf h},{\bf k},{\bf l})) - n_2({\bf hk},{\bf l})(n_2+\omega_2)({\bf h},{\bf k})\}}  \nonumber \\
& \times (-1)^{n_1({\bf g}) n_1({\bf h}) n_2({\bf k},{\bf l})(1+\omega_2({\bf k},{\bf l}))}~.
\end{align}

The full expression for $\mathcal{O}_4^{(0)}$ is
\begin{align}\label{o4,partial}
\mathcal{O}_4^{(0)}({\bf g},{\bf h},{\bf k},{\bf l}) &= (-1)^{(n_2 \cup n_2 + \omega_2 \cup n_2)({\bf g},{\bf h},{\bf k},{\bf l}) + dn_2({\bf gh}, {\bf k},{\bf l}) (n_2+\omega_2)({\bf g},{\bf h})} \times (-1)^{n_1({\bf g}) n_1({\bf h})n _2({\bf k},{\bf l})(n_2+\omega_2)({\bf k},{\bf l})} \nonumber \\
& \times (-1)^{n_1({\bf g})\{n_2({\bf h},{\bf kl}) ((n_2+\omega_2)({\bf k},{\bf l}) + dn_2({\bf h},{\bf k},{\bf l})) - n_2({\bf hk},{\bf l})(n_2+\omega_2)({\bf h},{\bf k})\}} \nonumber \\
& \times (-1)^{\frac{c_-}{4} ((n_2 \cup \omega_2 + \omega_2 \cup n_2 + \omega_2 \cup \omega_2)({\bf g},{\bf h},{\bf k},{\bf l}) + \omega_2({\bf g},{\bf h}) dn_2({\bf gh},{\bf k},{\bf l}) ) }~.
\end{align}

Now, we can use the equivalence relation
$n_2 \sim n_2 + \omega_2$ (recall this follows from a gauge transformation $n_1\rightarrow n_1+d\phi_0$, $n_2\rightarrow n_2+\phi_0 \cup \omega_2$ due to $dn_2=n_1\cup\omega_2$; equivalently, it follows from relabeling the  $e$ and $m$ fluxes).
This gives
\begin{align}
\mathcal{O}_4^{(0)}({\bf g},{\bf h},{\bf k},{\bf l})&\sim (-1)^{(n_2 \cup (n_2 + \omega_2))({\bf g},{\bf h},{\bf k},{\bf l}) + dn_2({\bf gh}, {\bf k},{\bf l}) n_2({\bf g},{\bf h})} \times (-1)^{n_1({\bf g}) n_1({\bf h})n _2({\bf k},{\bf l})(n_2+\omega_2)({\bf k},{\bf l})}\nonumber \\
& \times (-1)^{n_1({\bf g})\{(n_2+\omega_2)({\bf h},{\bf kl}) (n_2({\bf k},{\bf l}) + dn_2({\bf h},{\bf k},{\bf l})) - (n_2+\omega_2)({\bf hk},{\bf l})n_2({\bf h},{\bf k})\}} \nonumber \\
& \times (-1)^{\frac{c_-}{4} ((n_2 \cup \omega_2 + \omega_2 \cup n_2+ \omega_2 \cup \omega_2)({\bf g},{\bf h},{\bf k},{\bf l}) + \omega_2({\bf g},{\bf h}) dn_2({\bf gh},{\bf k},{\bf l}) ) } \\
&= (-1)^{(n_2 \cup (n_2 + \omega_2))({\bf g},{\bf h},{\bf k},{\bf l})  + n_1({\bf g}) (n_2 \cup_1 n_2)({\bf h},{\bf k},{\bf l}) +  dn_2({\bf gh}, {\bf k},{\bf l}) n_2({\bf g},{\bf h}) +  dn_2({\bf g}, {\bf hk},{\bf l}) n_2({\bf h},{\bf k}) +  dn_2({\bf g}, {\bf h},{\bf kl}) n_2({\bf k},{\bf l})} \nonumber \\
& \times (-1)^{n_1({\bf g})(n_2+\omega_2)({\bf h},{\bf kl})dn_2({\bf h},{\bf k},{\bf l}) + n_1({\bf g}) n_1({\bf h})n _2({\bf k},{\bf l})(n_2+\omega_2)({\bf k},{\bf l})} \nonumber \\ &\times (-1)^{\frac{c_-}{4} ((n_2 \cup \omega_2 + \omega_2 \cup n_2+ \omega_2 \cup \omega_2)({\bf g},{\bf h},{\bf k},{\bf l}) + \omega_2({\bf g},{\bf h}) dn_2({\bf gh},{\bf k},{\bf l}) ) }. \\
&= (-1)^{(n_2 \cup (n_2 + \omega_2) + dn_2 \cup_1 n_2)({\bf g},{\bf h},{\bf k},{\bf l}) + dn_2({\bf g},{\bf h},{\bf kl}) dn_2({\bf h},{\bf k},{\bf l}) } \nonumber \\
& \times (-1)^{ n_1({\bf g}) ((n_2 \cup_1 n_2)({\bf h},{\bf k},{\bf l}) + n_2({\bf h},{\bf kl})dn_2({\bf h},{\bf k},{\bf l})) + n_1({\bf g}) n_1({\bf h})n _2({\bf k},{\bf l}) + n_1({\bf g}) dn_2({\bf h},{\bf k},{\bf l})n_2({\bf k},{\bf l})} \nonumber \\
& \times (-1)^{\frac{c_-}{4} ((n_2 \cup \omega_2 + \omega_2 \cup n_2+ \omega_2 \cup \omega_2)({\bf g},{\bf h},{\bf k},{\bf l}) + \omega_2({\bf g},{\bf h}) dn_2({\bf gh},{\bf k},{\bf l}) ) }~,
\end{align}
where we repeatedly used $dn_2 = n_1 \cup \omega_2$.
To simplify the expression for ${\cal O}_4$, we
used the definition
\begin{equation}
(dn_2 \cup_1 n_2)({\bf g},{\bf h},{\bf k},{\bf l}) = dn_2({\bf gh},{\bf k},{\bf l})n_2({\bf g},{\bf h}) + dn_2({\bf g},{\bf hk},{\bf l})n_2({\bf h},{\bf k}) + dn_2({\bf g},{\bf h},{\bf kl})n_2({\bf k},{\bf l})~,
\end{equation}
and we add a coboundary $du$ to ${\cal O}_4$ by redefining $\nu_3\rightarrow \nu_3u$, where
$u({\bf g},{\bf h},{\bf k}) = i^{[n_1({\bf g}) n_2({\bf h},{\bf k})]_2}$. It satisfies
\begin{align}
&du({\bf g},{\bf h},{\bf k},{\bf l})= i^{([n_1({\bf g})]_2 + [n_1({\bf h})]_2 - [n_1({\bf gh})]_2) [n_2({\bf k},{\bf l})]_2 + [n_1({\bf g})]_2 ([n_2({\bf h},{\bf k})]_2 + [n_2({\bf hk},{\bf l})]_2 - [n_2({\bf h},{\bf kl})]_2 - [n_2({\bf k},{\bf l})]_2)} \\
&\quad = (-1)^{n_1({\bf g}) n_1({\bf h}) n_2({\bf k},{\bf l}) + (n_1 \cup (n_2 \cup_1 n_2))({\bf g},{\bf h},{\bf k},{\bf l})}\times i^{[n_1({\bf g})]_2 ( [n_2({\bf h},{\bf k}) + n_2({\bf hk},{\bf l})]_2 - [n_2({\bf h},{\bf kl}) + n_2({\bf k},{\bf l})]_2 )  } \\
&\quad = (-1)^{n_1({\bf g}) n_1({\bf h}) n_2({\bf k},{\bf l}) + (n_1 \cup (n_2 \cup_1 n_2))({\bf g},{\bf h},{\bf k},{\bf l})}\times (-1)^{n_1({\bf g}) dn_2({\bf h},{\bf k},{\bf l})(n_2({\bf h},{\bf kl}) + n_2({\bf k},{\bf l}))} \times  i^{[n_1({\bf g}) dn_2({\bf h},{\bf k},{\bf l})]_2}~.
\end{align}
Replacing $\mathcal{O}_4$ by $\mathcal{O}_4 du$ gives
\begin{align}
\mathcal{O}_4[c_-=4k,n_1,n_2]({\bf g},{\bf h},{\bf k},{\bf l}) &\sim (-1)^{(n_2 \cup (n_2 + \omega_2) + dn_2 \cup_1 n_2)({\bf g},{\bf h},{\bf k},{\bf l}) + dn_2({\bf g},{\bf h},{\bf kl}) dn_2({\bf h},{\bf k},{\bf l}) } \times  i^{[n_1({\bf g}) dn_2({\bf h},{\bf k},{\bf l})]_2} \nonumber \\
& \times (-1)^{\frac{c_-}{4} ((n_2 \cup \omega_2 + \omega_2 \cup n_2+ \omega_2 \cup \omega_2)({\bf g},{\bf h},{\bf k},{\bf l}) + \omega_2({\bf g},{\bf h}) dn_2({\bf gh},{\bf k},{\bf l}) ) \label{Eq:O4_c0}}~.
\end{align} 
We can further simplify the expression by introducing the Cartan coboundary $\zeta$, defined as \cite{Cartan_coboundary_2020}:
\begin{eqs}
    Sq^2(x_1 \cup x_2) = Sq^1 (x_1) \cup Sq^1 (x_2) + x_1 \cup Sq^2(x_2) + d \zeta~,
\label{eq: Cartan formula}
\end{eqs}
where $x_1$ and $x_2$ are arbitrary 1-cocycle and 2-cocycle ($x_1 \in \H^1(G,\ZZ_2)$ and $x_2 \in \H^2(G,\ZZ_2)$) and the solution of $\zeta$ is given by $\zeta ({\bf g},{\bf h},{\bf k},{\bf l})=x_1 ({\bf g}) x_1({\bf h}) x_2({\bf h},{\bf kl}) x_2 ({\bf k},{\bf l})$, which is equivalent to $\zeta = x_1 \cup [(x_1 \cup x_2) \cup_2 x_2 + x_1 \cup x_2]$.
When evaluated on 4-simplex $(01234)$ it is $\zeta(01234) = x_1(01) x_1(12) x_2(124) x_2 (234)$.
For even $c_-$ this is the unique solution for $\zeta$, since the other solution can be obtained by changing $\zeta$ with a coboundary $\zeta\rightarrow \zeta+n_1\cup\omega_2=\zeta+dn_2$.
Choosing $x_1 = n_1$ and $x_2 = \omega_2$, and using $dn_2 = n_1 \cup \omega_2$, $\zeta$ becomes
\begin{eqs}
    \zeta ({\bf g},{\bf h},{\bf k},{\bf l}) &= n_1({\bf g}) \omega_2({\bf h},{\bf kl}) n_1({\bf h}) \omega_2({\bf k},{\bf l}) \\
    &= dn_2 ({\bf g},{\bf h},{\bf kl}) dn_2 ({\bf h},{\bf k},{\bf l})~.
\label{eq: Cartan coboundary in Appendix}
\end{eqs}
The $c_-$ terms in \eqref{Eq:O4_c0} can be simplified as
\begin{equation}
    \mathcal{A}_{c_-} \equiv \frac{c_-}{4} \left( n_2 \cup \omega_2 + \omega_2 \cup n_2 + \omega_2 \cup \omega_2 + (n_1 \cup_1 \omega_2 ) \cup \omega_2 \right)~,
\end{equation}
where we have used $dn_2 = n_1 \cup \omega_2$ and $(n_1 \cup_1 \omega_2) ({\bf g},{\bf h})= n_1({\bf gh}) \omega_2({\bf g},{\bf h})$.
We can verify that ${\cal A}_{c_-}$ is closed.
$\mathcal{A}_{c_-}$ can be expressed in another form:
\begin{eqs}
    \mathcal{A}_{c_-} &= \frac{c_-}{4} \left( (n_1\cup \omega_2) \cup_1 \omega_2 + \omega_2 \cup \omega_2 + (n_1 \cup_1 \omega_2 ) \cup \omega_2 + d(n_2 \cup_1 \omega_2) \right), \\
    &= \frac{c_-}{4} \left( n_1 \cup (\omega_2 \cup_1 \omega_2) + \omega_2 \cup \omega_2 + d(n_2 \cup_1 \omega_2) \right)~,
\end{eqs}
which follows from the Hirsch identity.
Notice that $n_1 \cup (\omega_2 \cup_1 \omega_2)$ is also a coboundary:
\begin{eqs}
    n_1 \cup (\omega_2 \cup_1 \omega_2) &= n_1 \cup Sq^1 \omega_2 \\
    &= (Sq^1 n_1) \cup \omega_2 + db',\quad b'=\frac{1}{2} (n_1 \cup \omega_2) \\
    &= db'',\quad b''=b'+n_1 \cup n_2~,
\end{eqs}
where in the last line we used $Sq^1n_1=n_1\cup n_1$ and $dn_2=n_1\cup \omega_2$. 
Therefore, the term with $c_-$ in the obstruction is simply
\begin{eqs}
    (-1)^{\frac{c_-}{4}(\omega_2 \cup \omega_2)}.
\end{eqs}
To summarize, the anomaly ${\cal O}_4$ for $c_-=4k$ can be expressed as
\begin{equation}\label{eqn:O4c=4n}
    \mathcal{O}_4[c_-=4k,n_1,n_2] = (-1)^{{\cal P}(n_2) + n_2 \cup \om + \zeta(n_1,\omega_2) + \frac{1}{2} n_1 \cup n_1 \cup \omega_2} i^{\frac{c_-}{2} {\cal P}(\om)}~.
\end{equation}
Notice that $\mathcal{O}_4$ changes by a coboundary when we replace $n_2$ by $n_2 + \om$, which is consistent with the equivalence $n_2 \sim n_2 + \om$ we used in the derivation.

\subsection{$c_-=0$ and $s_1 \ne 0$}

In this subsection we consider the $\H^4$ anomaly computation with antiunitary symmetries. We fix $c_-=0$. When the symmetry is nonpermuting, i.e. when $n_1 = 0$, the absolute anomaly can be obtained as usual, and the resulting expression coincides with Eq.~\eqref{Eq:O4_c0_t} with $c_-=0$.

When $n_1 \ne 0$, the main problem is to pick a suitable reference state which is non-anomalous. The reference must satisfy
\begin{equation}
    dn_2^{\text{ref}} = n_1 \cup (\om^{\text{ref}} + s_1 \cup n_1)~.
\end{equation}
Notice that there exists a solution with 
\begin{equation}
    \om^{\text{ref}} = s_1 \cup n_1; \quad n_2^{\text{ref}} = 0~.
\end{equation}
Such a reference has $\eta_a^{\text{ref}}({\bf g},{\bf h}) = (-1)^{a_e a_m s_1({\bf g}) n_1({\bf h})}$. Since it has a simple form with $\eta_e = \eta_m = 1$, we believe that it should be non-anomalous, although we cannot prove this. The rest of the calculation will proceed by assuming that this reference is indeed non-anomalous.

The reference is related to the state with the desired $(\omega_2, n_2)$ by the following 2-cocycle in $Z^2_{n_1}(G_b,\Z_2\times\Z_2)$:
\begin{equation}
    \mathfrak{t} = m^{\omega_2 + s_1 \cup n_1} \times \psi^{n_2}~.
\end{equation}
This is indeed a cocycle, by virtue of $dn_2 = n_1 \cup (\omega_2 + s_1 \cup n_1)$. We now use the relative anomaly formula with this choice of reference state and $\mathfrak{t}$ cocycle. We set $\omega_2' := \omega_2 + s_1 \cup n_1$. Proceeding as in the previous section, we obtain
\begin{align}
    R &= (-1)^{((n_2 + \omega_2') \cup n_2)({\bf g},{\bf h},{\bf k},{\bf l}) + dn_2({\bf gh},{\bf k},{\bf l}) (n_2+\omega'_2)({\bf g},{\bf h})}\nonumber \\
    \eta^{\text{ref}} &= (-1)^{s_1 \cup n_1 \cup f}\\
    ``\frac{U}{U}" &= (-1)^{n_1({\bf g})[(n_2 \cup_1 (n_2 + \omega'_2))({\bf h},{\bf k},{\bf l}) + n_2({\bf h},{\bf kl})dn_2({\bf h},{\bf k},{\bf l})]} \times i^{n_1({\bf g})[f({\bf h},{\bf k}) + f({\bf hk},{\bf l}) -f({\bf h},{\bf kl}) -f({\bf k},{\bf l})]} \\
    ``\frac{FFF}{FFF}" &= 1, \nonumber \\
    f({\bf h},{\bf k}) &:= n_2({\bf h},{\bf k})(1 + \omega'_2({\bf h},{\bf k})) \mod 2~.
\end{align}
For $b = i^{n_1 \cup f}$, we find that
\begin{equation}
    db({\bf g},{\bf h},{\bf k},{\bf l}) = (-1)^{(n_1 \cup n_1 \cup f + s_1 \cup n_1 \cup f)({\bf g},{\bf h},{\bf k},{\bf l})} i^{n_1({\bf g})[f({\bf h},{\bf k}) + f({\bf hk},{\bf l}) -f({\bf h},{\bf kl}) -f({\bf k},{\bf l})]}~.
\end{equation}
The term proportional to $s_1$ arises because we are using twisted coboundaries in $B^4(G_b,U(1)_T)$. Upon multiplying the above terms with $db$, and simplifying, we get
\begin{align}
    \mathcal{O}_4({\bf g},{\bf h},{\bf k},{\bf l}) &\sim (-1)^{((n_2 + \omega_2') \cup n_2 )({\bf g},{\bf h},{\bf k},{\bf l}) + dn_2({\bf gh},{\bf k},{\bf l}) (n_2+\omega'_2)({\bf g},{\bf h})} \times (-1)^{n_1 \cup n_1 \cup f}\nonumber \\ &\times (-1)^{n_1({\bf g})[(n_2 \cup_1 (n_2 + \omega'_2))({\bf h},{\bf k},{\bf l}) + n_2({\bf h},{\bf kl})dn_2({\bf h},{\bf k},{\bf l})]}~. 
\end{align}
This expression is identical to Eq.~\eqref{o4,partial} with $\om$ replaced by $\omega'_2$.
By defining a new variable $n'_2 \equiv n_2 + \omega'_2$ and following a similar calculation in the previous section, we get
\begin{align}
    \mathcal{O}_4({\bf g},{\bf h},{\bf k},{\bf l}) &\sim \nonumber (-1)^{(n'_2 \cup (n'_2 + \omega'_2) + dn'_2 \cup_1 n'_2 + \zeta)({\bf g},{\bf h},{\bf k},{\bf l}) }\nonumber \\ &\times (-1)^{(n_1 \cup n_1 \cup n'_2 + n_1 \cup (n'_2 \cup_1 n'_2))({\bf g},{\bf h},{\bf k},{\bf l}) + (n'_2({\bf h},{\bf kl}) + n'_2({\bf k},{\bf l})) dn'_2({\bf h},{\bf k},{\bf l})}. 
\end{align}
Here $\zeta$ is the same Cartan coboundary Eq.~\eqref{eq: Cartan coboundary in Appendix} defined in the previous section. The second line can be written as $db' \times i^{n_1 \cup dn'_2} \times (-1)^{s_1 \cup n_1 \cup n'_2}$, with $b' = i^{n_1 \cup n'_2}$. Once again, the term with $s_1$ arises because we are working with twisted cochains. Finally we obtain
\begin{align}
    \mathcal{O}_4 &\sim (-1)^{n'_2 \cup (n'_2 + \omega'_2) + dn'_2 \cup_1 n'_2 + \zeta(n_1,\omega'_2) } \times (-1)^{s_1 \cup n_1 \cup n'_2} \times i^{n_1 \cup n_1 \cup \omega'_2}~.
\end{align}
Finally, we substitute $n'_2 = n_2 + \omega'_2$ back and obtain the $\mathcal{O}_4$ in terms of $n_2$ (up to a coboundary):
\begin{align}
    \mathcal{O}_4 &\sim (-1)^{n_2 \cup (n_2 + \omega'_2) + dn_2 \cup_1 n_2 + \zeta(n_1,\omega'_2) } \times (-1)^{s_1 \cup n_1 \cup n_2} \times i^{n_1 \cup n_1 \cup \omega'_2}~,
\label{eq: O4 for s1 in Appendix}
\end{align}
where we have used $n_1 \cup \omega'_2 = dn_2$ and therefore $s_1 \cup n_1 \cup \omega'_2$ is a coboundary.
Let us check that $\mathcal{O}_4$ is closed. When we compute $d\mathcal{O}_4$, the terms that appear when $s_1 = 0$ also appear in this case; they all cancel out. The new contributions which explicitly depend on $s_1$ come from taking coboundaries of the last two terms.
The definition of the coboundary operator of the anti-unitary case is
\begin{eqs}
    d \mathcal{O}_4 ({\bf g} ,{\bf h},{\bf k},{\bf l}, {\bf m}) = \frac{{\cal K}^{s_1({\bf g})}\mathcal{O}_4 ({\bf h},{\bf k},{\bf l}, {\bf m}){\cal K}^{-s_1({\bf g})}
    \mathcal{O}_4 ({\bf g},{\bf h}{\bf k},{\bf l},{\bf m})
    \mathcal{O}_4 ({\bf g},{\bf h},{\bf k},{\bf l}{\bf m})
    }
    {\mathcal{O}_4 ({\bf g}{\bf h} ,{\bf k},{\bf l}, {\bf m}) 
    \mathcal{O}_4 ({\bf g},{\bf h},{\bf k}{\bf l}, {\bf m})
    \mathcal{O}_4 ({\bf g},{\bf h},{\bf k},{\bf l}) }~,
\end{eqs}
where ${\cal K}$ is the complex conjugation, with ${\cal K}^0=1$.
In Eq.~\eqref{eq: O4 for s1 in Appendix}, the cobundary of the penultimate term and the last terms are
\begin{eqs}
    d (-1)^{s_1 \cup n_1 \cup n_2} = (-1)^{s_1 \cup n_1 \cup dn_2}
\label{eq: s1 O4 additional 1}
\end{eqs}
and
\begin{eqs}
    d (i^{n_1 \cup n_1 \cup \om^\prime}) = d_{0} (i^{n_1 \cup n_1 \cup \om^\prime}) \times (-1)^{s_1 \cup n_1 \cup n_1 \cup \om^\prime},
\label{eq: s1 O4 additional 2}
\end{eqs}
where $d_0$ is the untwisted coboundary operator, and the first piece $d_{0} (i^{n_1 \cup n_1 \cup \om^\prime})$ already exists in the previous calculation for $s_1 = 0$, while $(-1)^{s_1 \cup n_1 \cup n_1 \cup \om^\prime} = (-1)^{s_1 \cup n_1 \cup dn_2}$ is additional. We can see that additional terms in Eqs.~\eqref{eq: s1 O4 additional 1} and \eqref{eq: s1 O4 additional 2} cancel out.
This means that $d\mathcal{O}_4=0$.  

\subsection{$c_-=4k+2$}

In this case we have, in vector notation, the following relations:
\begin{align}
    \mathfrak{t}({\bf g},{\bf h}) &= (n_2({\bf g},{\bf h}),(n_2 + \omega_2)({\bf g},{\bf h})) \\
    {^{\bf g}}\mathfrak{t}({\bf h},{\bf k}) &= (n_2({\bf h},{\bf k}) + n_1({\bf g})\omega_2({\bf h},{\bf k}),(n_2 + \omega_2)({\bf h},{\bf k}) + n_1({\bf g})\omega_2({\bf h},{\bf k}))~.
\end{align}

\subsubsection{$n_1 = 0$}

In this case the anomaly ${\cal O}_4$ only contains the $R$ part and the ``$FFF/FFF$" part. 
Let us compute the 
$R$ part, $R^{\mathfrak{t}({\bf g},{\bf h})\mathfrak{t}({\bf k},{\bf l})}$.
Its value depends on the 4 arguments $n_2({\bf g},{\bf h}),n_2({\bf k},{\bf l}),\omega_2({\bf g},{\bf h}),\omega_2({\bf k},{\bf l})$. Depending on whether these quantities take the values $0$ or 1, there are a total of 16 different input arguments for a fixed $c_-$. By examining the values attained by $R$ as a function of these arguments, we find that 
\begin{equation}
R^{\mathfrak{t}({\bf g},{\bf h})\mathfrak{t}({\bf k},{\bf l})} 
= e^{2\pi i \left(\frac{1}{2} n_2 \cup n_2 + \frac{c_-}{8} ([\omega_2]_2 \cup [\omega_2]_2 + [n_2]_2 \cup [\omega_2]_2 + [\omega_2]_2 \cup [n_2]_2  )\right)({\bf g},{\bf h},{\bf k},{\bf l})}~.
\end{equation} 

Next we evaluate the $F$ symbols. The computation is as follows. First we consider the two $F$ symbols with argument $\mathfrak{t}({\bf ghk},{\bf l})$. The contribution from these $F$ symbols is 
\begin{align}
    \frac{F^{\mathfrak{t}({\bf h},{\bf k})\mathfrak{t}({\bf g},{\bf hk})\mathfrak{t}({\bf ghk},{\bf l})}}{F^{\mathfrak{t}({\bf g},{\bf h})\mathfrak{t}({\bf gh},{\bf k})\mathfrak{t}({\bf ghk},{\bf l})}} &= (-1)^{n_2({\bf ghk},{\bf l}) (n_2({\bf g},{\bf h}) n_2({\bf gh},{\bf k}) + n_2({\bf h},{\bf k}) n_2({\bf g},{\bf hk}) ) + (n_2 \leftrightarrow n_2 + \omega_2)} \\
    &= (-1)^{n_2({\bf ghk},{\bf l}) ((n_2 \cup_1 n_2)({\bf g},{\bf h},{\bf k})) + (n_2+\omega_2)({\bf ghk},{\bf l}) ((n_2 +\omega_2) \cup_1 (n_2+\omega_2))({\bf g},{\bf h},{\bf k})}~.
\end{align}
The symbol $(n_2 \leftrightarrow n_2 + \omega_2)$ in the first line means that the entire preceding expression is repeated after replacing $n_2$ by $n_2 + \omega_2$. Next, we consider the two $F$ symbols with argument $\mathfrak{t}({\bf g},{\bf hkl})$. The contribution from these $F$ symbols is
\begin{equation}
    \frac{F^{\mathfrak{t}({\bf k},{\bf l})\mathfrak{t}({\bf h},{\bf kl})\mathfrak{t}({\bf g},{\bf hkl})}}{F^{\mathfrak{t}({\bf h},{\bf k})\mathfrak{t}({\bf hk},{\bf l})\mathfrak{t}({\bf g},{\bf hkl})}} = (-1)^{n_2({\bf g},{\bf hkl}) ((n_2 \cup_1 n_2)({\bf h},{\bf k},{\bf l})) + (n_2+\omega_2)({\bf g},{\bf hkl}) ((n_2 +\omega_2) \cup_1 (n_2+\omega_2))({\bf h},{\bf k},{\bf l})}~.
\end{equation}
The final pair of $F$ symbols vanishes by symmetry. To combine the above expressions we use the fact that the cup-1 product of a 2-cochain $n_2$ with a 3-cochain $f_3$ is
\begin{equation}
    (n_2 \cup_1 f_3)({\bf g},{\bf h},{\bf k},{\bf l}) = n_2({\bf ghk},{\bf l}) f_3({\bf g},{\bf h},{\bf k}) + n_2({\bf g},{\bf hkl}) f_3({\bf h},{\bf k},{\bf l})~.
\end{equation}

Using this, we obtain the following result for the ``$FFF/FFF$" part of the anomaly ${\cal O}_4$,
\begin{align}
``\;\frac{FFF}{FFF}\;"&= (-1)^{n_2 \cup_1(n_2 \cup_1 n_2) + (n_2 + \omega_2) \cup_1((n_2+\omega_2) \cup_1 (n_2+\omega_2))}~.
\end{align}
This implies that
\begin{align}
  \mathcal{O}_4 = e^{2\pi i \left(\frac{1}{2} n_2 \cup n_2 + \frac{c_-}{8} ([\omega_2]_2 \cup [\omega_2]_2 + [n_2]_2 \cup [\omega_2]_2 + [\omega_2]_2 \cup [n_2]_2  )\right)} \times (-1)^{n_2 \cup_1(n_2 \cup_1 n_2) + (n_2 + \omega_2) \cup_1((n_2+\omega_2) \cup_1 (n_2+\omega_2))}~. 
\end{align}
It can be simplified as:
\begin{eqs}
    \mathcal{O}_4 = \exp \left [ 2 \pi i \left(\frac{c_-}{8} n_2 \cup n_2  + \frac{c_-}{8}  (n_2 + \omega_2) \cup (n_2 + \omega_2) \right) \right]
    \times
    \exp \left [ 2 \pi i \left( -\frac{1}{4} n_2 \cup_1 d n_2  -\frac{1}{4} (n_2+\omega_2) \cup_1 d (n_2 + \omega_2) \right)\right]~,
\end{eqs}
where we have used the identify $\frac{d B_2}{2} = B_2 \cup_1 B_2$ for any 2-cocycle $B_2$ (see Eq.~\eqref{eqn:dXidentity}). Using the definition of the Pontryagin square
\begin{eqs}
    {\cal P}(a) \equiv a \cup a - a \cup_1 d a~
\end{eqs}
we can express the anomaly ${\cal O}_4$ as
\begin{eqs}
    \mathcal{O}_4 = \exp \left[ 2\pi i ( \frac{c_-}{8}  ({\cal P}(n_2) + {\cal P}(n_2 + \omega_2)) ) \right]~.
\end{eqs}
We can further simplify it using
\begin{eqs}
    {\cal P}(B_2 + B'_2) - {\cal P}(B_2) - {\cal P}(B'_2) &= 2 B_2 \cup B'_2 + d z + 2w + d B_2 \cup_2 d B'_2~,
\label{eq: diff of Pontry}
\end{eqs}
where $w = - B_2 \cup_1 d B'_2$ and $z = B_2 \cup_1 B'_2 + d B_2 \cup_2 B'_2$.
\begin{eqs}
    {\cal P}(n_2) + {\cal P}(n_2 + \omega_2) &=  {\cal P}(2n_2 + \om) - 2 n_2 \cup (n_2 + \om) + \text{coboundary}\quad \text{mod } 4 \\
    &= 2n_2 \cup (n_2 + \om) + {\cal P}(\om) +\text{coboundary} \quad  \text{mod } 4~.
\end{eqs}
The expression of $\mathcal{O}_4$ finally becomes
\begin{eqs}
    \mathcal{O}_4[c_-= 4k + 2,n_1=0,n_2] = (-1)^{(n_2 + \omega_2) \cup n_2} i^{ \frac{c_-}{2} {\cal P}(\omega_2)}.
\end{eqs}

\subsubsection{$n_1 \ne 0$}
Each part in the anomaly ${\cal O}^{(0)}_4$ is given by
\begin{align}
R &= \theta^{[n_2({\bf g},{\bf h})]_2 [n_2({\bf k},{\bf l}) + dn_2({\bf gh},{\bf k},{\bf l})]_2 + [(n_2+\omega_2)({\bf g},{\bf h})]_2[(n_2+\omega_2)({\bf k},{\bf l}) + dn_2({\bf gh},{\bf k},{\bf l})]_2} \nonumber\\
\eta &= 1 \nonumber\\
``\;\frac{U}{U}\;" &= 1 \nonumber\\
``\;\frac{FFF}{FFF}\;" &= (-1)^{n_2({\bf ghk},{\bf l})[ n_2({\bf g},{\bf hk}) (n_2({\bf h},{\bf k})+n_1({\bf g}) \om({\bf h},{\bf k}))
+ n_2({\bf gh},{\bf k}) n_2({\bf g},{\bf h})]} \nonumber \\
& \quad \times (-1)^{n_2({\bf g},{\bf hkl}) [  
(n_2({\bf h},{\bf kl})+ n_1({\bf g}) \om({\bf h},{\bf kl})) (n_2({\bf k},{\bf l})+n_1({\bf gh}) \om({\bf k,l}))
+ (n_2({\bf hk},{\bf l}) + n_1({\bf g}) \om({\bf hk},{\bf l}))(n_2({\bf h},{\bf k}) + n_1({\bf g}) \om({\bf h},{\bf k}))]
} \nonumber \\
& \quad \times ( n_2 \leftrightarrow n_2 + \om)
\label{Eq:O4_c2}~, 
\end{align}
where $\theta = e^{i \pi c_-/4}$, and $( n_2 \leftrightarrow n_2 + \om)$ represents the same terms above with replacement of $n_2$ by $n_2 + \om$. Multiplying these together gives $\mathcal{O}_4^{(0)}$.
To simplify the formula, we first use the following identities
\begin{eqs}
    [n_2(234) + (n_1 \cup \om)(0234)]_2 &= n_2(234) + (n_1 \cup \om)(0234) - 2 n_2(234) (n_1 \cup \om)(0234) \\
    [n_2(012) + \om (012)]_2 &= n_2(012) + \om (012) - 2 n_2(012) \om (012) \\
    [(n_2+\om)(234) + (n_1 \cup \om)(0234)]_2&= n_2(234) + \om(234) - 2 n_2(234)\om(234) + (n_1 \cup \om)(0234) \\
    & - 2 (n_2(234) + \om(234) - 2 n_2(234)\om(234)) (n_1 \cup \om)(0234) ~.
\end{eqs}
Let us first consider the part involving $R$:
\begin{eqs}
    & \frac{c_-}{4}\Big[n_2 \cup n_2 - (n_1 \cup_1 n_2)\cup \om
    +(n_2+\om-2n_2 \cup_2 \om) \cup (n_2+\om-2n_2 \cup_2 \om) 
    - (n_1 \cup_1 (n_2+\om-2n_2 \cup_2 \om))\cup \om \\
    &\qquad + 2(n_1 \cup_1 n_2) \cup (n_2 \cup_2 \om) 
    + 2[n_1 \cup_1 (n_2+\om-2n_2 \cup_2 \om)] \cup [(n_2+\om-2n_2 \cup_2 \om) \cup_2 \om]\Big] \\
    &= \frac{c_-}{4}\Big[n_2 \cup n_2 - (n_1 \cup_1 n_2)\cup \om+(n_2+\om) \cup (n_2 +\om) - (n_1 \cup_1 (n_2 + \om))\cup \om \Big] \\
    &\qquad + (n_2 \cup_2 \om) \cup (n_2 + \om) + (n_2 + \om) \cup (n_2 \cup_2 \om) + [n_1 \cup_1 (n_2 \cup_2 \om)] \cup \om \\
    &\qquad + (n_1 \cup_1 n_2) \cup (n_2 \cup_2 \om) 
    + [n_1 \cup_1 (n_2+\om)] \cup [(n_2+\om) \cup_2 \om] \\
    &= \frac{c_-}{4}\Big[n_2 \cup n_2 - (n_1 \cup_1 n_2)\cup \om+(n_2+\om) \cup (n_2 +\om) - (n_1 \cup_1 (n_2 + \om))\cup \om \Big] \\
    &\qquad + (n_2 \cup_2 \om) \cup (n_2 + \om) + (n_2 + \om) \cup (n_2 \cup_2 \om) + [n_1 \cup_1 (n_2 \cup_2 \om)] \cup \om \\
    &\qquad + (n_1 \cup_1 \om) \cup (n_2 \cup_2 \om) + (n_1 \cup_1 (\om+n_2)) \cup \om \\
    &= \frac{c_-}{4}\Big[n_2 \cup n_2 - (n_1 \cup_1 n_2)\cup \om+(n_2+\om) \cup (n_2 +\om) - (n_1 \cup_1 (n_2 + \om))\cup \om \Big] \\
    &
    \qquad + (n_2 + \om) \cup_1 d(n_2 \cup_2 \om)
    + dn_2 \cup_1 (n_2 \cup_2 \om)
    + [n_1 \cup_1 (n_2 \cup_2 \om)] \cup \om \\
    &\qquad + (n_1 \cup_1 \om) \cup (n_2 \cup_2 \om) + (n_1 \cup_1 (\om+n_2)) \cup \om ~.
\end{eqs}
The first line in the last expression equals
\begin{eqs}
    &\frac{c_-}{4}(n_2 \cup n_2 
    - (n_1 \cup_1 n_2)\cup \om
    +(n_2+\om) \cup (n_2 +\om) 
    - (n_1 \cup_1 (n_2 + \om))\cup \om) \\
    &= \frac{c_-}{4} (2 n_2 \cup n_2 + 2 (n_1 \cup_1 n_2) \cup \om 
    + \om \cup \om +n_2 \cup \om+ \om \cup n_2 - (n_1 \cup_1 \om) \cup \om) \\
    &\sim  \frac{c_-}{4} (2 n_2 \cup (n_2 +\om) 
    + 2 (n_1 \cup_1 n_2) \cup \om 
    +\om \cup \om - dn_2 \cup_1 \om 
    - n_2 \cup_1 d\om 
    - (n_1 \cup_1 \om) \cup \om ) \\
    &= n_2 \cup (n_2 +\om) +  (n_1 \cup_1 n_2) \cup \om 
    + n_2 \cup_1 (\om \cup_1 \om)  
    + \frac{c_-}{4} (\om \cup \om -dn_2 \cup_1 \om 
    - (n_1 \cup_1 \om) \cup \om) \\
    &= n_2 \cup (n_2 +\om) +  (n_1 \cup_1 n_2) \cup \om
    + n_2 \cup_1 (\om \cup_1 \om) 
    + dn_2 \cup_1 \om\cr
    &\quad + \frac{c_-}{4} ( \om \cup \om  + (dn_2 - n_1 \cup \om) \cup_1 \om 
    -  n_1 \cup (\om \cup_1 \om) ) \\
    &= n_2 \cup (n_2 +\om) +  (n_1 \cup_1 n_2) \cup \om
    + n_2 \cup_1 (\om \cup_1 \om) 
     + (n_2 \cup_1 n_2 + n_2 \cup_2 dn_2) \cup_1 \om\cr
    &\quad  + dn_2 \cup_1 \om
     + \frac{c_-}{4} ( \om \cup \om  +  n_1 \cup (\om \cup_1 \om) )~,\label{eqn:R0}
\end{eqs}
where $\sim$ denotes equals up to a coboundary that can be absorbed into redefinition of $\nu_3$, and we have used the Hirsch identity \cite{Hirsch:1955}:
\begin{eqs}
    (a\cup b) \cup_1 c = (-1)^{\deg a} a \cup (b\cup_1 c) + (-1)^{\deg b \deg c} (a\cup_1 c) \cup b~.
\end{eqs}
Since $(-1)^{n_1 \cup n_1 \cup \om}=d((-1)^{n_1\cup n_2})$ is a coboundary (due to $dn_2=n_1\cup \omega_2$ mod 2), and $n_1 \cup (\om \cup_1 \om) $ equals $n_1 \cup n_1 \cup \om$ up to a coboundary,  Eq.~\eqref{eqn:R0} for $c_-=4k+2$ equals up a coboundary to
\begin{eqs}
    &n_2 \cup (n_2 +\om) +  (n_1 \cup_1 n_2) \cup \om + n_2 \cup_1 (\om \cup_1 \om)  + \frac{1}{2} n_1 \cup n_1 \cup \om \\
    &+ (n_2 \cup_1 n_2 + n_2 \cup_2 dn_2) \cup_1 \om
    + dn_2 \cup_1 \om 
    + \frac{c_-}{4} ( \om \cup \om  )~.
\end{eqs}
The total $R$ part is:
\begin{eqs}
    &n_2 \cup (n_2 +\om) +  (n_1 \cup_1 n_2) \cup \om + n_2 \cup_1 (\om \cup_1 \om)  + \frac{1}{2} n_1 \cup n_1 \cup \om + \frac{c_-}{4} ( \om \cup \om  ) \\
    &+ (n_2 \cup_1 n_2 + n_2 \cup_2 dn_2) \cup_1 \om
    + dn_2 \cup_1 \om \\
    &+ (n_2 + \om) \cup_1 d(n_2 \cup_2 \om)
    + dn_2 \cup_1 (n_2 \cup_2 \om)
    + [n_1 \cup_1 (n_2 \cup_2 \om)] \cup \om \\
    &+ (n_1 \cup_1 \om) \cup (n_2 \cup_2 \om) + (n_1 \cup_1 (\om+n_2)) \cup \om ~.
\end{eqs}
For the remaining parts in ${\cal O}_4$,
the $\eta$ and $U/U$ part equals 1 up to a coboundary.
The ``$FFF/FFF$" part without the $(n_2 \rightarrow n_2 + \om)$ piece equals
\begin{eqs}
    &\quad[n_2(123)+dn_2(0123)] n_2(013) n_2(034) + n_2(012) n_2(023) n_2 (034) \\
    &\quad + [n_2(234) + dn_2(0234)] [n_2(124) + dn_2(0124)] n_2(014) 
     + [n_2(123) + dn_2 (0123)] [n_2(134) + dn_2 (0134)]n_2(014) \\
    &=~n_2 \cup_1 (n_2 \cup_1 n_2) + n_2(034) n_1(01) n_2(013) \om(123) \\
    &\quad + n_2(014)[n_2(234) n_1(01) \om(124) + n_2(124) (n_1(01)+n_1(12)) \om(234) + n_1(01) n_1(02) \om(234) \om(124) \\
    & ~~~~ ~~~~ ~~~~ ~~~~ ~~ + n_2(123) n_1(01) \om(134) + n_2(134) n_1(01) \om(123) + n_1(01) \om (123) \om(134)]\\
    &= ~n_2 \cup_1 (n_2 \cup_1 n_2) 
    + n_2 \cup_1 [(n_1 \cup \om) \cup_2 n_2 + n_1 \cup (\om \cup_2 n_2)] \\
    &\quad + n_1(01) n_2(014)[ \om(124) \om(234) + n_2(124) \om(234) +\om(124) \om(234) + n_1(12) \om(124) \om(234)\\
    & ~~~~ ~~~~ ~~~~ ~~~~ ~~~~ ~~~~ +\om(134) n_2(123) +n_2(134) \om(123) +\om(134) \om(123)]~,
\end{eqs}
where we have used $[(n_1 \cup \om) \cup_2 n_2 + n_1 \cup (\om \cup_2 n_2)](1234) = n_1(12) n_2(124) \om(234)$.
The last two lines can be written using the 4-cochain valued function $f$ in (\ref{eqn:fandg}), $f(a_1, b_2, c_3) = a_1(01) b_2(014) c_3(1234)$:
\begin{eqs}
    &f(n_1,n_2, n_2 \cup_1 \om + \om \cup_1 n_2 + \om \cup_1 \om + (n_1 \cup \om) \cup_2 \om + n_1 \cup \om) \\
    &\quad = f(n_1,n_2, d(n_2 \cup_2 \om) + \om \cup_1 \om + dn_2)~. 
\end{eqs}
The last entry of $f$ does not depend on $n_2 \rightarrow n_2 + \om$, and since $f$ is linear in each entry, the sum of the $n_2$ part and the $n_2 + \om$ part is simply:
\begin{eqs}
    f(n_1,\om, d(n_2 \cup_2 \om) + \om \cup_1 \om + n_1 \cup \om)~.
\end{eqs}
The total ``$FFF/FFF$" part is
\begin{eqs}
    &~n_2 \cup_1 (n_2 \cup_1 n_2) + (n_2+\om) \cup_1 ((n_2+\om) \cup_1 (n_2+\om)) \\
    &\quad + n_2 \cup_1 [(n_1 \cup \om) \cup_2 \om + n_1 \cup \om ] 
    + \om \cup_1 [(n_1 \cup \om) \cup_2 n_2 + n_1 \cup (\om \cup_2 n_2)] \\
    &\quad + \om \cup_1 [(n_1 \cup \om) \cup_2 \om + n_1 \cup \om ] 
    + f(n_1,\om, d(n_2 \cup_2 \om) + \om \cup_1 \om + n_1 \cup \om)~.
\end{eqs}

The 4-cochain valued function $f$ has the following properties:
\begin{eqs}
    f(n_1, \om, d(n_2 \cup_2 \om)) &=(n_1 \cup_1 \om) \cup (n_2 \cup_2 \om)
    + \om \cup_1 (n_1 \cup (n_2 \cup_2 \om))
    + n_1 \cup (\om \cup_1 (n_2 \cup_2 \om)) \\
    f(n_1, \om, n_1 \cup \om) &= g(n_1 \cup n_1, \om , \om) + \zeta (n_1,\om) \\
    &= (n_1^2 \cup \om) \cup_2 \om + n_1^2 \cup \om + \zeta(n_1, \om) \\
    &\sim   (n_1^2 \cup \om) \cup_2 \om + \zeta(n_1, \om) \\
    f(n_1, \om, \om \cup_1 \om) 
    &= [n_1 \cup (\om \cup_1 \om)] \cup_2 \om + n_1 \cup [(\om \cup_1 \om) \cup_2 \om]\\
    &=  \frac{1}{2}[n_1 \cup d\om] \cup_2 \om\\
    &= \frac{1}{2} 
    (dn_1 \cup \om) \cup_2 \om - \frac{1}{2} d(n_1 \cup \om) \cup_2 \om \\
    &\sim (n_1^2 \cup \om) \cup_2 \om  + \frac{1}{2} d[dn_2 - n_1 \cup \om] \cup_2 \om\\
    &= (n_1^2 \cup \om) \cup_2 \om  
    + d(n_2 \cup_1 n_2 + n_2 \cup_2 dn_2) \cup_2 \om~,
\end{eqs}
where $\sim$ means equal up to a coboundary, and the first line follows from substituting $X = n_2 \cup_2 \om$ in (\ref{eqn:f(dX)}).
Using the above properties, the ``$FFF/FFF$" part can be expressed as
\begin{eqs}
    &~n_2 \cup_1 (n_2 \cup_1 n_2) + (n_2+\om) \cup_1 ((n_2+\om) \cup_1 (n_2+\om)) 
    + n_2 \cup_1 [(n_1 \cup \om) \cup_2 \om + n_1 \cup \om ] 
     \\
     &
    + \om \cup_1 [(n_1 \cup \om) \cup_2 n_2] 
    + \om \cup_1 [(n_1 \cup \om) \cup_2 \om + n_1 \cup \om ] \\
    & + (n_1 \cup_1 \om) \cup (n_2 \cup_2 \om)
    + n_1 \cup (\om \cup_1 (n_2 \cup_2 \om)) 
    + d(n_2 \cup_1 n_2 + n_2 \cup_2 dn_2) \cup_2 \om
    + \zeta(n_1, \om) \\
    =& ~ n_2 \cup_1 (\om \cup_1 \om + d(n_2 \cup_2 \om) + dn_2) \\
    &+ \om \cup_1 (\om \cup_1 \om + d(n_2 \cup_2 \om) + dn_2 + n_2 \cup_1 n_2 + dn_2 \cup_2 n_2)\\
    & + (n_1 \cup_1 \om) \cup (n_2 \cup_2 \om)
    + n_1 \cup (\om \cup_1 (n_2 \cup_2 \om))
    + d(n_2 \cup_1 n_2 + n_2 \cup_2 dn_2) \cup_2 \om
    + \zeta(n_1, \om) \\
    =& ~ n_2 \cup_1 (\om \cup_1 \om + d(n_2 \cup_2 \om) + dn_2) \\
    &+ \om \cup_1 (\om \cup_1 \om + d(n_2 \cup_2 \om) + n_2 \cup_1 n_2 + n_2 \cup_2 dn_2)\\
    & + (n_1 \cup_1 \om) \cup (n_2 \cup_2 \om)
    + n_1 \cup (\om \cup_1 (n_2 \cup_2 \om)) + d(n_2 \cup_1 n_2 + n_2 \cup_2 dn_2) \cup_2 \om 
    + \zeta(n_1, \om) ~.
\end{eqs}
The total anomaly is ${\cal O}_4=e^{i \pi \phi_4}$ with
\begin{eqs}\label{eqn:anomaly00}
    \phi_4 = & n_2 \cup (n_2 +\om) +  (n_1 \cup_1 n_2) \cup \om 
    + 
    n_2 \cup_1 (\om \cup_1 \om)
    + \frac{1}{2} n_1 \cup n_1 \cup \om + \frac{c_-}{4} ( \om \cup \om  ) \\
    &\quad + (n_2 \cup_1 n_2 + n_2 \cup_2 dn_2) \cup_1 \om
    + dn_2 \cup_1 \om 
     \\
     &\quad
    + 
    (n_2 + \om) \cup_1 d(n_2 \cup_2 \om)
    + dn_2 \cup_1 (n_2 \cup_2 \om)
    + [n_1 \cup_1 (n_2 \cup_2 \om)] \cup \om \\
    &\quad + 
    (n_1 \cup_1 \om) \cup (n_2 \cup_2 \om)
    + (n_1 \cup_1 (\om+n_2)) \cup \om 
    + n_2 \cup_1 (
    \om \cup_1 \om
    + 
    d(n_2 \cup_2 \om)
    + dn_2) \\
    &\quad + \om \cup_1 (\om \cup_1 \om 
    + 
    d(n_2 \cup_2 \om)
    + n_2 \cup_1 n_2 + n_2 \cup_2 dn_2)\\
    &\quad  + 
    (n_1 \cup_1 \om) \cup (n_2 \cup_2 \om)
    + n_1 \cup (\om \cup_1 (n_2 \cup_2 \om)) + dn_2 \cup_1 \om 
    + \zeta(n_1, \om) \\
    &= n_2 \cup (n_2 +\om) + n_2 \cup_1 dn_2 
      + \zeta(n_1, \om) + \frac{1}{2} n_1 \cup n_1 \cup \om + \frac{c_-}{4} ( \om \cup \om  - \om \cup_1 d\om)\\
    &\quad + 
    (n_1 \cup_1 n_2) \cup \om
    + (n_2 \cup_1 n_2 + n_2 \cup_2 dn_2) \cup_1 \om + \om \cup_1 (n_2 \cup_1 n_2 + n_2 \cup_2 dn_2) \\
    & \quad +d(n_2 \cup_1 n_2 + n_2 \cup_2 dn_2) \cup_2 \om 
    + 
    (n_1 \cup \om) \cup_1 (n_2 \cup \om)
    + 
    [n_1 \cup_1 (n_2 \cup_2 \om)] \cup \om
    \\
    &\quad + dn_2 \cup_1 \om + (n_1 \cup_1 (\om+ 
    n_2
    )) \cup \om + 
    n_1 \cup (\om \cup_1 (n_2 \cup_2 \om))
    ~.
\end{eqs}
To simplify the expression for the anomaly, we note that
the following expression is a coboundary:
\begin{eqs}
    &d\left((n_2 \cup_1 n_2 + n_2 \cup_2 dn_2)\cup_2\omega_2\right)\cr
    &\quad =
    (n_2 \cup_1 n_2 + n_2 \cup_2 dn_2) \cup_1 \om + \om \cup_1 (n_2 \cup_1 n_2 + n_2 \cup_2 dn_2) + d(n_2 \cup_1 n_2 + n_2 \cup_2 dn_2) \cup_2 \om \sim 0~,
\end{eqs}
where $\sim 0$ means it equals to zero up to a coboundary.
We also use the identity
\begin{eqs}
    (n_1 \cup_1 \om) \cup \om &= n_1 \cup (\om \cup_1 \om) + (n_1 \cup \om) \cup_1 \om \\
    &\sim  dn_2 \cup_1 \om~,
\end{eqs}
which cancels out the first term in the last line of (\ref{eqn:anomaly00}).
Thus the anomaly simplifies into ${\cal O}_4\sim e^{i \pi \phi_4}$ with
\begin{eqs}
    \phi_4=n_2 \cup (n_2 +\om) + n_2 \cup_1 dn_2 
      + \zeta(n_1, \om) + \frac{1}{2} n_1 \cup n_1 \cup \om + \frac{c_-}{4} ( \om \cup \om  - \om \cup_1 d\om)~.
\end{eqs}

\subsection{$c_-=2k+1$}\label{sec:coddanomaly}
In this case we have, in mod 4 notation, the following relations:
\begin{align}\label{eqn:Z4Z2}
    \mathfrak{t}({\bf g},{\bf h}) &= ([\omega_2]_2 + 2n_2)({\bf g},{\bf h}) \mod 4 \\
    {^{\bf g}}\mathfrak{t}({\bf h},{\bf k}) &= ([\omega_2]_2 + 2n_2)({\bf h},{\bf k}) + 2 n_1({\bf g}) \omega_2({\bf h},{\bf k}) \mod 4~.
\end{align}
We will write $\omega_2,n_2$ as $[\omega_2]_2, [n_2]_2$ unless it is clear that the overall expression is invariant under a shift of $\omega_2,n_2$ by even integers.
\subsubsection{$n_1 = 0$}
Here the anomaly has the form ``$\mathcal{O}_4 = R \frac{FFF}{FFF}$". It is perhaps simplest to leave the expression in terms of the cocycle $\mathfrak{t} \in Z^2(G_b,\Z_4)$: we obtain

\begin{align}
    R &= \theta^{[\mathfrak{t]}_4 \cup [\mathfrak{t}]_4} \\
   ``\; \frac{FFF}{FFF}\;" &= \theta^{[\mathfrak{t}]_4 \cup_1 d[\mathfrak{t}]_4}\\
    \implies \mathcal{O}_4[c_-=2k+1,n_1=0,n_2] &= \theta^{[\mathfrak{t]}_4 \cup [\mathfrak{t}]_4 + [\mathfrak{t}]_4 \cup_1 d[\mathfrak{t}]_4} = e^{i \pi 
\frac{c_-}{4} \mathcal{P}(\mathfrak{t})}~, \label{Eq:O4_c1_t}
\end{align}
where we used the property that $\theta^{ [\mathfrak{t}]_4 \cup_1 d[\mathfrak{t}]_4}=\theta^{ -[\mathfrak{t}]_4 \cup_1 d[\mathfrak{t}]_4}$, since $d\mathfrak{t}=0$ mod 4, 
and thus $e^{\pi i c_- [\mathfrak{t}]_4 \cup_1 {d[\mathfrak{t}]_4\over 4}}=\pm1=e^{-\pi i c_- [\mathfrak{t}]_4 \cup_1 {d[\mathfrak{t}]_4\over 4}}$.
Thus the anomaly is completely described in terms of the Pontryagin square of $\mathfrak{t}$. 
We can also express the obstruction ${\cal O}_4$ using $n_1, n_2, \om$ instead of $\mathfrak{t}$. Substituting (\ref{eqn:Z4Z2}) into ${\cal O}_4 = e^{i \pi \phi_4}$:
\begin{eqs}
    \phi_4 & = \frac{c_-}{4}  (4 n_2 \cup n_2 + 2n_2 \cup \om + 2 \om \cup n_2 + \om \cup \om - 4 n_2 \cup_1 dn_2 - 2 n_2 \cup_1 d\om - 2 \om \cup_1 d n_2 - \om \cup_1 d\om) \\
    &= n_2 \cup n_2 - n_2\cup_1 dn_2 + \frac{c_-}{2} (n_2 \cup \om + \om \cup n_2 - n_2 \cup_1 d\om  - \om \cup_1 dn_2) + \frac{c_-}{4}(\om \cup \om - \om \cup_1 d\om)\\
    &\sim  n_2 \cup n_2 - n_2 \cup_1 dn_2 + n_2 \cup \om + \frac{c_-}{2} (-dn_2 \cup_1 \om - \om \cup_1 dn_2) + \frac{c_-}{4}(\om \cup \om - \om \cup_1 d\om) \\
    &\sim {\cal P}(n_2) + n_2 \cup \om + \frac{c_-}{2} (dn_2 \cup_2 d\om) +\frac{c_-}{4} {\cal P}(\om) \\
    &= {\cal P}(n_2) + n_2 \cup \om + c_- (\frac{d\om}{2}) \cup_2 (\frac{d\om}{2}) + \frac{c_-}{4} {\cal P}(\om)\\
    & \sim {\cal P}(n_2) + n_2 \cup \om + \frac{c_-}{4} {\cal P}(\om)~,
\end{eqs}
where $\sim$ means equal up to a coboundary, which can be absorbed into a redefinition of $\nu_3$.

\subsubsection{$n_1 \ne 0$}

In the general case, we have the relative anomaly expression that takes the form $\mathcal{O}^{(0)}_4 =``\; R \eta \frac{U}{U} \frac{FFF}{FFF}\;"$. 
Each part in the anomaly ${\cal O}_4$ is given by
\begin{align}
R &= \theta^{[ ([\omega_2]_2 + 2n_2)({\bf g},{\bf h}) ]_4 [([\omega_2]_2 + 2n_2)({\bf k},{\bf l}) + 2 n_1({\bf gh}) \omega_2({\bf k},{\bf l})]_4} \nonumber \\
\eta &=
\begin{cases}
    (-1)^{n_1 \cup n_1 \cup \om},& \text{if } c_-=3 ~\text{(mod $4$)}\\
    1,              & \text{if } c_-=1 ~\text{(mod $4$)}
\end{cases} \nonumber \\
``\;\frac{U}{U}\;" &= (-1)^{n_1({\bf g}) ( \omega_2({\bf hk},{\bf l}) \omega_2({\bf h},{\bf k}) n_2({\bf h},{\bf k}) + \omega_2({\bf h},{\bf kl}) \omega_2({\bf k},{\bf l}) n_2({\bf k},{\bf l}) )}  (-1)^{n_1 \cup (\om \cup_1 (n_2 + \om))} 
\nonumber \\
``\;\frac{FFF}{FFF}\;" &= \theta^{\omega_2({\bf ghk},{\bf l}) ( [([\omega_2]_2 + 2 n_2)({\bf gh},{\bf k})]_4 + [([\omega_2]_2 + 2 n_2)({\bf g},{\bf h})]_4 - [([\omega_2]_2 + 2 n_2)({\bf g},{\bf hk})]_4 - [([\omega_2]_2 + 2 n_2)({\bf h},{\bf k}) + 2 n_1({\bf g}) \omega_2({\bf h},{\bf k})]_4)} \nonumber \\
&\quad  \times \theta^{\omega_2({\bf g},{\bf hkl}) ( [([\omega_2]_2 + 2 n_2)({\bf hk},{\bf l}) + 2 n_1({\bf g}) \omega_2({\bf hk},{\bf l})]_4 + [([\omega_2]_2 + 2 n_2)({\bf h},{\bf k}) + 2 n_1({\bf g}) \omega_2({\bf h},{\bf k})]_4)} \nonumber \\
& \quad \times \theta^{- \omega_2({\bf g},{\bf hkl}) ([([\omega_2]_2 + 2 n_2)({\bf h},{\bf kl}) + 2 n_1({\bf g}) \omega_2({\bf h},{\bf kl})]_4 + [([\omega_2]_2 + 2 n_2)({\bf k},{\bf l}) + 2 n_1({\bf gh}) \omega_2({\bf k},{\bf l})]_4) ~,\label{Eq:O4_c1}}
\end{align}
where $\theta = e^{i \pi c_-/4}$, and $[\cdot]_4$ denotes mod 4.
Let us simplify the expression of each part in the anomaly ${\cal O}_4^{(0)}$.
The ``$U/U$" part can be shown to equal
\begin{equation}
    n_1 \cup (\omega_2\cup_1 (\omega_2 \cup_2 n_2)) + n_1 \cup (\om \cup_1 (n_2 + \om))~.
\end{equation}
To simplify the $R$ part, we use the identity $[(\om + 2n_2)(g,h)]_4 = (\om + 2n_2)(g,h)$ and $[(\om + 2n_2)(k,l)+2n_1(gh) \om (k,l)]_4 = (\om + 2n_2)(k,l)+2n_1(gh) \om (k,l) - 4 n_1(gh) n_2 (k,l) \om (k,l)$. We find the $R$ part equals
\begin{eqs}
    &\theta^{(\om+2n_2) \cup (\om+2n_2)
    -2( (n_1\cup_1 (\om+ 2 n_2)) \cup \om) -4(n_1 \cup_1 \om) \cup (\om \cup_2 n_2)}\\
    &=(-1)^{n_2 \cup n_2 + (n_1 \cup_1 n_2) \cup \om + (n_1 \cup_1 \om) \cup (\om \cup_2 n_2) +\frac{c_-}{2}(n_2 \cup \om + \om \cup n_2 - (n_1 \cup_1 \om) \cup \om) + \frac{c_-}{4}(\om \cup \om)}~.
\end{eqs}
To simplify the ``$FFF/FFF$" part, 
we use the identity
$[(\omega_2 + 2 n_2)({\bf h},{\bf k}) + 2 n_1({\bf g}) \omega_2({\bf h},{\bf k})]_4= (\omega_2 + 2 n_2)({\bf h},{\bf k}) + 2 n_1({\bf g}) \omega_2({\bf h},{\bf k})-4n_1({\bf g}) n_2({\bf h},{\bf k}) \om ({\bf h},{\bf k})$, which simplifies the first line of the exponent in the ``$FFF/FFF$" part into
\begin{equation}
    \om (034) ( -d(\om+2n_2)[0123] - 2 n_1 (01) \om(123) + 4 n_1(01) \om (123) n_2 (123))~.
\end{equation}
The second and the third lines of the exponent in the ``$FFF/FFF$" part equal to (when evaluated on the simplex $(01234)$)
\begin{eqs}
    \om(014)& \Big( -d(\om+ 2n_2)[1234] + 2n_1(01) \om(134)[1-2n_2(134)]+ 2n_1(01) \om(123) (1-2n_2(123))  \nonumber\\
    & \quad  -2n_1(01) \om(124) [1-2 n_2(124)] -2n_1 (02) \om(234)[1-2n_2(234)] \Big)~.
\end{eqs}
Combining all terms, we find the exponent of the ``$FFF/FFF$" part on 4-simplex $(01234)$ equals
\begin{align}
  &-\om \cup_1 (d(\om+2n_2))(01234) + 4 f(n_1,\om, d(n_2 \cup_2 \om))(01234) + 4 f(n_1, \om, \frac{d \om}{2})(01234) \cr
    &\quad + 2 \om(014) dn_1 (012) \om (234)+ 4 \om \cup_1 [n_1 \cup (n_2 \cup_2 \om)](01234) - 2 \om \cup_1 (n_1 \cup \om)(01234) \cr
    &=\left(-\om \cup_1 (d(\om+2n_2)) + 4 f(n_1,\om, d(n_2 \cup_2 \om)) + 4 f(n_1, \om, \frac{d \om}{2})\right)(01234) \cr
    &\quad + 4 n_1 (01) \om(014)  [n_1 (12) \om (234)]+ \left(4 \om \cup_1 [n_1 \cup (n_2 \cup_2 \om)] - 2 \om \cup_1 (n_1 \cup \om) \right)(01234)\cr
    &= \big(-\om \cup_1 (d(\om+2n_2)) + 4 f(n_1,\om, d(n_2 \cup_2 \om)) + 4 f(n_1, \om, \frac{d \om}{2}) \cr
    &\quad + 4 f(n_1,\om, n_1 \cup \om) + 4 \om \cup_1 [n_1 \cup (n_2 \cup_2 \om)] - 2 \om \cup_1 (n_1 \cup \om)\big)(01234)~,  
\end{align}
which can be simplified into
\begin{equation}
    -\om \cup_1 (d(\om+2n_2)) + 4 f(n_1,\om, d(n_2 + n_2 \cup_2 \om)) + 4 \om \cup_1 [n_1 \cup (n_2 \cup_2 \om)] - 2 \om \cup_1 (n_1 \cup \om)~.
\end{equation}
We will use the property of the 4-cochain valued function $f$ in (\ref{eqn:fandg}) to simplify the above expression.
Take $X = n_2 \cup_2 \om$ in (\ref{eqn:f(dX)}),
\begin{equation}
    f(n_1, \om ,d(n_2 \cup_2 \om))=
    (n_1 \cup_1 \om) \cup (n_2 \cup_2 \om)
    + \om \cup_1 (n_1 \cup (n_2 \cup_2 \om))
    + n_1 \cup (\om \cup_1 (n_2 \cup_2 \om))~.
\end{equation}
We find the exponent of the ``$FFF/FFF$" part simplifies into
\begin{eqs}
    &-\om \cup_1 (d(\om+2n_2)) - 2 \om \cup_1 (n_1 \cup \om) 
    +4 f(n_1,\om,dn_2)\nonumber\\
    &\quad + 4(n_1 \cup_1 \om) \cup (n_2 \cup_2 \om)
    + 4 n_1 \cup (\om \cup_1 (n_2 \cup_2 \om))~.
\end{eqs}
Now, we add up all terms except the $\eta$ part:
\begin{eqs}
    &  
    n_1 \cup (\omega_2\cup_1 (\omega_2 \cup_2 n_2))
    + n_1 \cup (\om \cup_1 (n_2 + \om)) 
    + n_2 \cup n_2 + (n_1 \cup_1 n_2) \cup \om 
    + 
    (n_1 \cup_1 \om) \cup (\om \cup_2 n_2)
    \\
    & +\frac{c_-}{2}(n_2 \cup \om + \om \cup n_2 - (n_1 \cup_1 \om) \cup \om) + \frac{c_-}{4}(\om \cup \om) 
    - \frac{c_-}{4} \om \cup_1 (d(\om+2n_2)) 
    - \frac{c_-}{2} \om \cup_1 (n_1 \cup \om) \\
    & +f(n_1,\om,dn_2)
    + 
    (n_1 \cup_1 \om) \cup (n_2 \cup_2 \om)
    +  
    n_1 \cup (\om \cup_1 (n_2 \cup_2 \om))
    \\
    =& n_1 \cup (\om \cup_1 (n_2 + \om)) 
     +n_2 \cup n_2 + (n_1 \cup_1 n_2) \cup \om + f(n_1,\om,dn_2)\\
    & +\frac{c_-}{2}(n_2 \cup \om + \om \cup n_2 - (n_1 \cup_1 \om) \cup \om) 
    + \frac{c_-}{4}(\om \cup \om) 
    -\frac{c_-}{4} \om \cup_1 (d(\om+2n_2)) - \frac{c_-}{2} \om \cup_1 (n_1 \cup \om) ~.\\
\label{eq: 2k+1 without eta 1}
\end{eqs}
The first term in the last line equals
\begin{eqs}
    &\frac{c_-}{2}(n_2 \cup \om + \om \cup n_2 - (n_1 \cup_1 \om) \cup \om) \\
    \sim & ~ n_2 \cup \om + \frac{c_-}{2}(-(n_1 \cup_1 \om) \cup \om - dn_2 \cup_1 \om - n_2 \cup_1 d\om) \\
    \sim & ~ n_2 \cup \om + \frac{c_-}{2}(-(n_1 \cup_1 \om) \cup \om + dn_2 \cup_2 d\om  + \om \cup_1 dn_2 - n_2 \cup_1 d\om) \\
    \sim & ~ n_2 \cup \om + c_- (n_1 \cup \om) \cup (\om \cup_1 \om) + \frac{c_-}{2}(-(n_1 \cup_1 \om) \cup \om + \om \cup_1 dn_2 - n_2 \cup_1 d\om)~,
\end{eqs}
where $\sim$ means equal up to a coboundary, and
we have used $dn_2 \cup_2 (\om \cup_1 \om) = (n_1 \cup \om) \cup_2 (\om \cup_1 \om)$ since $Sq^1 (Sq^1 \om) = 0$.
Eq.~\eqref{eq: 2k+1 without eta 1} becomes
\begin{eqs}
    & n_2 \cup n_2  + n_2 \cup \om + c_- (n_1 \cup \om) \cup_2 (\om \cup_1 \om) 
    + \frac{c_-}{4}(\om \cup \om - \om \cup_1 d \om) \\
    &\quad\quad + \frac{c_-}{2}(-(n_1 \cup_1 \om) \cup \om - n_2 \cup_1 d\om) - \frac{c_-}{2} \om \cup_1 (n_1 \cup \om) 
     \\
     &\quad\quad
    + (n_1 \cup_1 n_2) \cup \om + n_1 \cup (\om \cup_1 (n_2 + \om)) + f(n_1,\om,dn_2) ~.
\end{eqs}
To simplify the expression, we use the following identities
\begin{align}
    f(n_1,\om,dn_2) =
     & f(n_1,\om, n_1 \cup \om) + f(n_1,\om, \om \cup_1 \om) \cr
    =& g(n_1 \cup n_1, \om, \om) + \zeta (n_1,\om) + f(n_1,\om, \om \cup_1 \om) \cr
    f(n_1,\om, \om \cup_1 \om) =& -c_- f(n_1,\om, \frac{d\om}{2}) \text{ (for odd $c_-$)}\cr
    \sim & -\frac{c_-}{2} [ -(n_1 \cup \om) \cup_1 \om - \om \cup_1 (n_1 \cup \om) - 2n_1 \cup (\om \cup_1 \om)] \cr
    & + (n_1 \cup n_1 \cup \om) \cup_2 \om - (n_1 \cup \om) \cup_2 (\om \cup_1 \om) \cr
    &+ n_1 \cup n_1 \cup \om - n_1 \cup (\om \cup_2 (\om \cup_1 \om))\cr
    \sim & -\frac{c_-}{2} [ -(n_1 \cup_1 \om) \cup \om - \om \cup_1 (n_1 \cup \om) ] \cr
    & + (n_1 \cup n_1 \cup \om) \cup_2 \om + (n_1 \cup \om) \cup_2 (\om \cup_1 \om) \cr
    & - \frac{c_-}{2} n_1 \cup (\om \cup_1 \om) \quad \text{mod }2 ~,
\label{eq: 2n+1 function f}
\end{align}
where the first identity can be proven by evaluating $ f(n_1,\om,dn_2) (01234)$ on 4-simplex $(01234)$, \footnote{
Explicitly, 
\begin{align*}
    & f(n_1,\om, n_1 \cup \om)(01234) \cr
    =& n_1(01) \om(014) n_1(12) \om(234) \cr
    =& n_1(01) [\om(014)+\om(124)] n_1(12) \om(234) \cr
    &+ n_1(01) \om(124) n_1(12) \om(234) \cr
    =& \left[ g(n_1 \cup n_1, \om, \om) + \zeta (n_1,\om) \right](01234).
\end{align*}
}
and we used $\om \cup_2 (\om\cup_1 \om) = 0 ~(\text{mod 2})$.
The Hirsch identity $(n_1 \cup_1 \om) \cup \om = (n_1 \cup \om) \cup_1 \om + n_1 \cup (\om \cup_1 \om)$ is also used.\\
The last term of \eqref{eq: 2n+1 function f} is $-\frac{c_-}{2} n_1 \cup (\om \cup_1 \om)$, which is cohomologous to $\frac{c_-}{2} n_1 \cup n_1 \cup \om$ (using $\frac{d\om}{2} = - \om \cup_1 \om$ and $\frac{dn_1}{2} = n_1 \cup n_1$). Now we include the $\eta$ part, which modifies this term into $\frac{1}{2} n_1 \cup n_1 \cup \om$.
The anomaly ${\cal O}_4=e^{i \pi \phi_4}$ is equal to
\begin{eqs}
     \phi_4 =& ~n_2 \cup n_2  + n_2 \cup \om + c_- (n_1 \cup \om) \cup_2 (\om \cup_1 \om) \\
    &+ \frac{c_-}{4}(\om \cup \om - \om \cup_1 d \om) + \zeta(n_1,\om) 
    +\frac{1}{2} n_1 \cup n_1 \cup \om\\
    &+ n_2 \cup_1 (\om \cup_1 \om) + (n_1 \cup_1 n_2) \cup \om + n_1 \cup (\om \cup_1 (n_2 + \om))
    +g(n_1\cup n_1, \om, \om) \\
    & + (n_1 \cup n_1 \cup \om) \cup_2 \om + (n_1 \cup \om) \cup_2 (\om \cup_1 \om) \\
    =&  ~ n_2 \cup n_2  + n_2 \cup_1 dn_2+ n_2 \cup \om + c_- (n_1 \cup \om) \cup_2 (\om \cup_1 \om) \\
    &+ \frac{c_-}{4}(\om \cup \om - \om \cup_1 d \om) + \zeta(n_1,\om) 
    +\frac{1}{2} n_1 \cup n_1 \cup \om\\
    &+n_2 \cup_1 (n_1 \cup \om) + (n_1 \cup_1 n_2) \cup \om + n_1 \cup (\om \cup_1 (n_2 + \om))
    +g(n_1\cup n_1, \om, \om) \\
    & + (n_1 \cup n_1 \cup \om) \cup_2 \om + (n_1 \cup \om) \cup_2 (\om \cup_1 \om)~.
\end{eqs}
We use $(n_1 \cup_1 n_2) \cup \om = (n_1 \cup \om) \cup_1 n_2 + n_1 \cup (\om \cup_1 n_2)$. The penultimate line is
\begin{eqs}
    &n_2 \cup_1 (n_1 \cup \om) + (n_1 \cup \om) \cup_1 n_2 + n_1 \cup (\om \cup_1 \om) + g(n_1\cup n_1, \om, \om) \\
    \sim &  (n_1 \cup \om) \cup_2 dn_2  + n_1 \cup (\om \cup_1 \om) + (n_1^2 \cup \omega_2) \cup_2 \omega_2 + n_1^2 \cup \omega_2  \\
    \sim & (n_1 \cup \om) \cup_2 dn_2 + (n_1^2 \cup \om) \cup_2 \om~.
\end{eqs}
We note that
 $(n_1 \cup \om) \cup_2 (n_1 \cup \om)=Sq^1(n_1\cup\omega_2)=\frac{d}{2}(n_1\cup\omega_2) \sim 0$ is a coboundary.
The final expression for the anomaly ${\cal O}_4$ is
\begin{eqs}
    O_4 =(-1)^{{\cal P}(n_2)+n_2\cup \omega_2
    + \zeta(n_1,\omega_2)+c_-(n_1\omega_2)\cup_2Sq^1(\omega_2)
    + \frac{1}{2} n_1^2\cup\omega_2 + {c_-\over 4}{\cal P}(\omega_2)}~.
\end{eqs}

\subsection{$c_-=k+1/2$}
In this case there are no permutations. Also we have $\mathfrak{t}({\bf g},{\bf h}) = \psi^{n_2({\bf g},{\bf h})}$. Since $F^{\psi \psi \psi} = 1$, and $R^{\psi \psi} = -1$, we obtain the simple result
\begin{equation}\label{Eq:O4_chalf}
    \mathcal{O}_4[c_-=k+1/2,n_1,n_2]({\bf g},{\bf h},{\bf k},{\bf l}) = R^{\mathfrak{t}({\bf g},{\bf h}) \mathfrak{t}({\bf k},{\bf l})} = (-1)^{n_2({\bf g},{\bf h})n_2({\bf k},{\bf l})}.
\end{equation}
This is the Gu-Wen supercohomology equation \cite{PhysRevB.90.115141,Gaiotto:2015zta}.

\subsection{Computation of stacking rule for $\nu_3$ with integer $c_-$}
\label{sec:stackingnu3}

\subsubsection{$s_1=0$}

Here we will complete the derivation of the stacking rules for the invertible phases characterized by the data $(n_1^a,n_2^a,\omega_2^a,\nu_3^a)$ and $(n_1^b,n_2^b,\omega_2^b,\nu_3^b)$. 
We will first consider the case when $G_b$ is a unitary symmetry, {\it i.e.} $s_1=0$.
As discussed in Section~\ref{Sec:stacking}, the stacking laws for $n_1, n_2,\omega_2, c_-$ are
\begin{eqs}\label{eqn:stackingn1n2}
    \om^{\text{tot}} &= \om^a = \om^b \\
    n_1^{\text{tot}} &= n_1^a + n_1^b \\
    n_2^{\text{tot}} &= n_2^a + n_2^b + n_1^a \cup n_1^b,\\
    c_-^{\text{tot}} &= c_-^a + c_-^b~.
\end{eqs}
(While the stacking rule for $n_2$ is a conjecture, it is comparible with our expressions for $\mathcal{O}_4$, as we will see below.) The stacking rule for $\nu_3$ is computed as follows. We use the property that ${\cal O}_4(n_1^{\text{tot}},n_2^{\text{tot}},\omega_2^{\text{tot}},c_-^{\text{tot}})$ differs from ${\cal O}_4(n_1^a,n_2^a,\omega_2^a,c_-^a){\cal O}_4(n_1^b,n_2^b,\omega_2^b,c_-^b)$ by a coboundary. Let us discuss each term in ${\cal O}_4$:
\begin{enumerate}
    \item ${\cal P}(n_2)$:
    \begin{eqs}
        &{\cal P}(n_2^a + n_2^b + n_1^a \cup n_1^b) - {\cal P}(n^a_2) - {\cal P}(n^b_2) \\
        =& dn_2^a \cup_2 dn_2^b + {\cal P}(n_1^a \cup n_1^b) + d[(n_1^a \cup n_1^b) \cup_1 (n_2^a + n_2^b) + n_2^a \cup_1 n_2^b + dn_2^a \cup_2 n_2^b]~.
    \label{eq: stacking 1}
    \end{eqs}
    where we have used \eqref{eq: diff of Pontry} to compute the difference of Pontryagin squares. Notice that ${\cal P}(n_1^a \cup n_1^b) = d [ \frac{1}{2} n_1^a \cup n_1^b \cup n_1^b + n_1^a \cup (n_1^a \cup_1 n_1^b) \cup n_1^b]$.
    \item $n_2 \cup \om$:
    \begin{equation}
        (n_2^a + n_2^b + n_1^a \cup n_1^b) \cup \om - n_2^a \cup \om - n_2^b \cup \om 
        = n_1^a \cup n_1^b \cup \om~.
    \label{eq: stacking 2}
    \end{equation}
    \item $\zeta(n_1, \om) [01234] = n_1 (01) n_1 (12) \om(124) \om(234)$:
    \begin{eqs}
        &\left(\zeta(n_1^a + n_1^b, \om) - \zeta(n_1^a, \om) - \zeta( n_1^b, \om)\right)(01234) \\
        =& n_1^b (01) n_1^a (12) \om (124) \om(234) + n_1^a (01) n_1^b (12) \om(124) \om(234) \\
        =& n_1^b \cup [(n_1^a \cup \om) \cup_2 \om + n_1^a \cup \om]+ n_1^a \cup [(n_1^b \cup \om) \cup_2 \om + n_1^b \cup \om]~.
    \label{eq: stacking 3}
    \end{eqs}
    \item $\frac{1}{2} [n_1]_2 \cup [n_1]_2 \cup \om$:
    \begin{eqs}
        &\frac{1}{2}[ (n_1^a + n_1^b - 2 n_1^a \cup_1 n_1^b)^2 \cup \om - (n_1^a)^2\cup  \om -(n_1^b)^2 \cup \om] \\
        =& \frac{1}{2}(n_1^a \cup n_1^b + n_1^b \cup n_1^a) \cup \om 
        + [(n_1^a + n_1^b) \cup (n_1^a \cup_1 n_1^b) + (n_1^a \cup_1 n_1^b) \cup (n_1^a + n_1^b)]\cup\omega_2 \text{(mod $2$)} \\
        =& \frac{1}{2} d(n_1^a \cup_1 n_1^a) \cup \om + [n_1^a \cup_1 \frac{dn_1^b}{2} + \frac{dn_1^a}{2} \cup_1 n_1^b] \cup \om 
        + [(n_1^a + n_1^b) \cup_1 d(n_1^a \cup_1 n_1^b)] \cup \om \\
        =& \frac{1}{2} d[(n_1^a \cup_1 n_1^b) \cup \om] + ((n_1^a \cup_1 n_1^b) \cup \frac{d\om}{2} \\
        &+ [n_1^a \cup_1 (n_1^b \cup n_1^b) + (n_1^a \cup n_1^a) \cup_1 n_1^b] \cup \om \\
        &+ [ (n_1^a + n_1^b) \cup_1 (n_1^a \cup n_1^b + n_1^b \cup n_1^a)] \cup \om \\
        =& \frac{1}{2} d[(n_1^a \cup_1 n_1^b) \cup \om] + ((n_1^a \cup_1 n_1^b) \cup \frac{d\om}{2}~.
    \label{eq: stacking 4}
    \end{eqs}
    \item $c_- (n_1 \cup \om) \cup_2 (\om \cup_1 \om)$:
    \begin{eqs}
        c_-^a (n_1^b \cup \om) \cup_2 (\om \cup_1 \om) + c_-^b (n_1^a \cup \om) \cup_2 (\om \cup_1 \om)~.
    \label{eq: stacking 5}
    \end{eqs}
\end{enumerate}
The total contribution of \eqref{eq: stacking 1}, \eqref{eq: stacking 2}, \eqref{eq: stacking 3}, \eqref{eq: stacking 4}, \eqref{eq: stacking 5} gives a coboundary:
\begin{eqs}
    &d[ \frac{1}{2} n_1^a \cup n_1^b \cup n_1^b + n_1^a \cup (n_1^a \cup_1 n_1^b) \cup n_1^b ]
    + d[c^a (\om \cup_1 \om) \cup_3 (n_1^b \cup \om)]
    \cr
    &\quad
    + d[(n_1^a \cup n_1^b) \cup_1 (n_2^a + n_2^b) + n_2^a \cup_1 n_2^b + dn_2^a \cup_2 n_2^b] 
    + d[\frac{1}{2} (n_1^a \cup_1 n_1^b) \cup \om]~.
\end{eqs}
Therefore, up to a possible 3-cocycle term, the stacking law for $\nu_3$ is
\begin{equation}\label{eqn:stackingnu3}
    \nu_3^{\text{tot}}=\nu_3^a\nu_3^b\,
    (-1)^{(n_1^a \cup n_1^b) \cup_1 (n_2^a + n_2^b) + n_2^a \cup_1 n_2^b + dn_2^a \cup_2 n_2^b + n_1^a \cup (n_1^a \cup_1 n_1^b) \cup n_1^b + c^a (\om \cup_1 \om) \cup_3 (n_1^b \cup \om)}
    \,i^{ (n_1^a \cup_1 n_1^b) \cup \om + n_1^a \cup n_1^b \cup n_1^b}~.
\end{equation}
The fact that we have obtained a reasonable stacking rule for $\nu_3$ which reduces to the known correct result when $\om = 0$ supports our conjectured stacking rule for $n_2$. Equations (\ref{eqn:stackingn1n2}) and (\ref{eqn:stackingnu3}) are our conjectured stacking rules for invertible phases with unitary symmetry.

\end{widetext}
 
\section{Equivalence of ${\cal O}_4$ obstructions for $c_-=0$ from $G$-crossed braided tensor categories and the fixed-point wave function}\label{sec: equivalence of O4}

In Ref.~\onlinecite{Wang2020fSPT}, the ${\cal O}_4$ obstruction, Eq.~(136), is obtained by the fixed-point wave function construction:
\begin{eqs}
    &{\cal O}_4(01234) \\
    =& (-1)^{[n_2 \cup n_2 + n_2 \cup_1 dn_2 + \om \cup n_2](01234) + \om(013) dn_2(1234)} \\
    & + (-1)^{dn_2(0124) dn_2(0234)} (-i)^{dn_2(0123)[1-dn_2(0124)] (\text{mod }2)}.
\label{eq: WangGuO4}
\end{eqs}
We first expand the last few terms which are not cup products. Notice that, in Ref.~\onlinecite{Wang2020fSPT}, the convention  is $dn_2 = \om \cup n_1$, which is different from $dn_2 = n_1 \cup \om$ defined in this paper.
\begin{eqs}
    &\om(013) dn_2(1234) + dn_2(0124) dn_2(0234) \\
    =& \om(013) \om(123) n_1 (34) + \om(012) \om(023) n_1(24) n_1(34).
\end{eqs}
To match the formula Eq.~\eqref{eqn: O4 c=k} with $c_-=0$, we reverse the branching structure $\lr{01234} \ra \lr{43210}$ and the terms above becomes
\begin{eqs}
    &n_1(01) \om(123) \om(134) + n_1(01) n_1(02) \om(124) \om(234) \\
    =& n_1(01) \om(123) \om(134) \\
    &+ n_1(01) (n_1(01) + n_1(12)) \om(124) \om(234) ~(\text{mod }2) \\
    =& n_1(01) ( \om(123) \om(134) + \om(124) \om(234) ) \\
    &+ n_1(01) n_1(12) \om(124) \om(234) \\
    =& [n_1 \cup (\om \cup_1 \om) + \zeta(n_1, \om)](01234).
\end{eqs}
The first term $n_1 \cup (\om \cup_1 \om)$ is a coboundary since it is $n_1 \cup \frac{d\om}{2} \sim \frac{dn_1}{2} \cup \om \sim d(n_1 \cup n_2) \sim 0$.
The last term in Eq.~\eqref{eq: WangGuO4} (after reversing the branching structure) is
\begin{eqs}
    &-\frac{1}{2}[ dn_2(1234) (1-dn_2 (0234)) ~(\text{mod }2)] \\
    =& -\frac{1}{2} [n_1 (12) \om(234) - n_1(02) n_1(12) \om(234)],
\end{eqs}
where we have used $dn_2 = n_1 \cup \om$ since the ordering of vertices are reversed. The expression in the bracket is automatically $0, 1$ if all $n_1$ and $\om$ are $0,1$, so the mod 2 action is unnecessary. It can be further simplified as:
\begin{eqs}
    & -\frac{1}{2} [n_1 (12) \om(234) - (n_1(01) + n_1(12)) n_1(12) \om(234) \\
    & \quad \quad + dn_1(012) n_1(12) \om(234)]\\
    =& \frac{1}{2} [ n_1(01) n_1(12) \om(234) ] - \frac{dn_1(012)}{2} n_1(12) \om(234) \\
    =&  \frac{1}{2} [n_1 \cup n_1 \cup \om](01234) - n_1(01) n_1(12 ) n_1(12) \om(234) \\
    =&  \frac{1}{2} [n_1 \cup n_1 \cup \om](01234) - [n_1 \cup n_1 \cup \om](01234) \\
    \sim &  \frac{1}{2} [n_1 \cup n_1 \cup \om](01234),
\end{eqs} 
where we have used $n_1 \cup n_1 \cup \om = d(n_1 \cup n_2) ~(\text{mod }2)$. In the cup product part, $n_2 \cup n_2 + n_2 \cup_1 dn_2$ is invariant under reversing branching structure, and $\om \cup n_2$ is transformed to $n_2 \cup \om$. Hence, the total ${\cal O}_4$ obstruction is:
\begin{eqs}
    {\cal O}_4 = (-1)^{{\cal P}(n_2) + n_2 \cup \om + \zeta(n_1,\omega_2) }i^{ n_1 \cup n_1 \cup \om},
\end{eqs}
which agrees with the obstruction obtained from $G$-crossed braided tensor categories Eq.~\eqref{eqn: O4 c=k} with $c_-=0$.
 
\section{A Non-anomalous Reference State for integer $c_-$}\label{sec:refstate}
 
Let us provide a description for a non-anomalous ``reference" bosonic shadow theory enriched by $G_b$ symmetry, with general $n_1$, $n_2=0$, $\omega_2=0$ and we assume the symmetry is unitary $s_1=0$.
When $c_-$ is a half integer, $n_1=0$, and we can take the $G_b$ symmetry to act trivially, {\it i.e.} the background $G_b$ gauge field decouples, and thus the symmetry is non-anomalous. Thus we will focus on the case of integer $c_-$.

To describe the data $n_1$, we note it is related to the gauge field of the charge conjugation symmetry in the Chern-Simons theory $Spin(2c_-)_1$ description of the bosonic shadow theory,
\begin{equation}\label{eqn:n1Gb}
    B_1=A_b^*n_1~.
\end{equation}
Gauging the $G_b$ symmetry is equivalent to orbifold of the $Spin(2c_-)_1$ chiral algebra by $G_b$ symmetry with action specified by $n_1$ \cite{Dijkgraaf:1989hb}. In the following we will provide another description of the resulting theory, which is manifestly well-defined. Furthermore, since
the theory couples to the $G_b$ gauge field only through (\ref{eqn:n1Gb}), and
the unitary $\mathbb{Z}_2$ charge conjugation symmetry does not act projectively on the anyons, the $G_b$ symmetry does not fractionalize on the anyons. In the language of $G_b$-crossed BTCs, thts means that the theory has all reference $\eta$ symbols equal to 1.

To obtain the theory coupled to the gauge field $A_b$ of the $G_b$ symmetry, we can start with the partition function that couples to $B_1$, denoted by $Z[B_1]$, and then perform the following manipulation
\begin{equation}\label{eqn:Z}
    \sum_{B_2}\left(\sum_{B_1}Z[B_1](-1)^{\int (B_1+A_b^*n_1)\cup B_2}\right)=Z[A_b^*n_1]~,
\end{equation}
where $B_2$ is the two-form $\mathbb{Z}_2$ gauge field that gauges the one-form symmetry generated by $(-1)^{\oint B_1+A_b^*n_1}$ in the theory whose partition function is given by the bracket on the left hand side of (\ref{eqn:Z}), denoted as
\begin{equation}
    {\cal T}\times (G_b)_0~,
\end{equation}
where ${\cal T}$ is the theory obtained by gauging the $\mathbb{Z}_2$ charge conjugation symmetry, with partition function $Z_{\cal T}=\sum_{B_1}Z[B_1]$, and $(G_b)_0$ is the untwisted $G_b$ gauge theory with vanishing Dijkgraaf-Witten action \cite{Dijkgraaf1989pz}, and it can described by the quantum double model based on group $G_b$ \cite{Kitaev:1997wr}.

Let us give an explicit description of the theory ${\cal T}$.
If we gauge the $\mathbb{Z}_2$ charge conjugation 0-form symmetry, 
the bosonic shadow theory $Spin(2c_-)_1$ becomes \cite{Cordova:2017vab}
\begin{equation}
    {\cal T}=Pin^+(2c_-)_1\leftrightarrow Spin(1)_1\times Spin(2c_--1)_1~,
\end{equation}
where $Spin(1)_1$ is the non-spin Ising TQFT with $c_-=1/2$ and three anyons $I, \psi,\sigma$.
If we gauge the dual $\mathbb{Z}_2$ one-form symmetry generated by the boson $(\psi,\psi)$ of $Spin(1)_1\times Spin(2c_--1)_1$, we recover the original bosonic shadow theory $Spin(2c_-)_1$.

Thus we find that gauging the $G_b$ symmetry in the bosonic shadow theory with $n_1$ gives
\begin{equation}\label{eqn:ref}
    {{\cal T}\times (G_b)_0\over\mathbb{Z}_2}=
    {Spin(1)_1\times Spin(2c_--1)_1\times (G_b)_0\over \mathbb{Z}_2}~,
\end{equation}
where the $\mathbb{Z}_2$ quotient denotes gauging the non-anomalous $\mathbb{Z}_2$ one-form symmetry generated by the boson
\begin{equation}\label{eqn:condense}
    (\psi,\psi, e^{i\oint A_b^*n_1})~,
\end{equation}
where $e^{i\oint A_b^*n_1}$ is the bosonic Wilson line of the $G_b$ gauge theory that carries the one-dimensional representation specified by $n_1$. Since the generator of the diagonal $\mathbb{Z}_2$ one-form symmetry is a boson, it can be gauged, giving a well-defined non-anomalous theory.
If the $G_b$ gauge field is not dynamical but a background gauge field, this gives a non-anomalous $G_b$ symmetry-enriched theory based on the topological order $Spin(2c_-)_1$.
The topological data of the theory (\ref{eqn:ref}) can be obtained using anyon condensation of the boson (\ref{eqn:condense}) as in Ref.~\onlinecite{Eliens:2013epa}.

\section{Absolute anomaly for nonpermuting symmetries and the Pontryagin square}\label{Sec:Psquare}
Consider a bosonic SET phase with anyons given by the UMTC $\mathcal{C}$, and with a general symmetry group $G$. Suppose the symmetry does not permute the anyons. Then, Ref.~\onlinecite{Barkeshli2019} shows that the absolute anomaly of this phase is given by Eq. \eqref{Eq:AbsAnomNoPerm}. This equation has the form ``$\mathcal{O}_4 = R \frac{FFF}{FFF}$''. Here we will show that this formula can be reproduced from the anomaly of the one-form symmetry \cite{Hsin:2018vcg} as discussed in \cite{Benini2018}.

It will be useful to have a concrete representation for the $F$ and $R$ symbols of the abelian anyons in $\mathcal{C}$. If the abelian anyons in $\mathcal{C}$ (which are given by a group $\A$) themselves form a modular category, we can describe their $F$ and $R$ symbols by a nondegenerate symmetric $K$ matrix with even diagonal entries:
\begin{align}
    F^{a b c} &= e^{i \pi \vec{a}^T K^{-1}(\vec{b} + \vec{c} - \overrightarrow{b+c})} \\
    R^{a b} &= e^{i \pi \vec{a}^T K^{-1} \vec{b}}.
\end{align}
If the abelian anyons in $\A$ do not form a modular category, we can use the same formulas by taking the $K$ matrix corresponding to the Drinfeld center $\mathcal{Z}(\A)$ of $\A$. In the $K$ matrix formalism, let the symmetry fractionalization anyon $\mathfrak{t}({\bf g},{\bf h})$ be denoted by $\overrightarrow{\mathfrak{t}({\bf g},{\bf h})}$. Substituting the relevant expressions into Eq. \eqref{Eq:AbsAnomNoPerm}, and simplifying, we obtain
\begin{widetext}
    \begin{align}
        \mathcal{O}_4({\bf g},{\bf h},{\bf k},{\bf l}) &= e^{i \pi \overrightarrow{\mathfrak{t}({\bf g},{\bf h})}^T K^{-1} \overrightarrow{\mathfrak{t}({\bf k},{\bf l})}} \times e^{i \pi \overrightarrow{\mathfrak{t}({\bf ghk},{\bf l}) }^T K^{-1} [\overrightarrow{\mathfrak{t}({\bf gh},{\bf k})} + \overrightarrow{\mathfrak{t}({\bf g},{\bf h})} - \overrightarrow{\mathfrak{t}({\bf g},{\bf hk})} - \overrightarrow{\mathfrak{t}({\bf h},{\bf k})}] } \times e^{i \pi \overrightarrow{\mathfrak{t}({\bf g},{\bf hkl}) }^T K^{-1} [\overrightarrow{\mathfrak{t}({\bf h},{\bf k})} + \overrightarrow{\mathfrak{t}({\bf hk},{\bf l})} - \overrightarrow{\mathfrak{t}({\bf h},{\bf kl})} - \overrightarrow{\mathfrak{t}({\bf k},{\bf l})}] } \\
        &= e^{i \pi \overrightarrow{\mathfrak{t}({\bf g},{\bf h})}^T K^{-1} \overrightarrow{\mathfrak{t}({\bf k},{\bf l})}} \times e^{i \pi \overrightarrow{\mathfrak{t}({\bf ghk},{\bf l}) }^T K^{-1} d\overrightarrow{\mathfrak{t}({\bf g},{\bf h},{\bf k})} } \times e^{i \pi \overrightarrow{\mathfrak{t}({\bf g},{\bf hkl}) }^T K^{-1} d\overrightarrow{\mathfrak{t}({\bf h},{\bf k},{\bf l})}}. 
    \end{align}
\end{widetext}

Let the components of $\vec{\mathfrak{t}}$ be given by integers $\mathfrak{t}_i$, $i = 1,2,\dots , \dim K$. Then, using the definition of the cup-1 product (Appendix \ref{sec:highercup}) we can decompose the anomaly into the following terms:
\begin{widetext}
    \begin{align}
        \mathcal{O}_4 &= e^{i \pi \sum\limits_i (K^{-1})_{ii} (\mathfrak{t}_i \cup \mathfrak{t}_i - \mathfrak{t}_i \cup_1 d\mathfrak{t}_i)} \times e^{i 2 \pi \sum\limits_{i < j} (K^{-1})_{ij} (\mathfrak{t}_i \cup \mathfrak{t}_j - \mathfrak{t}_i \cup_1 d\mathfrak{t}_j) } \\
        &= e^{i \pi \sum\limits_i (K^{-1})_{ii} \mathcal{P}(\mathfrak{t}_i)} \times e^{i 2 \pi \sum\limits_{i < j} (K^{-1})_{ij} (\mathfrak{t}_i \cup \mathfrak{t}_j - \mathfrak{t}_i \cup_1 d\mathfrak{t}_j) }.
    \end{align}
\end{widetext}
To obtain the second term in this result, we used $K^{-1}_{ij} = K^{-1}_{ji}$. For a 2-cochain $X$, the Pontryagin square is defined as $\mathcal{P}(X) = X \cup X - X \cup_1 dX$. 
The last expression can be obtained from the anomaly of one-form symmetry \cite{Hsin:2018vcg} by setting the background of the intrinsic one-form symmetry equals $A^*\mathfrak{t}$ for background $G$ gauge field $A$ \cite{Benini2018}.

\section{Equivalences in the $G_b$-crossed description}\label{sec:gtappendix}
\subsection{$s_1=0$}
In Section \ref{Sec:Equivs} we showed that for unitary symmetries, relabelling the fermion parity fluxes leads to an equivalence $(n_2, \nu_3) \simeq (n_2 + \om, \nu'_3)$. In this section, we show that this is the only allowed equivalence upon relabelling $e$ and $m$. We argue as follows. Any additional equivalences can be thought of as a superposition of the one which takes $n_2 \rightarrow n_2 + \om$, and another gauge transformation which preserves the $F,R$ and $U$ symbols but modifies the $\eta$ symbols, and thus $n_2$, by some additional terms. Here we show that such additional gauge transformations cannot transform $\eta$. We consider $c_-=0$ for simplicity; a stacking argument given below shows that this is also true when $c_- \ne 0$.  

The desired gauge transformation is some combination of a vertex basis transformation $\Gamma$ and a symmetry action gauge transformation $\gamma$ (defined in Appendix \ref{Sec:Gxreview}). By demanding that the $F$ and $R$ symbols are invariant under $\Gamma$, we obtain the constraints
\begin{align}
    \Gamma^{a b} \Gamma^{a\times b, c} &= \Gamma^{b c} \Gamma^{a, b\times c} \nonumber \\
    \Gamma^{a b} &= \Gamma^{b a}.
\end{align}
where $a,b,c$ are anyons. Using this, we can express each $\Gamma^{a b}$ in terms of three independent variables, say $\Gamma^{e m} = \Gamma^{m e}, \Gamma^{e \psi} = \Gamma^{\psi e}, \Gamma^{m \psi} = \Gamma^{\psi m}$, the rest being obtained as follows: 
\begin{align}
    \Gamma^{\psi \psi} &= \Gamma^{\Psi e} \Gamma^{\Psi m} \\
    \Gamma^{e e} &= \Gamma^{e \psi} \Gamma^{e m} \\
    \Gamma^{m m} &= \Gamma^{m \psi} \Gamma^{e m}.
\end{align}
For the $U$ symbols to also be invariant under the gauge transformation, we require
\begin{align}
    \frac{\gamma_{a\times b}}{\gamma_a \gamma_b} &= \frac{\Gamma^{^{\bf g}a ^{\bf g}b}}{\Gamma^{a b}} \nonumber \\
    &= \left(\frac{\Gamma^{a' b'}}{\Gamma^{a b}}\right)^{n_1({\bf g})}
\end{align}
where a permutation of $e$ and $m$ takes $a \rightarrow a', b \rightarrow b'$. We impose the physical constraint $\gamma_{\psi} = 1$. From invariance of $U$ we find that
\begin{align}
    \gamma_e^2 = \frac{1}{\gamma_m^2} &= \left(\frac{\Gamma^{e e}}{\Gamma^{mm}}\right)^{n_1} \\
    \gamma_e \gamma_m = 1. 
\end{align}
Thus we can set $\gamma_m = \frac{1}{\gamma_e} = \theta$ for some complex function $\theta: G_b \rightarrow U(1)$. Such a $\gamma$ changes $\eta_m$ by the amount
\begin{align}
    \frac{\gamma_m({\bf gh})}{\gamma_m({\bf g})\gamma_{^{\bf g}m}({\bf h})} &= \frac{\gamma_m({\bf gh})}{\gamma_m({\bf g})\gamma_m({\bf h})} \left(\frac{\gamma_m({\bf h})}{\gamma_e({\bf h})}\right)^{n_1({\bf g})} \\
    &= \theta^{n_1({\bf gh})-n_1({\bf g})-n_1({\bf h}) + 2n_1({\bf g}) n_1({\bf h})} \\
    &= 1,
\end{align}
the last line being an identity. Thus, gauge transformations that preserve $F,R,U$ cannot change $\eta_m$; hence they cannot change $n_2$.

We now argue that there cannot be additional equivalences even when $c_-=0$. Suppose there was an equivalence of the form $(c_-,n_1, n_2, \nu_3) \simeq (c_-,n_1, n_2 + \lambda_2, \nu'_3)$ for some 2-cocycle $\lambda_2$, with $c_- \ne 0$. Now we can consider a stack of 3 systems: a system with $(c_-=0,n_1^0, n_2^0, \nu_3^0)$, the state with data $(c_-,n_1, n_2, \nu_3)$, and the inverse of the state with data $(c_-,n_1, n_2 + \lambda_2, \nu'_3)$. Upon stacking, we would obtain a new state with $c_-=0$ but with $n_2^0 \rightarrow n_2^0 + \lambda_2$. But this state should be equivalent to the one with $(c_-=0,n_1^0, n_2^0, \nu_3^0)$, since the other two states in the stack were inverses of each other, by assumption. Thus the equivalence under shifting by $\lambda_2$ would hold even for $c_-=0$. This is a contradiction.

\subsection{$s_1 \ne 0$}
In the main text we showed that there is a relabelling of $e$ and $m$ which, along with a gauge transformation, takes $n_2 \rightarrow n_2 + \om$. We ask if there are other gauge transformations which preserve $F,R,U$ but also change $\eta$. When $s_1 \ne 0$ the constraints on $F$ and $R$ are identical to those given above, however the invariance condition on $U$ now depends on $s_1$:
\begin{equation}
    \frac{\gamma_{a\times b}}{\gamma_a \gamma_b} =  \left(\frac{\Gamma^{a' b'}}{\Gamma^{a b}}\right)^{n_1} (\Gamma^{a b})^{2 s_1}.
\end{equation}
Taking $\gamma_{\psi} = 1$, and $a=b=\psi$, we obtain $(\Gamma^{\psi \psi})^{2s_1} =1$, thus $\Gamma^{\psi \psi} = \pm 1$. Taking $a=b=e,m$ separately we obtain
\begin{align}
    \gamma_m^2 &= (\Gamma^{m m})^{-2 s_1} \left(\frac{\Gamma^{m m}}{\Gamma^{e e}}\right)^{n_1} \nonumber \\
     \gamma_e^2 &= (\Gamma^{e e})^{-2 s_1} \left(\frac{\Gamma^{e e}}{\Gamma^{m m}}\right)^{n_1}. \nonumber
\end{align}

Under this $\gamma$, $\eta_m({\bf g},{\bf h})$ changes by an amount
\begin{align}
    & \frac{\gamma_m({\bf gh})}{\gamma_m({\bf g}) (\gamma_{^{\bf g}m}({\bf h}))^{1-2s_1({\bf g})}} \nonumber \\
    &= d\gamma^{-1}_m({\bf g},{\bf h}) (\gamma_m({\bf h}))^{2 s_1({\bf g})} \left(\frac{\gamma_m}{\gamma_e}({\bf h})\right)^{n_1({\bf g}) - 2 n_1({\bf g}) s_1({\bf g})}. 
\end{align}
Note that $\eta_m$ can only change by $\pm 1$. Thus the square of the above quantity must equal 1. By plugging in the expressions for $\gamma_m^2$ and $\gamma_e^2$ and simplifying, we obtain
\begin{align}
    1 &= \left(\frac{\Gamma^{e e}}{\Gamma^{m m}}\right)^{2 \alpha}, \nonumber \\
    \alpha &= -s_1 n_1 + 2 n_1 s_1 + 2 (n_1 \cup_1 s_1)(n_1 + 2 s_1). 
\end{align}
This equality should hold for arbitrary $n_1,s_1$. Therefore we demand that
\begin{equation}
    \left(\frac{\Gamma^{m m}}{\Gamma^{e e}}\right)^{2} = 1.
\end{equation}
We also have $\frac{\Gamma^{m m}}{\Gamma^{e e}} = \frac{\Gamma^{m \psi}}{\Gamma^{e \psi}} = \Gamma^{\psi \psi} (\Gamma^{m \psi})^2$. The above equation then implies that
\begin{equation}
    (\Gamma^{m \psi})^4 = (\Gamma^{e \psi})^4 =1.
\end{equation}
Now we can obtain the change in $\eta_m$. We use the parameterization
\begin{align}
    \gamma_m &= (\Gamma^{m \psi})^{n_1 - s_1} (\Gamma^{e m})^{-s_1} (\Gamma^{\psi \psi})^{\frac{n_1}{2}}\\
    \gamma_e &= (\Gamma^{m \psi})^{s_1 - n_1} (\Gamma^{e m})^{-s_1} (\Gamma^{\psi \psi})^{-\frac{n_1}{2}}.
\end{align}
After simplifying with the help of the above constraints, the total shift in $\eta_m$ equals 
\begin{equation}
    \eta_m \rightarrow \eta_m \times (\Gamma^{\psi \psi})^{s_1 n_1} (\Gamma^{m \psi})^{2(s_1 n_1 + n_1 s_1)}. 
\end{equation}
We see that gauge transformations with $\Gamma^{\psi \psi} = -1$ shift $\eta_m$ by $(-1)^{s_1 n_1}$. The corresponding shift in $n_2$ is
\begin{equation}
    n_2 \rightarrow n_2 + s_1 n_1.
\end{equation}
In Ref. \cite{bulmashSymmFrac}, it is noted that the condition $\Gamma^{\psi \psi} = -1$ leads to problems in the classification of SET phases with antiunitary symmetries. In our case, this condition introduces an extra equivalence when $n_1 \ne 0$ compared to when $n_1 = 0$. Thus, following  Ref. \cite{bulmashSymmFrac}, we impose the restriction $\Gamma^{\psi \psi} = 1$ by hand. Now if $(\Gamma^{m \psi})^2 = -1$, $n_2$ is only shifted by a 2-coboundary $n_1 s_1 + s_1 n_1$; this equivalence is already present in the theory. We conclude that all gauge transformations with $\gamma_{\psi} = 1$ and $\Gamma^{\psi \psi} = 1$ which preserve the $F,R$ and $U$ symbols must also preserve $\eta$ up to 2-coboundaries. Thus the only other equivalence on $n_2$ is the one that takes $n_2 \rightarrow n_2 + \om$. 

\bibliographystyle{utphys}
\bibliography{fInvertible_draft.bib}

\providecommand{\href}[2]{#2}\begingroup\raggedright\begin{thebibliography}{10}

\bibitem{hasan2010}
M.~Z. Hasan and C.~L. Kane, ``Colloquium: Topological insulators,''
  \href{http://dx.doi.org/10.1103/RevModPhys.82.3045}{{\em Rev. Mod. Phys.}
  {\bfseries 82} (Nov, 2010) 3045--3067}.
  \url{https://link.aps.org/doi/10.1103/RevModPhys.82.3045}.

\bibitem{qi2011}
X.-L. Qi and S.-C. Zhang, ``Topological insulators and superconductors,''
  \href{http://dx.doi.org/10.1103/RevModPhys.83.1057}{{\em Rev. Mod. Phys.}
  {\bfseries 83} (Oct, 2011) 1057--1110}.
  \url{https://link.aps.org/doi/10.1103/RevModPhys.83.1057}.

\bibitem{Chen2013}
X.~Chen, Z.-C. Gu, Z.-X. Liu, and X.-G. Wen, ``Symmetry protected topological
  orders and the group cohomology of their symmetry group,''
  \href{http://dx.doi.org/10.1103/PhysRevB.87.155114}{{\em Phys. Rev. B}
  {\bfseries 87} (Apr, 2013) 155114}.
  \url{https://link.aps.org/doi/10.1103/PhysRevB.87.155114}.

\bibitem{Senthil2015SPT}
T.~Senthil, ``{Symmetry Protected Topological phases of Quantum Matter},''
  \href{http://dx.doi.org/10.1146/annurev-conmatphys-031214-014740}{{\em Ann.
  Rev. Condensed Matter Phys.} {\bfseries 6} (2015) 299},
  \href{http://arxiv.org/abs/1405.4015}{{\ttfamily arXiv:1405.4015
  [cond-mat.str-el]}}.

\bibitem{Kapustin2014}
A.~Kapustin, ``Bosonic topological insulators and paramagnets: a view from
  cobordisms,'' \href{http://arxiv.org/abs/arXiv:1404.6659}{{\ttfamily
  arXiv:1404.6659}}.

\bibitem{Kapustin:2014dxa}
A.~Kapustin, R.~Thorngren, A.~Turzillo, and Z.~Wang, ``{Fermionic Symmetry
  Protected Topological Phases and Cobordisms},''
  \href{http://dx.doi.org/10.1007/JHEP12(2015)052}{{\em JHEP} {\bfseries 12}
  (2015) 052}, \href{http://arxiv.org/abs/1406.7329}{{\ttfamily arXiv:1406.7329
  [cond-mat.str-el]}}.

\bibitem{Freed:2016rqq}
D.~S. Freed and M.~J. Hopkins, ``{Reflection positivity and invertible
  topological phases},'' \href{http://dx.doi.org/10.2140/gt.2021.25.1165}{{\em
  Geom. Topol.} {\bfseries 25} (2021) 1165--1330},
  \href{http://arxiv.org/abs/1604.06527}{{\ttfamily arXiv:1604.06527
  [hep-th]}}.

\bibitem{yonekura2019}
K.~Yonekura, ``On the cobordism classification of symmetry protected
  topological phases,''
  \href{http://dx.doi.org/10.1007/s00220-019-03439-y}{{\em Communications in
  Mathematical Physics} {\bfseries 368} no.~3, (2019) 1121--1173}.
  \url{https://doi.org/10.1007/s00220-019-03439-y}.

\bibitem{gaiotto2017}
D.~Gaiotto and T.~Johnson-Freyd, ``Symmetry protected topological phases and
  generalized cohomology,''
  \href{http://arxiv.org/abs/arXiv:1712.07950}{{\ttfamily arXiv:1712.07950}}.

\bibitem{kitaev2009}
A.~Kitaev, ``Periodic table for topological insulators and superconductors,''
  {\em AIP Conf. Proc.} {\bfseries 1134} (2009) 22.

\bibitem{klitzing1980}
K.~v. Klitzing, G.~Dorda, and M.~Pepper, ``New method for high-accuracy
  determination of the fine-structure constant based on quantized hall
  resistance,'' \href{http://dx.doi.org/10.1103/PhysRevLett.45.494}{{\em Phys.
  Rev. Lett.} {\bfseries 45} (Aug, 1980) 494--497}.
  \url{https://link.aps.org/doi/10.1103/PhysRevLett.45.494}.

\bibitem{Barkeshli2019}
M.~Barkeshli, P.~Bonderson, M.~Cheng, and Z.~Wang, ``Symmetry
  fractionalization, defects, and gauging of topological phases,''
  \href{http://dx.doi.org/10.1103/PhysRevB.100.115147}{{\em Phys. Rev. B}
  {\bfseries 100} (Sep, 2019) 115147}.
  \url{https://link.aps.org/doi/10.1103/PhysRevB.100.115147}.

\bibitem{Manjunath2020fqh}
N.~Manjunath and M.~Barkeshli, ``Classification of fractional quantum hall
  states with spatial symmetries,'' {\em arXiv preprint arXiv:2012.11603}
  (2020) .

\bibitem{guo2018}
M.~Guo, P.~Putrov, and J.~Wang, ``Time reversal, su(n) yang–mills and
  cobordisms: Interacting topological superconductors/insulators and quantum
  spin liquids in 3+1d,''
  \href{http://dx.doi.org/10.1016/j.aop.2018.04.025}{{\em Annals of Physics}
  {\bfseries 394} (Jul, 2018) 244–293}.
  \url{http://dx.doi.org/10.1016/j.aop.2018.04.025}.

\bibitem{guo2020}
M.~Guo, K.~Ohmori, P.~Putrov, Z.~Wan, and J.~Wang, ``Fermionic finite-group
  gauge theories and interacting symmetric/crystalline orders via cobordisms,''
  \href{http://dx.doi.org/10.1007/s00220-019-03671-6}{{\em Communications in
  Mathematical Physics} {\bfseries 376} no.~2, (Jan, 2020) 1073–1154}.
  \url{http://dx.doi.org/10.1007/s00220-019-03671-6}.

\bibitem{kapustin2021}
A.~Kapustin, N.~Sopenko, and B.~Yang, ``A classification of invertible phases
  of bosonic quantum lattice systems in one dimension,''
  \href{http://dx.doi.org/10.1063/5.0055996}{{\em Journal of Mathematical
  Physics} {\bfseries 62} no.~8, (Aug, 2021) 081901}.
  \url{http://dx.doi.org/10.1063/5.0055996}.

\bibitem{sopenko2021}
N.~Sopenko, ``An index for two-dimensional spt states,''
  \href{http://arxiv.org/abs/2101.00801}{{\ttfamily arXiv:2101.00801
  [math-ph]}}.

\bibitem{ogata2021}
Y.~Ogata, ``A $h^{3}(g,t)$-valued index of symmetry protected topological
  phases with on-site finite group symmetry for two-dimensional quantum spin
  systems,'' \href{http://arxiv.org/abs/2101.00426}{{\ttfamily arXiv:2101.00426
  [math-ph]}}.

\bibitem{bourne2021}
C.~Bourne and Y.~Ogata, ``The classification of symmetry protected topological
  phases of one-dimensional fermion systems,''
  \href{http://dx.doi.org/10.1017/fms.2021.19}{{\em Forum of Mathematics,
  Sigma} {\bfseries 9} (2021) }. \url{http://dx.doi.org/10.1017/fms.2021.19}.

\bibitem{Wang2020fSPT}
Q.-R. Wang and Z.-C. Gu, ``Construction and classification of
  symmetry-protected topological phases in interacting fermion systems,''
  \href{http://dx.doi.org/10.1103/PhysRevX.10.031055}{{\em Phys. Rev. X}
  {\bfseries 10} (Sep, 2020) 031055}.
  \url{https://link.aps.org/doi/10.1103/PhysRevX.10.031055}.

\bibitem{Gu2014Supercoh}
Z.-C. Gu and X.-G. Wen, ``Symmetry-protected topological orders for interacting
  fermions: Fermionic topological nonlinear $\ensuremath{\sigma}$ models and a
  special group supercohomology theory,''
  \href{http://dx.doi.org/10.1103/PhysRevB.90.115141}{{\em Phys. Rev. B}
  {\bfseries 90} (Sep, 2014) 115141}.
  \url{https://link.aps.org/doi/10.1103/PhysRevB.90.115141}.

\bibitem{Wang2017Supercoh}
Q.-R. Wang and Z.-C. Gu, ``Towards a complete classification of
  symmetry-protected topological phases for interacting fermions in three
  dimensions and a general group supercohomology theory,''
  \href{http://dx.doi.org/10.1103/PhysRevX.8.011055}{{\em Phys. Rev. X}
  {\bfseries 8} (Mar, 2018) 011055}.
  \url{https://link.aps.org/doi/10.1103/PhysRevX.8.011055}.

\bibitem{kitaev2006}
A.~Kitaev, ``Anyons in an exactly solved model and beyond,''
  \href{http://dx.doi.org/https://doi.org/10.1016/j.aop.2005.10.005}{{\em
  Annals of Physics} {\bfseries 321} no.~1, (2006) 2--111}.
  \url{https://www.sciencedirect.com/science/article/pii/S0003491605002381}.
  January Special Issue.

\bibitem{Barkeshli2020Anomaly}
M.~Barkeshli and M.~Cheng, ``Relative anomalies in (2+1)d symmetry enriched
  topological states,''
  \href{http://dx.doi.org/10.21468/scipostphys.8.2.028}{{\em SciPost Physics}
  {\bfseries 8} no.~2, (Feb, 2020) }.
  \url{http://dx.doi.org/10.21468/SciPostPhys.8.2.028}.

\bibitem{Bulmash2020}
D.~Bulmash and M.~Barkeshli, ``Absolute anomalies in (2+1)d symmetry-enriched
  topological states and exact (3+1)d constructions,''
  \href{http://arxiv.org/abs/arXiv:2003.11553}{{\ttfamily arXiv:2003.11553}}.

\bibitem{Cheng2018fSPT}
M.~Cheng, Z.~Bi, Y.-Z. You, and Z.-C. Gu, ``Classification of
  symmetry-protected phases for interacting fermions in two dimensions,''
  \href{http://dx.doi.org/10.1103/PhysRevB.97.205109}{{\em Phys. Rev. B}
  {\bfseries 97} (May, 2018) 205109}.
  \url{https://link.aps.org/doi/10.1103/PhysRevB.97.205109}.

\bibitem{Chen2019freeinteracting}
Y.-A. Chen, A.~Kapustin, A.~Turzillo, and M.~You, ``Free and interacting
  short-range entangled phases of fermions: Beyond the tenfold way,''
  \href{http://dx.doi.org/10.1103/PhysRevB.100.195128}{{\em Phys. Rev. B}
  {\bfseries 100} (Nov, 2019) 195128}.
  \url{https://link.aps.org/doi/10.1103/PhysRevB.100.195128}.

\bibitem{Bhardwaj_2017}
L.~Bhardwaj, D.~Gaiotto, and A.~Kapustin, ``State sum constructions of
  spin-tfts and string net constructions of fermionic phases of matter,''
  \href{http://dx.doi.org/10.1007/jhep04(2017)096}{{\em Journal of High Energy
  Physics} {\bfseries 2017} no.~4, (Apr, 2017) }.
  \url{http://dx.doi.org/10.1007/JHEP04(2017)096}.

\bibitem{brumfiel2018pontrjagin}
G.~Brumfiel and J.~Morgan, ``The pontrjagin dual of 3-dimensional spin
  bordism,'' \href{http://arxiv.org/abs/1612.02860}{{\ttfamily arXiv:1612.02860
  [math.AT]}}.

\bibitem{Bischoff_2019}
M.~Bischoff, C.~Jones, Y.-M. Lu, and D.~Penneys, ``Spontaneous symmetry
  breaking from anyon condensation,''
  \href{http://dx.doi.org/10.1007/jhep02(2019)062}{{\em Journal of High Energy
  Physics} {\bfseries 2019} no.~2, (Feb, 2019) }.
  \url{http://dx.doi.org/10.1007/JHEP02(2019)062}.

\bibitem{Jiang2017}
S.~Jiang and Y.~Ran, ``Anyon condensation and a generic tensor-network
  construction for symmetry-protected topological phases,''
  \href{http://dx.doi.org/10.1103/PhysRevB.95.125107}{{\em Phys. Rev. B}
  {\bfseries 95} (Mar, 2017) 125107}.
  \url{https://link.aps.org/doi/10.1103/PhysRevB.95.125107}.

\bibitem{Altland1997}
A.~Altland and M.~R. Zirnbauer, ``Nonstandard symmetry classes in mesoscopic
  normal-superconducting hybrid structures,''
  \href{http://dx.doi.org/10.1103/PhysRevB.55.1142}{{\em Phys. Rev. B}
  {\bfseries 55} (Jan, 1997) 1142--1161}.
  \url{https://link.aps.org/doi/10.1103/PhysRevB.55.1142}.

\bibitem{Ryu_2010}
S.~Ryu, A.~P. Schnyder, A.~Furusaki, and A.~W.~W. Ludwig, ``Topological
  insulators and superconductors: tenfold way and dimensional hierarchy,''
  \href{http://dx.doi.org/10.1088/1367-2630/12/6/065010}{{\em New Journal of
  Physics} {\bfseries 12} no.~6, (Jun, 2010) 065010}.
  \url{https://doi.org/10.1088/1367-2630/12/6/065010}.

\bibitem{gu2014}
Z.-C. Gu and M.~Levin, ``Effect of interactions on two-dimensional fermionic
  symmetry-protected topological phases withz2symmetry,''
  \href{http://dx.doi.org/10.1103/physrevb.89.201113}{{\em Physical Review B}
  {\bfseries 89} no.~20, (May, 2014) }.
  \url{http://dx.doi.org/10.1103/PhysRevB.89.201113}.

\bibitem{Benini2018}
F.~Benini, C.~C\'ordova, and P.-S. Hsin, ``{On 2-Group Global Symmetries and
  their Anomalies},'' \href{http://dx.doi.org/10.1007/JHEP03(2019)118}{{\em
  JHEP} {\bfseries 03} (2019) 118},
  \href{http://arxiv.org/abs/1803.09336}{{\ttfamily arXiv:1803.09336
  [hep-th]}}.

\bibitem{Kapustin2014anomaly}
A.~Kapustin and R.~Thorngren, ``Anomalous discrete symmetries in three
  dimensions and group cohomology,''
  \href{http://dx.doi.org/10.1103/PhysRevLett.112.231602}{{\em Phys. Rev.
  Lett.} {\bfseries 112} (Jun, 2014) 231602}.
  \url{https://link.aps.org/doi/10.1103/PhysRevLett.112.231602}.

\bibitem{fidkowski2018surface}
L.~Fidkowski, A.~Vishwanath, and M.~A. Metlitski, ``Surface topological order
  and a new 't hooft anomaly of interaction enabled 3+1d fermion spts,''
  \href{http://arxiv.org/abs/1804.08628}{{\ttfamily arXiv:1804.08628
  [cond-mat.str-el]}}.

\bibitem{Gaiotto:2015zta}
D.~Gaiotto and A.~Kapustin, ``{Spin TQFTs and fermionic phases of matter},''
  \href{http://dx.doi.org/10.1142/S0217751X16450445}{{\em Int. J. Mod. Phys. A}
  {\bfseries 31} no.~28n29, (2016) 1645044},
  \href{http://arxiv.org/abs/1505.05856}{{\ttfamily arXiv:1505.05856
  [cond-mat.str-el]}}.

\bibitem{wang2019spt}
Q.-R. Wang, Y.~Qi, and Z.-C. Gu, ``Anomalous symmetry protected topological
  states in interacting fermion systems,''
  \href{http://dx.doi.org/10.1103/physrevlett.123.207003}{{\em Physical Review
  Letters} {\bfseries 123} no.~20, (Nov, 2019) }.
  \url{http://dx.doi.org/10.1103/PhysRevLett.123.207003}.

\bibitem{Cartan_coboundary_2020}
A.~M. Medina-Mardones, ``An effective proof of the cartan formula: The even
  prime,'' {\em Journal of Pure and Applied Algebra} {\bfseries 224} no.~12,
  (2020) 106444.

\bibitem{Whitehead1949}
J.~H.~C. Whitehead, ``On simply connected, 4-dimensional polyhedra,''
  \href{http://dx.doi.org/10.1007/BF02568048}{{\em Commentarii Mathematici
  Helvetici} {\bfseries 22} no.~1, (Dec, 1949) 48--92}.
  \url{https://doi.org/10.1007/BF02568048}.

\bibitem{aasen2021characterization}
D.~Aasen, P.~Bonderson, and C.~Knapp, ``Characterization and classification of
  fermionic symmetry enriched topological phases,''
  \href{http://arxiv.org/abs/2109.10911}{{\ttfamily arXiv:2109.10911
  [cond-mat.str-el]}}.

\bibitem{Morimoto2015}
T.~Morimoto, A.~Furusaki, and C.~Mudry, ``Breakdown of the topological
  classification $\mathbb{Z}$ for gapped phases of noninteracting fermions by
  quartic interactions,''
  \href{http://dx.doi.org/10.1103/PhysRevB.92.125104}{{\em Phys. Rev. B}
  {\bfseries 92} (Sep, 2015) 125104}.
  \url{https://link.aps.org/doi/10.1103/PhysRevB.92.125104}.

\bibitem{Barkeshli:2014cna}
M.~Barkeshli, P.~Bonderson, M.~Cheng, and Z.~Wang, ``{Symmetry
  Fractionalization, Defects, and Gauging of Topological Phases},''
  \href{http://dx.doi.org/10.1103/PhysRevB.100.115147}{{\em Phys. Rev. B}
  {\bfseries 100} no.~11, (2019) 115147},
  \href{http://arxiv.org/abs/1410.4540}{{\ttfamily arXiv:1410.4540
  [cond-mat.str-el]}}.

\bibitem{barkeshli2019tr}
M.~Barkeshli, P.~Bonderson, M.~Cheng, C.-M. Jian, and K.~Walker, ``Reflection
  and time reversal symmetry enriched topological phases of matter: Path
  integrals, non-orientable manifolds, and anomalies,''
  \href{http://dx.doi.org/10.1007/s00220-019-03475-8}{{\em Communications in
  Mathematical Physics} (2019) },
  \href{http://arxiv.org/abs/arXiv:1612.07792}{{\ttfamily arXiv:1612.07792}}.
  \url{https://doi.org/10.1007/s00220-019-03475-8}.

\bibitem{Steenrod1947}
N.~Steenrod, ``Products of cocycles and extensions of mappings,'' {\em Ann.
  Math.} {\bfseries 48} (1947) 290--320.

\bibitem{ENO2010}
P.~Etingof, D.~Nikshych, and V.~Ostrik, ``Fusion categories and homotopy
  theory,'' {\em Quantum Topology} {\bfseries 1} (2010) 209--273,
  \href{http://arxiv.org/abs/arXiv:0909.3140}{{\ttfamily arXiv:0909.3140}}.

\bibitem{bulmashSymmFrac}
D.~Bulmash and M.~Barkeshli, ``Symmetry fractionalization in fermionic
  topological phases,''.

\bibitem{aasen2021torsorial}
D.~Aasen, P.~Bonderson, and C.~Knapp, ``Torsorial actions on g-crossed braided
  tensor categories,'' \href{http://arxiv.org/abs/2107.10270}{{\ttfamily
  arXiv:2107.10270 [math.QA]}}.

\bibitem{Freed:2004yc}
D.~S. Freed and G.~W. Moore, ``{Setting the quantum integrand of M-theory},''
  \href{http://dx.doi.org/10.1007/s00220-005-1482-7}{{\em Commun. Math. Phys.}
  {\bfseries 263} (2006) 89--132},
  \href{http://arxiv.org/abs/hep-th/0409135}{{\ttfamily arXiv:hep-th/0409135}}.

\bibitem{Cordova:2017vab}
C.~Cordova, P.-S. Hsin, and N.~Seiberg, ``{Global Symmetries, Counterterms, and
  Duality in Chern-Simons Matter Theories with Orthogonal Gauge Groups},''
  \href{http://dx.doi.org/10.21468/SciPostPhys.4.4.021}{{\em SciPost Phys.}
  {\bfseries 4} no.~4, (2018) 021},
  \href{http://arxiv.org/abs/1711.10008}{{\ttfamily arXiv:1711.10008
  [hep-th]}}.

\bibitem{Goddard:1976qe}
P.~Goddard, J.~Nuyts, and D.~I. Olive, ``{Gauge Theories and Magnetic
  Charge},'' \href{http://dx.doi.org/10.1016/0550-3213(77)90221-8}{{\em Nucl.
  Phys. B} {\bfseries 125} (1977) 1--28}.

\bibitem{Kapustin:2014tfa}
A.~Kapustin, ``{Symmetry Protected Topological Phases, Anomalies, and
  Cobordisms: Beyond Group Cohomology},''
  \href{http://arxiv.org/abs/1403.1467}{{\ttfamily arXiv:1403.1467
  [cond-mat.str-el]}}.

\bibitem{Yonekura:2018ufj}
K.~Yonekura, ``{On the cobordism classification of symmetry protected
  topological phases},''
  \href{http://dx.doi.org/10.1007/s00220-019-03439-y}{{\em Commun. Math. Phys.}
  {\bfseries 368} no.~3, (2019) 1121--1173},
  \href{http://arxiv.org/abs/1803.10796}{{\ttfamily arXiv:1803.10796
  [hep-th]}}.

\bibitem{Seiberg:2016rsg}
N.~Seiberg and E.~Witten, ``{Gapped Boundary Phases of Topological Insulators
  via Weak Coupling},'' \href{http://dx.doi.org/10.1093/ptep/ptw083}{{\em PTEP}
  {\bfseries 2016} no.~12, (2016) 12C101},
  \href{http://arxiv.org/abs/1602.04251}{{\ttfamily arXiv:1602.04251
  [cond-mat.str-el]}}.

\bibitem{Hsin:2016blu}
P.-S. Hsin and N.~Seiberg, ``{Level/rank Duality and Chern-Simons-Matter
  Theories},'' \href{http://dx.doi.org/10.1007/JHEP09(2016)095}{{\em JHEP}
  {\bfseries 09} (2016) 095}, \href{http://arxiv.org/abs/1607.07457}{{\ttfamily
  arXiv:1607.07457 [hep-th]}}.

\bibitem{Hastings_2008}
M.~B. Hastings, ``Topology and phases in fermionic systems,''
  \href{http://dx.doi.org/10.1088/1742-5468/2008/01/l01001}{{\em Journal of
  Statistical Mechanics: Theory and Experiment} {\bfseries 2008} no.~01, (Jan,
  2008) L01001}. \url{https://doi.org/10.1088/1742-5468/2008/01/l01001}.

\bibitem{RevModPhys.88.035001}
E.~Witten, ``Fermion path integrals and topological phases,''
  \href{http://dx.doi.org/10.1103/RevModPhys.88.035001}{{\em Rev. Mod. Phys.}
  {\bfseries 88} (Jul, 2016) 035001}.
  \url{https://link.aps.org/doi/10.1103/RevModPhys.88.035001}.

\bibitem{Atiyah:1975jf}
M.~F. Atiyah, V.~K. Patodi, and I.~M. Singer, ``{Spectral asymmetry and
  Riemannian Geometry 1},''
  \href{http://dx.doi.org/10.1017/S0305004100049410}{{\em Math. Proc. Cambridge
  Phil. Soc.} {\bfseries 77} (1975) 43}.

\bibitem{Hsin:2019gvb}
P.-S. Hsin and S.-H. Shao, ``{Lorentz Symmetry Fractionalization and Dualities
  in (2+1)d},'' \href{http://dx.doi.org/10.21468/SciPostPhys.8.2.018}{{\em
  SciPost Phys.} {\bfseries 8} (2020) 018},
  \href{http://arxiv.org/abs/1909.07383}{{\ttfamily arXiv:1909.07383
  [cond-mat.str-el]}}.

\bibitem{Kapustin:2017jrc}
A.~Kapustin and R.~Thorngren, ``{Fermionic SPT phases in higher dimensions and
  bosonization},'' \href{http://dx.doi.org/10.1007/JHEP10(2017)080}{{\em JHEP}
  {\bfseries 10} (2017) 080}, \href{http://arxiv.org/abs/1701.08264}{{\ttfamily
  arXiv:1701.08264 [cond-mat.str-el]}}.

\bibitem{Hsin:2020nts}
P.-S. Hsin and H.~T. Lam, ``{Discrete theta angles, symmetries and
  anomalies},'' \href{http://dx.doi.org/10.21468/SciPostPhys.10.2.032}{{\em
  SciPost Phys.} {\bfseries 10} no.~2, (2021) 032},
  \href{http://arxiv.org/abs/2007.05915}{{\ttfamily arXiv:2007.05915
  [hep-th]}}.

\bibitem{Gaiotto:2014kfa}
D.~Gaiotto, A.~Kapustin, N.~Seiberg, and B.~Willett, ``{Generalized Global
  Symmetries},'' \href{http://dx.doi.org/10.1007/JHEP02(2015)172}{{\em JHEP}
  {\bfseries 02} (2015) 172}, \href{http://arxiv.org/abs/1412.5148}{{\ttfamily
  arXiv:1412.5148 [hep-th]}}.

\bibitem{Hsin:2018vcg}
P.-S. Hsin, H.~T. Lam, and N.~Seiberg, ``{Comments on One-Form Global
  Symmetries and Their Gauging in 3d and 4d},''
  \href{http://dx.doi.org/10.21468/SciPostPhys.6.3.039}{{\em SciPost Phys.}
  {\bfseries 6} no.~3, (2019) 039},
  \href{http://arxiv.org/abs/1812.04716}{{\ttfamily arXiv:1812.04716
  [hep-th]}}.

\bibitem{Kapustin:2014gua}
A.~Kapustin and N.~Seiberg, ``{Coupling a QFT to a TQFT and Duality},''
  \href{http://dx.doi.org/10.1007/JHEP04(2014)001}{{\em JHEP} {\bfseries 04}
  (2014) 001}, \href{http://arxiv.org/abs/1401.0740}{{\ttfamily arXiv:1401.0740
  [hep-th]}}.

\bibitem{Hsin:2019fhf}
P.-S. Hsin and A.~Turzillo, ``{Symmetry-enriched quantum spin liquids in (3 +
  1)$d$},'' \href{http://dx.doi.org/10.1007/JHEP09(2020)022}{{\em JHEP}
  {\bfseries 09} (2020) 022}, \href{http://arxiv.org/abs/1904.11550}{{\ttfamily
  arXiv:1904.11550 [cond-mat.str-el]}}.

\bibitem{milnor1974characteristic}
J.~Milnor, J.~Stasheff, J.~Stasheff, and P.~University, {\em Characteristic
  Classes}.
\newblock Annals of mathematics studies. Princeton University Press, 1974.
\newblock \url{https://books.google.com/books?id=5zQ9AFk1i4EC}.

\bibitem{kapustin2013}
A.~Kapustin and R.~Thorngren, ``Higher symmetry and gapped phases of gauge
  theories,'' \href{http://arxiv.org/abs/arXiv:1309.4721}{{\ttfamily
  arXiv:1309.4721}}.

\bibitem{Cordova:2018cvg}
C.~C\'ordova, T.~T. Dumitrescu, and K.~Intriligator, ``{Exploring 2-Group
  Global Symmetries},'' \href{http://dx.doi.org/10.1007/JHEP02(2019)184}{{\em
  JHEP} {\bfseries 02} (2019) 184},
  \href{http://arxiv.org/abs/1802.04790}{{\ttfamily arXiv:1802.04790
  [hep-th]}}.

\bibitem{Tachikawa:2017gyf}
Y.~Tachikawa, ``{On gauging finite subgroups},''
  \href{http://dx.doi.org/10.21468/SciPostPhys.8.1.015}{{\em SciPost Phys.}
  {\bfseries 8} no.~1, (2020) 015},
  \href{http://arxiv.org/abs/1712.09542}{{\ttfamily arXiv:1712.09542
  [hep-th]}}.

\bibitem{PhysRevLett.123.207003}
Q.-R. Wang, Y.~Qi, and Z.-C. Gu, ``Anomalous symmetry protected topological
  states in interacting fermion systems,''
  \href{http://dx.doi.org/10.1103/PhysRevLett.123.207003}{{\em Phys. Rev.
  Lett.} {\bfseries 123} (Nov, 2019) 207003}.
  \url{https://link.aps.org/doi/10.1103/PhysRevLett.123.207003}.

\bibitem{PhysRevB.95.195147}
C.~Wang, C.-H. Lin, and Z.-C. Gu, ``Interacting fermionic symmetry-protected
  topological phases in two dimensions,''
  \href{http://dx.doi.org/10.1103/PhysRevB.95.195147}{{\em Phys. Rev. B}
  {\bfseries 95} (May, 2017) 195147}.
  \url{https://link.aps.org/doi/10.1103/PhysRevB.95.195147}.

\bibitem{PhysRevB.103.235101}
M.~Levin, A.~Kapustin, and L.~Spodyneiko, ``Nernst and ettingshausen effects in
  gapped quantum materials,''
  \href{http://dx.doi.org/10.1103/PhysRevB.103.235101}{{\em Phys. Rev. B}
  {\bfseries 103} (Jun, 2021) 235101}.
  \url{https://link.aps.org/doi/10.1103/PhysRevB.103.235101}.

\bibitem{Gu2014interaction}
Z.-C. Gu and M.~Levin, ``Effect of interactions on two-dimensional fermionic
  symmetry-protected topological phases with ${Z}_{2}$ symmetry,''
  \href{http://dx.doi.org/10.1103/PhysRevB.89.201113}{{\em Phys. Rev. B}
  {\bfseries 89} (May, 2014) 201113}.
  \url{https://link.aps.org/doi/10.1103/PhysRevB.89.201113}.

\bibitem{Ryu2012interaction}
S.~Ryu and S.-C. Zhang, ``Interacting topological phases and modular
  invariance,'' \href{http://dx.doi.org/10.1103/PhysRevB.85.245132}{{\em Phys.
  Rev. B} {\bfseries 85} (Jun, 2012) 245132}.
  \url{https://link.aps.org/doi/10.1103/PhysRevB.85.245132}.

\bibitem{Qi_2013}
X.-L. Qi, ``A new class of (2 + 1)-dimensional topological superconductors with
  $\mathbb {Z}_8$ topological classification,''
  \href{http://dx.doi.org/10.1088/1367-2630/15/6/065002}{{\em New Journal of
  Physics} {\bfseries 15} no.~6, (Jun, 2013) 065002}.
  \url{http://dx.doi.org/10.1088/1367-2630/15/6/065002}.

\bibitem{Lu2016}
Y.-M. Lu and A.~Vishwanath, ``Classification and properties of
  symmetry-enriched topological phases: Chern-simons approach with applications
  to ${Z}_{2}$ spin liquids,''
  \href{http://dx.doi.org/10.1103/PhysRevB.93.155121}{{\em Phys. Rev. B}
  {\bfseries 93} (Apr, 2016) 155121}.
  \url{https://link.aps.org/doi/10.1103/PhysRevB.93.155121}.

\bibitem{Wu:1954_3}
W.-T. Wu, ``{On Pontrjagin classes III},'' {\em Acta Math. Sinica} {\bfseries
  4} (1954) 323--347.

\bibitem{WL15}
C.~Wang and M.~Levin, ``Topological invariants for gauge theories and
  symmetry-protected topological phases,''
  \href{http://dx.doi.org/10.1103/PhysRevB.91.165119}{{\em Phys. Rev. B}
  {\bfseries 91} (Apr, 2015) 165119}.
  \url{https://link.aps.org/doi/10.1103/PhysRevB.91.165119}.

\bibitem{T17}
N.~Tantivasadakarn, ``Dimensional reduction and topological invariants of
  symmetry-protected topological phases,''
  \href{http://dx.doi.org/10.1103/PhysRevB.96.195101}{{\em Phys. Rev. B}
  {\bfseries 96} (Nov, 2017) 195101}.
  \url{https://link.aps.org/doi/10.1103/PhysRevB.96.195101}.

\bibitem{CET21}
Y.-A. Chen, T.~D. Ellison, and N.~Tantivasadakarn, ``Disentangling
  supercohomology symmetry-protected topological phases in three spatial
  dimensions,'' \href{http://dx.doi.org/10.1103/PhysRevResearch.3.013056}{{\em
  Phys. Rev. Research} {\bfseries 3} (Jan, 2021) 013056}.
  \url{https://link.aps.org/doi/10.1103/PhysRevResearch.3.013056}.

\bibitem{Chiu2016review}
C.-K. Chiu, J.~C.~Y. Teo, A.~P. Schnyder, and S.~Ryu, ``Classification of
  topological quantum matter with symmetries,''
  \href{http://dx.doi.org/10.1103/RevModPhys.88.035005}{{\em Rev. Mod. Phys.}
  {\bfseries 88} (Aug, 2016) 035005}.
  \url{https://link.aps.org/doi/10.1103/RevModPhys.88.035005}.

\bibitem{Qi2009TRS-TSC}
X.-L. Qi, T.~L. Hughes, S.~Raghu, and S.-C. Zhang, ``Time-reversal-invariant
  topological superconductors and superfluids in two and three dimensions,''
  \href{http://dx.doi.org/10.1103/PhysRevLett.102.187001}{{\em Phys. Rev.
  Lett.} {\bfseries 102} (May, 2009) 187001}.
  \url{https://link.aps.org/doi/10.1103/PhysRevLett.102.187001}.

\bibitem{Zhang2013MKP}
F.~Zhang, C.~L. Kane, and E.~J. Mele, ``Time-reversal-invariant topological
  superconductivity and majorana kramers pairs,''
  \href{http://dx.doi.org/10.1103/PhysRevLett.111.056402}{{\em Phys. Rev.
  Lett.} {\bfseries 111} (Aug, 2013) 056402}.
  \url{https://link.aps.org/doi/10.1103/PhysRevLett.111.056402}.

\bibitem{Liu2013}
Z.-X. Liu and X.-G. Wen, ``Symmetry-protected quantum spin hall phases in two
  dimensions,'' \href{http://dx.doi.org/10.1103/PhysRevLett.110.067205}{{\em
  Phys. Rev. Lett.} {\bfseries 110} (Feb, 2013) 067205}.
  \url{https://link.aps.org/doi/10.1103/PhysRevLett.110.067205}.

\bibitem{ning2021enforced}
S.-Q. Ning, Y.~Qi, Z.-C. Gu, and C.~Wang, ``Enforced symmetry breaking by
  invertible topological order,''
  \href{http://arxiv.org/abs/2109.15307}{{\ttfamily arXiv:2109.15307
  [cond-mat.str-el]}}.

\bibitem{Wang_Gu_unpublished}
Q.-R. Wang and Z.-C. Gu {\em to appear} (to appear) .

\bibitem{tata2020}
S.~Tata, ``{Geometrically Interpreting Higher Cup Products, and Application to
  Combinatorial Pin Structures},''
  \href{http://arxiv.org/abs/2008.10170}{{\ttfamily arXiv:2008.10170
  [hep-th]}}.

\bibitem{Hirsch:1955}
G.~Hirsch, ``Quelques propri\'et\'es des produits de steenrod,'' {\em C. R.
  Acad. Sci. Paris} {\bfseries 241} (1955) .

\bibitem{PhysRevB.90.115141}
Z.-C. Gu and X.-G. Wen, ``Symmetry-protected topological orders for interacting
  fermions: Fermionic topological nonlinear $\ensuremath{\sigma}$ models and a
  special group supercohomology theory,''
  \href{http://dx.doi.org/10.1103/PhysRevB.90.115141}{{\em Phys. Rev. B}
  {\bfseries 90} (Sep, 2014) 115141}.
  \url{https://link.aps.org/doi/10.1103/PhysRevB.90.115141}.

\bibitem{Dijkgraaf:1989hb}
R.~Dijkgraaf, C.~Vafa, E.~P. Verlinde, and H.~L. Verlinde, ``{The Operator
  Algebra of Orbifold Models},''
  \href{http://dx.doi.org/10.1007/BF01238812}{{\em Commun. Math. Phys.}
  {\bfseries 123} (1989) 485}.

\bibitem{Dijkgraaf1989pz}
R.~Dijkgraaf and E.~Witten, ``{Topological Gauge Theories and Group
  Cohomology},'' \href{http://dx.doi.org/10.1007/BF02096988}{{\em Commun. Math.
  Phys.} {\bfseries 129} (1990) 393}.

\bibitem{Kitaev:1997wr}
A.~Y. Kitaev, ``{Fault tolerant quantum computation by anyons},''
  \href{http://dx.doi.org/10.1016/S0003-4916(02)00018-0}{{\em Annals Phys.}
  {\bfseries 303} (2003) 2--30},
  \href{http://arxiv.org/abs/quant-ph/9707021}{{\ttfamily
  arXiv:quant-ph/9707021}}.

\bibitem{Eliens:2013epa}
I.~S. Eli\"ens, J.~C. Romers, and F.~A. Bais, ``{Diagrammatics for Bose
  condensation in anyon theories},''
  \href{http://dx.doi.org/10.1103/PhysRevB.90.195130}{{\em Phys. Rev. B}
  {\bfseries 90} no.~19, (2014) 195130},
  \href{http://arxiv.org/abs/1310.6001}{{\ttfamily arXiv:1310.6001
  [cond-mat.str-el]}}.

\end{thebibliography}\endgroup

\end{document}